%% file: article.tex
\documentclass{jfm}

\usepackage{graphicx}
\usepackage{newtxtext}
\usepackage[varvw]{newtxmath}
\usepackage{natbib}
\usepackage{longtable}
\usepackage{hyperref}
\hypersetup{
    colorlinks = true,
    urlcolor   = blue,
    citecolor  = black,
}

\usepackage[normalem]{ulem}

\usepackage{comment}
\usepackage{subcaption}

\newcommand\omegas{\omega_s}

\title{Magnetohydrodynamic drag on an oscillating sphere in a rotating spherical cavity}

\author{David Cébron\aff{1}
  \corresp{\email{david.cebron@univ-grenoble-alpes.fr}},
 \and Paolo Personnettaz\aff{1,2}}
\affiliation{
\aff{1} Univ. Grenoble Alpes, Univ. Savoie Mont Blanc, CNRS, IRD, Univ. Gustave Eiffel, ISTerre, 38000 Grenoble, France
\aff{2}{CNES - Centre National d’Etudes Spatiales, 2 Place Maurice Quentin, 75039, Paris, France}
}

\begin{document}
\maketitle

\begin{abstract}
Oscillating-sphere drag is a classical fluid-mechanics problem, yet no theory simultaneously accounts for confinement, rotation, viscosity and magnetic fields. We consider a conducting sphere undergoing translational oscillations inside a rotating spherical cavity, modelling confined magnetohydrodynamic flows relevant to planetary interiors and liquid-metal experiments. Planetary applications include polar (rectilinear oscillations) and equatorial (circular trajectories) Slichter modes of solid inner cores or subsurface-ocean interiors relative to outer ice shells. Existing theories are restricted to separate asymptotic regimes, including viscous drag in bounded fluids \citep{stokes1851effect}, rotational effects in inviscid cavities \citep{busse1974free}, and magnetic coupling through oscillatory boundary layers \citep{buffett1995magnetic}. We derive a unified asymptotic framework for oscillatory drag in confined spherical shells with conductivity and permeability contrasts between the inner sphere, fluid shell and outer solid. Closed-form force laws and dissipation predictions are obtained for polar and equatorial oscillations at weak rotation, with finite-rotation corrections for the polar mode. The framework accounts for viscous and magnetic pressure contributions, magnetic tension, Alfv\'en-wave radiation and magnetohydrodynamic Stokes-Ekman layers. Recovering classical boundary-layer theories as limiting cases, the analysis yields a local Alfv\'en-wave drag law for spherical shells, analogous to bounded oscillatory Stokes drag. Beyond the thin boundary-layer approximation, complementary asymptotics yield a closed-form representation of bounded Stokes flow and extend inductionless theory to spherical shells. Direct numerical simulations validate the predictions across a broad parameter range. The framework predicts oscillatory coupling, added-mass, drag and dissipation in planetary cores, subsurface oceans and liquid-metal experiments.
\end{abstract}

\begin{keywords}
Authors should not enter keywords: they are chosen by the author during the submission process and will be added during the typesetting (see \href{https://www.cambridge.org/core/journals/journal-of-fluid-mechanics/information/list-of-keywords}{Keyword PDF} for the list).  
\end{keywords}

{\bf MSC Codes }  {\it(Optional)} Please enter your MSC Codes here
\newpage
\section{Introduction}
\label{sec:intro}
\subsection{Geophysical background}
The Earth's solid inner core can undergo small translational oscillations relative to the surrounding liquid outer core and solid mantle. In the presence of seismic or impact events, the inner-core centre of mass behaves as a damped harmonic oscillator whose dominant restoring force is Archimedean \citep{rosat2011review}. Earth’s rotation splits these oscillations into one polar and two equatorial Slichter modes \citep{slichter1961fundamental}, corresponding to axial and orbital-like motions, respectively. Their dynamics remain far less constrained than those of rotational modes driven by external interactions \cite[e.g.][]{bars2015flows}.

The oscillation period follows from Newton's second law applied to the inner
core, including the Archimedean restoring force and the fluid added mass
\citep{slichter1961fundamental}. Rotational, inertial and viscous effects were
examined by \cite{won1973oscillation}. Beyond lumped models,
\cite{busse1974free}, hereafter B74, considered a homogeneous, incompressible and inviscid rotating outer core while assuming the mantle to remain stationary. For the polar mode, B74 showed that confinement and fluid Coriolis effects nearly compensate, yielding only small corrections to previous period estimates, and derived a closed-form inviscid velocity field. The linear hydrodynamic problem was later revisited numerically by \cite{rieutord2002slichter}, who retained a fixed rigid outer boundary, thereby excluding translational mantle motion, while allowing for a neutrally stratified fluid with radially varying density and treating viscosity non-perturbatively.

Related rotating viscous formulations by \cite{smylie1998viscous,smylie2000the} allowed the outer solid shell to translate under conservation of total linear momentum, while the influence of mantle recoil on the Slichter period was isolated by \cite{grinfeld2005motion}. For a non-rotating planet with homogeneous rigid inner core and mantle and an inviscid incompressible outer core, neglecting magnetic effects, they found mantle motion to have a negligible effect for the Earth and an unexpectedly small effect even for Mercury despite appreciable mantle displacement. In a complementary homogeneous three-layer elasto-gravitational model with an elastic mantle, inviscid liquid outer core and deformable inner core, \cite{grefflefftz2007fluid} showed that, in a centre-of-mass frame, the translational components of pressure and traction at the fluid boundaries are constrained by the global force balance. These results distinguish the relative-motion fluid problem considered below from the global dynamics determining the absolute translations of the planetary layers. Analogous relative translations arise in icy bodies with subsurface oceans, for which \citet{escapa2011free} derived the free oscillations of two rigid solids bounding an ideal-fluid ocean and related their periods and relative amplitudes to the internal structure.

Viscous and Ohmic dissipation have received less attention. Earlier studies of electromagnetic core–mantle coupling had already demonstrated that magnetic stresses can modify the frequencies and damping of global core motions \citep{macdonald1961study,crossley1974electromagnetic,toomre1974diurnal}. Neglecting magnetic-pressure effects, \cite{buffett1995magnetic} used a boundary-layer model to derive the magnetic-tension contribution to the Lorentz force for polar oscillations in an inviscid, unbounded, and non-rotating fluid. \cite{smylie1998viscous} analysed the three modes in a non-magnetic, bounded, viscous, and rotating fluid, constructing the basic flow from products of Legendre functions \cite[inertial-mode approach of][]{greenspan1968theory} and including Stokes--Ekman boundary layers. They obtained the inviscid pressure force and viscous tangential stresses, but neglected viscous corrections to the pressure required to recover the exact solution of \cite{stokes1851effect}. This formulation was used by \cite{smylie1999viscosity} to infer an unrealistically large viscosity near the inner core \cite[see also][]{smylie2000the}. Despite these efforts, the relative importance of dissipation mechanisms remains uncertain \citep{rosat2011review}.

After more than two decades of limited attention, the dissipation of Earth's core Slichter modes was revisited by \cite{personnettaz2026ohmic}, hereafter referred to as P26. Combining magnetohydrodynamic (MHD) simulations with scaling arguments, they estimated both Ohmic and viscous dissipation and inferred quality factors and decay times for Earth’s core. Beyond its geophysical motivation, however, the problem is appealing due to its simple geometry, which is amenable to analytical approaches such as potential-flow theory and boundary-layer analysis. In the following, we review the relevant contributions from both fluid-dynamical and MHD perspectives.

\subsection{Fluid dynamics context}
The flow and forces generated by an oscillating sphere form a canonical problem in fluid mechanics, relevant from laboratory experiments to planetary interiors. Classical solutions describe viscous effects in simple configurations, whereas many geophysical systems involve rotation, confinement and magnetic fields. In such environments, oscillations of electrically conducting fluids generate coupled viscous, rotational and magnetic boundary layers.

\begin{table}
  \begin{center}
\def~{\hphantom{0}}
  \begin{tabular}{lccccccc}
     &  Viscous  & Confinement  & Rotation & Magnetic & ${\eta}_s$ & Equatorial \\
 \cite{stokes1851effect}    & $\checkmark$   & $\checkmark$ & $\times$ & $\times$ & $\times$&  $\times$  \\ 
\cite{reitz1961the}    & $\times$   & $\times$    & $\times$
    & $L_\eta \gg a_s$ $^\mathrm{(b)}$ & any   & $\checkmark$  \\
  \cite{singh1965drag}$^\mathrm{(a)}$  & $\checkmark$   & $\times$ & $\omegas \gg \varOmega_o$  & $L_\eta \gg a_s$  $^\mathrm{(b)}$    & any & $\times$  \\ 
    \cite{motz1966magnetohydrodynamic}  & $\times$   & $\times$ & $\times$  & $L_\eta \gg a_s$ & $0$ & $\times$  \\
  \cite{busse1974free}  & $\times$   & $\checkmark$ & $\omegas > 2 \varOmega_o$ & $\times$ & $\times$ & $\times$  \\
   \cite{buffett1995magnetic}  & $\times$   & $\times$ & $\times$ & $L_\eta \ll \ell$ & $\eta_f$  & $\times$  \\
      \cite{smylie1998viscous} $^\mathrm{(c)}$  & $L_\nu \ll \ell$   & $\checkmark$ & $\omegas > 2 \varOmega_o$ $^\mathrm{(d)}$ & $\times$ & $\times$ & $\checkmark$  \\
  This work & $L_\nu \ll \ell$   & $\checkmark$ & $\omegas > 2 \varOmega_o$ $^\mathrm{(e)}$ & $\checkmark$ $^\mathrm{(f)}$ & any & $\checkmark$  
  \end{tabular}
  \caption{Main theoretical works for a sphere (radius $a_s$, magnetic diffusivity $\eta_s$) oscillating at frequency $\omegas$ in a fluid rotating at $\varOmega_o$, with viscous and magnetic skin depths $L_\nu=\sqrt{2 \nu/\omegas}$ and $L_\eta=\sqrt{2 \eta_f/\omegas}$ ($\nu$ and $\eta_f$ are the fluid viscous and magnetic diffusivities). (a) Compressibility included, (b) weak magnetic fields, (c) spheroidal shape considered, (d) $\omegas \gg \varOmega_o$ for the drag, (e) polar mode only, (f) $L_\eta \ll a_s$ and $L_\eta \gg a_s$. Length scales $\ell=\min(a_s,D)$ with the fluid gap $D=a_m-a_s$.}
  \label{tab:mainTheoryContributions}
  \end{center}
\end{table}

The drag on an oscillating sphere has been studied for more than two centuries, initially in connection with pendulum experiments aimed at determining the Earth's figure \citep{darrigol2002between}. Unless stated otherwise, the studies reviewed below concern rectilinear oscillations (polar-mode forcing). Early measurements by \cite{buat1779principles} and \cite{bessel1828ueber} attributed the fluid resistance primarily to inertial effects, while \cite{poisson1831memoire} proposed a mathematical solution consistent with added-mass estimates. Experiments by \cite{edward1829xviii} subsequently demonstrated the influence of viscosity on the oscillation period, motivating the seminal analyses of Stokes. In Art.~9 of \citet{stokes1843on}, the inviscid flow generated by an oscillating sphere in a concentric spherical enclosure was derived, correcting Poisson's result, while \citet{stokes1851effect} later incorporated viscous effects; the resulting force was subsequently tested experimentally in unbounded and concentric bounded configurations \citep{meyer1871ueber,mcconnell1965added,motz1966magnetohydrodynamic,dolfo2020experimental}. The spherical-container solution was also used by \citet{transontay1990motion} to interpret the wall and inertial effects measured for a sphere oscillating in a cylindrical tube. These results were extended to arbitrary linear accelerations by \cite{basset1888motion} and to transient motions by \cite{bromwich1929motion}. Stokes' solution remains one of the earliest successful applications of the Navier--Stokes equations with boundary conditions \citep{darrigol2002between}.

Extensions to non-spherical geometries followed. Using the potential theory of \cite{green1828essay}, \cite{kirchhoff1876vorlesungen} derived irrotational solutions for ellipsoids \citep[e.g.\ pp.~141, 144 of][for prolate and oblate cases]{lamb1924hydrodynamics}, while \cite{oberbeck1876ueber} incorporated viscous effects. Spherical confinement for prolate spheroids was later approximated by B74, and analytical drag expressions for oscillating spheroids were obtained by \cite{lai1972stokes}. These low-Reynolds-number hydrodynamics results are reviewed by \cite{happel1960low}.

The influence of rotation on the flow generated by a moving sphere was first studied by \cite{taylor1922motion} and \cite{goertler1944einige}. For steady axial translation at small Reynolds number, \cite{childress1964slow} obtained the weak-rotation correction to Stokes drag. In an unbounded fluid, \cite{tanzosh1994motion} determined the steady drag for arbitrary rotation, whereas \cite{candelier2008time} derived the unsteady force for arbitrary particle motion at small Reynolds and Taylor numbers. Experiments by \cite{maxworthy1965experimental,maxworthy1968observed,maxworthy1970flow} explored the associated Taylor-column structure and rotation-modified drag. For a sphere translating transversely through an inviscid fluid confined between planes normal to the rotation axis, \citet{stewartson1967slow} described the establishment of a Taylor column and the asymptotic two-dimensional exterior flow. At small Rossby number, \cite{hocking1979drag} showed that axial end-wall confinement enhances the steady drag through Taylor-column blocking; this mechanism, distinct from the spherical oscillatory confinement considered below, was recently quantified by \cite{auregan2023flow} and shown to explain the long-standing discrepancy between Maxworthy's measurements and unbounded-domain predictions; see \cite{bush1994particle} for a historical review and \cite{ungarish2024drag} for a recent reassessment of the short-container problem. For forcing frequencies below twice the rotation rate, including steady translation, inertial-wave radiation also contributes to the drag, as analysed by \cite{stewartson1953slow} and investigated experimentally for transverse motions by \cite{mason1975forces,mason1977forces}. By contrast, planetary oscillations typically occur at frequencies exceeding twice the rotation rate (figure~\ref{fig:planets}), precluding inertial-wave radiation. In this regime, \cite{singh1964oscillation} showed that rotation induces an azimuthal flow without affecting the leading-order drag, whereas B74 determined the inviscid drag in a spherical fluid cavity. Despite its relevance to planetary applications, this regime remains comparatively unexplored, motivating our study.

Beyond the linear response, oscillatory motion generates steady streaming through nonlinear advection \citep{riley2001steady}. This second-order flow was first analysed for a transversely oscillating sphere by \citet{wang1965flow,riley1966sphere}, and extended to superposed transverse and torsional oscillations of an isolated sphere by \citet{gopinath1994steady}. High-frequency streaming was subsequently formulated in terms of the effective slip generated by oscillatory boundary layers \citep{rednikov2004steady,rednikov2011acoustic}, with corresponding spheroidal flows visualised experimentally by \citet{kotas2007visualization}. Confinement was later treated for separate transverse and torsional forcing between concentric hemispheres \citep{kong2015oscillatory}, and for their superposition in a concentric spherical shell, where \citet{kong2017oscillatory} obtained an explicit leading-order field and the associated steady streaming for unrestricted Womersley number. In rotating systems, steady streaming generates mean zonal flows \citep{cebron2021mean} and has recently been investigated experimentally in oscillating spherical shells \citep{kozlov2017steady,subbotin2020inertial,subbotin2021linear}. These nonlinear effects lie outside the scope of the present study.

Magnetic effects on oscillatory and bluff-body flows have long been investigated in canonical MHD configurations. Hydromagnetic extensions of Stokes' oscillating-boundary problem established the structure of oscillatory conducting boundary layers \citep{kakutani1958effect,kakutani1960effect,hide1960hydromagnetic}, while Ludford and co-workers examined the electromagnetic forces acting on translating conducting bodies \citep{ludford1959rayleigb,ludford1960inviscid,ludford1960on}. Building on these developments, \citet{stewartson1956motion} formulated the canonical inductionless problem of a conducting sphere translating through an unbounded conducting fluid, which was subsequently developed by \citet{chester1957effect} and \citet{reitz1961the} and extended to viscous creeping flow by \citet{gotoh1960stokes}. \citet{singh1965drag} later generalised the theory to oscillatory motion, deriving analytical expressions for the viscomagnetic drag and added-mass forces. Subsequently, \cite{motz1966wave,motz1966magnetohydrodynamic} extended the classical theory to finite interaction parameters. Strong-field experiments subsequently documented magnetic drag on conducting spheres over broad ranges of interaction parameter \citep{yonas1967measurements,maxworthy1968experimental,vives1974sphere}; \citet{lielausis1975liquid} reviewed the liquid-metal literature. The steady inductionless sphere problem has been revisited \citep{sekhar2005magnetohydrodynamic,haverkort2009magnetohydrodynamics}, providing a more complete description of the flow while extending the classical theory beyond its original asymptotic limits \citep{delacroix2018drag,liu2025analysis}. In parallel, space-plasma studies showed that conducting obstacles moving through magnetised plasmas generate Alfv\'en wings, providing the complementary ideal-MHD analogue of electromagnetic drag \citep{drell1965drag,neubauer1980nonlinear}. This picture has recently been confirmed by global numerical simulations \citep{de2025numerical,corso2026simulations}.

Taken together, these studies established the principal asymptotic descriptions of electromagnetic drag on translating and oscillating bodies in conducting fluids, but no existing theory combines spherical confinement, rotation, viscosity and electromagnetic property contrasts for an oscillating sphere. Although the main analytical developments below concern thin magnetic boundary layers, the present framework also covers the opposite large-diffusivity regime, extending the inductionless oscillatory theory of \citet{singh1965drag} from an unbounded fluid to spherical shells. It further generalises the magnetic boundary-layer theory of \citet{buffett1995magnetic} to confined, viscous and rotating configurations, for both polar and equatorial oscillations. The resulting predictions for electromagnetic drag, added mass and Ohmic dissipation are validated against direct numerical simulations (\textsc{dns}) and establish, in the strong-field limit, an explicit asymptotic connection with the Alfv\'en-wing drag law. The principal theoretical developments and their domains of validity are summarised in table~\ref{tab:mainTheoryContributions}.

\subsection{Motivations of this work}

In P26, various regimes were revealed for an oscillating sphere, spanning parameter ranges described by distinct asymptotic laws. While these laws capture well their respective limits, a comprehensive and robust theoretical framework is still lacking. The aim of this work is to establish a unified asymptotic description connecting the previously disconnected regimes within their respective validity domains. We therefore consider a solid conducting sphere undergoing rapid, small-amplitude oscillations within a rotating, viscous and electrically conducting fluid-filled spherical shell bounded by a solid exterior and permeated by an imposed magnetic field. The analysis is developed for arbitrary radius ratios in concentric spherical shells and electromagnetic property contrasts between the three domains, encompassing Earth-like configurations ($a=0.35$) as a particular case. The oscillation amplitude is assumed small relative to the shortest geometric scale, $\epsilon_s/\ell\ll1$ with $\ell=\min(a_s,D)$ and $D=a_m-a_s$, corresponding to the high-Strouhal-number, or small-excursion, limit \citep{wang1965flow} relevant to planetary interiors, where typical amplitudes are only $1$--$5$~mm \citep{rosat2011review,coyette2014slichter}.

With the characteristic velocity $\epsilon_s\omega_s$, the ratio of convective to local acceleration is $\epsilon_s/\ell\ll1$. Small excursion therefore controls the linearisation: nonlinear advection and the associated steady streaming enter at higher order in amplitude, without requiring that the viscous ($\epsilon_s\omega_s a_s/\nu$) and magnetic ($\epsilon_s\omega_s a_s/\eta_f$) Reynolds numbers to be small. The induction equation is linearised consistently about the imposed field $\boldsymbol{B}_0$. Viscous and magnetic diffusion remain independently controlled by the skin-depth ratios $L_\nu/a_s$ and $L_\eta/a_s$, with the Womersley number $\mathrm{Wo}=a_s\sqrt{2}/L_\nu$ providing an equivalent measure of oscillatory viscous effects in linear-response problems.

\begin{figure}
\centering
\includegraphics[width=1\textwidth]{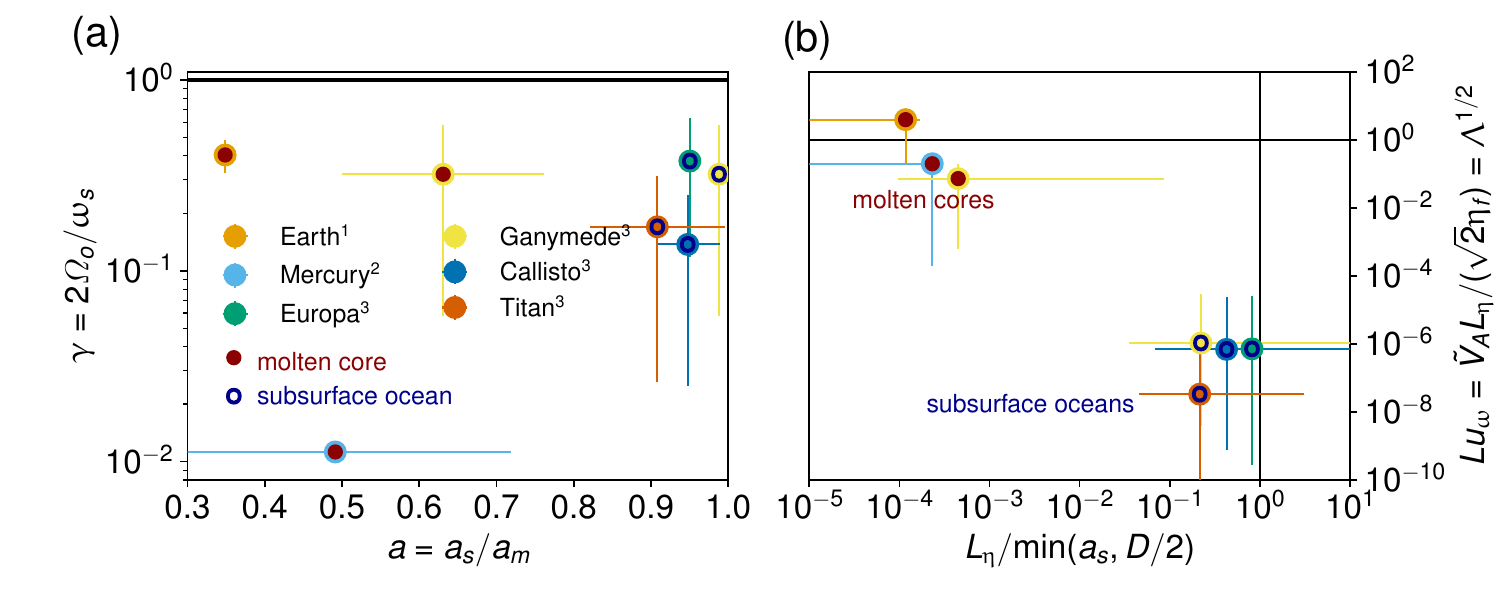}
\caption{(a) Radius ratio $a$ and frequency ratio $\gamma$. (b) Normalised magnetic skin depth and oscillatory Lundquist number, in planetary settings. Periods from $^1$\cite{rosat2011review}, $^2$\cite{grinfeld2005motion} and $^3$\cite{coyette2014slichter}. Conductivities $\sigma_f=0.2$-$20~\mathrm{S\,m^{-1}}$ for subsurface oceans \citep{psarakis2024electrical}, and $\sigma_f=10^6~\mathrm{S\,m^{-1}}$ for liquid cores \citep{pozzo2014thermal}. In (a) and (b), horizontal lines mark emission of inertial ($\gamma>1$) and Alfv\'en ($Lu_\omega>1$) waves; boundary-layer regime for $L_\eta/\min(a_s,D/2)<1$ (vertical line in b).} 
\label{fig:planets}
\end{figure}

Planetary liquid layers span depths ranging from shallow subsurface oceans (kilometres) to deep molten cores (thousands of kilometres). Bounding effects are conveniently characterised by the inner-to-outer radius ratio $a$ (figure~\ref{fig:planets}a), whose wide range leads to diverse confinement regimes, particularly pronounced in icy-moon oceans, motivating extensions beyond unbounded models \cite[e.g.][BG95 hereafter]{buffett1995magnetic}. The viscous and magnetic skin depths, based on their respective diffusivities and oscillation or rotation timescales, remain larger than $\epsilon_s$ but smaller than both the inner radius $a_s$ and the gap $D=a_m-a_s$. For deep liquid cores,  $D$ is typically of the order of $a_s$ and the hierarchy is (notations of table~\ref{tab:mainTheoryContributions})
\begin{equation}
    \epsilon_s < L_\nu < \sqrt{2\nu\varOmega_o^{-1}} < L_\eta < \sqrt{2\eta_f\varOmega_o^{-1}} < \min(a_s,D) ,
\end{equation}
with the skin depths $L_\nu=(2\nu\omega_s^{-1})^{1/2}$ and
$L_\eta=(2\eta_f\omega_s^{-1})^{1/2}$. In subsurface oceans, the magnetic skin depth can be comparable to the half-gap (figure~\ref{fig:planets}b), beyond the validity of boundary-layer models; corresponding asymptotic estimates of the Ohmic dissipation are also derived here. More broadly, the large-diffusivity limit studied in appendix~\ref{appA} completes the regime map with bulk-diffusive solutions, in which viscous or magnetic diffusion extends over scales comparable to the smallest system length scale $\min(a_s,D)$.

The rotation parameter $\gamma=2\varOmega_o/\omegas<1$ ranges from $0.33$-$0.5$ for Earth \citep{rosat2011review} to $\sim10^{-2}$ for Mercury (figure~\ref{fig:planets}a), precluding inertial-wave excitation. Rotation may thus be treated perturbatively for Mercury but not for Earth.
Both approaches are considered below for the polar mode; the equatorial
formulae are restricted to zero or weak rotation. Although the oscillation period $T_s=2\pi/\omegas$ is short relative to the viscous timescale, it can match the magnetic timescale in subsurface oceans, allowing magnetic perturbations to extend well beyond the boundary layer.

Planetary magnetic fields originate from internal dynamos (e.g.\ Earth’s core) or external forcing (e.g.\ Jupiter’s field on Europa) and often display complex spatial and temporal variability. On the short oscillation timescales considered here (hours), the imposed field is taken steady. Its strength is characterised by the Lundquist number $Lu_\omega=\tilde{V}_A L_\eta/(\sqrt{2}\eta_f)$, where $\tilde{V}_A=\tilde{B}_0/\sqrt{\rho_f \mu_f}$ is the Alfv\'en velocity related to the typical amplitude $\tilde{B}_0$ of the field through the fluid density $\rho_f$ and magnetic permeability $\mu_f$ \citep{dormy2007mathematical}. This number compares the resistive timescale $L_\eta^2/\eta_f$ with the Alfv\'en timescale $L_\eta/\tilde{V}_A$ over a magnetic skin depth $L_\eta$. BG95 instead use $\Lambda=Lu_\omega^2$, with $\Lambda<1$ corresponding to the magnetic skin layer regime, in which magnetic diffusion remains confined to a thin boundary layer, and $\Lambda>1$ to the diffusion-free Alfv\'en-wave regime, in which momentum is transported away by propagating Alfv\'en waves. Unlike the interaction (Stuart) parameter $N=\sigma_f\tilde{B}_0^2a_s/(\rho_f\mathcal{V})$ commonly used for steadily translating bodies at velocity $\mathcal V$, the present linear oscillatory problem is governed by $\Lambda=\sigma_f\tilde{B}_0^2/(\rho_f\omega_s)=N \epsilon_s/a_s$, which compares Lorentz forces with the oscillatory inertia. Since $\mathcal{V}=\epsilon_s \omega_s$ gives $\mathcal{V}/(\omega_s a_s)=\epsilon_s/a_s \ll 1$, convective inertia is asymptotically smaller than oscillatory inertia, so that $\Lambda$, rather than $N$, governs the leading-order dynamics. Most planets and moons lie in the weak $\Lambda$ regime (figure~\ref{fig:planets}b), especially subsurface oceans with low conductivity ($\approx1\,\mathrm{S/m}$) and weak external forcing, while Earth lies near the transition (see P26). Electrical properties may differ between fluid and solid domains: metallic cores are comparably conducting in both phases, outer solid regions may be insulating or conducting, and subsurface-ocean boundaries are typically insulating. Accordingly, the magnetic field experienced by the fluid can differ markedly from an externally imposed field owing to magnetic shielding and permeability contrasts. Our boundary-layer framework accommodates an arbitrary axisymmetric internal field; for a uniform externally imposed field, we relate the internal magnetostatic field to the applied field through permeability contrasts, while both conductivity and permeability contrasts enter the oscillatory response at the solid-fluid interfaces.

The problem combines six physical ingredients: oscillation type, confinement, viscosity, rotation, electromagnetic contrasts and magnetic forcing. Existing theories describe only restricted regions of this parameter space (table~\ref{tab:mainTheoryContributions}). The objective is to derive an analytical framework providing explicit analytical predictions across the asymptotic regimes relevant to planetary and laboratory MHD systems (figure~\ref{fig:reg}). We additionally derive an explicit closed-form representation of the complete bounded oscillatory Stokes solution (appendix~\ref{sec:ExpStoU}), valid for arbitrary gap width and Womersley number, and directly suited to force asymptotics and \textsc{dns} validation.

Even in the absence of magnetic fields, the drag remains subtle because viscous effects arise from both boundary stresses and pressure corrections generated by boundary-layer pumping. For stress-free and no-slip spheres, \cite{moore1959rise} and \cite{smylie1998viscous,smylie2000the} retained only viscous stresses, recovering two-thirds of the force obtained by \cite{levich1949motion} and \cite{stokes1851effect}. The missing contribution arises from the viscous pressure correction identified by \cite{kang1988drag}, which can also be recovered using the power approach \citep[\S5.13 of][]{batchelor1967introduction}; see also \cite{joseph2004dissipation}. Revisiting the complete bounded problem of \citet{stokes1851effect}, we derive a representation of the oscillatory field and the associated force expansions in the slow- and rapid-oscillation limits, providing the basis for extensions to rotating and magnetohydrodynamic fluids. We thereby generalise the theories of \cite{stokes1851effect} and \cite{smylie1998viscous} to bounded rotating fluids, including viscously modified pressure forces, and extend the MHD boundary-layer theory of BG95 from polar oscillations in an unbounded inviscid fluid to bounded, viscous configurations with conductivity and permeability contrasts, treating polar and equatorial oscillations at zero or weak rotation, as well as finite-rotation for the polar mode. The extension yields asymptotic predictions that agree quantitatively with \textsc{dns} both in their functional dependence and absolute magnitude.

\subsection{Structure of the work}
The paper is organised as follows. Section~\ref{sec:model} formulates the problem and introduces the governing equations, while \S\ref{sec:methods} presents the analytical and numerical methods. Section~\ref{sec:bulkBr} establishes the hydrodynamic framework, including rotational corrections and viscous dissipation. Building on these results, \S\ref{sec:BTtang} and \S\ref{sec:BTtangvisc} analyse magnetohydrodynamic boundary layers, drag and dissipation in inviscid and viscous fluids, respectively. Finally, \S\ref{sec:discon} discusses the physical implications of the theory and its relevance to planetary and laboratory systems.

\section{Description of the problem} \label{sec:model}
This section derives the governing equations, characterises the forces acting on the inner sphere, and concludes with a concise discussion of the dissipation.

\begin{figure}
\centering
\includegraphics[width=1\linewidth]{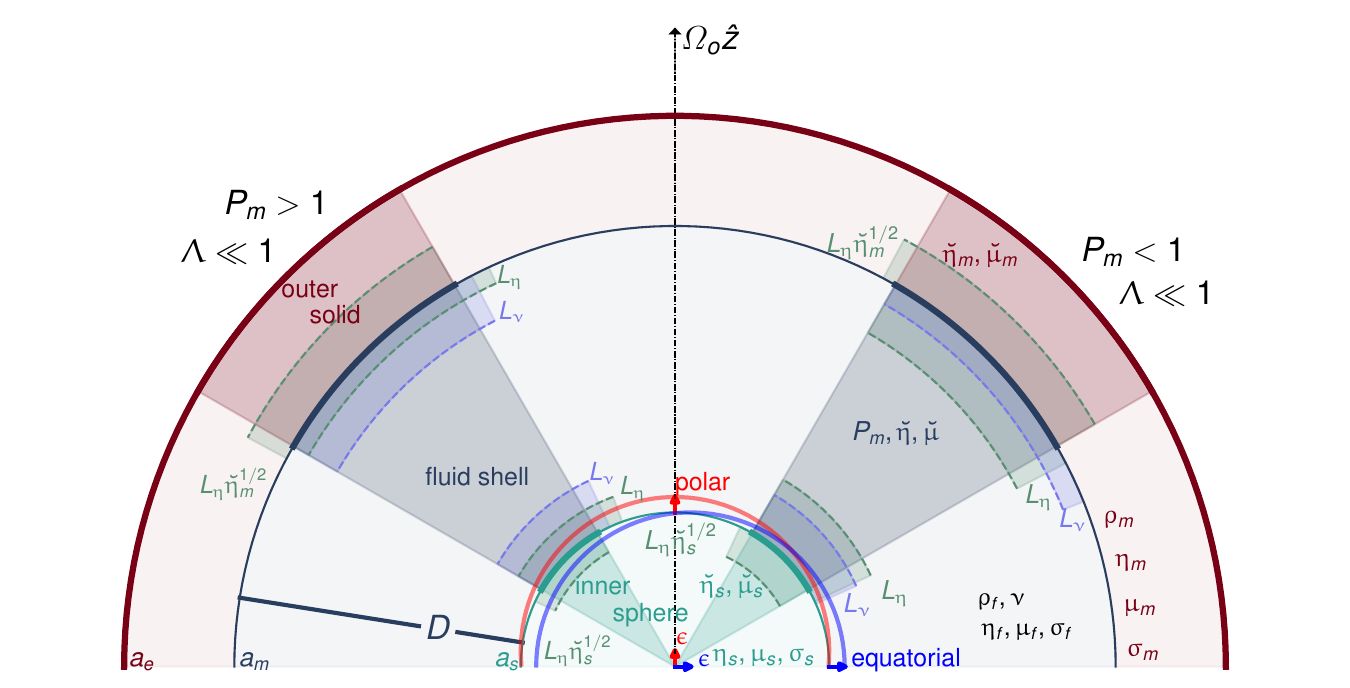}
    \caption{Three-domain configuration and associated viscous and magnetic skin depths in the weak-rotation ($\gamma\ll1$) and weak-field ($\Lambda\ll1$) regime: $P_m>1$ (left) and $P_m<1$ (right), with magnetic Prandtl number $P_m=\nu/\eta_f$.}
    \label{fig:sketch}
\end{figure}

\subsection{Governing equations}\label{sec:phymodel}
We consider a fluid-filled spherical shell of inner and outer radii ${a}_s$ and ${a}_m$, bounded by inner and outer solid domains (figure~\ref{fig:sketch}). The forced problem is formulated in the non-inertial frame attached to the outer solid of radius $a_e$, with origin $O$ at the outer-sphere centre and axes rotating at ${\boldsymbol{\varOmega}}_o=\varOmega_o\hat{\boldsymbol{z}}$, so that the outer solid has no translational motion in this frame. The incompressible Newtonian fluid has uniform electrical conductivity $\sigma_f$, density $\rho_f$, kinematic viscosity $\nu$, and magnetic permeability $\mu_f$. The spherical interfaces separate the fluid from inner and outer solid domains with uniform electrical conductivities $\sigma_s$ and $\sigma_m$, and magnetic permeabilities $\mu_s$ and $\mu_m$, respectively.
The fluid velocity ${\boldsymbol{V}}$ is governed by
\begin{subequations}
\label{syst:eqn_sl}
\begin{equation}
\partial_t  {\boldsymbol{V}} +  ({\boldsymbol{V}} \boldsymbol{\cdot} \boldsymbol{\nabla}) \, {\boldsymbol{V}} +2 { \boldsymbol{\varOmega}}_o \times {\boldsymbol{V}}  = -\nabla {\varPi} + \nu \, \boldsymbol{\nabla}^2 {\boldsymbol{V}} +(\rho_f \mu_f)^{-1} ({\boldsymbol{B}}  \cdot \nabla ) {\boldsymbol{B}} , \qquad \boldsymbol{\nabla} \boldsymbol{\cdot} {\boldsymbol{V}} = 0,
\tag{\theequation~\emph{a,b}}
\end{equation}
\end{subequations}
in the fluid region $\Omega_f$ of volume $(4 \pi/3) ({a}_m^3-{a}_s^3)$, and the magnetic field ${\boldsymbol{B}}$ is governed by
\begin{subequations}
\label{syst:eqn_slB0}
\begin{equation}
   \partial_t {\boldsymbol{B}} = \nabla \times ({\boldsymbol{V}} \times {\boldsymbol{B}}) + \eta \nabla^2 {\boldsymbol{B}} , \qquad \boldsymbol{\nabla} \boldsymbol{\cdot} {\boldsymbol{B}} = 0 ,
\tag{\theequation~\emph{a,b}}
\end{equation}
\end{subequations}
where the magnetic diffusivity $\eta$ is $\eta_f=(\mu_f\sigma_f)^{-1}$, $\eta_s=(\mu_s\sigma_s)^{-1}$, and $\eta_m=(\mu_m\sigma_m)^{-1}$ in, respectively, the fluid, the solid inner and outer domains. The reduced pressure is defined as $\varPi=\varPi_h+B^2/(2\rho_f\mu_f)$, where $\varPi_h$ includes the centrifugal contribution, such that $\varPi$ contains both hydrodynamic and magnetic pressures.

For a free planetary mode, our frame of reference would additionally translate relative to an inertial centre-of-mass frame, as required by the global linear-momentum balance \citep{grinfeld2005motion,grefflefftz2007fluid}. The resulting spatially uniform inertial acceleration is a gradient and can therefore be absorbed into the reduced pressure; the local equations for prescribed relative motion consequently retain the same form, while the physical pressure force and planetary eigenfrequency must be reconstructed together with the global translational dynamics. Let $\boldsymbol{\xi}_s(t)$ and $\boldsymbol{\xi}_m(t)$ denote the positions of the centres of the inner and outer solid domains, respectively, in an inertial frame. In our working frame of reference, the prescribed mechanical forcing is their relative displacement
\begin{equation} 
{\boldsymbol{\delta}}(t) ={\boldsymbol{\xi}}_s(t)-{\boldsymbol{\xi}}_m(t) =\Real \left({\boldsymbol{\epsilon}}_s \mathrm{e}^{\mathrm{i}{\omegas}t}\right), \end{equation} 
with relative velocity ${\boldsymbol{V}}_s=\mathrm{d}_t{\boldsymbol{\delta}}$, with the oscillation frequency $\omegas$ and the time-independent displacement vector $\boldsymbol{\epsilon}_s$. We use the Cartesian basis $(\hat{\boldsymbol{x}},\hat{\boldsymbol{y}},\hat{\boldsymbol{z}})$ and spherical-coordinate basis $(\hat{\boldsymbol{r}},\hat{\boldsymbol{\theta}},\hat{\boldsymbol{\phi}})$, centred on $O$. The oscillation can be along the rotation axis, with
\begin{eqnarray} \label{eq:ppolDC}
 {\boldsymbol{\delta}}=\Real \left[ {\epsilon_s} \hat{\boldsymbol{z}} \mathrm{e}^{\mathrm{i} {\omegas} t} \right] = \Real \left[ {\epsilon_s} (\cos \theta \hat{\boldsymbol{r}}-\sin \theta \hat{\boldsymbol{\theta}}) \mathrm{e}^{\mathrm{i} {\omegas} t} \right] ,  
\end{eqnarray}
which is the so-called polar mode, or along a circular trajectory, with
\begin{eqnarray} \label{eq:eqqDC}
 {\boldsymbol{\delta}}=\Real \left[{\epsilon_s} (\hat{\boldsymbol{x}}\pm \mathrm{i}\hat{\boldsymbol{y}}) \mathrm{e}^{\mathrm{i} {\omegas} t} \right]  = \Real \left[ {\epsilon_s} (\sin \theta \hat{\boldsymbol{r}}+\cos \theta \hat{\boldsymbol{\theta}}\pm\mathrm{i} \hat{\boldsymbol{\phi}}) \mathrm{e}^{\mathrm{i}( {\omegas} t \pm \phi)} \right] ,
\end{eqnarray}
which are the prograde and retrograde equatorial modes (equations 4 and 5 of BG95). The three modes have distinct frequencies $\omegas$ and amplitudes $\epsilon_s$ in planetary cores and are analysed independently below (using the generic notation $\omegas$ and $\epsilon_s$). The amplitude $\epsilon_s$ is assumed to be the smallest length scale, so the inner-boundary position is fixed at leading order. From (\ref{eq:ppolDC})-(\ref{eq:eqqDC}), the polar mode corresponds to a rectilinear harmonic translation, whose velocity magnitude oscillates in time. By contrast,
the equatorial modes correspond to a uniform circular translation: the
Cartesian velocity components oscillate harmonically with a phase shift of $\pi/2$, but the velocity magnitude remains constant. Although the maximum kinetic energy is identical for all three modes, the time-averaged kinetic energy per unit mass is $\langle (\mathrm{d}_t {\boldsymbol{\delta}})^2/2 \rangle=\mathcal{E}(\epsilon_s\omega_s)^2/2$, where $\mathcal{E}$ is the kinetic energy factor: $\mathcal{E}=1/2$ for the polar mode and $\mathcal{E}=1$ for equatorial modes (twice as much). 

To close the problem, boundary conditions are needed. With outward normal $\hat{\boldsymbol{n}}$, the velocity of inviscid fluids verifies the non-penetration condition $\hat{\boldsymbol{n}} \cdot {\boldsymbol{V}}=\hat{\boldsymbol{n}} \cdot \mathrm{d}_t{\boldsymbol{\delta}}$ and $\hat{\boldsymbol{n}} \cdot {\boldsymbol{V}}=0$ at the inner and outer boundaries, respectively. For viscous fluids, conditions on tangential velocities are also required: no-slip boundaries impose ${\boldsymbol{V}}=\mathrm{d}_t{\boldsymbol{\delta}}$ and ${\boldsymbol{V}}=\boldsymbol{0}$ at the inner and outer boundaries, respectively, whereas stress-free boundaries impose
\begin{subequations}
\label{syst:eqn_slB}
\begin{equation}
  ( {\boldsymbol{\tau}}_v \cdot \hat{\boldsymbol{n}}) \times \hat{\boldsymbol{n}} =\boldsymbol{0} , \qquad     {\boldsymbol{\tau}}_v=\rho_f \nu (\nabla {\boldsymbol{V}}+(\nabla {\boldsymbol{V}})^\top),
\tag{\theequation~\emph{a,b}}
\end{equation}
\end{subequations}
where ${\boldsymbol{\tau}}_v$ is the viscous stress tensor. For the magnetic field, equations (\ref{syst:eqn_slB0}) are solved in the solid domains, using ${\boldsymbol{V}}=\mathrm{d}_t {\boldsymbol{\delta}}$ and ${\boldsymbol{V}}= \boldsymbol{0}$ in the inner and outer domains, respectively. The outer solid domain extends with uniform properties up to external radius ${a}_e$.

\subsection{Forces from pressure, viscous and magnetic stresses}\label{sec:forces}
The total fluid force ${\boldsymbol{F}}$ on the sphere results from the combined action of tangential (${\boldsymbol{F}}_v$) and normal (${\boldsymbol{F}}_{vn}$) viscous stresses, pressure forces ${\boldsymbol{F}}_p$, and Lorentz force $\boldsymbol{F}_\eta={\boldsymbol{F}}_{mt}+{\boldsymbol{F}}_{mp}$ that splits into magnetic tension ${\boldsymbol{F}}_{mt}$ and magnetic pressure ${\boldsymbol{F}}_{mp}$ forces. This gives
\begin{eqnarray}
{\boldsymbol{F}}={\boldsymbol{F}}_v+{\boldsymbol{F}}_{vn}+{\boldsymbol{F}}_p+{\boldsymbol{F}}_{mt}+{\boldsymbol{F}}_{mp} ,
\end{eqnarray}
with ${\boldsymbol{F}}_v+{\boldsymbol{F}}_{vn}=\int_S {\boldsymbol{\tau}}_v\cdot\mathrm{d}\boldsymbol{S}$, and ${\boldsymbol{F}}_p=-\int_S \rho_f{\varPi}_h\,\mathrm{d}\boldsymbol{S}$, where $\mathrm{d}\boldsymbol{S}=\hat{\boldsymbol{n}}\mathrm{d}S$ is related to the sphere outward unit normal $\hat{\boldsymbol{n}}=\hat{\boldsymbol{r}}$. Writing the Lorentz stress as
\begin{eqnarray}
\mu_f^{-1}(\nabla\times\boldsymbol{B})\times\boldsymbol{B}=-\mu_f^{-1}\nabla(\boldsymbol{B}^2/2)+\mu_f^{-1}(\boldsymbol{B}\cdot\nabla)\boldsymbol{B}
\end{eqnarray}
gives $\boldsymbol{F}_\eta=\mu_f^{-1}\int_V(\nabla\times\boldsymbol{B})\times\boldsymbol{B}\,\mathrm{d}V$ as the sum of ${\boldsymbol{F}}_{mt}=\mu_f^{-1}\int_S\boldsymbol{B}(\boldsymbol{B}\cdot\mathrm{d}\boldsymbol{S})$ and ${\boldsymbol{F}}_{mp}=-(2\mu_f)^{-1}\int_S\boldsymbol{B}^2\,\mathrm{d}\boldsymbol{S}$. As viscous effects modify the pressure, and thus ${\boldsymbol{F}}_p$, we calculate the viscous contribution $\boldsymbol{F}_\nu$ as $\boldsymbol{F}_\nu=\boldsymbol{F}_v+{\boldsymbol{F}}_{n}-\boldsymbol{F}_n(\nu=0)$, where ${\boldsymbol{F}}_n={\boldsymbol{F}}_p+{\boldsymbol{F}}_{vn}$ is the total normal force, and $\boldsymbol{F}_n(\nu=0)=\boldsymbol{F}_p(\nu=0)$ is the inviscid fluid force. Hence $\boldsymbol{F}_\nu$, unlike $\boldsymbol{F}_v$, can remain non-zero for stress-free viscous boundaries (and both vanish for inviscid flows).

Neglecting magnetic and rotational effects, the oscillating no-slip sphere was analysed by \cite{stokes1851effect}, who showed that ${\boldsymbol{F}}_n+{\boldsymbol{F}}_v$ is the sum of a viscous drag proportional to ${\boldsymbol{V}}_s$ (vanishing as $\nu\to0$) and an inertial contribution. Noting the fluid mass $M_{fs}=4\pi a_s^3 \rho_f/3$ displaced by the inner sphere, this latter inertial contribution is proportional to $\mathrm{d}_t{\boldsymbol{V}}_s$, written $-C_a M_{fs} \mathrm{d}_t{\boldsymbol{V}}_s$, where $C_a\to C_a^i\neq0$ in the inviscid limit owing to ${\boldsymbol{F}}_p$. Although the steady result ${\boldsymbol{F}}_n+{\boldsymbol{F}}_v=-6\pi\rho_f\nu a_s{\boldsymbol{V}}_s$ for $\omegas\to0$ is well known, the relative roles of pressure and viscous stresses, and the rapidly oscillating regime, have received much less attention. Moreover, \cite{stokes1851effect} also considered an outer spherical boundary, deriving implicit bounded-flow solutions and an explicit force on the sphere. Revisiting these results, we derive an explicit velocity field (appendix~\ref{sec:ExpStoU}) and asymptotic expansions of the force for slow and rapid oscillations, quantifying confinement effects (appendix~\ref{sec:StokEq59}).

For stress-free boundary conditions, the force differs but remains relevant to the no-slip configuration, as it directly reflects viscous dissipation in the fluid bulk. This contribution becomes significant when viscosity is not negligible; we therefore recall the corresponding stress-free bulk-force expressions (appendix~\ref{app:steLeal}).

In the presence of rotation or magnetic effects, the inviscid-fluid force does not necessarily retain the form ${\boldsymbol{F}}_p=-C_a^i {m}_{fs} \mathrm{d}_t {\boldsymbol{V}}_s$ ; in general ${\boldsymbol{F}}_p$ is not even opposite to the acceleration (B74). However, for the three Slichter modes, B74 showed that this property persists when rotation is included (without magnetic or viscous effects). In this inviscid rotating limit, the pressure force is parallel to $\hat{\boldsymbol{z}}$ for the polar mode and antiparallel to the cylindrical-radius unit vector $\hat{\boldsymbol{s}}$ for equatorial modes.

\subsection{Ohmic and viscous dissipation}\label{sec:dissipation}
In the fluid, the Ohmic and viscous dissipations are
\begin{eqnarray}
{\mathcal{D}}_\eta= \eta_f \mu_f^{-1} \int_{\Omega_f} (\nabla \times {\boldsymbol{B}})^2\mathrm{d}V \quad , \quad {\mathcal{D}}_\nu=\int_{\Omega_f} {\boldsymbol{\tau}}_v \boldsymbol{:} \nabla {\boldsymbol{V}} \mathrm{d}V ,
\end{eqnarray}
with ${\boldsymbol{\tau}}_v=\rho_f \nu [\nabla {\boldsymbol{V}}+(\nabla {\boldsymbol{V}})^\top]$. Ohmic dissipation in solids follows from analogous expressions with the corresponding volumes and material properties. Although viscous dissipation is often expressed using the vorticity $\boldsymbol{\mathcal{W}}=\nabla\times\boldsymbol{V}$ alone, the oscillating inner boundary prevents this simplification. Using the Bobyleff-Forsyth identity \citep{buresti2009notes},
\begin{eqnarray}
 {\mathcal{D}}_\nu=\int_{\Omega_f} {\boldsymbol{\tau}}_v \boldsymbol{:} \nabla {\boldsymbol{V}} \mathrm{d}V =  \rho_f \nu \int_{\Omega_f} \boldsymbol{\mathcal{W}}^2\mathrm{d}V + 2 \rho_f \nu  \int_{ {\Omega_f}} \nabla \cdot(\mathrm{D}_t \boldsymbol{V} ) \mathrm{d}V, \label{eq:ensDnu0}
\end{eqnarray}
with $\mathrm{D}_t \boldsymbol{V}=\partial_t \boldsymbol{V}+(\boldsymbol{V}\cdot\nabla)\boldsymbol{V}$, an additional acceleration term must be retained. Since $\nabla\cdot\partial_t\boldsymbol{V}=0$,
\begin{eqnarray}
 \int_{ {\Omega_f}} \nabla \cdot(\mathrm{D}_t \boldsymbol{V} ) \mathrm{d}V = \int_{\partial {\Omega_f}} \hat{\boldsymbol{n}} \cdot (\mathrm{D}_t \boldsymbol{V}) \mathrm{d}S = \int_{\partial {\Omega_f}} \hat{\boldsymbol{n}} \cdot \left(  \boldsymbol{\mathcal{W}} \times  \boldsymbol{V} + \nabla( \boldsymbol{V}^2/2) \right) \mathrm{d}S , \label{eq:ensDnu}
\end{eqnarray}
where $\boldsymbol{\mathcal{W}}\times\boldsymbol{V}$ is the Lamb vector and
\begin{eqnarray}
\int_{\partial {\Omega_f}} \hat{\boldsymbol{n}}\cdot(\mathrm{D}_t \boldsymbol{V})\mathrm{d}S=\int_{r=a_m} (\mathrm{D}_t {V}_r) \mathrm{d}S-\int_{r=a_s} (\mathrm{D}_t {V}_r) \mathrm{d}S .
\end{eqnarray}
The surface stress work rate (power input from the boundary) is then \citep{wu1999note}
\begin{eqnarray}
   \rho_f \nu \int_{\Omega_f} \boldsymbol{V} \cdot (\nabla^2 \boldsymbol{V}) \mathrm{d}V+  {\mathcal{D}}_\nu=\rho_f \nu \int_{\partial \Omega_f} \hat{\boldsymbol{n}} \cdot \left(\boldsymbol{\mathcal{W}} \times  \boldsymbol{V} + \nabla \boldsymbol{V}^2 \right)  \mathrm{d}S .
\end{eqnarray}

\section{Analytical and numerical methods} \label{sec:methods}

Dimensionless variables are defined using the gap
$D=a_m-a_s$ as length scale, the magnetic diffusion time
$D^2/\eta_f$ as time scale, $\rho_f$ as density scale, and
$\eta_f D^{-1}\sqrt{\rho_f\mu_f}$ as magnetic-field scale
(nomenclature in \S\ref{sec:nomTab} of the supplementary material). Hereafter all quantities are dimensionless, but retain their dimensional symbols unless stated otherwise.  With the solenoidal constraints $\boldsymbol{\nabla}\boldsymbol{\cdot}\boldsymbol{V}=\boldsymbol{\nabla}\boldsymbol{\cdot}\boldsymbol{B}=0$, the governing equations read
\begin{align}
\mathrm{\partial}_t \boldsymbol{V} +(\boldsymbol{V} \cdot \nabla) \boldsymbol{V} +\gamma \omega \, \hat{\boldsymbol{z}}  \times \boldsymbol{V}  &= -\nabla \varPi + P_m \boldsymbol{\nabla}^2 \boldsymbol{V} +(\boldsymbol{B}  \cdot \nabla ) \boldsymbol{B}    , \label{eq:dc0} \\
\mathrm{\partial}_t  \boldsymbol{B} &= \breve{\eta} \nabla^2 \boldsymbol{B} + (\boldsymbol{B} \cdot \nabla) \boldsymbol{V} - (\boldsymbol{V} \cdot \nabla) \boldsymbol{B} ,    \label{eq:dc1}
\end{align}
with forcing displacement $\boldsymbol{\delta}= \boldsymbol{\epsilon} \mathrm{e}^{\mathrm{i}( \omega t + m \phi)}$ of amplitude $\epsilon$ and frequency $\omega=\omega_s D^2/\eta_f$, magnetic Prandtl number $P_m=\nu/\eta_f$, and frequency ratio $\gamma=2\varOmega_o/\omega_s$. The azimuthal wavenumber is $m=-1$ and $m=1$ for the prograde and retrograde equatorial modes, respectively, while the $\phi$ dependency is removed with $m=0$ for the polar mode. The dimensionless diffusivity is $\breve{\eta}=1$ in the fluid and $\breve{\eta}_s=\eta_s/\eta_f$, $\breve{\eta}_m=\eta_m/\eta_f$ in the inner and outer solids, whose velocities are $\boldsymbol{V}_s=\mathrm{d}_t\boldsymbol{\delta}$ and $\boldsymbol{V}=\boldsymbol{0}$, respectively; only the magnetic equation~is solved in solids. Defining the permeability ratios $\breve{\mu}_s=\mu_s/\mu_f$ and $\breve{\mu}_m=\mu_m/\mu_f$, electrically insulating inner or outer domains correspond to $\breve{\eta}_s\to\infty$ and $\breve{\eta}_m\to\infty$. 

\subsection{Numerical modelling}\label{sec:modelling}
Equations (\ref{eq:dc0})-(\ref{eq:dc1}) are solved using the finite-difference/pseudo-spectral code \textsc{xshells} \citep[][]{schaeffer2017turbulent}, or the finite-element code \textsc{comsol} Multiphysics\textsuperscript{\textregistered}. \textsc{xshells} employs a poloidal-toroidal decomposition of the velocity field in spherical harmonics with degree $l\le l_{\max}$ and azimuthal wavenumber $m\le m_{\max}$ via the \textsc{shtns} library \citep{schaeffer2013efficient}, combined with second-order finite differences on $N_r$ radial points. Its spectral formulation enables access to extreme viscosity and magnetic-field regimes and has been extensively validated \citep[e.g.][]{marti2014full}. In the weak-forcing limit $\epsilon\ll1$, the nonlinear equations are integrated in a fixed geometry, with translational forcing imposed through the boundary velocity $\mathrm{d}_t\boldsymbol{\delta}$ at the inner boundary \cite[as in][]{rieutord2002slichter}. Only diffusion, discarding any motion, is solved in the solid regions. The region surrounding the outer solid shell is treated as a perfect electrical insulator, with the magnetic boundary conditions described in \cite{ChristensenWicht2015}. Implementation details are given in P26. Simulations use a semi-implicit third-order backward-difference scheme \citep[SBDF3,][]{ascher1995implicit}. A typical resolution at $L_\nu=10^{-3}$, $\Lambda=0.5$, and $P_m=10^{-2}$ is $N_r=511$, $l_{\max}=127$, and $m_{\max}=8$, with sufficient refinement in the nested boundary layers and, when Alfv\'en waves are present, in the interior. For laminar flows with simple forcing (e.g. axial magnetic field), only a few azimuthal modes are required: polar mode is axisymmetric ($m=0$), whereas equatorial modes require the single non-zero azimuthal order
 ($m=\pm1$). Simulations with \textsc{xshells} are restricted here to no-slip boundaries. The system starts from rest and is allowed to evolve until it reaches a converged regime over a few diffusive time scales. The initial transient is discarded. Dissipation is computed from volume integrals (\S\ref{sec:dissipation}) and time-averaged in the converged regime. The vorticity formulation avoids solving for the pressure, reducing computational cost but precluding direct evaluation of pressure forces and therefore of the added-mass coefficient. For these quantities, \textsc{comsol} is used instead (\S\ref{sec:COMSim}).

\subsection{Perturbation approach and boundary-layer theory (BLT)} \label{sec:BLp}
We restrict attention to the weak-forcing regime, $\epsilon=\epsilon_s/D\ll1$. Analytical solutions are derived using boundary-layer theory (BLT), for which the boundary-layer components must decay over distances small compared with the local radii of curvature and the separation between neighbouring boundaries. In the fluid, the asymptotic expansion is therefore organised by $L_\nu/\ell$ and $L_\eta/\ell$, with $\ell=\min(a_s,D)$; as shown below by deriving their thicknesses, the magnetic layers in the inner and outer solids additionally require $L_\eta\ll a_s \breve{\eta}_s^{-1/2}$ and $L_\eta\ll \breve{\eta}_m^{-1/2} \min(a_m,a_e-a_m)$, respectively. Except for this last condition, the boundary layer problem is fully independent with respect to $a_e$. These restrictions do not apply to the exact bounded Stokes hydrodynamic solution of \citet{stokes1851effect} in appendix~\ref{sec:ExpStoU}, which remains valid for arbitrary $L_\nu/\ell$ within the linear problem.

\begin{figure}
\centering
\includegraphics[width=0.9\linewidth]{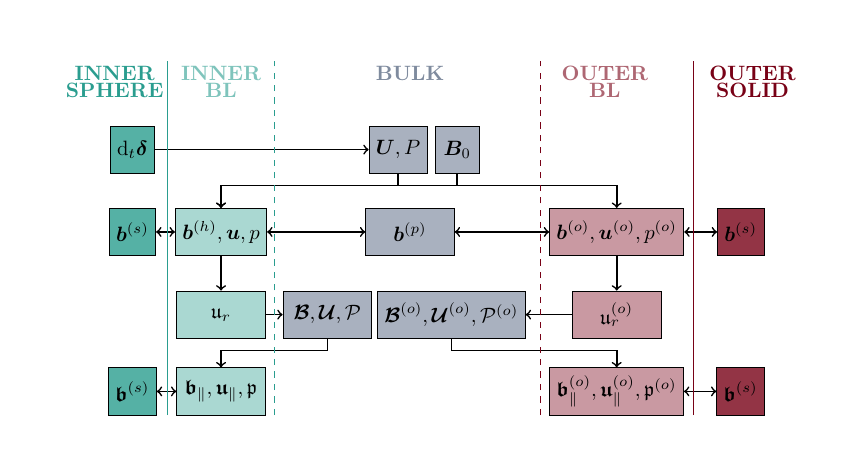}
    \caption{Hierarchy and spatial distribution of velocity and magnetic base fields and perturbations. All fields are of order $\epsilon$ in this work, except the zeroth-order field $\boldsymbol{B}_0$, which interacts with the hydrodynamic solution $(\boldsymbol{U},P)$ forced by the inner sphere velocity $\mathrm{d}_t \boldsymbol{\delta}$, to generate corrections in the whole space (e.g.\ $\boldsymbol{u}$). At next order in the boundary-layer thickness, ($\boldsymbol{u}_\parallel,\boldsymbol{u}^{(o)}_\parallel$) drive boundary-layer radial flows $(\mathfrak{u}_r,\mathfrak{u}_r^{(o)})$, which force secondary bulk fields (e.g.\ $\boldsymbol{\mathcal{B}}$) and their associated boundary-layer corrections (e.g.\ $\boldsymbol{\mathfrak{b}}_\parallel$).}
    \label{fig:decomposition}
\end{figure}

With the dependence $\mathrm{e}^{\mathrm{i}(\omega t+m\phi)}$, the polar and equatorial modes correspond to $m=0$ and $m=\pm1$, respectively. The analysis starts from the imposed magnetic field
$\boldsymbol{B}_0$ and the inviscid non-magnetic solution
$(\boldsymbol{U}, P)$. Boundary layers generated at the inner and outer spheres then induce successive magnetic, velocity and pressure
corrections, yielding the hierarchy
\begin{eqnarray} \label{syst:ansstz}
\boldsymbol{B} &=& \boldsymbol{B}_0+\epsilon 
 \Real  \left[(\boldsymbol{b}+\boldsymbol{\mathcal{B}}+\boldsymbol{\mathfrak{b}}+\boldsymbol{b}^{(o)}+\boldsymbol{\mathcal{B}}^{(o)}+\boldsymbol{\mathfrak{b}}^{(o)}) \mathrm{e}^{\mathrm{i}( \omega t + m \phi)} \right] , 
\end{eqnarray}
\begin{eqnarray}
\boldsymbol{V} &=&\epsilon 
 \Real  \left[(\boldsymbol{U}+ \boldsymbol{u}+\boldsymbol{\mathcal{U}}+\boldsymbol{\mathfrak{u}}+ \boldsymbol{u}^{(o)}+\boldsymbol{\mathcal{U}}^{(o)}+\boldsymbol{\mathfrak{u}}^{(o)}) \mathrm{e}^{\mathrm{i}( \omega t + m \phi)} \right] , \label{eq:uuUB}  
\end{eqnarray}
and
\begin{eqnarray}
 \varPi&=&\epsilon 
 \Real  \left[(P+p+\mathcal{P}+\mathfrak{p}+p^{(o)}+\mathcal{P}^{(o)}+\mathfrak{p}^{(o)}) \mathrm{e}^{\mathrm{i}( \omega t + m \phi)} \right] ,   \label{syst:ansstzP}
\end{eqnarray}
where $\boldsymbol{B}_0(r,\theta)$ is an imposed steady magnetic field, whose normal component $B_{0r}$ at the solid boundary primarily controls boundary-layer magnetic effects (e.g.\ BG95). Figure~\ref{fig:decomposition} provides the roadmap of the analytical
construction adopted throughout the paper, indicating
the successive bulk and boundary-layer corrections and
their coupling. 

The boundary-layer analysis holds for an arbitrary $\phi$-independent $\boldsymbol{B}_0(r,\theta)$, while a uniform axial field is used in \textsc{dns}. For $r\le a_s$, the field remains uniform when a uniform external field $\boldsymbol{B}_e$ is applied, and can reach up to three times $\boldsymbol{B}_e$ for large $\mu_s \gg \mu_0$, or can be strongly reduced for large $\mu_f$ due to magnetic shielding. The analytical derivation of $\boldsymbol{B}_0$, including the effective permeability formulation and shielding limits, is given in \S\ref{sec:B00}.

Considering the hydrodynamic inviscid solution $(\boldsymbol{U},P)$ for $\boldsymbol{B}_0=\boldsymbol{0}$, the non-zero $\boldsymbol{B}_0$ and viscous no-slip conditions generate a hierarchy of BLT corrections from inner and outer boundaries, the latter denoted by ${}^{(o)}$. The perturbation hierarchy relies on the thin boundary-layer assumption, introducing a second small
parameter in addition to $\epsilon$. Starting from the irrotational solution $(\boldsymbol{U}, P)$, successive boundary-layer and bulk corrections are determined order by order. We first consider the leading-order corrections $(\boldsymbol{b},\boldsymbol{u},p)$ and $(\boldsymbol{b}^{(o)},\boldsymbol{u}^{(o)},p^{(o)})$ to the basic state $(\boldsymbol{B}_0,\boldsymbol{U},P)$, arising from the inner and outer boundary layers, respectively (figure~\ref{fig:decomposition}). The fields $(\boldsymbol{b},\boldsymbol{u})=(\boldsymbol{b}^{(p)}+\boldsymbol{b}^{(h)},\,\boldsymbol{u}^{(p)}+\boldsymbol{u}^{(h)})$ decompose into a bulk contribution $(\boldsymbol{b}^{(p)},\boldsymbol{u}^{(p)})$, driven by induction and vanishing in the non-magnetic limit where $^{(p)}$ stands for particular, and a boundary-layer correction $(\boldsymbol{b}^{(h)},\boldsymbol{u}^{(h)})$ enforcing the boundary conditions at $r=a_s$, where $^{(h)}$ stands for homogeneous. Through higher-order boundary-layer radial flow $\mathfrak{u}_r$, required to ensure mass conservation, these corrections, in turn, force higher-order bulk fields $(\boldsymbol{\mathcal{B}},\boldsymbol{\mathcal{U}},\mathcal{P})$ and $(\boldsymbol{\mathcal{B}}^{(o)},\boldsymbol{\mathcal{U}}^{(o)},\mathcal{P}^{(o)})$, smaller by a factor of the order of the thickness of the boundary-layer. The resulting secondary bulk fields generally violate the tangential boundary conditions and therefore generate a final boundary-layer readjustment $(\boldsymbol{\mathfrak{b}}_\parallel,\boldsymbol{\mathfrak{u}}_\parallel)$, and similarly at the outer boundary (see \S\ref{sec:BLp2} and \S\ref{sec:highP}).

\subsection{Calculation of the boundary-layer solution}
Boundary-layer homogeneous solutions decay exponentially into the bulk, where diffusion is negligible, so that the magnetic particular solution $\boldsymbol{b}^{(p)}$ approaches the diffusionless limit $\boldsymbol{b}^{(p)}_0$ induced by $\boldsymbol{U}$. In the weak-Lorentz-force regime, $(\boldsymbol{b}^{(p)},\boldsymbol{u}^{(p)}) \approx (\boldsymbol{b}^{(p)}_0,\boldsymbol{u}^{(p)}_0)$ can be obtained analytically (appendix~\ref{appA}). Outer-boundary corrections (superscript ${}^{(o)}$) satisfy the same equations as those at $r=a_s$ and vanish in the unbounded limit $a_m \gg a_s$; they are thus deduced from the inner solutions, and only $(\boldsymbol{b},\boldsymbol{u},p)$ is detailed below. 

With $\boldsymbol{U}$ satisfying non-penetration, the leading-order boundary-layer perturbation verifies $u_r=0$. Within BLT, radial derivatives $\partial_r \boldsymbol{u}$ and $\partial_r \boldsymbol{b}$ dominate over the fields and over derivatives in other directions. Radial variations of $\boldsymbol{B}_0$ are neglected relative to those of $\boldsymbol{b}$, so $\partial_r \boldsymbol{B}_0$ is omitted when $\partial_r \boldsymbol{b}$ is present; this is exact for the uniform axial fields considered below. For $\boldsymbol{B}_0=[B_{0r},B_{0\theta},B_{0\phi}]$ depending only on $\theta$, the tangential components of $\boldsymbol{u}=(0,\boldsymbol{u}_\parallel)$ and $\boldsymbol{b}=(b_r,\boldsymbol{b}_\parallel)$ satisfy the BLT form of (\ref{eq:dc0})-(\ref{eq:dc1}),
\begin{eqnarray}
\mathrm{i} \omega \boldsymbol{u}_\parallel +\gamma\omega  \hat{\boldsymbol{z}} \times  \boldsymbol{u}_\parallel &=& - \nabla_\parallel p+P_m \partial_r^2 \boldsymbol{u}_\parallel+ B_{0r}\partial_r \boldsymbol{b}_\parallel , \label{eq:dc3} \\
\mathrm{i} \omega \boldsymbol{b}_\parallel &=&  \breve{\eta} \partial_r^2 \boldsymbol{b}_\parallel + B_{0r} \partial_r \boldsymbol{u}_\parallel + \mathcal{T}(\boldsymbol{U}, \boldsymbol{B}_0)  , \label{eq:dc2} 
\end{eqnarray}
where $(\boldsymbol{u}_\parallel,\boldsymbol{b}_\parallel)$ also denote $\boldsymbol{u}_\parallel=[0,u_\theta,u_\phi]^\top$ and $\boldsymbol{b}_\parallel=[0,b_\theta,b_\phi]^\top$ for simplicity. Combining (\ref{eq:dc2}) with the curl of (\ref{eq:dc3}) yields a homogeneous system for $(\boldsymbol{u}_\parallel,\boldsymbol{b}_\parallel)$, except for 
\begin{eqnarray} \label{eq:inhoTT}
\mathcal{T} &=& B_{0r} \partial_r \boldsymbol{U}- \boldsymbol{B}_0 (\partial_r U_r) + r^{-1} (B_{0r} \boldsymbol{U}-U_r \boldsymbol{B}_0)+ r^{-1} \mathcal{F} , \\
     \mathcal{F} &=& \mathrm{i} m ( \sin \theta)^{-1} (B_{0\phi} U_\theta-B_{0\theta}  U_\phi ) \hat{\boldsymbol{\theta}}  + \mathcal{F}_\phi \hat{\boldsymbol{\phi}} , \\
 \mathcal{F}_\phi&=&B_{0\theta} \partial_\theta U_\phi-B_{0\phi} \partial_\theta U_\theta+U_\phi \partial_\theta B_{0\theta}-U_\theta \partial_\theta B_{0\phi} ,
\end{eqnarray}
related to the induction forcing. Neglecting rotation and viscosity recovers equations (20)–(21) of BG95 when $B_{0r}\partial_r\boldsymbol{U}$ is assumed dominant in (\ref{eq:inhoTT}); in their subsequent analysis all forcing terms are then neglected, i.e.\ $\mathcal{T}=0$. Here, all inhomogeneous contributions in (\ref{eq:inhoTT}) are retained, showing that $B_{0r}\partial_r\boldsymbol{U}$ is not generally dominant (see \S\ref{sec:BTtang}).

Since $b_r$ does not enter (\ref{eq:dc3})–(\ref{eq:dc2}), it is determined independently within the BLT framework. At leading order, the forcing $\nabla\times(\boldsymbol{U}\times\boldsymbol{B}_0)$ leads to
\begin{eqnarray}
 \mathrm{i} \omega b_r&=& \breve{\eta} \partial_r^2 b_r+  r^{-1 } [(U_r B_{0 \theta} - U_\theta B_{0r} ) \cot \theta +B_{0 \theta} \partial_\theta U_r - B_{0r} \partial_\theta U_\theta] \nonumber \\
    & & + \mathrm{i} m(r \sin \theta)^{-1}(B_{0 \phi}  U_r - B_{0r}  U_\phi)+ r^{-1}(U_r \partial_\theta B_{0\theta}-U_\theta \partial_\theta B_{0r}) , \label{eq:tracBr}
\end{eqnarray}
as detailed in \S\ref{sec:Br}. All theoretical evaluations, from basic flows \citep[e.g.\ from][]{busse1974free} to boundary-layer solutions or surface integrations, are carried out with the arbitrary-precision library \textit{mpmath} ensuring the high accuracy required.

\subsection{Boundary and interface conditions} \label{sec:BINTCBC}
Boundary conditions at $r=a_s$ or $r=a_m$ require continuity of (i) $b_r$, (ii) the tangential components $\breve{\mu}^{-1}\boldsymbol{b}_\parallel$  of the $\boldsymbol{H}$-field $\boldsymbol{B}= \mu \boldsymbol{H}$, and (iii) the tangential electric field $\breve{\eta} \nabla\times\boldsymbol{B}-\boldsymbol{V}\times\boldsymbol{B}$ \citep[e.g.][]{stratton2007electromagnetic}. In the BLT limit, condition (iii) reduces to
\begin{eqnarray} \label{eq:Bddic}
[ \breve{\eta} \partial_r \boldsymbol{b}_\parallel]_-^+ + B_{0r} [\boldsymbol{U}_\parallel+\boldsymbol{u}_\parallel]_-^+ = \boldsymbol{0},
\end{eqnarray}
as in BG95. Using $\nabla\cdot\boldsymbol{B}=0$, conditions (i)–(ii) also imply continuity of $\breve{\mu}^{-1}\partial_r (r^2 b_r)$. For inviscid fluids, the above boundary conditions are sufficient, as the velocity field is slaved to the magnetic field. Velocity discontinuities may then occur at the boundaries, introducing forcing through the term $B_{0r}[\boldsymbol{U}_\parallel+\boldsymbol{u}_\parallel]_-^+\neq\boldsymbol{0}$ in (\ref{eq:Bddic}). At $r=a_s$, $[\boldsymbol{U}_\parallel+\boldsymbol{u}_\parallel]_-^+=\boldsymbol{U}_\parallel+\boldsymbol{u}_\parallel-\mathrm{d}_t\boldsymbol{\delta}_\parallel$, whereas at $r=a_m$, $[\boldsymbol{U}_\parallel+\boldsymbol{u}_\parallel]_-^+=\boldsymbol{U}_\parallel+\boldsymbol{u}_\parallel$ since the outer boundary is at rest. This boundary forcing can then generate a non-zero homogeneous solution $(\boldsymbol{b}_\parallel^{(h)},\boldsymbol{u}_\parallel^{(h)})$, even for a negligible particular solution $(\boldsymbol{b}_\parallel^{(p)},\boldsymbol{u}_\parallel^{(p)})$.

For viscous fluids, additional boundary conditions are required. The stress-free condition in the BLT formulation (\ref{syst:eqn_slB}) imposes $\partial_r(\boldsymbol{U}_\parallel+\boldsymbol{u}_\parallel)=\boldsymbol{0}$, whereas the no-slip condition enforces $\boldsymbol{U}+\boldsymbol{u}=\mathrm{d}_t\boldsymbol{\delta}$, reducing (\ref{eq:Bddic}) to $\left[\breve{\eta}\partial_r\boldsymbol{b}_\parallel\right]_-^+=\boldsymbol{0}$.

\subsection{Higher-order MHD solutions} \label{sec:BLp2}
Equations (\ref{eq:dc3})-(\ref{eq:dc2}) give the boundary-layer solution but do not enforce $\nabla \cdot \boldsymbol{u}=0$. To do so, the radial flow  $\epsilon 
 \Real (\mathfrak{u}_r\mathrm{e}^{\mathrm{i}( \omega t + m \phi)} ) $ is needed at the next order (in the small boundary layer thickness). With the non-penetration condition, the BLT approach gives \cite[][]{greenspan1968theory}
\begin{equation}
r \partial_r \mathfrak{u}_r =- \cot (\theta){u}_{\theta}-\mathrm{i} m (\sin \theta)^{-1} {u}_{\phi}-\partial_{\theta} {u}_{\theta} , \label{eq:consmass}
\end{equation} 
where the factor $r$ on the left-hand side of equation~(\ref{eq:consmass}) is set constant to $r=a_s$ or $r=a_m$. The solution $\mathfrak{u}_r $ is then obtained by integrating equation~(\ref{eq:consmass}) with the condition that $\mathfrak{u}_r $ vanishes away from the boundary, and is non-zero at the boundary ($\mathfrak{u}_r$ being called secondary flow, or Ekman pumping in rotating fluids). The non-penetration boundary condition $\mathfrak{u}_r+\mathcal{U}_r=0$ then forces a secondary bulk flow $\boldsymbol{\mathcal{U}}$ (the so-called Ekman circulation for rotating fluids). Neglecting diffusive effects in the bulk, $\boldsymbol{\mathcal{U}}$ and the pressure field $\mathcal{P}$ are governed by 
\begin{eqnarray}
    \mathrm{i} \omega \boldsymbol{\mathcal{U}}   &=& -  \nabla \mathcal{P}+ \boldsymbol{B}_0 \cdot \nabla \boldsymbol{\mathcal{B}}+\boldsymbol{\mathcal{B}}\cdot \nabla \boldsymbol{B}_0-\gamma\omega \hat{\boldsymbol{z}} \times \boldsymbol{\mathcal{U}} , \label{eq:dc40} \\
    \mathrm{i} \omega \boldsymbol{\mathcal{B}} &=&  (\boldsymbol{B}_0 \cdot \nabla) \boldsymbol{\mathcal{U}} - (\boldsymbol{\mathcal{U}} \cdot \nabla) \boldsymbol{B}_0 ,    \label{eq:dc41}
\end{eqnarray}
with the secondary bulk field $\boldsymbol{\mathcal{B}}$. In contrast to the exponentially decaying $\mathfrak{u}_r$, $\boldsymbol{\mathcal{U}}$ is nearly uniform across the boundary layer, with $\mathcal{U}_r(r=a_s)=-\mathfrak{u}_r(r=a_s)$, so that the radial flow is $\mathfrak{u}_r+\mathcal{U}_r \approx \mathfrak{u}_r-\mathfrak{u}_r(r=a_s)$ in the boundary layer. This higher-order contribution can be significant (e.g.\ for recovering the Stokes force in appendix~\ref{sec:stoBL}). 

Although explicit determination of $(\boldsymbol{\mathcal{U}},\boldsymbol{\mathcal{B}})$ is difficult, their contribution to mean dissipation can be obtained indirectly via power arguments (\S\ref{sec:dragduissp2}). When used below, $\boldsymbol{\mathcal{U}}$ is approximated by neglecting magnetic and rotational effects in (\ref{eq:dc40})-(\ref{eq:dc41}) and assuming an irrotational form; weak rotational corrections can nevertheless be derived a posteriori for the polar mode (\S\ref{sec:InvBulk}). Once the secondary bulk fields are known, their residual tangential mismatch at the boundaries is removed by the higher-order boundary-layer corrections $(\boldsymbol{\mathfrak{u}}_\parallel,\boldsymbol{\mathfrak{b}}_\parallel)$. As it will be illustrated in \S\ref{sec:highP}, they are obtained by solving the same homogeneous boundary-layer problem as $(\boldsymbol{u}^{(h)}_\parallel,\boldsymbol{b}^{(h)}_\parallel)$ with amplitudes fixed by the interface conditions of \S\ref{sec:BINTCBC}, but where $\boldsymbol{U}$ is replaced by $\boldsymbol{\mathcal{U}}+\boldsymbol{\mathcal{U}}^{(o)}$, and with a stationary inner boundary (as its motion is already accounted for in $\boldsymbol{u}$).

\subsection{Scaled forces and dissipation from the boundary-layer approach} \label{sec:dragduissp2}

We consider the forces on the inner solid domain using leading-order corrections to the basic state; those on the outer solid domain are obtained similarly, using the relevant fields at  $r=a_m$. At the order $\epsilon$, the viscous force $\boldsymbol{F}_{v}$, the hydrodynamic pressure force $\boldsymbol{F}_p$, and the magnetic tension $\boldsymbol{F}_{mt}$ are obtained from $\boldsymbol{F}_{j}= \epsilon  \mathrm{e}^{\mathrm{i} \omega t} \int_S  \boldsymbol{f}_{j} \, \mathrm{e}^{\mathrm{i} m \phi} \mathrm{d}S$, with $\boldsymbol{f}_{v} = P_m \partial_r \boldsymbol{u}_\parallel$, $\boldsymbol{f}_p = - (P+p_h) \hat{\boldsymbol{r}}$, and $\boldsymbol{f}_{mt} =   B_{0r} \boldsymbol{b}+b_r \boldsymbol{B}_0$, where $p_h$ and $p_m$ are the hydrodynamic and magnetic parts of $p=p_h+p_m$. In BG95, $\boldsymbol{f}_{mt} $ is further simplified to its BLT expression $\boldsymbol{f}_{mt} =   B_{0r} \boldsymbol{b}_\parallel $. The magnetic pressure force $\boldsymbol{F}_{mp}$ is similarly obtained from $\boldsymbol{f}_{mp} =  {f}_{mp}  \hat{\boldsymbol{r}}$, with ${f}_{mp} = - \boldsymbol{B}_0 \cdot \boldsymbol{b} $ since $B^2 \approx 2  \boldsymbol{B}_0 \cdot \boldsymbol{b} $  at this order. 
We also note the torque $\boldsymbol{\Gamma}_j=\epsilon  \mathrm{e}^{\mathrm{i} \omega t }\int_S \boldsymbol{r} \times  \boldsymbol{f}_{j} \, \mathrm{e}^{\mathrm{i} m \phi} \mathrm{d}S$ for each local force $\boldsymbol{f}_{j}$. The magnetic and viscous torques originate from the magnetic tension stress $\boldsymbol{f}_{mt}$ and tangential viscous stress $\boldsymbol{f}_{v}$, respectively.

Following BG95, we characterise each force contribution by the scaled force
\begin{eqnarray}
 \frac{\Delta \boldsymbol{\omega}_j}{\omega}
 =
 - \frac{1}{2 M_s \omega^2}
 \int_S \boldsymbol{f}_j \, \mathrm{e}^{\mathrm{i} m \phi} \mathrm{d}S
 =
 - \frac{1}{2 M_s \omega^2}
 \frac{\boldsymbol{F}_j}
 {\epsilon \mathrm{e}^{\mathrm{i} \omega t }}
 ,
 \label{eq:detun}
\end{eqnarray}
which measures the force relative to the inertial-force amplitude
$M_s\epsilon\omega^2$. Although introduced here for a forced problem,
this notation follows directly from the corresponding
fixed-outer-boundary free-oscillation dynamics governed by $M_s \mathrm{d}_t^2 \boldsymbol{\delta}=  \boldsymbol{F} $, where $M_s=(4\pi/3)\rho_s a_s^3$ is the solid-sphere mass and
$\boldsymbol{F}$ is the total force
\cite[as in the fixed-outer-boundary linear formulation of][]{rieutord2002slichter}.
Indeed, a weak frequency perturbation gives
$M_s\epsilon(\omega+\Delta\omega_j)^2
\simeq M_s\epsilon\omega^2(1+2\Delta\omega_j/\omega)$,
so that projection of (\ref{eq:detun}) onto the oscillation direction
yields the first-order complex frequency correction associated with
$\boldsymbol{F}_j$ for weak forces. In the absence of rotation, the
polar-mode forces considered by BG95 are parallel to
$\hat{\boldsymbol{z}}$, such that
$\Delta\boldsymbol{\omega}_j=\Delta\omega_j\hat{\boldsymbol{z}}$;
the force response can then be characterised by the scalar
$\Delta\omega_j$. Within this fixed-outer-boundary free-oscillation
interpretation, the real and imaginary parts of the resulting detuning
$\Delta\omega_j/\omega$ quantify, respectively, the eigenfrequency correction and modal damping-rate contribution, while
$-2(\rho_s/\rho_f)\Real(\Delta\omega_j/\omega)$
gives the associated relative correction to the effective inertia of
the solid domain. For a finite-mass freely translating outer solid, however, $\boldsymbol{\delta}$ is the relative displacement and
$\Delta\boldsymbol{\omega}_j/\omega$ retains its meaning as a
fluid-force response, while its planetary eigenfrequency interpretation requires the global translational dynamics
\citep{grinfeld2005motion,grefflefftz2007fluid}.

Decomposing the viscous force $\boldsymbol{F}_\nu$ and the Lorentz force $\boldsymbol{F}_\eta$ into non-dissipative components $[\boldsymbol{F}_\nu^{nd},\boldsymbol{F}_\eta^{nd}]\propto \mathrm{d}_t^2\boldsymbol{\delta}$ (in phase with acceleration) and dissipative components $[\boldsymbol{F}_\nu^d,\boldsymbol{F}_\eta^d]\propto \mathrm{d}_t\boldsymbol{\delta}$ (in phase with velocity), we can write
\begin{eqnarray}
[\boldsymbol{F}_\nu^d,\boldsymbol{F}_\eta^d] = - \frac{\mathrm{d}_t \boldsymbol{\delta}}{\omega\epsilon} [F_\nu^d,F_\eta^d] 
=-\frac{\mathrm{d}_t\boldsymbol{\delta}}{\omega}\int_S[f_\nu^d,f_\eta^d] \mathrm{e}^{\mathrm{i}m\phi} \mathrm{d}S .
\end{eqnarray}
Time-averages follow by equating the work required to maintain the motion to viscous and Ohmic losses \cite[][]{motz1966magnetohydrodynamic,batchelor1967introduction}. Projecting onto the velocity $\Real(\mathrm{d}_t\boldsymbol{\delta})$ gives
$\omega^{-1}\langle \Real(\mathrm{d}_t\boldsymbol{\delta})^2\rangle\int_S[f_\nu^d,f_\eta^d] \,\mathrm{e}^{\mathrm{i}m\phi} \mathrm{d}S
=\langle[\mathcal{D}_\nu,\mathcal{D}_\eta]\rangle$, where $\langle \Real(\mathrm{d}_t\boldsymbol{\delta})^2\rangle/(\epsilon\omega)^2=\mathcal{E} $ is the kinetic energy factor (\S\ref{sec:phymodel}). Using BLT, these forces are given by \cite[e.g.][]{batchelor1967introduction}
\begin{eqnarray}
 \int_S f_\nu^d \mathrm{d}S &=&   \frac{ P_m}{2 \mathcal{E} \omega} \int_{\Omega_f} \left(\Real \left( \partial_r  \boldsymbol{u}_\parallel \right)\right)^2+\left(\Imag \left( \partial_r  \boldsymbol{u}_\parallel \right)\right)^2\mathrm{d}V \implies {f}_\nu^d = -  \frac{ P_m   {\mathcal{G}_\nu}{(a_s)}}{2 \mathcal{E}\omega}  ,  \label{eq:totPowoh0}  \qquad \\
 \int_S f_\eta^d \mathrm{d}S  &=& \frac{1}{2 \mathcal{E} \omega}   \int_{\Omega_f}  \left(\Real \left( \partial_r  \boldsymbol{b}_\parallel \right)\right)^2+\left(\Imag \left( \partial_r  \boldsymbol{b}_\parallel \right)\right)^2  \mathrm{d}V \implies {f}_\eta^d= - \frac{ {\mathcal{G}_\eta}{(a_s)} }{2 \mathcal{E}\omega}  \label{eq:totPowoh}
\end{eqnarray}
for the (thin) inner fluid boundary layer of velocity amplitude $\omega \epsilon$. Because the boundary-layer perturbations decay exponentially,
the thin-layer limit yields $\int_{\Omega_f}\cdots\,\mathrm dV=\int_S\cdots\,\mathrm dS \int_{a_s}^{\infty}\mathrm dr$, so the radial integration reduces to the opposite of the primitives $(\mathcal{G}_\nu,\mathcal{G}_\eta)$ evaluated at $r=a_s$. The same procedure applies to the outer boundary layer by evaluating the primitive at $r=a_m$. Ohmic dissipation in solid domains and the associated forces follow from analogous expressions using the appropriate volume, magnetic diffusivity, and permeability.

Whereas (\ref{eq:detun}) is based on instantaneous-force amplitudes,
mean dissipation follows from the power balance. We define
\begin{eqnarray}
 \widehat{\mathcal D}_j
 \equiv
 \frac{\langle\mathcal D_j\rangle}{M_s\epsilon^2\omega^3}
 =
 \frac{\mathcal E}{M_s\omega^2}\int_S f_j^d\,\mathrm{e}^{\mathrm{i}m\phi} \mathrm{d}S
 =
 \frac{\mathcal E}{M_s\omega^2}\frac{F_j^d}{\epsilon},
 \qquad j\in\{\nu,\eta\}.
 \label{eq:detun2}
\end{eqnarray}
Since the modal energy is $M_s\mathcal E(\epsilon\omega)^2$,
equations~(\ref{eq:detun}) and~(\ref{eq:detun2}) give, for the
dissipative component of the same force, $\Imag(\Delta\omega_j)/\omega = \widehat{\mathcal D}_j/(2\mathcal E)$.
Thus, the normalised amplitude-damping rate equals
$\widehat{\mathcal D}_j$ for the polar mode and
$\widehat{\mathcal D}_j/2$ for the equatorial modes.

In (\ref{eq:detun})–(\ref{eq:detun2}), surface integration over $S$ is generally cumbersome and, when required, performed numerically. Analytical simplifications follow from the BG95 assumptions: (i) retaining only the homogeneous solution, which is accurate for sufficiently thin boundary layers, and (ii) neglecting spatial variations of boundary-layer thickness, i.e.\ assuming a uniform layer. The latter is exact in specific limits (e.g.\ no rotation with a monopole field at $r=a_s$), and otherwise relies on a characteristic thickness. This approach is hereafter termed \textit{huBL} (homogeneous uniform Boundary Layer).

\section{Non-magnetic basic flows}\label{sec:bulkBr}

The hydrodynamic base flow provides the starting point for the
subsequent MHD boundary-layer analysis. We first derive the potential
flow (\S\ref{sec:InvBulk}) and its associated bulk dissipation
(\S\ref{sec:bulkrot}), before incorporating rotational effects
(\S\ref{sec:InvBulkBusse}).

\subsection{Potential flows, added-mass and weak rotational perturbations} \label{sec:InvBulk}
For $\gamma\ll1$, rotational effects are negligible and the basic flow reduces to an irrotational oscillatory motion. For $\boldsymbol{B}_0=\boldsymbol{0}$, an irrotational bulk flow $\boldsymbol{U}$ can then be sought \cite[with finite viscosity, see][]{buresti2009notes}, so that $\boldsymbol{U}=\nabla \varPhi$ with $\nabla^2 \varPhi=0$. In the unbounded limit $a_s \ll a_m$ used by BG95 \cite[or problem 92 in][]{lamb1924hydrodynamics}, the velocity potential $\varPhi$ can be obtained for arbitrary displacements at small velocities. In the frame of the sphere, the flow reduces to dipolar contributions with internal sources of the form $(k_1 x + k_2 y + k_3 z) r^{-3}$, where the coefficients $k_i$ are determined from the non-penetration condition $(U_r - \mathrm{d}_t \boldsymbol{\delta} \cdot \hat{\boldsymbol{r}})\vert_{r=a_s} = 0$. Extension to bounded domains follows by superposing dipolar terms with external sources, as in Art.~9 of \cite{stokes1843on} or problem 93 of \cite{lamb1924hydrodynamics}. In the frame of the outer boundary at rest, additional terms ${k}_4 x + {k}_5 y + {k}_6 z$ enforce non-penetration at $r=a_m$. The flow $\epsilon \boldsymbol{U} \mathrm{e}^{\mathrm{i}(\omega t + m \phi)} = \epsilon \nabla[\varPhi \mathrm{e}^{\mathrm{i}(\omega t + m \phi)}]$ is given by
\begin{equation}
    [\varPhi_0,\varPhi_\pm]=-\frac{\mathrm{i}}{2} \left( \frac{a_s}{r} \right)^2 \omega a_s \frac{1+2(r/a_m)^3}{1-a^3} [ \cos \theta, \sin \theta ] , \label{syst:Lamb}
\end{equation}
for the polar and equatorial modes ($a=a_s/a_m$). The associated velocity fields are
\begin{equation}
    \boldsymbol{U}= \mathrm{i} \omega \frac{1}{1-a^3} \frac{[a_s^3-(a r)^3] \cos (\theta)\, \hat{\boldsymbol{r}} +[(a r)^3+a_s^3/2] \sin (\theta)\, \hat{\boldsymbol{\theta}}}{r^3},
    \label{eq:LambPolar}
\end{equation}
for the polar mode, while the velocity field of equatorial modes is
\begin{equation}
    \boldsymbol{U}= \mathrm{i} \omega \frac{1}{1-a^3} \frac{[a_s^3-(a r)^3] \sin (\theta)\, \hat{\boldsymbol{r}} -[(a r)^3+a_s^3/2][\cos (\theta)\, \hat{\boldsymbol{\theta}} + \mathrm{i} m \hat{\boldsymbol{\phi}}]}{r^3}.
     \label{eq:LambEquatorial}
\end{equation}
These flows are exact solutions of the viscous Navier--Stokes equation~(\ref{eq:dc0}) in the absence of magnetic and rotational effects \cite[as for any potential flow, see][]{buresti2009notes}.

\begin{figure}
\centering
\begin{subfigure}[b]{0.48\textwidth}
\centering
\includegraphics[width=1\linewidth]{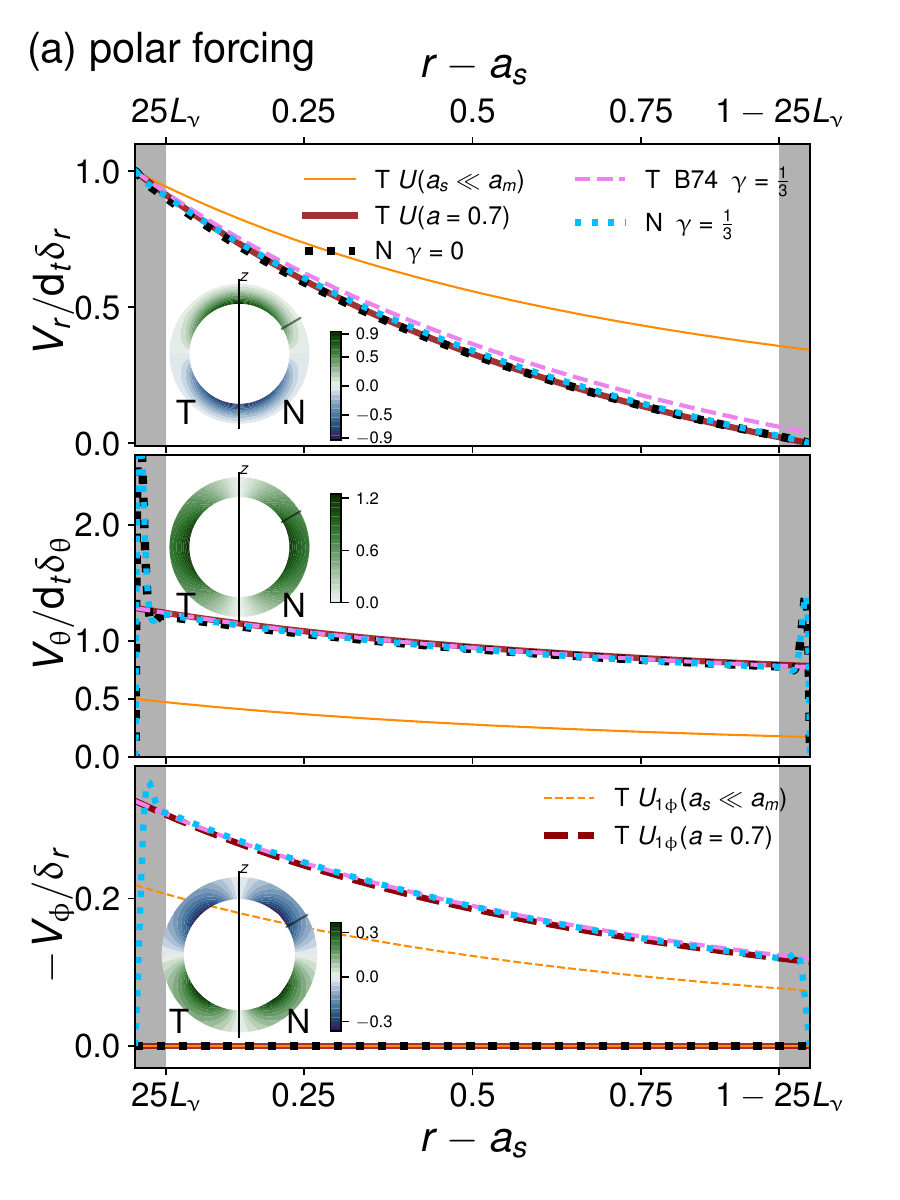}
\end{subfigure}
\begin{subfigure}[b]{0.48\textwidth}
\centering
\includegraphics[width=1\linewidth]{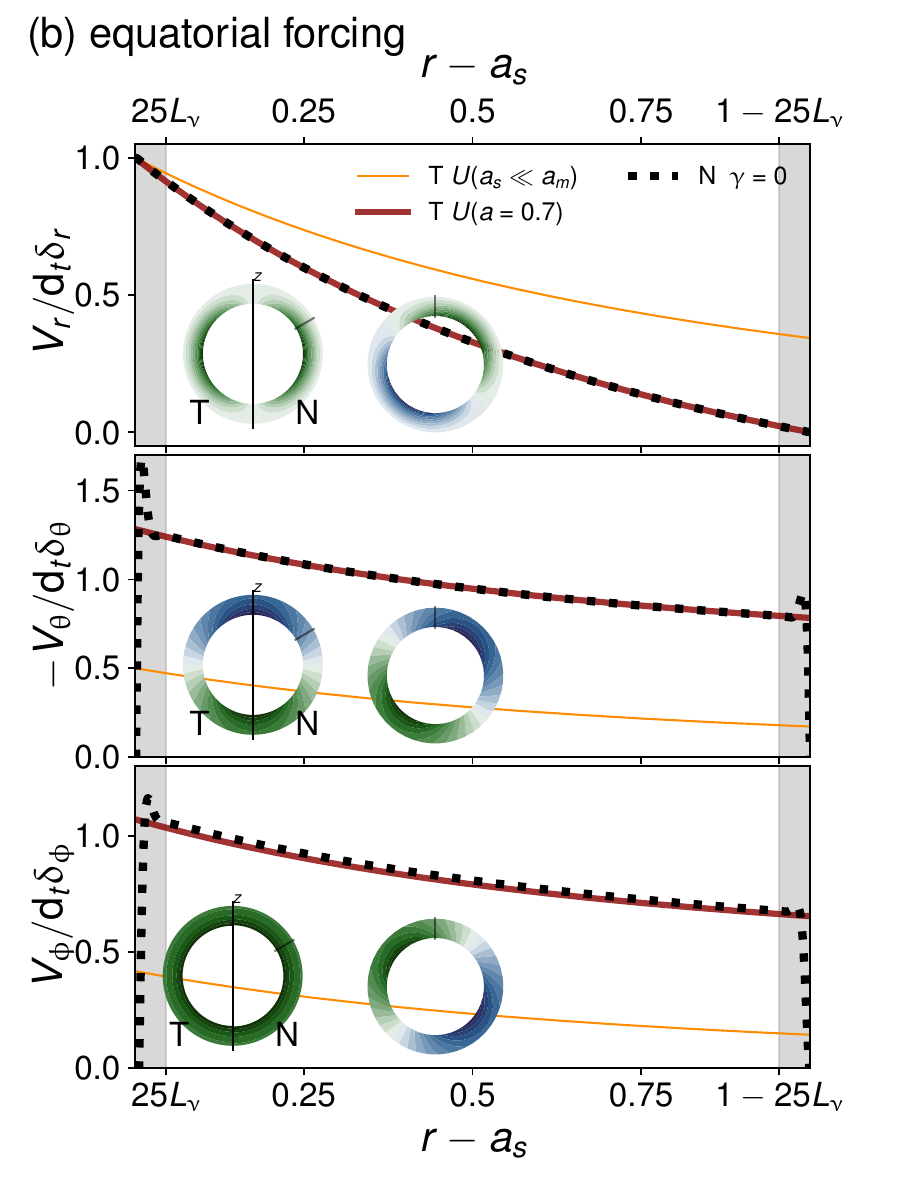}
\end{subfigure}
\caption{Theoretical basic flow (T) vs. \textsc{xshells} (N, dotted, black for $\gamma=0$, and cyan for the polar mode at $\gamma=1/3$) for polar (a) and prograde equatorial (b) forcing ($\theta=60^{\circ}$, $a=0.7$, $L_\nu=6\times10^{-3}$). Potential flow: unbounded (orange) and bounded (red) at $\gamma=0$ (solid, eq.~\ref{eq:LambPolar}-\ref{eq:LambEquatorial}), and rotation perturbation of the polar mode (non-zero azimuthal flow~\ref{syst:Lamb2}, bottom plot of a). B74 solution (pink dashed line, \S\ref{sec:InvBulkBusse}), using $\zeta_m$ from equation~(\ref{syst:zet_est}). Surface plots compare bounded analytical (T) and numerical (N) velocity in a meridional plane (equatorial section of T also shown in b).} 
\label{fig:BaseFlow}
\end{figure}

For all three modes, we have $\nabla^2\boldsymbol{U}=\boldsymbol{0}$,
and the inviscid pressure force follows from the
linearised unsteady Bernoulli equation. With $\partial_t[\epsilon\varPhi\mathrm{e}^{\mathrm{i}(\omega t+m\phi)}]
=\mathrm{i}\omega\epsilon\varPhi
\mathrm{e}^{\mathrm{i}(\omega t+m\phi)}$, we obtain
\begin{equation}
-\frac{\boldsymbol{F}_p}{M_{fs}}
= C_a^i\,\mathrm{d}_t^2\boldsymbol{\delta}
= -\frac{1}{4\pi a_s^3/3}
\int_{\tilde S}
\mathrm{i}\omega\epsilon\varPhi
\mathrm{e}^{\mathrm{i}(\omega t+m\phi)}
\,\mathrm{d}\boldsymbol{S}
= \frac{1}{2}\frac{1+2\tilde a^3}{1-a^3}\,
\mathrm{d}_t^2\boldsymbol{\delta}.
\label{eq:addedm}
\end{equation}
on any spherical surface $\tilde S$ of radius $\tilde a a_m$.
For the polar mode, the acceleration is axial and reverses during each
oscillation, whereas for the equatorial modes it is centripetal, along
$-\hat{\boldsymbol{s}}$. The inner-sphere force is obtained with
$\tilde a=a$; the corresponding radial pressure integral at the outer
boundary is obtained with $\tilde a=1$.

Weak rotational effects can be incorporated perturbatively. At leading order for $\gamma\ll1$, the rotation-induced azimuthal velocity of the polar-mode potential flow follows from the $\phi$ component of the Navier--Stokes equation, yielding $U_\phi/U_s=\mathrm{i}\gamma$ in cylindrical coordinates $(s,\phi,z)$. Perturbing the flow (\ref{syst:Lamb}), the first-order basic flow $\boldsymbol{U}=\boldsymbol{U}_1$ becomes
\begin{subequations}
\begin{equation}
  \boldsymbol{U}_1=\nabla \varPhi+U_{1\phi} \hat{\boldsymbol{\phi}} , \quad  U_{1\phi}=- \frac{3}{4} \frac{(a_s/r)^3}{1-a^3} \gamma \omega  \sin (2 \theta) ,
\tag{\theequation~\emph{a,b}}
\end{equation}
\label{syst:Lamb2}
\end{subequations}
for the polar mode (correcting a typo in equation~10 of BG95). 

Figure~\ref{fig:BaseFlow} compares these base flows with \textsc{xshells} (dotted profiles) for strong confinement ($a=0.7$). Dark grey bands near the boundaries indicate regions where viscous effects are significant, with thickness of $25$ viscous skin depths; these are excluded from the bulk-flow discussion and treated in the boundary-layer analysis. Outside the boundary layers, the irrotational approximation is excellent: the bounded base flow (dark-red; equations~\ref{eq:LambPolar} and~\ref{eq:LambEquatorial} in (a) and (b)) accurately reproduces the \textsc{dns}, remaining accurate for both radial and meridional components even under weak rotation. Conversely, the unbounded flow (orange) used by BG95 fails to capture the solution when confinement effects become significant for $a>0.3$. In the lowest panel of figure~\ref{fig:BaseFlow}a, the rotation-perturbed and bounded solution (\ref{syst:Lamb2}b) (dark-red dashed) agrees well with the \textsc{dns} for $\gamma=1/3$ (cyan dotted).

\begin{figure}
\centering
\begin{subfigure}[b]{0.48\textwidth}
\centering
\includegraphics[width=1\linewidth]{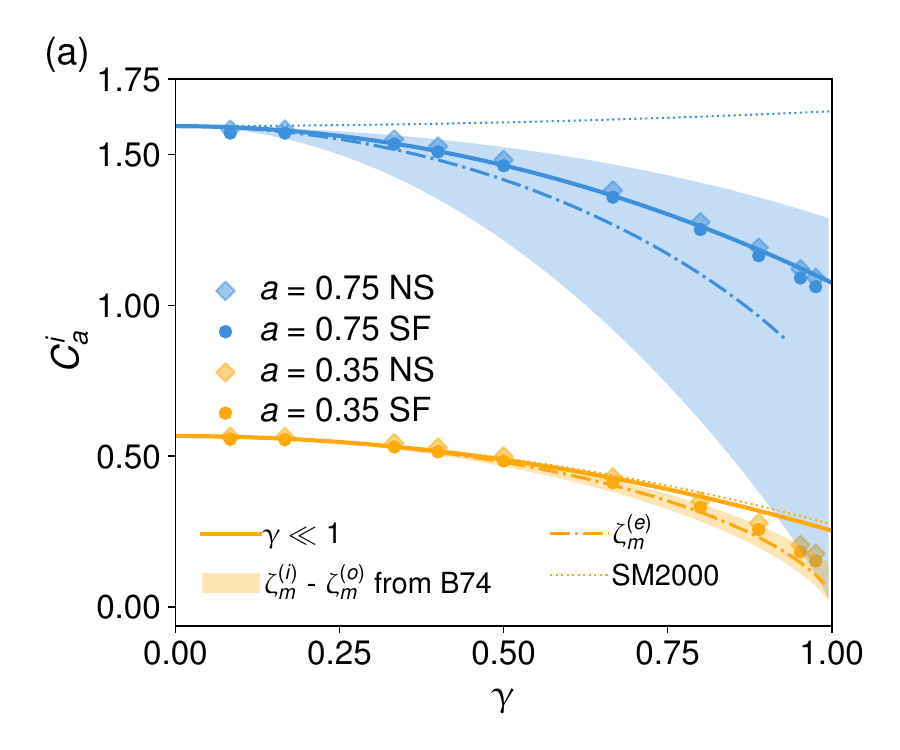}
\end{subfigure}
\begin{subfigure}[b]{0.48\textwidth}
\centering
\includegraphics[width=1\linewidth]{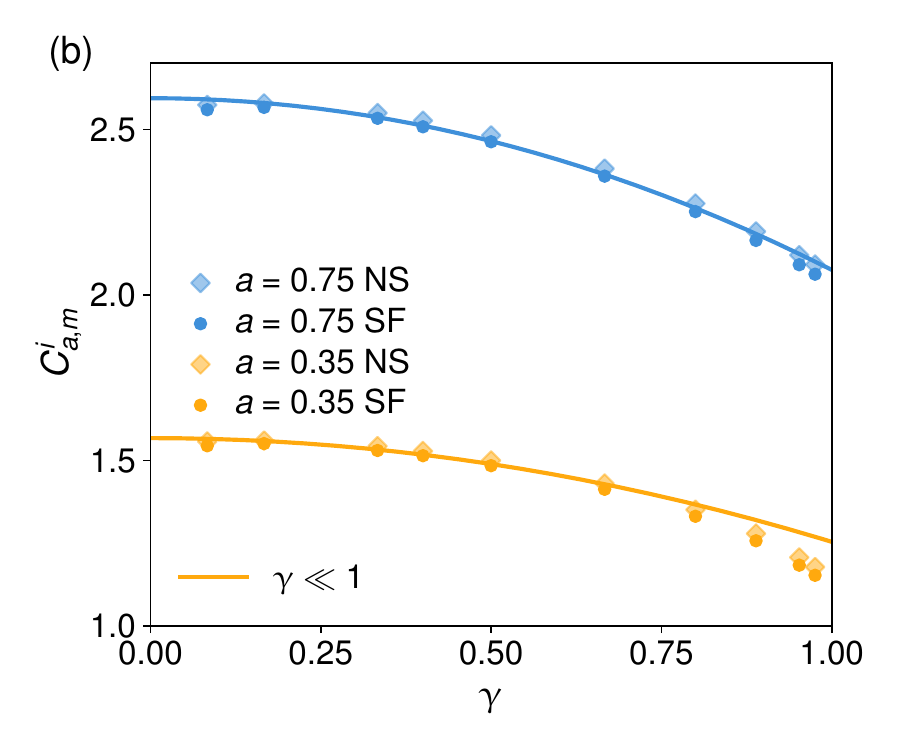}
\end{subfigure}
\caption{Inviscid added-mass coefficient $C_a^i$ at the inner (a) and outer (b) boundaries vs. $\gamma$ for two radius ratios, with no-slip (NS) and stress-free (SF) boundary conditions at $r=a_s$ (stress-free outer boundary) in \textsc{comsol}, at $(L_\nu/(a_m\sqrt{\gamma}))^2=10^{-5}$ and $\epsilon=0.5L_\nu$. Solid lines in (a,b) correspond to  (\ref{eq:OURCa}); dotted lines in (a) follow (\ref{eq:smh}). In (a), the colored area and dash-dotted lines represent the spread of the B74 solution (three estimates of $\zeta_m$). }
\label{fig:xGammayCa}
\end{figure}

Since (\ref{eq:addedm}) applies only to irrotational flows, the rotational correction to the pressure force is evaluated using the kinetic-energy approach \cite[e.g.][equation~6]{busse1974free}. Applying this approach to (\ref{syst:Lamb2}) yields the inviscid added-mass coefficient (for the inner sphere)
\begin{align}
C_a^i=\frac{1}{2} \frac{1+2a^3}{1-a^3} - \frac{3}{10} \frac{\gamma^2}{1-a^3} + \mathcal{O}(\gamma^4),
\label{eq:OURCa}
\end{align}
for the polar mode. The $\gamma^2$ correction originates from the centrifugal pressure associated with the azimuthal velocity $U_\phi$ and can equally be derived directly from the pressure field. Equation~(\ref{eq:OURCa}) accurately reproduces the rotational dependence of $C_a^i$ at small $\gamma$, in excellent agreement with the \textsc{comsol} \textsc{dns} (fig.~\ref{fig:xGammayCa}).

\subsection{Bulk dissipation of the rotation-perturbed potential flow} \label{sec:bulkrot}
In the absence of rotation, the real part of the flow $\nabla \varPhi$ can be used to calculate the viscous dissipation $\mathcal{D}_\nu$, that reduces here to $\mathcal{D}_\nu= 2 \rho_f \nu  \int_{\partial {\Omega_f}} \hat{\boldsymbol{n}} \cdot  \nabla( \boldsymbol{V}^2/2)  \mathrm{d}S$ (see \ref{eq:ensDnu0}-\ref{eq:ensDnu}),
 \begin{eqnarray}
 \frac{  \mathcal{D}_\nu}{P_m }  = 12 \pi  \frac{1-a^5}{(1-a^3)^2} a_s (\epsilon \omega)^2 (f_s \sin^2 \omega t + f_c \cos^2 \omega t ) \label{eq:sfPoJ0}
 \end{eqnarray}
with $(f_s,f_c)=(1,0)$ for the polar mode, and $(f_s,f_c)=(1,1)$ for the equatorial modes, and where $P_m$ comes from our unit choice (to be replaced by $\rho_f \nu$ in dimensional units). Considering the base flow $\boldsymbol{U}_1$ of equation~(\ref{syst:Lamb2}), equations (\ref{eq:ensDnu0})-(\ref{eq:ensDnu}) give the leading order rotation-induced correction $(f_s,f_c)=(1,2 \gamma^2/5)$ for the polar mode. As the $f_c$ correction is of order $\gamma^2$, we can wonder if higher-order $\gamma^2$ corrections of the basic flow could modify this result.
Equation~(\ref{eq:sfPoJ0}) gives 
  \begin{eqnarray}
F_\nu^d=
\frac{6\pi}{\mathcal E}
\frac{1-a^5}{(1-a^3)^2}
P_m a_s(f_s+f_c)\epsilon\omega,
\qquad
\widehat{\mathcal D}_\nu=
\frac{9}{4}\frac{\rho_f}{\rho_s}
\frac{L_\nu^2}{a_s^2}
\frac{1-a^5}{(1-a^3)^2}(f_s+f_c) , \label{eq:sfPoJ}
 \end{eqnarray}
 recovering the \cite{levich1949motion} drag, given by equation
(\ref{eq:FSto2}), and extending it to spherical confinement. We have successfully validated this prediction
of the bulk dissipation geometric dependence against \textsc{comsol}
\textsc{dns} (fig.~\ref{fig:x_Gamma_y_Fvratio}b in the Appendix).

This bulk-flow dissipation can be further enhanced by viscous boundary layers imposed by the boundary conditions. For a rapidly oscillating no-slip inner boundary and a no-slip outer boundary at rest, the boundary-layer contribution is given by (\ref{eq:Stob2}); its leading term, proportional to $a_s/L_\nu$, exceeds the bulk force (\ref{eq:sfPoJ}) when $a_s/L_\nu>2(1-a^5)/(1+a^4)$, i.e.\ for sufficiently thin boundary layers (the $a^4$ term is the outer boundary layer contribution). Comparing (\ref{eq:sfPoJ}) with the $L_\nu$-independent term of (\ref{eq:Stob2}) in the unbounded limit $a=0$ further shows that the viscous layer at the no-slip inner boundary reduces the contribution of this order to the force by a factor $1/2$.

\subsection{Large rotation effects: basic flow and added-mass coefficient}\label{sec:InvBulkBusse}
 
The regime of arbitrary rotation, i.e. $\gamma$, is analytically challenging, notably because inertial modes may be directly forced when $\gamma>1$. In the planetary-relevant range $\gamma<1$, the polar mode was analysed by B74, while all three modes were considered by \cite{smylie1998viscous,smylie2000the}. These studies include confinement by the outer boundary, and \cite{smylie1998viscous,smylie2000the} additionally account for the associated Stokes--Ekman layers. Their results relevant to this work are summarised in appendix~\ref{sec:appSmyli}. We outline below the formulation of B74 used to analyse rotational effects. Several typographical errors and implicit relations were identified in B74 in the course of our derivation; we provide explicit corrected forms below.

Using cylindrical coordinates and a stream function $\psi(s,z)$, the axisymmetric basic flow $\boldsymbol{U}=\boldsymbol{U}_{\mathrm{B74}}$ of B74 is written as $[U_s,U_\phi,U_z]=[s^{-1}\partial_z\psi,\,\mathrm{i}\gamma s^{-1}\partial_z\psi,\,-s^{-1}\partial_s\psi]$. Following B74, the stream function $\psi(s,z)$ of the polar mode reads
\begin{subequations}
\label{syst:psi}
\begin{equation}
\psi=\frac{k}{2}(1-\varsigma^2)(\zeta^2-1)[C_1 \mathcal{H}(\zeta)+C_2 ] , \qquad    \mathcal{H}(\zeta)= \frac{1}{2} \ln \left( \frac{\zeta+1}{\zeta-1} \right) -\frac{\zeta}{\zeta^2-1},
\tag{\theequation~\emph{a,b}}
\end{equation}
\end{subequations}
and the inviscid added-mass coefficient is
\begin{eqnarray}
C_a^i=\frac{\mathcal{H}(\zeta_s)-\mathcal{H}(\zeta_m)+\zeta_s^{-1}+\zeta_s(\zeta_s^2-1)^{-1}}{\mathcal{H}(\zeta_m)-\mathcal{H}(\zeta_s)}   , \label{eq:Ca_Busse}
\end{eqnarray}
where, correcting a sign error in B74's expression for $C_2$ (the correct sign is required to satisfy non-penetration at $\zeta = \zeta_m$),
\begin{subequations}
\label{syst:psi2}
\begin{equation}
C_1=\mathrm{i} \omega k [\mathcal{H}(\zeta_m)-\mathcal{H}(\zeta_s)]^{-1} , \qquad    C_2= -\mathrm{i} \omega k [1-\mathcal{H}(\zeta_s)/\mathcal{H}(\zeta_m)]^{-1},
\tag{\theequation~\emph{a,b}}
\end{equation}
\end{subequations}
with $\zeta_s=\gamma^{-1}$, $k=\gamma a_s (1-\gamma^2)^{-1/2}$, and where the ellipsoidal coordinates $(\varsigma,\zeta)$ are related to $(s,z)$ through $s=k\sqrt{(1-\varsigma^2)(\zeta^2-1)}$ and $z=k \varsigma \zeta \sqrt{1-\gamma^2}$. Here, $\zeta_s$ and $\zeta_m$ are the values of $\zeta$ at the inner and outer boundaries, respectively. Using these latter equations, four solutions are found for $(\varsigma,\zeta)$ in function of $(s,z)$, and the solution
\begin{subequations}
\label{syst:muzeta}
\begin{equation}
\varsigma= \frac{\varpi_+^{1/2}-\varpi_-^{1/2}}{2k}, \quad    \zeta= \frac{2z (1-\gamma^2)^{-1/2}}{\varpi_+^{1/2}-\varpi_-^{1/2}}, \quad \varpi_\pm= \left(\frac{z}{\sqrt{1-\gamma^2}} \pm k \right)^2+s^2
\tag{\theequation~\emph{a,b,c}}
\end{equation}
\end{subequations}
is found to recover the solution (\ref{eq:LambPolar}) of the non-rotating limit. The approach of B74 can only approximate the non-penetration condition at the outer boundary. 
The value of $\zeta=\zeta_m$ at the outer boundary can be bounded from below and above by $\zeta_m^{(i)}$ and $\zeta_m^{(o)}$, which are given by (correcting a typo in equation~16b of B74)
\begin{subequations}
\label{syst:zet_bd}
\begin{equation}
\zeta_m^{(i)}=\sqrt{1+(a_m/k)^2} , \qquad \zeta_m^{(o)}=(1-\gamma^2)^{-1/2} \, a_m/k ,
\tag{\theequation~\emph{a,b}}
\end{equation}
\end{subequations}
such that $[\zeta_m^{(i)}/\zeta_m^{(o)}]^2=1-\gamma^2(1-a^2) $. For small $\gamma^2(1-a^2) \ll 1$, the theory is thus expected to be accurate. An intermediate relevant estimate is given by equation~(15) of B74. Integrating the associated cubic polynomial, we obtain 
\begin{subequations}
\label{syst:zet_est}
\begin{equation}
 \zeta_m=\zeta_m^{(e)}= \frac{(12\mathcal{A} + 12 \sqrt{\mathcal{A}^2 - 12})^{2/3} + 12}{6(12\mathcal{A} + 12\sqrt{\mathcal{A}^2 - 12})^{1/3}} , \qquad \mathcal{A}=9 (a_m/k)^3(1-\gamma^2)^{-1/2} \ , 
\tag{\theequation~\emph{a,b}}
\end{equation}
\end{subequations}
which provides an explicit solution for $\zeta_m$. In the rapid-oscillation limit, all three estimates of $\zeta_m$ reduce the added-mass coefficient (\ref{eq:Ca_Busse}) to (\ref{eq:addedm}).

Figure~\ref{fig:xGammayCa}a compares the B74 prediction with the inviscid added-mass coefficient computed using \textsc{comsol}; the \textsc{dns} (dots) lie within the coloured bounds (\ref{syst:zet_bd}). While (\ref{eq:Ca_Busse}) computed with (\ref{syst:zet_est}) and (\ref{eq:OURCa}) coincide for $\gamma\ll1$, the former agrees better at moderate confinement ($a=0.35$), whereas (\ref{eq:OURCa}) is more accurate at strong confinement ($a=0.75$) owing to exact enforcement of non-penetration at $r=a_m$.

Higher-order corrections to the basic flow $\boldsymbol{U}_1$ of (\ref{syst:Lamb2}) can also be derived. In particular, the next-order contribution $\boldsymbol{U}_2$ (in $\gamma^2$, appendix~\ref{sec:Bus22}) is obtained explicitly. Although the three estimates of $\zeta_m$ yield different higher-order cross-terms in $a$ and $\gamma$, they coincide in the unbounded rotating limit ($a\ll1$), giving the same basic flow and added-mass coefficient. For example, $\zeta_m=\zeta_m^{(o)}$ yields, at order $\mathcal{O}(\gamma^4+a^5)$, 
\begin{eqnarray}
    \boldsymbol{U}_2^{a \to 0} &=& \boldsymbol{U}_1+ 3  \omega \left(\frac{\mathrm{i}}{5} f_r \cos \theta \hat{\boldsymbol{r}}+ \mathrm{i}  f_\theta \sin \theta \hat{\boldsymbol{\theta}}+\frac{(f_\phi+3a^3) \gamma \sin 2 \theta}{10} \hat{\boldsymbol{\phi}} \right) \, \frac{a_s^3}{r^3}\gamma^2 , \qquad \label{eq:Busse_ordre2} \\
    f_r&=&(5 \cos^2 \theta -3)(a_s^2 r^{-2}-1)(1+a^3) +2a^3(r^3a_s^{-3}-1) , \\
f_\theta &=& (\cos^2 \theta-1/5) (3a_s^2 r^{-2}-1)(1+a^3)/4-a^3(2r^3a_s^{-3}+1)/5 , \\
f_\phi&=&-[13-15(a_s/r)^2+(35a_s^2/r^2-25) \cos^2 \theta](1+a^3)/4 , \label{eq:fpjhi}
\end{eqnarray}
where $\epsilon \boldsymbol{U}_1=\epsilon (U_{1\phi} \hat{\boldsymbol{\phi}}+\nabla \varPhi)$ is given by (\ref{syst:Lamb})-(\ref{syst:Lamb2}), and where the unbounded rotation-corrected flow $\boldsymbol{U}_2(a = 0)$ is obtained by putting $a=0$ in equations (\ref{eq:Busse_ordre2})-(\ref{eq:fpjhi}).
At $\mathcal{O}(\gamma^3)$, this flow confirms the leading order rotation-induced correction of the viscous force and dissipation proposed in section \ref{sec:bulkrot}, and the $\mathcal{O}(\gamma^4)$ series expansion of the flow gives $f_s=1+4 \gamma^4/25$, and $f_c=2 \gamma^2/5+4\gamma^4/175$. 

In an unbounded rotating fluid, the added-mass coefficient (\ref{eq:Ca_Busse}) is found to be exactly
\begin{eqnarray}
    C_a^i&=&\frac{(\gamma^2-1)(2\gamma -\gamma_0)}{2\gamma + \gamma_0(\gamma^2-1)} , \label{eq:Busse_ordre2_Ca}
\end{eqnarray}
where $\gamma_0=\ln [(1+\gamma)/(1-\gamma)]$, providing a compact exact expression for the added-mass coefficient of a sphere oscillating along the rotation axis in an inviscid fluid (see the comparison with \cite{smylie2000the} and \textsc{dns} in appendix~\ref{sec:appSmyli}). For weak viscous and rotational effects, a perturbative approach can be used in which the viscous corrections to Stokes' result (\ref{eq:Stob1}) are combined with the rotational terms (from equation~\ref{eq:Busse_ordre2_Ca} in the unbounded limit, or from appendix~\ref{sec:appSmyli} in the bounded case). This approach is valid only up to order $\mathcal{O}(\gamma^2 L_\nu /a_s)$, since cross terms may arise at higher order.

\begin{figure}
\centering
\begin{subfigure}[b]{0.49\textwidth}
\centering
\includegraphics[width=1\linewidth]{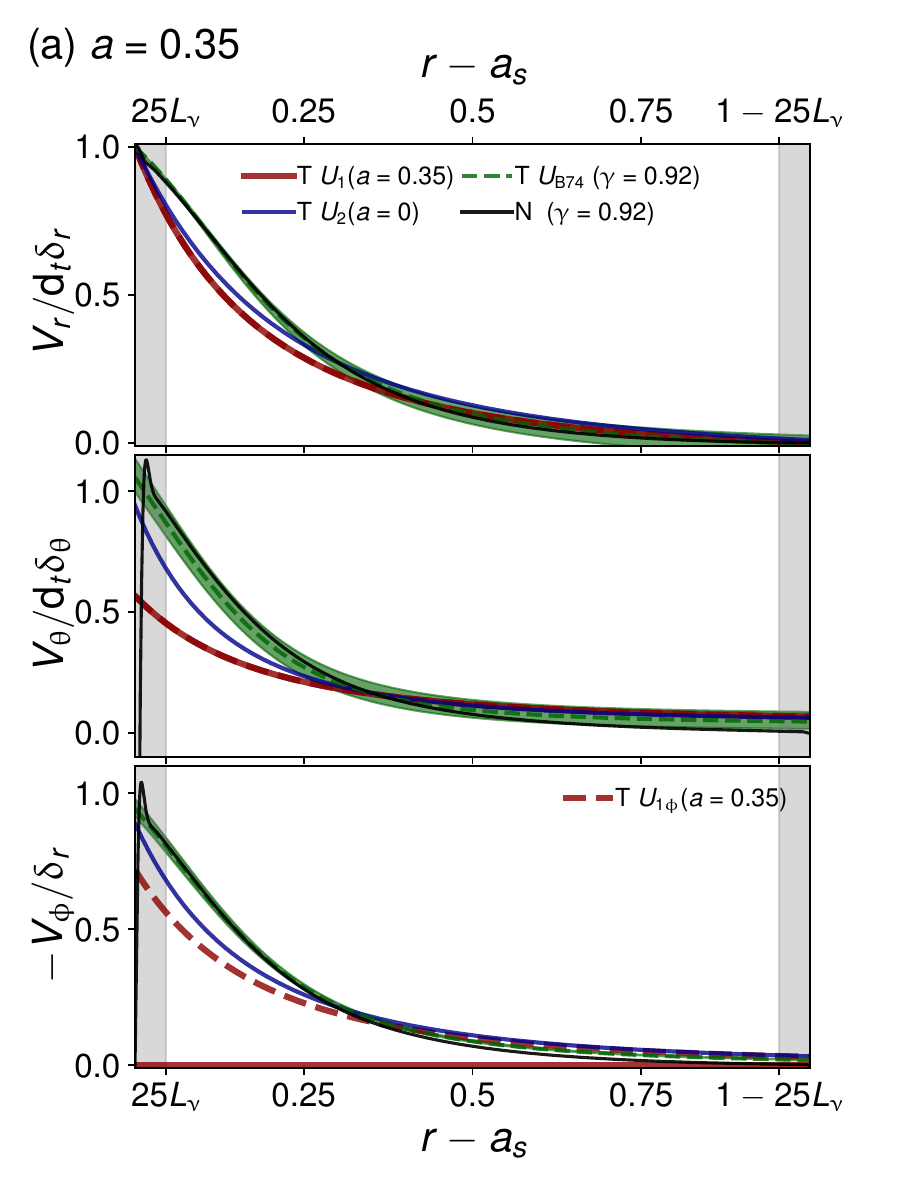}
\end{subfigure}
\begin{subfigure}[b]{0.49\textwidth}
\centering
\includegraphics[width=1\linewidth]{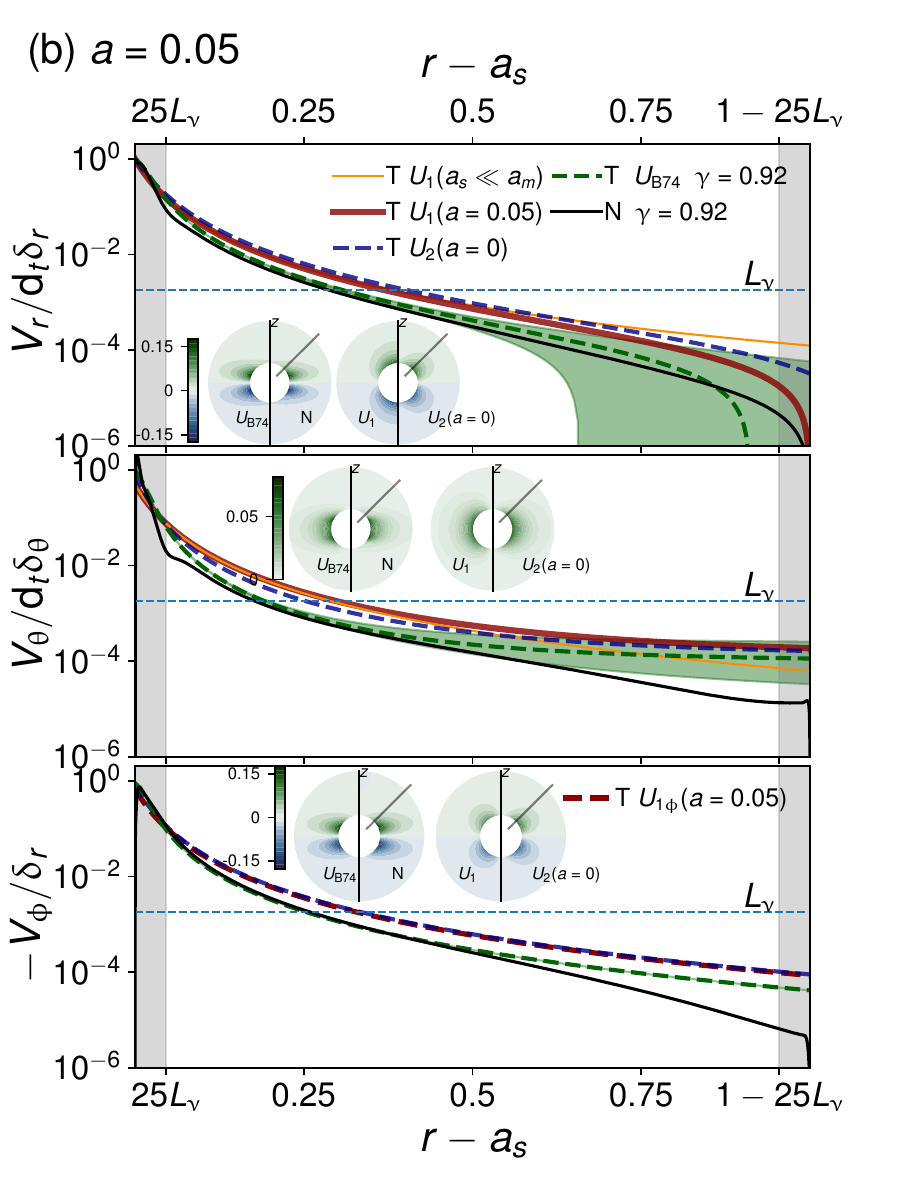}
\end{subfigure}
\caption{Theoretical (T) basic flow vs. \textsc{xshells} (N, black) for strong rotation ($\gamma=0.92$), at $a=0.35$ (left) and $a=0.05$ (right), polar mode ($\theta=45^{\circ}$, $L_\nu=0.002$). Potential flow: unbounded case (orange solid, eq.~\ref{eq:LambPolar} with $a=0$), bounded case at $\gamma=0$ (dark-red solid, eq.~\ref{eq:LambPolar}) and rotation perturbation (azimuthal flow~\ref{syst:Lamb2}, dark-red dashed). B74 flow (\ref{syst:psi}): $\zeta_m$ from (\ref{syst:zet_est}) (dashed), bounds (\ref{syst:zet_bd}) shown as a shaded region, and Taylor expansion (\ref{eq:Busse_ordre2}) for $a=0$ (blue). Grey side bands: viscous boundary layers. In (b), surface plots compare the B74 solution $\boldsymbol{U}_\mathrm{B74}$, the rotation-perturbed potential flow $\boldsymbol{U}_1$ (\ref{syst:Lamb2}), and its correction (\ref{eq:Busse_ordre2}) with the \textsc{xshells} (N) in the meridional plane (zoom to $r=6a_s$).} 
\label{fig:BaseFlowBusse}
\end{figure}

For strong confinement ($a=0.7$) and moderate rotation ($\gamma=1/3$), figure~\ref{fig:BaseFlow} shows that the B74 basic flow (dashed pink line) consistently recovers both the rotation-perturbed solution (\ref{syst:Lamb2}) and the \textsc{xshells} \textsc{dns}. At stronger rotation ($\gamma\sim1$), considered in figure~\ref{fig:BaseFlowBusse}a for $a=0.35$, the B74 solution (dashed), based on (\ref{syst:zet_est}), provides a markedly improved description of the velocity profiles outside the boundary layers (grey bands) compared with the rotation-perturbed potential flow (\ref{syst:Lamb2}) (dark-red). The influence of $\zeta_m$ is illustrated by the shaded region between the bounds in (\ref{syst:zet_bd}); the azimuthal velocity remains only weakly sensitive, with all formulations nearly coincident.

Figure~\ref{fig:BaseFlowBusse}b presents the same comparison for $a=0.05$ on a logarithmic scale, highlighting the spread of solutions. A horizontal dashed line at $L_\nu$ marks the threshold below which the basic flow becomes comparable to the viscously-driven secondary bulk flow $\boldsymbol{\mathcal{U}}$. The B74 solution again shows the best agreement with the \textsc{xshells} \textsc{dns}, whereas alternative models, including (\ref{syst:Lamb2}) (red) and (\ref{eq:Busse_ordre2}) (blue), overestimate the velocity, particularly near the inner boundary. Accurate representation of the bulk flow near $r=a_s$ is thus essential, as it sets the boundary-layer solution where dissipation is concentrated.

The meridional fields in figure~\ref{fig:BaseFlowBusse}b, in particular the comparison between $\boldsymbol{U}_1$ and the \textsc{xshells} (N), provide further evidence of the strong rotational constraint on the bulk flow. This behavior is captured only by the B74 solution $\boldsymbol{U}_\mathrm{B74}$, whereas its weakly bounded Taylor expansion $\boldsymbol{U}_2^{a \to 0}$, given by (\ref{eq:Busse_ordre2}), fails. Even in approximate form, accounting for confinement in B74 is therefore essential. Additional meridional cuts from numerical and analytical models at $\gamma \approx 1$ are provided in Supplementary Material~\ref{Supp:RotationPolar}. To our knowledge, this is the first validation and application of the B74 theoretical flow and added-mass coefficient. The hydrodynamic results of this section provide the reference solution for the magnetohydrodynamic extensions developed below.

\section{Induced radial field and inviscid MHD boundary layers} \label{sec:BTtang}

This section extends the hydrodynamic framework of \S\ref{sec:bulkBr} to inviscid magnetohydrodynamics. Magnetic perturbations generated by the oscillatory flow are first determined for the radial (\S\ref{sec:Br}) and tangential (\S\ref{sec:BGinv}) perturbations, as well as the higher-order bulk corrections (\S\ref{sec:highP}). The resulting magnetic forces and Ohmic dissipation are then derived (\S\ref{sec:maga}-\S\ref{sec:RelDisMKL}).

\subsection{Radial magnetic field: induction by the hydrodynamic base flow}  \label{sec:Br}
Our calculations of $b_r$ are performed in an identical way for the three modes. To illustrate the method, we detail explicitly this calculation below in the particular case of a uniform axial field $\boldsymbol{B}_0=[B_0 \cos \theta,-B_0 \sin \theta,0]^\top$, with the rotation-modified polar mode flow given by equations~(\ref{syst:Lamb})-(\ref{syst:Lamb2}). The evolution of $b_r$ in the fluid is then governed by 
\begin{eqnarray} \label{eq:brr}
   L_\eta^2 \partial_r^2 b_r= 2 \mathrm{i} b_r +3 \mathrm{i} \, B_0 \, a_s^3 r^{-4} (3 \cos^2 \theta -1) (1-a^3)^{-1}, 
\end{eqnarray}
recalling that the magnetic field perturbation is $ \epsilon \Real (\boldsymbol{b} )$. This equation corresponds to \ref{eq:tracBr} multiplied by $2\omega^{-1}$. 
In the diffusionless limit $L_\eta=0$, the solution $b^{(p)}_0$ of (\ref{eq:brr}) is readily obtained (appendix~\ref{sec:diffuWekL}). The induced radial field $b_r=b_r^{(p)}+b_r^{(h)}$ is written as the sum of a particular solution $b_r^{(p)}$ and a homogeneous solution $b_r^{(h)}$ satisfying $ L_\eta^{2} \partial_r^2 b_r=2 \mathrm{i} b_r$. The homogeneous solution gives rise to a magnetic skin layer of typical thickness $L_\eta$, ensuring the electromagnetic interface conditions at $r=a_s$ or $r=a_m$. In the illustrative case, the solution of (\ref{eq:brr}) reads
\begin{eqnarray}
    b_r^{(h)}&=&\tilde{A} \mathrm{e}^{\lambda_{\mathfrak{s}} r}+\tilde{B} \mathrm{e}^{-\lambda_{\mathfrak{s}} r}, \label{eq:solhomoBr}\\
        b_r^{(p)} &=& - \frac{B_0 a_s^3 }{2 L_\eta^4} \frac{3 \cos^2 \theta -1}{1-a^3} (\mathrm{Ei}_1(\lambda_{\mathfrak{s}} r) \mathrm{e}^{\lambda_{\mathfrak{s}} r}+\mathrm{Ei}_1(-\lambda_{\mathfrak{s}} r) \mathrm{e}^{-\lambda_{\mathfrak{s}} r} + 2 (\lambda_{\mathfrak{s}} r)^{-2}) , \label{eq:bpfl}
\end{eqnarray}
with the (complex) prefactors $(\tilde{A},\tilde{B})$, the exponential integral $\mathrm{Ei}_1(x)$, and the radial wavenumber $\lambda_{\mathfrak{s}}=(1+\mathrm{i})/L_\eta$ based on the magnetic skin depth $L_\eta$. Outside the boundary layer, we recover the diffusionless bulk solution (\ref{eq:brBdn1}) with equation~(\ref{eq:bpfl}) by considering the fifth-order series expansion $\mathrm{Ei}_1(x) \approx \mathrm{e}^{-x} (x^{-1}-x^{-2} +2x^{-3}-6x^{-4})$ for large $x$. 

In the inner and outer solid domains, the particular solution is zero for an imposed uniform axial field. The solution $b_r=b_r^{(h)}$ is thus of the form (\ref{eq:solhomoBr}), replacing $L_\eta$ by $\breve{\eta}_j^{1/2} L_\eta$. We enforce that the boundary-layer solution $b_r^{(h)}$ does not diverge when moving away from the boundary. This imposes the solution $b_r^{(h)}(r \leq a_s)$ in the inner solid domain, and the fluid solution $b_r^{(h)}(r \geq a_s)$ near $r=a_s$, to be of the form 
\begin{subequations}
\label{syst:Br_homo}
\begin{equation}
b_r(r\leq a_s) = \tilde{A}_+ \mathrm{e}^{ \lambda_{\mathfrak{s}}  \breve{\eta}_s^{-1/2} \, (r-a_s)} , \quad b_r^{(h)}(r\geq a_s) = \tilde{A}_- \mathrm{e}^{-\lambda_{\mathfrak{s}} (r-a_s)} ,
\tag{\theequation~\emph{a,b}}
\end{equation}
\end{subequations}
with two constants $\tilde{A}_\pm$. Similarly, the fluid solution $b_r^{(h)}(r  \leq a_m)$ near $r=a_m$, and the solution $b_r^{(h)}(r \geq a_m)$ in the outer solid domain, must take the form 
\begin{subequations}
\label{syst:Br_homo1}
\begin{equation}
b_r^{(h)}(r  \leq a_m) = \tilde{B}_+ \mathrm{e}^{\lambda_{\mathfrak{s}}  (r-a_m)} , \quad b_r(r\geq a_m) = \tilde{B}_- \mathrm{e}^{-\lambda_{\mathfrak{s}}  \breve{\eta}_m^{-1/2} \, (r-a_m)} ,
\tag{\theequation~\emph{a,b}}
\end{equation}
\end{subequations} 
with two constants $\tilde{B}_\pm$. Using the interface conditions, the four constants can then be analytically obtained (appendix~\ref{sec:CFbr}).

\begin{figure}
\centering
\begin{subfigure}[b]{0.8\textwidth}
\centering
\includegraphics[width=1\linewidth]{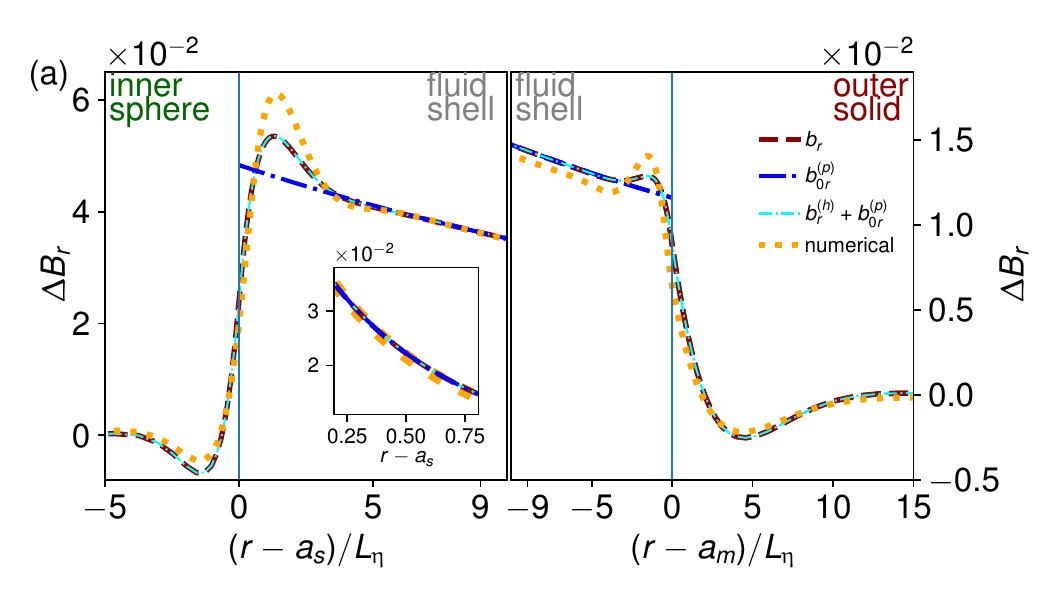}
\end{subfigure}
\begin{subfigure}[b]{1.0\textwidth}
\centering
\includegraphics[width=0.8\linewidth]{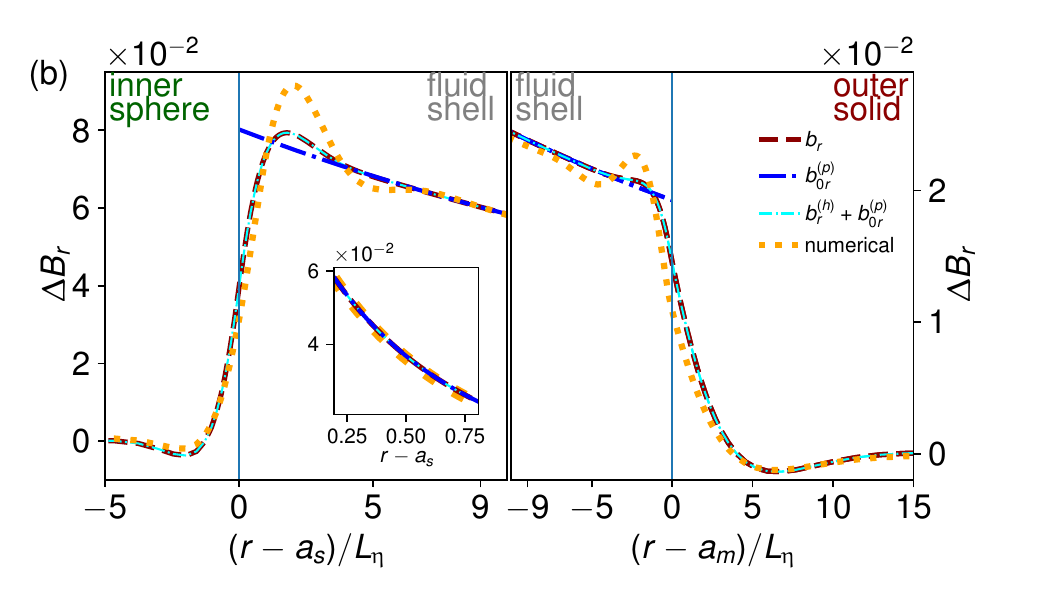} 
\end{subfigure}
\caption{Radial part of the magnetic perturbation $\Delta \boldsymbol{B}=(\boldsymbol{B}-\boldsymbol{B}_0)\epsilon^{-1}$ near the inner (left) and outer (right) boundaries: \textsc{xshells} (orange-dotted), induced field (dark-red dashed), induced field with diffusionless correction (cyan dash-dotted), and diffusionless bulk solution (blue dash-dotted,~\ref{eq:brBdn1}). (a) Polar and (b) equatorial forcing. Parameters:   $\Lambda_z=0.667$, $\Lambda_l=0.5$, $L_\eta=1.9\times 10^{-2}$, $P_m=0.1$, $a=0.7$, $a_e/a_m=1.1$, $\breve{\eta}_s=1, \breve{\eta}_m=10$, $\breve{\mu}_m=\breve{\mu}_s=1$, $\theta=30^{\circ}$, $\phi=0$, polar $\gamma=0.1$, equatorial $\gamma=10^{-3}$. Insets: bulk perturbation.} 
   \label{fig:xRadialMagneticBoundaryLayer}
\end{figure}
Unless stated otherwise, comparisons with \textsc{xshells} \textsc{dns} are performed for a confined configuration ($a = 0.7$), identical magnetic diffusivity in the inner sphere and fluid ($\breve{\eta}_s = 1$), and a solid exterior with $\breve{\eta}_m=10$. Magnetic permeability is uniform across the three domains ($\breve{\mu}_s=\breve{\mu}_m=1$). A uniform axial magnetic field of amplitude $B_0$ is imposed, with $\Lambda_z=B_{0}^2/\omega$. The fluid magnetic skin depth and magnetic Prandtl number are $L_\eta=1.9\times10^{-2}$ and $P_m=10^{-1}$, respectively, ensuring a clear separation of scales. We consider radial profiles at $\theta = 30^\circ$ and $\phi = 0$. The \textsc{dns} are shown in the converged regime at $t=(\tau_s+0.35)T_s$ for polar forcing and $t=(\tau_s+0.45)T_s$ for equatorial forcing, where $\tau_s$ is chosen such that $\tau_s T_s>10P_m^{-1}$, i.e.\ after more than ten viscous time scales.

Figure~\ref{fig:xRadialMagneticBoundaryLayer} shows excellent agreement between the analytical solution for $b_r$ (dark-red dashed line) and the \textsc{dns} (orange-dotted line) for both polar (a) and equatorial (b) modes at weak rotation, with $\gamma=0.1$ and $\gamma=10^{-3}$ respectively; similar agreement is obtained in the non-rotating case (not shown). Here and throughout, comparisons are performed using the magnetic perturbation $\Delta \boldsymbol{B}=(\boldsymbol{B}-\boldsymbol{B}_0)\epsilon^{-1}$. The interior radial field is accurately captured by the diffusionless correction (equation~\ref{eq:brBdn1}; dark-blue dash-dotted line; bottom-left inset). Therefore, it provides an effective particular solution, as shown by the sum of the homogeneous inviscid solution and diffusionless correction (cyan dash-dotted line). The magnetic skin layer extends over a few $L_\eta$ in both the inner sphere ($\breve{\eta}_s=1$) and the fluid, and is thicker in the outer solid ($\breve{\eta}_m=10$). The small discrepancies visible within the fluid boundary layer arise from higher-order radial-field corrections driven by the tangential boundary-layer components. Larger deviations are expected at higher $\gamma$, where magnetic induction by the rotation-modified basic flow yields a different particular solution.

\subsection{Tangential components in the inviscid MHD boundary layer} \label{sec:BGinv}
For rapid oscillations ($\gamma \ll 1$), we extend the study to cases where the Coriolis force becomes negligible while the magnetic induction from the rotation-corrected basic flow is not. This occurs when $\gamma^2$ is small compared with $\Lambda$, introduced by BG95 to estimate the magnetic-to-inertial force ratio (their equation~17). For $\gamma^2 \ll \Lambda$ and an inviscid fluid in a uniform axial field, (\ref{eq:dc3})-(\ref{eq:dc2}) become
\begin{eqnarray}
 B_{0r} \partial_r \boldsymbol{b}_\parallel &=& \mathrm{i} \omega \boldsymbol{u}_\parallel+ \nabla p ,
\label{eq:Lambi1}  \\ 
    L_\eta^2 \partial^2_r \boldsymbol{b}_\parallel
     &=& 2 \mathrm{i}  \boldsymbol{b}_\parallel - \frac{2 B_{0r}}{\omega } \partial_r \boldsymbol{u}_\parallel -\frac{6  B_{0\theta} a_s^3  }{r^4 (1-a^3)}  \begin{bmatrix}
 \mathrm{i}  \cos \theta \\ (1-5 \cos^2 \theta) \gamma / 2 
  \end{bmatrix}  ,  \label{eq:Lambi2}
\end{eqnarray}
using the rotation-modified polar base flow (\ref{syst:Lamb})-(\ref{syst:Lamb2}) and magnetic skin depth $L_\eta$. From (\ref{eq:Lambi2}), one may test whether $B_{0r}\partial_r\boldsymbol{U}$ dominates the forcing $\mathcal{T}$ in (\ref{eq:inhoTT}). Neglecting rotation gives $\mathcal{T}=-\boldsymbol{B}_0(\partial_r U_r)$, since $B_{0r}\partial_r\boldsymbol{U}=-\mathcal{T}/2$ is cancelled by $r^{-1}(B_{0r}\boldsymbol{U}-U_r\boldsymbol{B}_0)$.

Although a diffusionless solution of (\ref{eq:Lambi1})-(\ref{eq:Lambi2}) can be derived for weak Lorentz forces (appendix~\ref{sec:diffuWekL}), the general solution is written as $[\boldsymbol{u}_\parallel,\boldsymbol{b}_\parallel]=[\boldsymbol{u}_\parallel^{(p)},\boldsymbol{b}_\parallel^{(p)}]+[\boldsymbol{u}_\parallel^{(h)},\boldsymbol{b}_\parallel^{(h)}]$, i.e.\ the sum of a particular solution and the homogeneous solution. The latter, obtained by discarding the last term in (\ref{eq:Lambi2}), is required to satisfy the boundary conditions. Integration of (\ref{eq:Lambi1})-(\ref{eq:Lambi2}) then yields lengthy expressions for $[\boldsymbol{u}_\parallel^{(p)},\boldsymbol{b}_\parallel^{(p)}]$ in the inviscid fluid, e.g.
\begin{eqnarray}
b_\theta^{(p)} = \frac{L_\eta^2r^{-2}(\Lambda_l+\mathrm{i})+ (\mathrm{i} \pi  -2\, \mathrm{Shi}(\lambda r)) \mathrm{e}^{\lambda r}-(\mathrm{e}^{\lambda r}+\mathrm{e}^{-\lambda r}) \, \mathrm{Ei}_1(-\lambda r)}{B_{0r}^{-1} a_s^{-3} (1-a^3)(1-\mathrm{i} \Lambda_l)^2 L_\eta^4} \,  \sin \theta
\end{eqnarray}
where the radial wavenumber $\lambda=\lambda_{\mathfrak{s}}(1-\mathrm{i} \Lambda_l)^{-1/2}$ reduces to the magnetic skin depth $\lambda_{\mathfrak{s}}$ for $\Lambda_l \ll 1$. Here $\mathrm{Shi}(x)=\int_0^x t^{-1}\sinh(t)\,\mathrm{d}t$ denotes the hyperbolic sine integral, and $\Lambda_l=B_{0r}^2/\omega=\sigma_f \tilde{B}_{0r}^2/( \rho_f \omegas)$ is the local parameter based on $B_{0r}$ (dimensional counterpart $\tilde{B}_{0r}$). This parameter also appears in the BG95 model, where it governs the transition from a skin depth regime to one dominated by Alfv\'en-wave radiation. For $\Lambda_l \ll 1$, the diffusionless approximation $[\boldsymbol{u}^{(p)},\boldsymbol{b}^{(p)}]\approx[\boldsymbol{u}^{(p)}_0,\boldsymbol{b}^{(p)}_0]$ follows as simple analytical expressions for weak Lorentz forces (appendix~\ref{appA}).

\begin{figure}
\centering
\begin{subfigure}[b]{0.9\textwidth}
\centering
\includegraphics[width=0.8\linewidth]{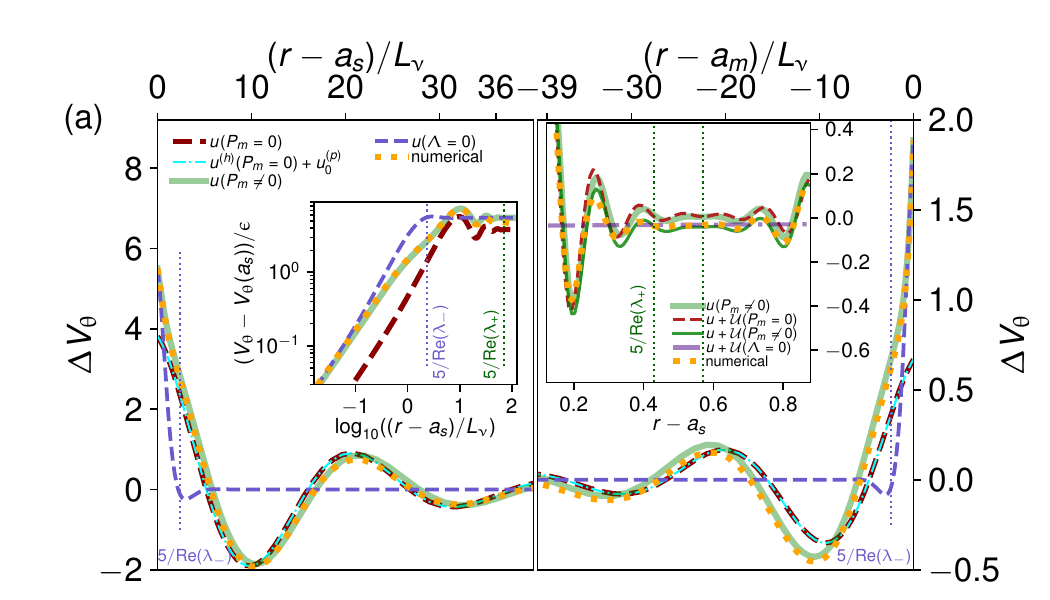}
\end{subfigure}
\begin{subfigure}[b]{0.9\textwidth}
\centering
\includegraphics[width=0.8\linewidth]{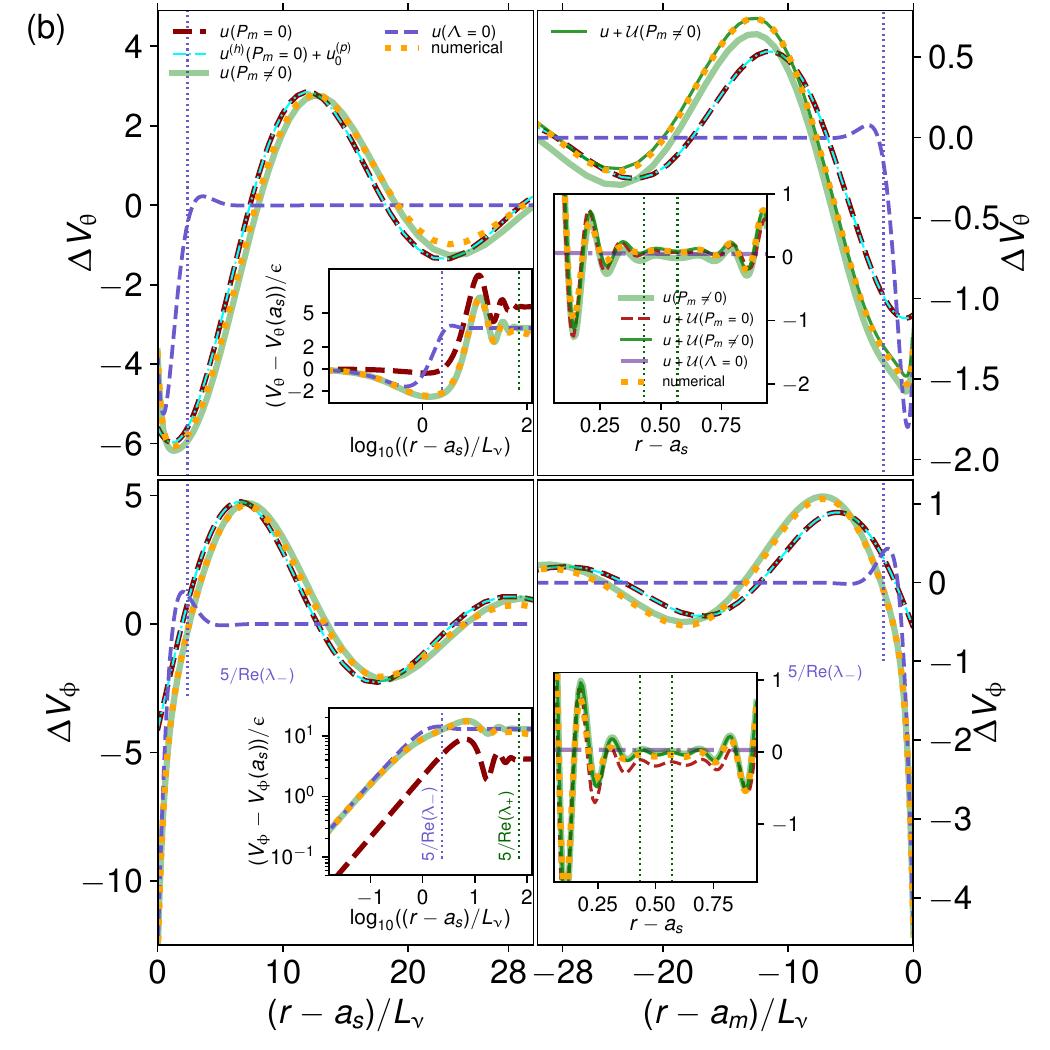}
\end{subfigure}
\caption{Tangential components of flow perturbation $\Delta \boldsymbol{V}=\boldsymbol{V}\epsilon^{-1}-\Real(\boldsymbol{U}_1\exp{(\mathrm{i}(\omega t+m\phi))})$ near the inner (left) and outer (right) boundaries: \textsc{xshells} (orange-dotted), inviscid MHD (dark-red dashed), inviscid MHD with diffusionless correction (cyan dash-dotted), viscomagnetic oscillatory (thick green), and purely viscous (purple dashed) solutions. Insets show bulk perturbations, including secondary bulk flows: inviscid MHD (red dashed), viscomagnetic oscillatory (dark-green), and purely viscous (purple dash-dotted); viscomagnetic oscillatory boundary layer (thick green). (a) Polar and (b) equatorial forcing. Parameters: as in figure~\ref{fig:xRadialMagneticBoundaryLayer}, but   $\Lambda_z=2.67$, $\Lambda_l=2$, $\gamma=0$.}
   \label{fig:xRyThetaPhiVelocityBLViscoMagWaves}
\end{figure}

From the homogeneous part of (\ref{eq:Lambi1})-(\ref{eq:Lambi2}), the solutions $\boldsymbol{b}_\parallel^{(h)}$ and $\boldsymbol{u}_\parallel^{(h)}$ in the fluid are
\begin{subequations}
\label{syst:hom_mag1}
\begin{equation}
\boldsymbol{b}_\parallel^{(h)}=\begin{bmatrix} A_\pm \\ B_\pm \end{bmatrix}   \mathrm{e}^{\pm \lambda \tilde{r}_\pm} , \, \boldsymbol{u}_\parallel^{(h)}= \pm \frac{ \omega \Lambda_l \lambda}{2 \mathrm{i} B_{0r}}  L_\eta^2 \boldsymbol{b}_\parallel^{(h)} , \, \boldsymbol{b}_\parallel^{(s)}= \begin{bmatrix} C_\pm \\ D_\pm \end{bmatrix}  \mathrm{e}^{\mp \breve{\eta}^{-1/2} \, \lambda_{\mathfrak{s}} \tilde{r}_\pm} ,
\tag{\theequation~\emph{a,b,c}}
\end{equation}
\end{subequations}
discarding exponentially growing terms, where $\boldsymbol{b}_\parallel^{(s)}$ is the solution in the solids. At $r\ge a_s$, the fluid solution is governed by $(A_-,B_-)$ with $\tilde r_-=r-a_s$, while it is described by $(A_+,B_+)$ at $r\le a_m$, with $\tilde r_+=r-a_m$. In the inner and outer solids, $\boldsymbol{b}_\parallel^{(s)}$ involves $(C_-,D_-)$ and $(C_+,D_+)$, respectively. The boundary-layer thicknesses in the solids are $\breve{\eta}_s^{1/2}\lambda_{\mathfrak{s}}^{-1}$ and $\breve{\eta}_m^{1/2}\lambda_{\mathfrak{s}}^{-1}$, while $\lambda=\lambda_{\mathfrak{s}}(1-\mathrm{i}\Lambda_l)^{-1/2}$ sets the fluid thickness $\lambda^{-1}\breve{\eta}^{1/2}$; the flow contribution vanishes as $\Lambda_l\to0$. For $\Lambda_l\gg1$, $\lambda$ becomes $\lambda \approx \sqrt{2}(\Lambda_l^{-3/2}/2+\mathrm{i}\Lambda_l^{-1/2})L_\eta^{-1}$, which recovers the Alfv\'en phase velocity $\omega/\Imag(\lambda)=\tilde{B}_{0r}/\sqrt{\rho_f\mu_f}=V_A$ and an attenuation length $L_A=\Real(\lambda)^{-1}=\sqrt{2}L_\eta\Lambda_l^{3/2}\gg L_\eta$, much larger than the skin depth $L_\eta$ (as noted by BG95). The inviscid MHD solution therefore transitions from a magnetic skin layer at $\Lambda_l\ll1$ to an attenuated Alfv\'en-wave regime. Our BLT approach requires $L_A$ small compared with the available propagation distance $D$, and absence of significant reflection, or boundary-layer interaction, on the opposite boundary (and thus absence of global resonance).

\begin{figure}
\centering
\begin{subfigure}[b]{0.8\textwidth}
\centering
\includegraphics[width=0.8\linewidth]{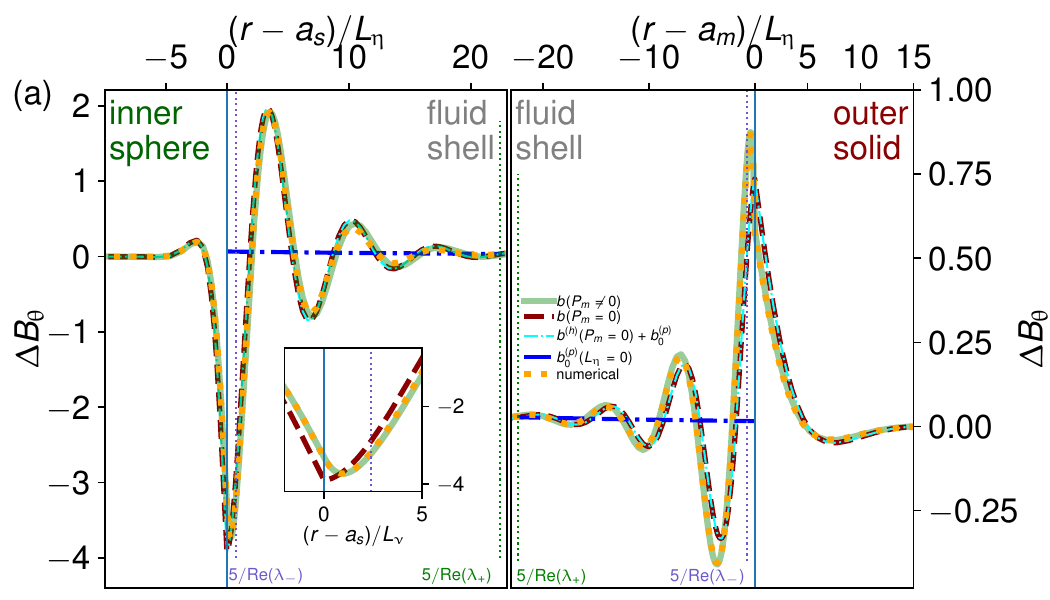}
\end{subfigure}
\begin{subfigure}[b]{0.8\textwidth}
\centering
\includegraphics[width=0.9\linewidth]{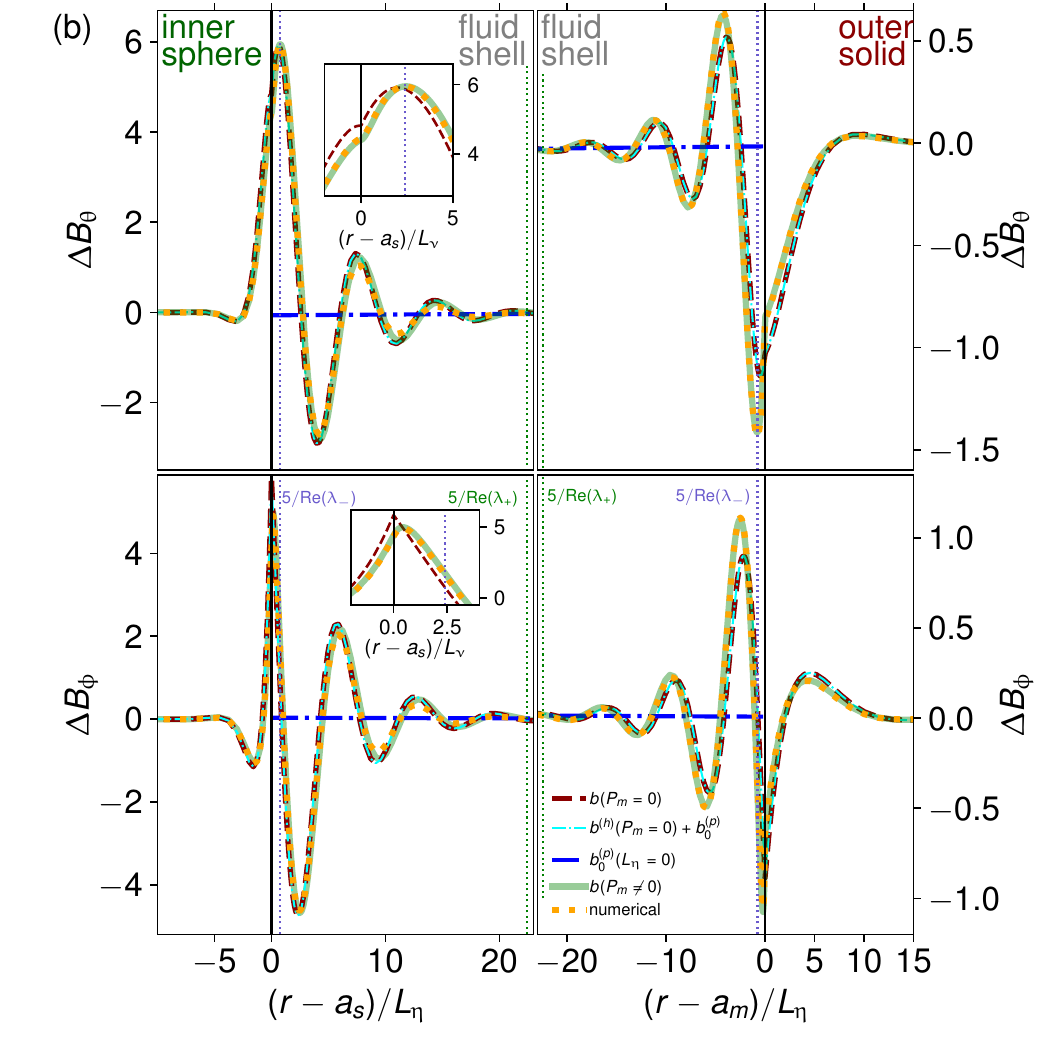}
\end{subfigure}
\caption{Tangential magnetic field perturbation near the inner (left) and outer (right) boundaries:  \textsc{xshells} (orange-dotted), inviscid MHD (dark-red dashed), inviscid MHD with diffusionless correction (cyan dash-dotted), viscomagnetic oscillatory (thick green), and purely viscous (purple dashed). Insets on the left show profile close to the interfaces. (a) Polar and (b) equatorial forcing. Parameters: as in figure~\ref{fig:xRadialMagneticBoundaryLayer}, but $\Lambda_z=2.67$, $\Lambda_l=2$, $\gamma=0$.}
   \label{fig:xRyThetaPhiMagneticBLViscoMagWaves}
\end{figure} 

The prefactors of (\ref{syst:hom_mag1}) are analytically obtained from the interface conditions (appendix~\ref{sec:CFbinv}); their complete symbolic expressions are available in the accompanying Zenodo repository \cite{zenodo}. These inviscid MHD boundary-layer solutions (including the particular contribution) are shown as dark-red dash-dotted lines in the following figures (e.g.\ figures~\ref{fig:xRyThetaPhiVelocityBLViscoMagWaves} and~\ref{fig:xRyThetaPhiMagneticBLViscoMagWaves} for velocity and magnetic fields, respectively).
The \textsc{dns} velocity perturbation is computed as scaled difference with respect to the potential base flow, $\Delta \boldsymbol{V}=\boldsymbol{V}\epsilon^{-1}-\Real(\boldsymbol{U}_1\exp{(\mathrm{i}(\omega t+m\phi))})$.
Without rotation, polar forcing produces vanishing azimuthal components in both theory and \textsc{dns}, so only the $\theta$ component is considered. The inviscid MHD solution (dark-red dashed) agrees with \textsc{dns} (orange-dotted) in both skin and Alfv\'en-wave regimes (figures~\ref{fig:xRyThetaPhiMagneticBLViscoMagWaves} at $\Lambda_l=2$ and~\ref{fig:xRyThetaPhiMagneticBLViscoMag} at $\Lambda_l=0.5$), except near interfaces where viscosity dominates within nested boundary layers (see insets), leading to slope changes absent from the inviscid model. In the Alfv\'en-wave regime, perturbations penetrate much deeper in the fluid (right insets showing the full liquid core) and exhibit wave-like structures (figure~\ref{fig:xRyThetaPhiMagneticBLViscoMagWaves}) compared with those in the skin-layer regime. In all cases, the diffusionless induced correction ($b^{(p)}_{0}$, $u^{(p)}_{0}$) is an efficient substitute for the inviscid MHD particular solution, as shown by the cyan dash-dotted line.

Larger discrepancies arise for velocity, especially near solid boundaries (figure~\ref{fig:xRyThetaPhiVelocityBLViscoMagWaves}), where viscous effects dominate. For large $\Lambda_l$, the velocity perturbation follows the inviscid MHD solution outside viscous layers, notably in the Alfv\'en-wave regime, whereas viscosity leads to marked deviations for small $\Lambda_l$ (figure~\ref{fig:xRyViscoMagRotVThetaPhiSmallLambda}). Profiles are shown at co-latitude $\theta=30^{\circ}$; agreement improves toward the poles where local $\Lambda_l$ is larger, and deteriorates at the equator where $\Lambda_l=0$. 
The above considerations are valid for both polar and equatorial modes, respectively indicated by (a) and (b) in the figures discussed.

\begin{figure}
\centering
\begin{subfigure}[b]{0.8\textwidth}
\centering
\includegraphics[width=1\linewidth]{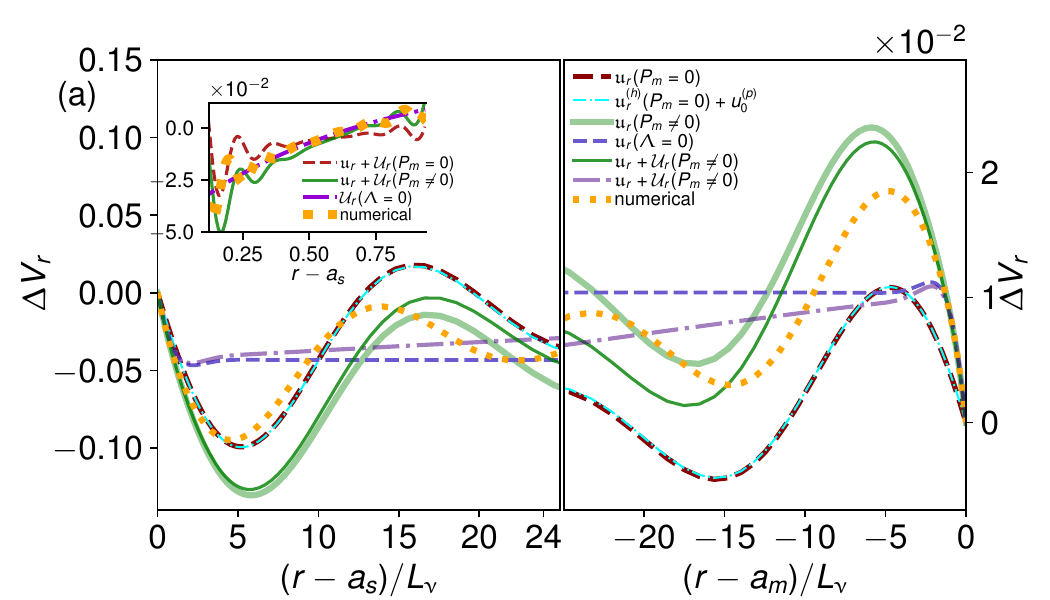}
\end{subfigure}
\begin{subfigure}[b]{0.8\textwidth}
\centering
\includegraphics[width=1\linewidth]{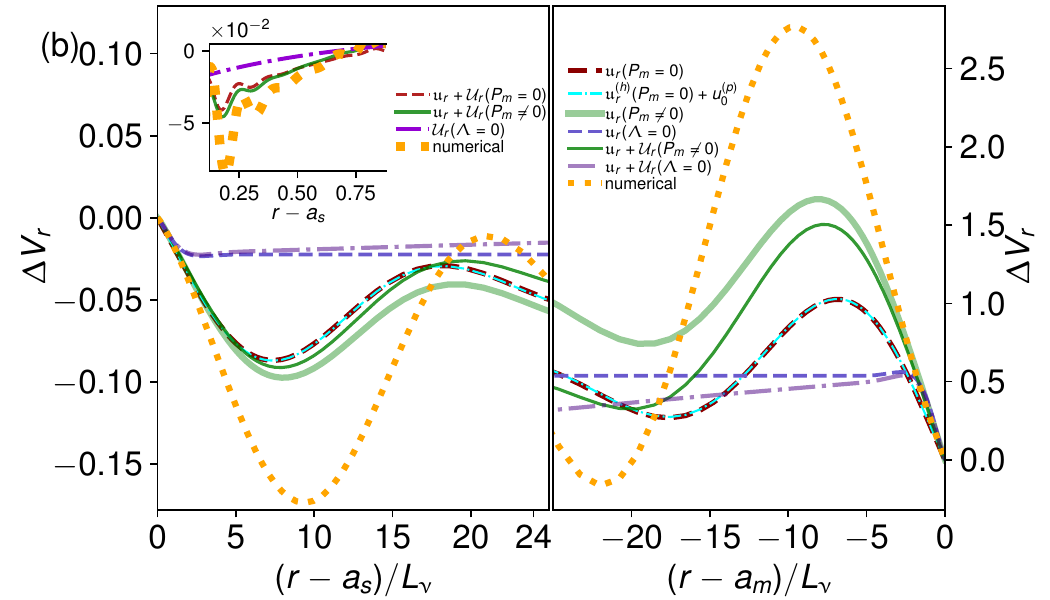}
\end{subfigure}
\caption{Radial flow perturbation $\Delta V_r$ near the inner (left) and outer (right) boundaries: \textsc{xshells} (orange-dotted), inviscid MHD (dark-red dashed), inviscid MHD with diffusionless correction (cyan dash-dotted), viscomagnetic oscillatory (thick green), and purely viscous (purple dashed). Viscomagnetic oscillatory (dark green) and purely viscous (purple dash-dotted) perturbation with secondary bulk flow.  (a) Polar and (b) equatorial forcing. Parameters: as in figure~\ref{fig:xRadialMagneticBoundaryLayer}, but   $\Lambda_z=2.67$, $\Lambda_l=2$, $\gamma=0$. Insets show the bulk perturbation.}
   \label{fig:xRadialVelocityBoundaryLayerWave}
\end{figure}

\subsection{Higher-order bulk fields driven by boundary-layer radial flow} \label{sec:highP}
The higher-order correction $\mathfrak{u}_r$ is obtained near both boundaries from $\boldsymbol{u}_\parallel$ using the BLT equation~(\ref{eq:consmass}). Discarding the particular solution and substituting $r\partial_r$ with $a_s \partial_r$ in (\ref{eq:consmass}) yields, in the fluid at $r=a_s$, and noting $\kappa_\lambda=\lambda^{-1}(r-a_s) \mp \lambda^{-2} \approx \mp \lambda^{-2}$,
\begin{eqnarray} \label{eq:brevBr}
    \mathfrak{u}_r  &=& \mp \frac{ \omega  \Lambda_l}{2 \mathrm{i} B_{0r} } L_\eta^2 \frac{A_\pm (\cot \theta-\kappa_\lambda \partial_\theta \lambda) +\partial_\theta A_\pm +\mathrm{i} m B_\pm}{a_s} \, \mathrm{e}^{\pm \lambda \tilde{r}_\pm}.
\end{eqnarray}
Within the \textit{huBL} framework, the $\theta$-dependence of $B_{0r}$, and hence of $\Lambda_l$ and $\lambda$, is neglected in the $\theta$-derivatives of (\ref{eq:consmass}) and (\ref{eq:brevBr}). Using the basic flow (\ref{syst:Lamb}), one obtains
\begin{eqnarray} \label{eq:vrBrfrK}
\frac{\mathfrak{u}_r}{\mathrm{e}^{- \lambda (r-a_s)}}&=& \frac{\mathfrak{u}_r^{(o)}}{-a^4 \mathrm{e}^{- \lambda (a_m-r)}}= - \frac{3}{2} \frac{ 1-\mathrm{i}}{\breve{c}_j+\sqrt{1-\mathrm{i} \Lambda_l}}  \frac{L_\eta \omega \Lambda_l}{a_s}  \frac{\mathfrak{c}}{1-a^3} , \qquad
\end{eqnarray}
noting material properties contrast $\breve{c}_j=\breve{\eta}_j^{1/2} \breve{\mu}_j$, and $\mathfrak{c}=\cos \theta$ for the polar mode and $\mathfrak{c}=\sin \theta$ for the equatorial modes. 

Figure~\ref{fig:xRadialVelocityBoundaryLayerWave} compares the numerical radial velocity perturbation, $\Delta V_r=V_r\epsilon^{-1}-\Real(U_{1r}\exp{(\mathrm{i}(\omega t+m\phi))})$ (orange-dotted), with the higher-order MHD inviscid solution $\mathfrak{u}_r$ (dark-red dashed). The asymptotic solution captures the spatial structure of the perturbation, reproducing both the boundary-layer behaviour and the bulk wavelength, but it underestimates the wave amplitude. This limitation is consistent with the comparatively weak radial velocity, which is more sensitive than the tangential components to higher-order effects neglected in the present approximation. As shown below, the inclusion of viscosity significantly improves the agreement between analytical predictions and \textsc{dns}.

Through the boundary conditions, the radial boundary-layer fluxes (\ref{eq:vrBrfrK}) drive bulk fields $( \boldsymbol{\mathcal{U}}, \boldsymbol{\mathcal{B}})$, whose Ohmic dissipation is subsequently evaluated using power arguments (\S\ref{sec:RelDisMKL}). Explicit expression of $(\boldsymbol{\mathcal{U}},\mathcal{P})$ and $(\boldsymbol{\mathcal{U}}^{(o)},\mathcal{P}^{(o)})$ can be obtained in the limit of weak Lorentz forces. By searching for irrotational flows of velocity potentials $(\varPsi,\varPsi^{(o)})$, with
\begin{eqnarray}
 \mathfrak{u}_r+\mathcal{U}_r|_{(r=a_s)}=\mathcal{U}_r|_{(r=a_m)}= \mathfrak{u}_r^{(o)}+\mathcal{U}_r^{(o)}|_{(r=a_m)}=\mathcal{U}_r^{(o)}|_{(r=a_s)} =0 ,
\end{eqnarray}
the method used in \S\ref{sec:InvBulk} for $\boldsymbol{U}$ shows that the pressure driven by the inner boundary layer is 
\begin{eqnarray} \label{eq:UcirBG}
\mathcal{P}&=&- \mathrm{i} \omega \varPsi =  \frac{3}{4} \frac{1+\mathrm{i}}{(1-a^3)^2} \frac{\omega^2 \Lambda_l L_\eta  a_s^2  (1+2 r^3/a_m^3) }{\breve{c}_s+\sqrt{1-\mathrm{i} \Lambda_l}} \frac{\cos \theta}{r^2 } , \label{eq:f9A} 
\end{eqnarray}
and the associated bulk secondary flow
\begin{eqnarray}
 \frac{\boldsymbol{\mathcal{U}}}{ \mathfrak{u}_r (r=a_s)}&=& - \frac{(a_s/r)^3}{1-a^3} \left[\left(1-\frac{r^3}{a_m^3} \right) \hat{\boldsymbol{r}} + \left(1+2\frac{r^3}{a_m^3} \right) \frac{\tan(\theta)}{2} \hat{\boldsymbol{\theta}} +3 \mathrm{i} \frac{\gamma}{2} \sin (\theta ) \hat{\boldsymbol{\phi}}   \right] ,  
\end{eqnarray}
for the polar mode, using the weak rotation-induced correction  $\mathcal{U}_\phi/ \mathcal{U}_s= \mathrm{i} \gamma $. For the equatorial modes, $\mathcal{P}$ is given by equation~(\ref{eq:f9A}) where $\cos \theta$ is replaced by $\sin \theta$, and
\begin{eqnarray}
  \boldsymbol{\mathcal{U}} &=& - \frac{3}{4} \frac{1-\mathrm{i}}{(1-a^3)^2} \frac{a_s^2 L_\eta  \omega \Lambda_l r^{-3}}{\breve{c}_s+\sqrt{1-\mathrm{i} \Lambda_l}} \left[ 2 \left(1-\frac{r^3}{a_m^3} \right)\sin (\theta) \hat{\boldsymbol{r}}+\left(1+2\frac{r^3}{a_m^3} \right) \boldsymbol{t}_\parallel \right] , \quad \label{eq:fcc2EQmag}
\end{eqnarray}
with $\boldsymbol{t}_\parallel=\cos (\theta) \hat{\boldsymbol{\theta}} + \mathrm{i} m  \hat{\boldsymbol{\phi}}$. For all modes, $\mathfrak{u}_r^{(o)}$ drives  $(\mathcal{P}^{(o)},\boldsymbol{\mathcal{U}}^{(o)})$, which can be obtained by multiplying $\mathcal{P}$ and $\boldsymbol{\mathcal{U}}$ by $a^4$, and by replacing $\breve{c}_s$ and $a_m$ by $\breve{c}_m$ and $a_s$, respectively. 

From the secondary bulk flow $\boldsymbol{\mathcal{U}}+\boldsymbol{\mathcal{U}}^{(o)}$, we can then obtain the bulk magnetic correction $\boldsymbol{\mathcal{B}}+\boldsymbol{\mathcal{B}}^{(o)} $ from equation~(\ref{eq:dc41}) by using $\boldsymbol{B}_0=B_0 \hat{\boldsymbol{z}}$. For the polar mode, we obtain
\begin{eqnarray} \label{eq:dfaB}
\frac{\boldsymbol{\mathcal{B}}}{B_0} &=& \frac{9}{4} \frac{1+\mathrm{i}}{(1-a^3)^2} \frac{\Lambda_l L_\eta a_s^2 }{\breve{c}_s+\sqrt{1-\mathrm{i} \Lambda_l}} \frac{(3\cos^2 \theta -1) \hat{\boldsymbol{r}}+\sin( 2 \theta )\hat{\boldsymbol{\theta}}+ \Xi \hat{\boldsymbol{\phi}} }{r^4} , \qquad
\end{eqnarray}
with the rotation-induced term $ \Xi=\mathrm{i} \gamma (5 \cos^2 \theta-1) \sin \theta$, and we obtain
\begin{eqnarray} \label{eq:dfaBeqi}
\frac{\boldsymbol{\mathcal{B}}}{B_0} &=& \frac{9}{4} \frac{1+\mathrm{i}}{(1-a^3)^2} \frac{\Lambda_l L_\eta a_s^2}{\breve{c}_s+\sqrt{1-\mathrm{i} \Lambda_l}} \left[  \frac{3 \sin 2\theta }{2 r^4} \hat{\boldsymbol{r}}+\frac{(1-2 \cos^2 \theta)\hat{\boldsymbol{\theta}}- \mathrm{i} m \cos (\theta) \hat{\boldsymbol{\phi}}}{r^4} \right], \qquad
\end{eqnarray}
for the equatorial modes. The field $\boldsymbol{\mathcal{B}}^{(o)}$ can then be obtained by multiplying  (\ref{eq:dfaB})-(\ref{eq:dfaBeqi}) by $a^4$, and by replacing $\breve{c}_s$ by $\breve{c}_m$. Magnetic corrections are thus negligible, preventing meaningful comparison with \textsc{dns}. The secondary flow forced by the boundary layer is also negligible, becoming significant in the bulk only when viscosity is included (see below).

As anticipated in \S\ref{sec:BLp2}, the secondary bulk fields $(\boldsymbol{\mathcal{U}}+\boldsymbol{\mathcal{U}}^{(o)},\boldsymbol{\mathcal{B}}+\boldsymbol{\mathcal{B}}^{(o)})$ do not in general satisfy the tangential boundary conditions, whose enforcement determines the final boundary-layer corrections $(\boldsymbol{\mathfrak{u}}_\parallel,\boldsymbol{\mathfrak{b}}_\parallel)$. These boundary-layer solutions are of the form (\ref{syst:hom_mag1}), using the prefactors (\ref{eq:A1}) where $ \mathfrak{h}=0$ since the boundary velocity has already been accounted for (at previous order). In these expressions, the velocity is only dependent on the bulk flow (and not on the bulk magnetic field). They can thus be used if $\boldsymbol{U}$ is replaced by $\boldsymbol{\mathcal{U}}+\boldsymbol{\mathcal{U}}^{(o)}$, giving
\begin{eqnarray} \label{eq:fgh1}
\frac{\boldsymbol{\mathfrak{u}}_\parallel}{ \mathrm{e}^{- \lambda (r-a_s)}}  &=&  \frac{3}{4} \frac{(1+\mathrm{i}) \omega \Lambda_l^2}{(1-a^3)^2} \frac{L_\eta \sin( \theta)}{a_s} \frac{(1+2 a^3+3a^4 )  \hat{\boldsymbol{\theta}}+3 \mathrm{i} \gamma (1+a^4) \cos  \theta \hat{\boldsymbol{\phi}}}{(\breve{c}_s+\sqrt{1-\mathrm{i} \Lambda_l})^2} , \qquad \quad  \\ \label{eq:fgh2}
\frac{\boldsymbol{\mathfrak{u}}_\parallel^{(o)}}{a^3 \mathrm{e}^{\lambda(r-a_m)} }  &=&  \frac{3}{4} \frac{(1+\mathrm{i}) \omega \Lambda_l^2}{(1-a^3)^2} \frac{L_\eta \sin( \theta)}{a_s} \frac{(3+2 a+a^4 ) \hat{\boldsymbol{\theta}}+3 \mathrm{i} \gamma (1+a^4) \cos  \theta \hat{\boldsymbol{\phi}}}{(\breve{c}_m+\sqrt{1-\mathrm{i} \Lambda_l})^2} ,
\end{eqnarray}
for the polar mode, while equatorial modes lead to
\begin{eqnarray} \label{eq:fgh1eq}
\frac{\boldsymbol{\mathfrak{u}}_\parallel}{ \mathrm{e}^{- \lambda (r-a_s)}}  &=&  -\frac{3}{4} \frac{(1+\mathrm{i}) \omega \Lambda_l^2}{(1-a^3)^2} \frac{L_\eta}{a_s} \frac{(1+2 a^3+3a^4 )  (\cos ( \theta) \hat{\boldsymbol{\theta}}+ \mathrm{i} m \hat{\boldsymbol{\phi}})}{(\breve{c}_s+\sqrt{1-\mathrm{i} \Lambda_l})^2} , \qquad \quad  \\ \label{eq:fgh2eq}
\frac{\boldsymbol{\mathfrak{u}}_\parallel^{(o)}}{a^3 \mathrm{e}^{\lambda(r-a_m)} }  &=& - \frac{3}{4} \frac{(1+\mathrm{i}) \omega \Lambda_l^2}{(1-a^3)^2} \frac{L_\eta}{a_s} \frac{(3+2 a+a^4 )( \cos (\theta) \hat{\boldsymbol{\theta}}+ \mathrm{i} m\hat{\boldsymbol{\phi}})}{(\breve{c}_m+\sqrt{1-\mathrm{i} \Lambda_l})^2} .
\end{eqnarray}
For all modes, these flows are associated to the boundary-layer magnetic fields $[\boldsymbol{\mathfrak{b}}_\parallel,\boldsymbol{\mathfrak{b}}_\parallel^{(o)}]=2 \mathrm{i} B_{0r}(\omega \Lambda_l \lambda L_\eta^2)^{-1} [-\boldsymbol{\mathfrak{u}}_\parallel,\boldsymbol{\mathfrak{u}}_\parallel^{(o)}]$, given by the induction equation and arising from the interaction of $\boldsymbol{B}_0$ with $\boldsymbol{\mathcal{U}}+\boldsymbol{\mathcal{U}}^{(o)}$. Interactions of $\boldsymbol{\mathcal{B}}$ with $\boldsymbol{U}$ or $\boldsymbol{\mathcal{U}}+\boldsymbol{\mathcal{U}}^{(o)}$ are of order $2$ or higher in $\epsilon$ and are thus neglected. In (\ref{eq:fgh1})–(\ref{eq:fgh2eq}), terms in $a^4$ or higher originate from the $\mathcal{U}^{(o)}$ contribution. These corrections are very small, and thus very difficult to benchmark against the \textsc{dns} results.

\subsection{Magnetic tension drag from the inviscid boundary layer} \label{sec:maga}
\begin{figure}
\centering
\includegraphics[width=1.0\linewidth]{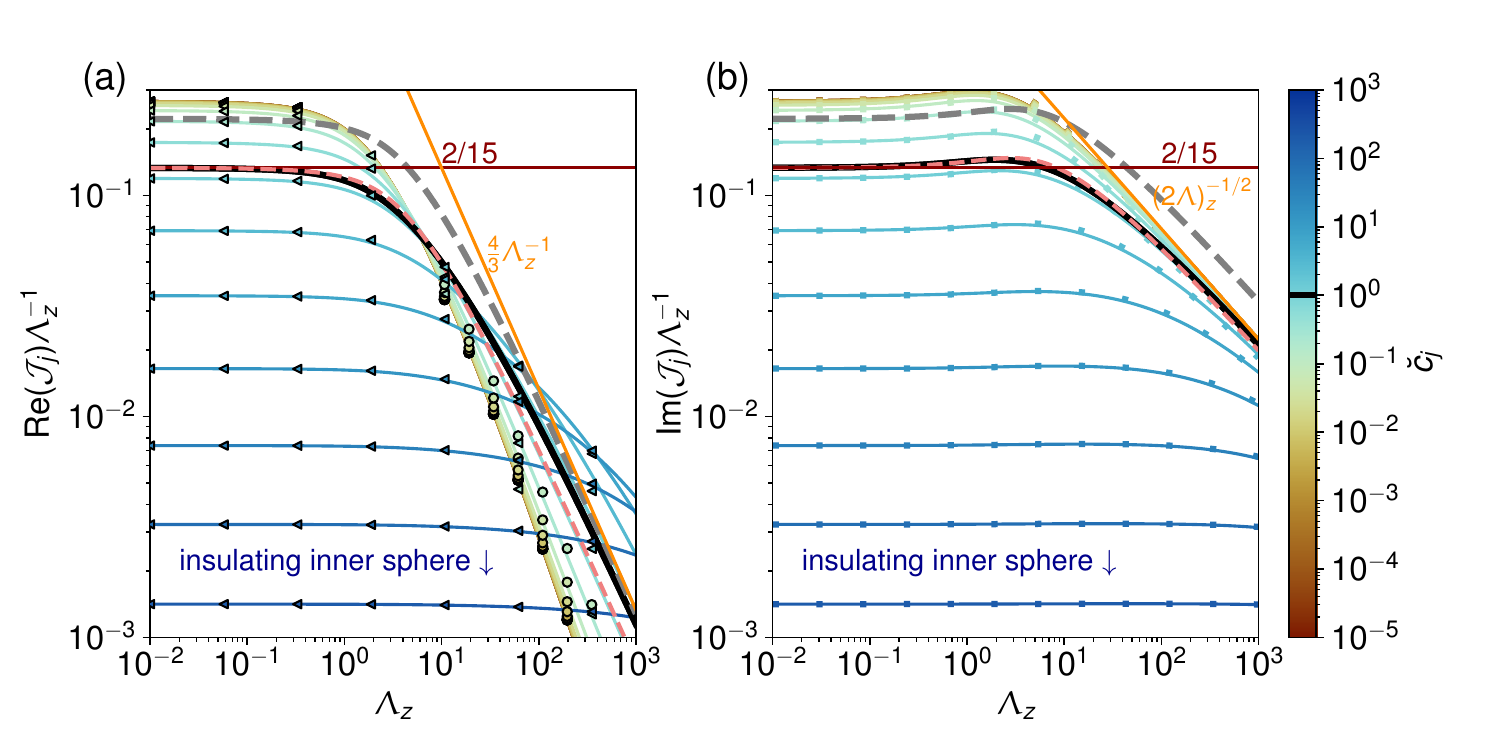}
\caption{Role of $\breve{c}_s$ and of $\Lambda_z$ on the real (left) and imaginary (right) part of the proxy $\mathcal{J}_j \Lambda_z^{-1}$ of the magnetic tension. For $\breve{c}_j=1$: (\ref{eq:exInt}) in black, with its asymptotic formula for $\Lambda_z \ll 1$ (red) and $\Lambda_z \gg 1$ (orange), compared to (\ref{eq:Bufextde}) with $\Lambda = \Lambda_z / 3$ (grey dashed line) and including the field-geometry correction factor $\mathcal{I}_{mt}^{\mathrm{pol}}=3/5$ (pink dashed line; defined in \S\ref{sec:maga}, eq.~\ref{eq:corrfactor}). The dots in (b) use the same correction factor $\mathcal{I}_{mt}^{\mathrm{pol}}=3/5$.
    For $\Real({\mathcal{J}_s})$, the analogous correction is $\mathcal{R}_{mt}^{\mathrm{pol}}=3/5$ for $\breve{c}_j\gg1$ or $\Lambda_z \ll 1$ (triangles) and $\mathcal{R}_{mt}^{\mathrm{pol}}=\sqrt{3} (\log(4 \Lambda_z) - 2)/ 4$ for $\breve{c}_j \ll 1$ and $\Lambda_z \gg 1$ (circles).}
    \label{fig:etamuDetuning}
\end{figure}

Integrating over the inner and outer boundaries yields the magnetic tension force $\boldsymbol{F}_{mt}$ and detuning $\Delta \omega_{mt}$. Neglecting the particular solution, $\Delta \omega_{mt}$ for the polar mode follows from (\ref{eq:A1pDC}). For an arbitrary axisymmetric imposed field within the BLT approximation, it depends only on the radial component $B_{0r}$ and reads
\begin{subequations}
\label{syst:exint}
\begin{equation}
\left. \frac{\Delta \omega_{mt}}{\omega} \right|_{r=a_s}=  \frac{9}{16} \frac{\rho_f}{\rho_s} \frac{ L_\eta}{ a_s }  \frac{\mathcal{J}_s}{1-a^3}  \quad , \quad \mathcal{J}_j=-\int_{0}^\pi  \frac{ (1+\mathrm{i}) \Lambda_l  (\hat{\boldsymbol{z}}\cdot\hat{\boldsymbol{\theta}} ) \sin^2 \theta}{ \breve{c}_j +\sqrt{1-\mathrm{i}\Lambda_l}  } \mathrm{d} \theta ,
\tag{\theequation~\emph{a,b}}
\end{equation}
\end{subequations}
with $\hat{\boldsymbol{z}}\cdot\hat{\boldsymbol{\theta}}=-\sin\theta$ from the projection $\boldsymbol{f}_{mt}\cdot\hat{\boldsymbol{z}}$. In the \textit{huBL} approximation, $\Lambda_l$ is constant, while it varies as $\Lambda_l=\Lambda_z\cos^2\theta$ for an axial field, with $\Lambda_z=B_{0}^2/\omega$. An insulating solid corresponds to $\breve{c}_j\gg1$, which reduces the imaginary part of the detuning for all $\Lambda_z$ relative to uniform electrical properties (figure~\ref{fig:etamuDetuning}). In the opposite limit of negligible $\breve{c}_j$, the integral (\ref{syst:exint}) saturates to a curve independent of the electrical properties ratio.

The detuning associated with the outer boundary-layer is also given by equations (\ref{syst:exint}), replacing $\mathcal{J}_s$ by $a\mathcal{J}_m$. The real part of this detuning provides the non-dissipative part of the magnetic tension force on the outer boundary, and its imaginary part is a dissipation source for the inner sphere dynamics. The total dissipation is thus
\begin{eqnarray}
  \frac{ \Imag \left(\Delta \omega_{mt}\right)}{\omega} =  \Imag \left(  \frac{9}{16} \frac{\rho_f}{\rho_s} \frac{ L_\eta}{ a_s }  \frac{\mathcal{J}_s+a \mathcal{J}_m}{1-a^3}   \right) . \label{eq:Im_maT}
\end{eqnarray}
For a uniform axial field, $\Lambda_z$ is constant and the integration can be analytically performed for $\breve{c}_j=1$. We have $\mathcal{J}_s=\mathcal{J}_m=\mathcal{J}_1$, and, noting $\chi=\sqrt{1- \mathrm{i} \Lambda_z} $,
\begin{eqnarray} \label{eq:exact}
\mathcal{J}_1 &=&  \frac{(1+\mathrm{i})\sqrt{\mathrm{-i}}}{12 \Lambda_z^{3/2}} [  (16 -6\chi)  \sqrt{-\mathrm{i} \Lambda_z^{3}}+3 \chi \sqrt{\mathrm{i} \Lambda_z} -(3-12 \mathrm{i}\Lambda_z) \tan^{-1}( \chi^{-1} \sqrt{\mathrm{i} \Lambda_z} ) ] \qquad \, \label{eq:exInt}
\end{eqnarray}
shown in black in figure~\ref{fig:etamuDetuning}. We thus have $\mathcal{J}_1 \approx 2 (1+\mathrm{i}) \Lambda_z/15 $ for $\Lambda_z \ll 1$, and $\mathcal{J}_1 \approx 4/3+\mathrm{i} \sqrt{\Lambda_z/2} $ for $\Lambda_z \gg 1$ (red and orange lines in figure~\ref{fig:etamuDetuning}, respectively).

Since evaluating the exact solution (\ref{eq:Im_maT}) requires the field geometry, BG95 rather used the approximate \textit{huBL} approach where $B_{0r}$ is replaced by a typical constant value at the boundary $r=a_s$. 
The local parameter $\Lambda_l$, based on $B_{0r}$ is then replaced by its surface average $\Lambda= \langle \Lambda_l \rangle$ at $r=a_s$. Note that the \textit{huBL} result is exact when the radial imposed field is constant at $r=a_s$. For the polar mode ($\boldsymbol{f}_{mt}$ along $\hat{\boldsymbol{z}}$), the estimate $\mathcal{J}_s \approx \mathcal{J}_s^{\mathrm{huBL}}$ reads
\begin{subequations}
\label{eq:Bufextde}
\begin{equation}
\frac{\mathcal{J}_s^{\mathrm{huBL}}}{ \Lambda} = \frac{4}{3} \frac{1+\mathrm{i}}{\breve{c}_s +\sqrt{1-\mathrm{i}\Lambda} }   , 
\left. \frac{\Delta \omega_{mt}}{\omega} \right|_{r=a_s}=    \frac{3 \Lambda}{4}  \frac{\rho_f}{\rho_s} \frac{ L_\eta}{ a_s } \frac{1+\mathrm{i}}{ \breve{c}_s +\sqrt{1-\mathrm{i}\Lambda} } \frac{1+\alpha \gamma^2}{1-a^3} ,
\tag{\theequation~\emph{a,b}}
\end{equation}
\end{subequations}
where $\alpha=0$ for the flow $\boldsymbol{U}_1$ given by (\ref{syst:Lamb2}), but re-doing the calculation gives $\alpha=-6a^3(1+a)  (1+a+a^2)^{-1}/5$ for the flow $\boldsymbol{U}_2$ given by equation~(\ref{eq:Busse_ordre4}). Since
$L_\eta^2/a_s^2= 2/(P_m\,\mathrm{Wo}^2)$, the magnetic skin depth is related to the Womersley number and the magnetic Prandtl number. 

Equation~(\ref{eq:Bufextde}) gives the magnetic-tension correction to $C_a$ as $C_a^i=C_a^i(\Lambda=0)-2 (\rho_s/\rho_f)\Real(\Delta \omega_{mt}/\omega)$ in the inviscid limit. Magnetic tension reduces the effective inertia of the inner solid domain, increasing the frequency of free oscillations (BG95). The influence of the particular solution on the magnetic tension in (\ref{eq:exInt}) or (\ref{eq:Bufextde}) can be assessed analytically using the diffusionless solution (\ref{eq:brBdn1}) for a uniform axial field and weak Lorentz forces. In the BLT limit $L_\eta \ll a_s$, equation~(\ref{eq:exInt}) is then multiplied by $1+(1-\mathrm{i})L_\eta/a_s$, showing that the particular solution introduces only a higher-order correction of order $(L_\eta/a_s)^2$.

Equations (\ref{eq:A1pDC}) and (\ref{eq:Bufextde}) extend equations (26)-(27) and (32) of BG95 to rotation-perturbed, bounded flows around an inner sphere with magnetic diffusivity and permeability. Equation~(\ref{eq:Bufextde}) shows that jumps in $\eta$ and $\mu$ are encapsulated through the ratio $\mu \eta^{1/2}$ across the boundary (unlike when the particular solution is included). In the limit $\mu_s \gg \mu_f$ of a high-permeability inner sphere, the tangential components of $\boldsymbol{b}$ vanish at the boundary, and hence $\boldsymbol{f}_{mt}=0$. The insulating limit $\sigma_s \ll \sigma_f$ also yields vanishing $\boldsymbol{f}_{mt}$, although the BLT approximation breaks down: the BLT approach requires $\omega a_s^2 \gg \max( \eta_f,\eta_s)$ to ensure thin boundary layers. In the opposite case $\sigma_s \gg \sigma_f$ of a highly conducting solid, where the tangential magnetic field vanishes at the boundary, $\breve{c}_s \to 0$ produces a slightly reduced $\boldsymbol{f}_{mt}$ relative to the geophysically relevant case $\breve{c}_s \approx 1$.

For a uniform axial field and varying electromagnetic properties contrast $\breve{c}_s$, figure~\ref{fig:etamuDetuning} shows how the field strength ($\Lambda_z$) controls the magnetic-tension contribution to the added-mass coefficient (panel~a, via $\Real(\mathcal{J}_i)$) and to dissipation (panel~b, via $\Imag(\mathcal{J}_i)$). For $\breve{c}_j=1$, the exact solution (\ref{eq:exact}) (solid black) is compared with the \textit{huBL} estimate (\ref{eq:Bufextde}) (grey dashed), obtained naively using simply the surface average value $\Lambda=\langle \Lambda_l \rangle=\Lambda_z/3$ for a uniform axial field. Although this averaging captures the overall field intensity, it does not reproduce the weaker detuning predicted by the exact integration (\ref{eq:exInt}). 
A further field-geometry correction factor can be estimated by considering the ratio of the exact formulation of (\ref{syst:exint}) and its \textit{huBL} approximation. The real and imaginary parts of the \textit{huBL} result should then be corrected by a factor, given by $\mathcal{R}_{mt}^{\mathrm{pol}}=\Real({\mathcal{J}_s})/\Real({\mathcal{J}_s^{\mathrm{huBL}}})$ and $\mathcal{I}_{mt}^{\mathrm{pol}}=\Imag({\mathcal{J}_s})/\Imag({\mathcal{J}_s^{\mathrm{huBL}}})$, respectively, for the polar mode. For this forcing, these definitions naturally give $\mathcal{R}_{mt}^{\mathrm{pol}}=\mathcal{I}_{mt}^{\mathrm{pol}}=1$ for a constant radial field at $r=a_s$, while an arbitrary field geometry can be handled through
\begin{equation}
    \mathcal{R}_{mt}^{\mathrm{pol}}(\Lambda \ll 1) = \mathcal{I}_{mt}^{\mathrm{pol}} (\Lambda \ll 1) = \frac{3}{2} \frac{\int_{0}^\pi \Lambda_l(\theta) \sin^3 \theta \mathrm{d} \theta}{\int_{0}^\pi \Lambda_l(\theta) \sin \theta \mathrm{d} \theta}
    \label{eq:corrfactor}
\end{equation} 
for $\Lambda \ll 1$. For a uniform axial field, $\Lambda_l(\theta)=\Lambda_z\cos^2(\theta)$, yielding $\mathcal{R}_{mt}^{\mathrm{pol}}=\mathcal{I}_{mt}^{\mathrm{pol}}=3/5$ in this limit. As shown in figure~\ref{fig:etamuDetuning}b, this single value already reproduces $\Imag(\mathcal{J}_s)$ almost exactly over the entire range of $\Lambda_z$ and $\breve{c}_j$. Moreover, in the opposite asymptotic limit $\Lambda\gg1$, we obtain $\mathcal{I}_{mt}^{\mathrm{pol}}=3\sqrt{3}/8\approx0.65$, sufficiently close to $3/5$ to justify adopting the approximation $\mathcal{I}_{mt}^{\mathrm{pol}}\approx3/5$ throughout the remainder of the paper. Although introduced here to account for the magnetic-tension drag, this correction factor also enters the Ohmic dissipation (\S\ref{sec:RelDisMKL}, \S\ref{sec:RelDisMKL2}) and the comparison with the \textsc{dns} of P26, making the choice $\mathcal{I}_{mt}^{\mathrm{pol}}\approx3/5$ a convenient and accurate approximation for a uniform axial field.

The real part varies more (figure~\ref{fig:etamuDetuning}a): $\mathcal{R}_{mt}^{\mathrm{pol}}\approx3/5$ remains a satisfactory approximation for $\breve{c}_j=1$ and becomes asymptotically exact for $\breve{c}_j\gg1$ or $\Lambda_z\ll1$, but is instead well captured by $\mathcal{R}_{mt}^{\mathrm{pol}}\approx\sqrt{3}\,(\ln(4\Lambda_z)-2)/4$ in the remaining regime $\breve{c}_j\ll1$, $\Lambda_z\gg1$. 

In the limit $\Lambda \to 0$, equation~(\ref{eq:Bufextde}) shows that the magnetic tension is linear in $\Lambda$, corresponding to the magnetic skin-layer regime (figure 2 of BG95). For large $\Lambda \gg 1$, Alfv\'en-wave propagation dominates and (\ref{eq:Bufextde}) reduces to
\begin{eqnarray}
 \left. \frac{\Delta \omega_{mt}}{\omega} \right|_{r=a_s}&\approx&  \frac{3}{4} \frac{\rho_f}{\rho_s} \frac{ L_\eta}{ a_s }  \frac{(1-\mathrm{i}) \breve{c}_s+ \mathrm{i} \, \sqrt{2 \Lambda }}{1-a^3}  , \label{eq:asymBG}
\end{eqnarray}
which, with the field geometry correction, gives the dimensional leading-order sum of the inviscid pressure force and magnetic-tension force
\begin{eqnarray}
    {\boldsymbol{F}_{p}}+  {\boldsymbol{F}_{mt}} &=& - {m}_{fs} C_a \mathrm{d}_t {\boldsymbol{V}}_s- 4 \pi \mathcal{I}_{mt}^{\mathrm{pol}} \sqrt{\frac{\rho_f}{\mu_f}} \frac{ a_s^2 \tilde{B}_0}{1-a^3}  {\boldsymbol{V}}_s, \label{eq:FStomagDC}  
\end{eqnarray}
with
\begin{eqnarray}
   C_a &=& \frac{1}{2} \frac{1+2a^3}{1-a^3}- \frac{3}{2} \frac{\mathcal{R}_{mt}^{\mathrm{pol}} \breve{c}_s}{1-a^3}\frac{L_\eta}{a_s}   ,
\end{eqnarray}
where ${\boldsymbol{F}_{p}}$ contributes only through the first term in $C_a$ (added-mass of the basic flow), and where magnetic diffusion no longer contributes to the leading-order drag. Instead, the oscillating sphere emits propagating Alfv\'en waves that transport momentum away through the mechanical and electromagnetic energy flux. Electromagnetic drag therefore remains finite although local Ohmic dissipation becomes asymptotically negligible, making this regime directly analogous to classical wave drag. Note also that (\ref{eq:FStomagDC}a) gives only the magnetic-tension contribution to the drag; the power approach of \S\ref{sec:RelDisMKL} additionally recovers the pressure-mediated
contribution and thereby the complete mechanical power budget. In a finite cavity, the Alfv\'en-wave attenuation length $L_A$ must satisfy $L_A\ll D$ for these formula to be valid. Otherwise, reflected waves couple the two boundaries and the local radiative law no longer describes the cavity response.

The coefficient $C_a$ of (\ref{eq:FStomagDC}), originating from the real part of (\ref{eq:asymBG}), shows how the magnetic tension modifies the added-mass, yielding a non-zero $\Lambda_l$-independent limit for the frequency shift, as noted by BG95 and here extended to arbitrary $\breve{c}_s$ and $a$. Whereas BG95 identified $\Lambda \approx 1$ as the transition between the magnetic skin layer and Alfv\'en-wave regimes for $\breve{c}_s=1$, the criterion is rather found to be $\Lambda \approx 2(\breve{c}_s + 1)^2$, with an additional transition at $\breve{c}_s=1$ in the magnetic skin layer regime (\S\ref{sec:roleEL} and fig.~\ref{fig:xetamuyLambdacI}).

In the non-rotating case, the prefactors (\ref{eq:A1pDC}) extend to equatorial modes by replacing $\sin \theta$ with $-\cos \theta$. Restoring the azimuthal dependence $\mathrm{e}^{\mathrm{i}\phi}$ and expressing the force $\boldsymbol{F}_{mt}$ in the Cartesian basis, the force components along $\hat{\boldsymbol{x}}$ and $\hat{\boldsymbol{y}}$ are $F_{mtz}$ and $\mathrm{i} m F_{mtz}$, where $\boldsymbol{F}_{mt}=F_{mtz}\hat{\boldsymbol{z}}=A_{mtz} \epsilon \mathrm{e}^{\mathrm{i}\omega t} \hat{\boldsymbol{z}}$ is the polar-mode force (such that the force is $A_{mtz} \boldsymbol{\delta}$). Magnetic torque vanishes for the equatorial modes, and for the rotation-perturbed polar mode. 

Considering the dissipative component of (\ref{eq:FStomagDC}a), the large-$\Lambda$ asymptotics recover the classical linear dependence on magnetic-field strength observed, for translating spheres, in the inductionless regime of low magnetic Reynolds number $Rm_\omega=\omega_s\epsilon_sa_s/\eta_f\ll1$ and large interaction parameter $N=\Lambda a_s/\epsilon_s\gg1$, as established experimentally by \cite{yonas1967measurements}, \cite{maxworthy1968experimental} and \cite{sekhar2005magnetohydrodynamic}. The same scaling also emerges in the complementary ideal-MHD limit, where momentum is transported away by the pair of Alfv\'en wings generated by a conducting sphere translating steadily across a uniform axial magnetic field \citep{drell1965drag,neubauer1980nonlinear}. Following the classical hydrodynamic definition of the drag coefficient, we define the magnetic drag coefficient $C_d^m$ by replacing the dynamic pressure $\rho_f \mathcal{V}^2/2$ with its magnetic counterpart $\tilde{B}_0^2/(2\mu_f)=\rho_f \tilde{V}_A^2/2$, such that the (dimensional) drag is written
\begin{equation}
C_d^m\,\pi a_s^2\mu_f^{-1}\tilde{B}_0^2/2=C_d^m \pi a_s^2 \rho_f \tilde{V}_A^2/2 ,
\end{equation}
with the sphere cross-section area $\pi a_s^2$. The usual drag coefficient $C_d$, based on the dynamic pressure, is then $C_d=C_d^m/A_l^2$, where the Alfv\'en number $A_l=\mathcal{V}/\tilde{V}_A$ (also known as Alfv\'en Mach number) reads here $A_l=(\epsilon/L_\eta) (2/\Lambda)^{1/2}<1$. For sub-Alfv\'enic motion ($A_l<1$), ideal-MHD Alfv\'en-wing theories predict the universal scaling $C_d^m/A_l=\mathcal{O}(1)$, implying the characteristic force scaling $\propto\tilde{B}_0^2A_l\propto\tilde{B}_0$, although the associated mechanical drag coefficient cannot be inferred uniquely from their equivalent electric-circuit formulation. In particular, equations~(54)-(55) of \citet{saur2013magnetic} and equation~(29) of \citet{bouvier2015protostellar} yield $C_d^m/A_l=4$, while finite-$A_l$ effects multiply this ratio by $(1+A_l^{-2})^{-1/2}$ \citep{neubauer1980nonlinear,zarka2007plasma}. By directly evaluating the Maxwell stresses acting on the sphere, equation~(\ref{eq:FStomagDC}a) not only recovers the Alfv\'en-wing scaling but also yields the explicit prefactor $C_d^m/A_l=8 \mathcal{I}_{mt}^{\mathrm{pol}}/(1-a^3)$. As for the classical Stokes drag coefficient (\S\ref{sec:StokEq59}), the total drag force is actually larger due to pressure effects: in the relevant large $\Lambda \gg 1$ limit, the energetic argument developed in \S\ref{sec:RelDisMKL} shows that this drag must be multiplied by $3 /[2(1-a^3)]$  giving the drag coefficient
\begin{equation}
C_d=\frac{12}{A_l} \frac{\mathcal{I}_{mt}^{\mathrm{pol}}}{(1-a^3)^2}.
\end{equation}
Normalised by the natural energy density of each problem, this asymptotic drag law recovers the mathematical form of the Stokes law (\ref{eq:Stob2}), where $A_l$ replaces the Reynolds number. In this sense, Alfv\'en-wave radiation transports momentum in ideal MHD just as viscous diffusion transports momentum in low-Reynolds-number hydrodynamics. These predictions remain in good agreement with the global ideal-MHD simulations of \citet{corso2026simulations} for a steady, compressible ideal-MHD flow perpendicular to the field past a non-magnetised conducting sphere. They recover the same asymptotic scaling in the incompressible low-$A_l$ regime (their equations~18--19). Their figure~6 at large value of $\beta=200 \gg 1$, approaching the incompressible limit, gives $C_A \approx 1.75$ in their notations, which corresponds to an effective coefficient $C_d^m/A_l\simeq 2 \times 1.75/\pi \approx 1.1$. The different prefactor is unsurprising since their simulations consider a steadily translating obstacle, whereas the present theory is derived for a rapidly oscillating sphere in the thin-boundary-layer regime $L_\eta \ll a_s$.

\subsection{Ohmic dissipation in inviscid MHD boundary layers} \label{sec:RelDisMKL}
The Ohmic dissipation does not arise solely from the magnetic-tension contribution (\ref{eq:Bufextde}). In both the solid and fluid domains it can be obtained from the volume integral of the squared radial derivatives of the magnetic perturbation, as specified by (\ref{eq:totPowoh}). For the thin inner boundary layer, the BLT reduces the volume integral to the surface element $a_s^2 \mathrm{d}\theta \mathrm{d}\phi$, retaining only leading-order terms in $L_\eta/a_s \ll 1$ and $\breve{\eta}_s^{1/2} L_\eta/a_s \ll 1$ (\S\ref{sec:dragduissp2}). 
To keep the analytical integration tractable, the \textit{huBL} framework is used and gives (for the polar mode)
\begin{eqnarray}
  \left.  \widehat{\mathcal D}_\eta^{\mathrm{pol}} \right|_{r \geq a_s}  &=&  \frac{9 }{8}  \frac{ \Lambda}{\vartheta_1 \vartheta_2^{1/4} |\breve{c}_s+\sqrt{1-\mathrm{i} \Lambda }|^{2}}  \frac{\rho_f}{\rho_s} \frac{ L_\eta}{ a_s }  \frac{1+\gamma^2/5+\alpha_2 \gamma^4}{(1-a^3)^2} , \label{eq:pmBatch21} 
\end{eqnarray}
in the fluid, noting $ \vartheta_1= \cos \left( \arg \left[ (1+\mathrm{i})(1-\mathrm{i} \Lambda )^{-1/2} \right] \right) $, and $ \vartheta_2=4(1+\Lambda^2) $, and where rotational effects modify the dissipation through the terms in $\gamma$ where $\alpha_2=0$ and $\alpha_2=8/175$ for the rotation-modified basic flows $\boldsymbol{U}_1$ (\ref{syst:Lamb2}) and $\boldsymbol{U}_2$ (\ref{eq:Busse_ordre4}), respectively. Within the solid inner sphere, the relative Ohmic contribution is given by (for the polar mode)
\begin{eqnarray}
       \left.  \widehat{\mathcal D}_\eta^{\mathrm{pol}} \right|_{r \leq a_s}  &=& \breve{\mu}_s  \vartheta_1 \vartheta_2^{1/4}   \left.  \widehat{\mathcal D}_\eta^{\mathrm{pol}} \right|_{r \geq a_s} . \label{eq:pmBatch22}
\end{eqnarray}
 The relative Ohmic contribution $\breve{\mu}_s \vartheta_1 \vartheta_2^{1/4}$ of the inner sphere decreases from $\breve{\mu}_s$ at $\Lambda\ll1$, where the flow-induced contribution is negligible, to $0$ for $\Lambda\gg1$, where the energy loss mainly arises from Alfv\'en-wave radiation in the fluid. The outer-boundary contribution follows from (\ref{eq:pmBatch21})-(\ref{eq:pmBatch22}) by multiplying by $a^4$ and replacing the index $s$ with $m$. Discarding rotation ($\gamma=0$), we simply obtain twice (\ref{eq:pmBatch21}) and (\ref{eq:pmBatch22}) for the equatorial modes. In any case, the total dissipation is obtained by summing the inner and outer solid domain contributions to the two fluid ones (at the two boundaries).

To assess the accuracy of the \textit{huBL} estimate, we can compute numerically the exact dissipation integrals, retaining the angular dependence of the imposed magnetic field rather than replacing $B_{0r}$ by its surface average, and form the ratio with the corresponding \textit{huBL} counterparts. For a uniform axial field where $\Lambda=\Lambda_z/3$, we then find slightly different ratios in each domain (\S\ref{sec:huBLaccu}). By analogy with the correction factor $\mathcal{I}_{mt}^{\mathrm{pol}}$ introduced in \S\ref{sec:maga} for magnetic tension, we thus introduce a global correction factor for the total dissipation (neglecting the outer solid contribution), denoted $\mathcal{I}_{\eta}^{\mathrm{pol}}$ for the polar mode, and $\mathcal{I}_{\eta}^{\mathrm{eq}}$ for the equatorial modes. For $\breve{c}=1$ and $\gamma=0$, we obtain that $\mathcal{I}_{\eta}^{\mathrm{pol}}= \mathcal{I}_{mt}^{\mathrm{pol}}$ at machine precision, and $\mathcal{I}_{\eta}^{\mathrm{eq}}\approx 2 \mathcal{I}_{\eta}^{\mathrm{pol}}$. While the \textit{huBL} approach shows that the equatorial-mode Ohmic dissipation is twice the polar one when $B_{0r}$ is constant at $r=a_s$, this factor thus increases to $2 \mathcal{I}_{\eta}^{\mathrm{eq}}/\mathcal{I}_{\eta}^{\mathrm{pol}}=4 $ for a uniform axial field  (figure~\ref{fig:huBL}). Note also that, while (\ref{eq:pmBatch21})--(\ref{eq:pmBatch22}) account for magnetic tension and pressure, the axial Lorentz force vanishes at leading order for such a field, owing to the cancellation between magnetic pressure and tension (\S\ref{sec:magPPOdc}).

\begin{figure}
\centering
\includegraphics[width=0.95\linewidth]{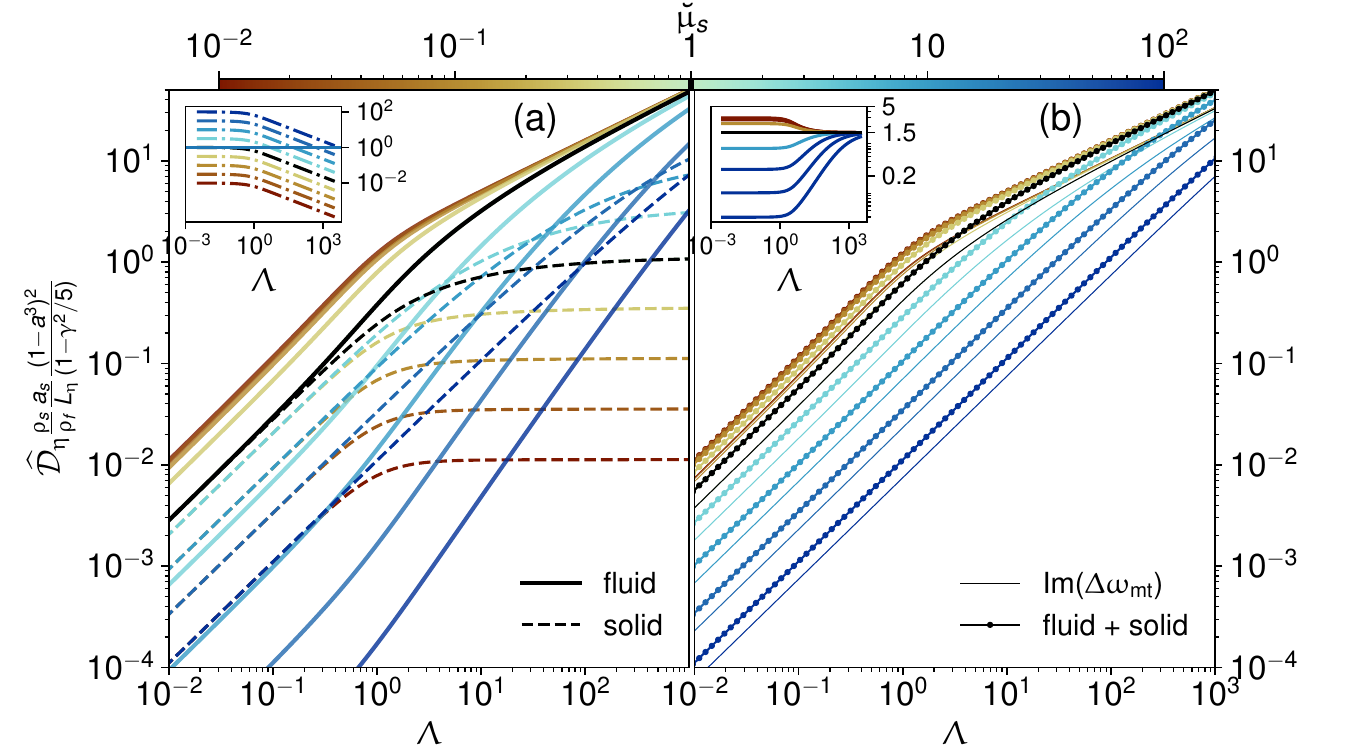}
    \caption{Scaled Ohmic dissipation for $\breve{\eta}_s=\mathcal{I}_{\eta}^{\mathrm{pol}}=1$, $\alpha_2=0$. (a) Fluid, from (\ref{eq:pmBatch21}), and solid, from (\ref{eq:pmBatch22}), regions. (b) Sum of fluid and solid contributions compared with the imaginary part of (\ref{eq:Bufextde}). Top left inset in (a): ratio between the scaled Ohmic dissipations in the solid and fluid for $\breve{\mu}_s=1$. Top left inset in (b): ratio between sum of the two contributions and the imaginary part of (\ref{eq:Bufextde}), colour of the curves is $\breve{\eta}_s$, assuming $\breve{\mu}_s=1$.}
    \label{fig:OhmicDetuningAnalytical}
\end{figure}

Figure~\ref{fig:OhmicDetuningAnalytical} shows the effect of magnetic-field strength $\Lambda$ and permeability ratio $\breve{\mu}_s$ on the scaled Ohmic dissipation: panel (a) compares fluid and solid contributions, and panel (b) contrasts total dissipation with that due to magnetic tension alone. For $\Lambda \gg 1$, increasing $\Lambda$ enhances fluid dissipation while leaving the solid contribution unchanged, so the fluid term dominates (inset of panel a). In figure~\ref{fig:OhmicDetuningAnalytical}b, the total dissipation (markers) is compared with its estimate (solid line) based on the dissipative part of the magnetic tension (as in BG95). 

For $\Lambda \ll 1$, equations (\ref{eq:pmBatch21})-(\ref{eq:pmBatch22}) converge towards the same \textit{huBL} estimate
\begin{eqnarray} \label{eq:plate}
  \left.  \widehat{\mathcal D}_\eta^{\mathrm{pol}} \right|_{r \geq a_s}  &=&  \breve{\mu}_s^{-1} \left.  \widehat{\mathcal D}_\eta^{\mathrm{pol}} \right|_{r \leq a_s}= \frac{9 }{8}  \frac{ \Lambda}{ (\breve{c}_s+1)^{2}}  \frac{\rho_f}{\rho_s} \frac{ L_\eta}{ a_s }  \frac{1+\gamma^2/5+\alpha_2 \gamma^4}{(1-a^3)^2} , \qquad
\end{eqnarray}
the total dissipation being thus twice (\ref{eq:plate}) at $\breve{\mu}_s = 1$. For $\Lambda \ll 1$, the fluid behaves as a solid heated by oscillatory currents, allowing (\ref{eq:plate}) to be related to the dissipation in a conducting sphere subjected to a field of magnitude $B$. Beyond the estimate $V_{ec} \omega B^2/2$, where $V_{ec}$ is the eddy-current volume \cite[e.g.][]{bouvier2015protostellar}, an exact analytical result exists for a uniform axial field of magnitude $B_{0}$ \cite[][who obtained identical shell dissipation for $L_\eta \ll a_s$]{chyba2021magnetic}. Matching equations (5) and (7) of \cite{chyba2021magnetic} also shows a regime transition at $L_\eta/a_s > \sqrt[3]{4/45} \approx 1/2$ for $a \ll 1$.

The opposite limit $\Lambda \gg 1$ gives a $\Lambda$-independent value in the solid inner domain, with
\begin{eqnarray}
 \breve{\mu}_s^{-1} \left.  \widehat{\mathcal D}_\eta^{\mathrm{pol}} \right|_{r \leq a_s}  &=& \frac{1}{\sqrt{2 \Lambda}} \left.  \widehat{\mathcal D}_\eta^{\mathrm{pol}} \right|_{r \geq a_s} =  \frac{9}{8}   \frac{\rho_f}{\rho_s} \frac{ L_\eta}{ a_s }  \frac{1+\gamma^2/5+\alpha_2 \gamma^4}{(1-a^3)^2} .
\end{eqnarray}
The full dissipation being the sum of (\ref{eq:pmBatch21})-(\ref{eq:pmBatch22}), we retrieve the scalings in $\Lambda$ and $\Lambda^{1/2}$ for $\Imag(\Delta \omega_{mt})$ that we found for $\Lambda \ll 1$ and $\Lambda \gg 1$, respectively. 

The dissipation from the power approach can now be compared with the one based on the magnetic tension. For the polar mode, $\mathcal{I}_{mt}^{\mathrm{pol}} \approx \mathcal{I}_{\eta}^{\mathrm{pol}} $ such that the ratio can be calculated as the sum of the fluid (equation~\ref{eq:pmBatch21}) and solid (equation~\ref{eq:pmBatch22}) contributions divided by the imaginary part of (\ref{eq:Bufextde}b). By neglecting outer-boundary dissipation and rotational effect at $ \breve{\eta}_s = 1 $, we obtain a gain of $3/(2(1-a^3))$, regardless of the values of $ \Lambda $ and $ \breve{\mu}_s $. This is illustrated by the black line in the top-left inset of (b), and justifies equation~20 in P26. For varying $\breve{\eta}_s$ (coloured lines in the top-left inset of figure~\ref{fig:OhmicDetuningAnalytical}b),  the gain becomes $1/(1-a^3)$ multiplied by $3(1+\breve{\mu}_s)/[2(1+\breve{c}_s)]$ for $\Lambda \ll 1$ and $3/2$ for $\Lambda \gg 1$.
For $\breve{c}_s=1$, these ratios coincide, recovering the same ratio known from the hydrodynamic viscous case (including rotation effects) between the total Stokes force and the tangential-stress contribution (see equations~\ref{eq:disSh2}-\ref{eq:Fv_pola} and appendix~\ref{sec:stoBL}). In the BG95 case $(\breve{c}_s,a)=(1,0)$, the disregarded factor $2/3$ between magnetic tension and total Ohmic dissipation is nearly compensated by the disregarded factor $\mathcal{I}_{mt}^{\mathrm{pol}} \approx 3/5$ from the \textit{huBL} approach (\S\ref{sec:maga}), explaining the good agreement of BG95 with $\textsc{dns}$ found by P26. 

Higher-order bulk-field estimates (\S\ref{sec:highP}) are uncertain for moderate Lorentz forces owing to the irrotational assumption, but allow comparison with the magnetic-tension force. Contributions from $\boldsymbol{\mathcal{B}}+\boldsymbol{\mathcal{B}}^{(o)}$ and $\boldsymbol{\mathfrak{b}}$ are higher order in $L_\eta$ and negligible, whereas the force from $\mathcal{P}+\mathcal{P}^{(o)}$ is comparable. Using the \textit{huBL} approach for the polar mode gives
\begin{eqnarray} \label{eq:ml35}
 \frac{\Delta \omega_{p}}{\Delta \omega_{mt}} = \frac{1}{2} \frac{1+2a^3+3a^4}{1-a^3} ,
\end{eqnarray}
with $\Delta\omega_{mt}$ from (\ref{eq:Bufextde}); the $a^4$ term arises solely from $\mathcal{P}^{(o)}$. We thus retrieve that the total magnetic force is $\sim 3/2$ larger than the magnetic tension (for $a \ll 1$). Although less accurate than the power-based estimate, (\ref{eq:ml35}) indicates comparable real and imaginary parts, yielding the magnetic correction to $C_a^i$, which is inaccessible from the power-based approach.

These results complete the theory and bring it into quantitative agreement with the \textsc{dns} of P26. There, the theoretical dissipation prefactor $d^\star_{\eta,2}$ was derived within the \textit{huBL} framework, which neglects the field-geometry correction. Using $\mathcal{I}_{\eta}^{\mathrm{pol}} \approx \mathcal{I}_{\eta}^{\mathrm{eq}}/2 \approx 3/5$, the improved estimates are now $d^\star_{\eta,2}=\pi\sqrt{2}\mathcal{I}_{\eta}^{\mathrm{pol}}/4\approx0.67$ for the polar mode and $d^\star_{\eta,2}=\pi\sqrt{2}\mathcal{I}_{\eta}^{\mathrm{eq}}/2\approx2.7$ for the equatorial modes. Relative to the \textsc{dns} prefactor of the polar mode and the average of the prograde and retrograde equatorial-mode prefactors (thereby removing rotational effects), the theoretical predictions agree to within $9\%$.

More broadly, the \textsc{dns} of P26 had already validated the asymptotic functional dependence over the geophysically relevant range $P_m\in[10^{-6},10^{-3}]$, including the previously unreported confinement scaling $(1-a^3)^{-2}$. The present correction of the analytical prefactors completes this agreement, demonstrating that the asymptotic theory reproduces both the scaling laws and the absolute magnitude of the Ohmic dissipation for all three translational modes. It also clarifies the physical origin of the confinement correction: estimating the dissipation from magnetic tension alone yields a confinement factor proportional to $(1-a^3)^{-1}$, whereas evaluating the volume-integrated Ohmic dissipation includes pressure effects and produces the stronger $(1-a^3)^{-2}$ dependence observed in the \textsc{dns}.

\subsection{Bulk-diffusive and boundary-layer Ohmic dissipation}

When $L_\eta \gtrsim a_s$, the boundary-layer approximation breaks down and the Ohmic dissipation is instead governed by bulk induction. In this bulk-diffusive regime, relevant to weakly conducting subsurface oceans (figure~\ref{fig:planets}b), we obtain the fluid dissipation (appendix~\ref{app:largeDeta})
\begin{subequations}
\label{eq:imB1}
\begin{equation}
\widehat{\mathcal D}_\eta|_{r \geq a_s}^{\mathrm{pol}}
    = \frac{3}{20}\frac{\rho_f}{\rho_s}\frac{1+\gamma^2}{1-a^3}  \Lambda_z f_\Lambda,       \quad \widehat{\mathcal D}_\eta|_{r \geq a_s}^{\mathrm{eq}}
    = \frac{\rho_f}{\rho_s}\frac{\Lambda_z}{1-a^3}
    \!\left(\frac{7}{20}+a^3\right),
\tag{\theequation~\emph{a,b}}
\end{equation}
\end{subequations}
for a uniform axial field with $\Lambda_z\ll1$, where $f_\Lambda=1$.
For the unbounded, non-rotating polar problem, \citet{motz1966magnetohydrodynamic} obtained and experimentally tested the finite-field correction $f_\Lambda=1-1.08\Lambda_z+1.15\Lambda_z^2$. We quote this correction for comparison, but do not interpret it as a large-$\Lambda_z$ asymptotic continuation of the present confined bulk-diffusive solution. Through its equivalence with the volume Ohmic dissipation (\S\ref{sec:dragduissp2}), equation~(\ref{eq:imB1}a) gives the exact quasi-static, weak-Lorentz-force contribution for the non-rotating polar mode and extends the magnetic drag of \citet{singh1965drag} from an unbounded fluid to confined spherical shells; its rotation-dependent term and the equatorial expression~(\ref{eq:imB1}b) are estimates obtained by neglecting the associated non-uniform electric potential (appendix~\ref{app:largeDeta}).

Equations~(\ref{eq:imB1}) are independent of $L_\eta$ and scale
linearly with $\Lambda_z$, in contrast to the Alfv\'en-wave scaling
proportional to $\Lambda_z^{1/2}$. For the polar mode
($\gamma=0$, $\Lambda_z=3\Lambda$), comparison of
(\ref{eq:imB1}) with (\ref{eq:plate}) shows that, within the controlled
weak-field regime, bulk Ohmic dissipation exceeds the magnetic
skin-layer contribution when $L_\eta/a_s>4(1-a^3)/5$.

\section{Magnetohydrodynamic boundary layers with viscous and rotational effects} \label{sec:BTtangvisc}
This section aims to include viscous and rotational effects in the theory, modifying both the boundary-layer structure and the pressure force. We first introduce viscosity, deriving the viscomagnetic oscillatory layers (\S\ref{sec:viscomag}), the tangential-stress drag (\S\ref{sec:fovisma}) and the dissipation (\S\ref{sec:RelDisMKL2}), before also including rotational effects (\S\ref{sec:rotvisc}).

\subsection{Tangential components in the viscomagnetic oscillatory layers} \label{sec:viscomag}
Viscosity enters through the additional diffusion term
$P_m\partial_r^2\boldsymbol{u}_\parallel$ in
(\ref{eq:Lambi1}). Analytical particular solutions of the three
Slichter modes (\ref{syst:Lamb})-(\ref{syst:Lamb2}) remain explicit, although their expressions are too lengthy to be reported here  without obscuring the presentation. All analytical expressions omitted from the manuscript, including these particular solutions, are nevertheless provided in the accompanying Zenodo repository
\citep{zenodo}, allowing direct evaluation, plotting and independent
verification.

The solution $\boldsymbol{b}_\parallel^{(s)}$ in the solid domains remains unchanged (equation~\ref{syst:hom_mag1}), while the homogeneous fluid solutions at both boundaries take the form
\begin{eqnarray} \label{eq:viscmag1} 
 \boldsymbol{b}_\parallel^{(h)} &=&  (A_{\pm 1}\hat{\boldsymbol{\theta}}+A_{\pm 2} \hat{\boldsymbol{\phi}}) \mathrm{e}^{\pm\lambda_+ \tilde{r}_\pm}+(B_{\pm 1}\hat{\boldsymbol{\theta}}+B_{\pm 2} \hat{\boldsymbol{\phi}}) \mathrm{e}^{\pm\lambda_- \tilde{r}_\pm} , \\
 B_{0r} \boldsymbol{u}_\parallel^{(h)} &=& \pm (\mathrm{i} \omega \lambda_+^{-1}-\lambda_+ ) (A_{\pm 1}\hat{\boldsymbol{\theta}}+A_{\pm 2} \hat{\boldsymbol{\phi}}) \mathrm{e}^{\pm\lambda_+ \tilde{r}_\pm}\pm(\mathrm{i} \omega \lambda_-^{-1} -\lambda_-) (B_{\pm 1}\hat{\boldsymbol{\theta}}+B_{\pm 2} \hat{\boldsymbol{\phi}})\mathrm{e}^{\pm\lambda_- \tilde{r}_\pm} , \qquad \label{eq:viscmag2}
\end{eqnarray}
since the induction equation~(\ref{eq:dc2}) gives $  B_{0r}  \boldsymbol{u}_\parallel=- \partial_r \boldsymbol{b}_\parallel + \mathrm{i} \omega \int \boldsymbol{b}_\parallel \mathrm{d}r$. The prefactors with the subscript $_-$ and $\tilde{r}_-$  (resp. $_+$ and $\tilde{r}_+$) are associated with the boundary layer at $r \geq a_s$ (resp. $r \leq a_m$). Here, the (complex) radial wavenumber $\lambda_\pm$ is
\begin{eqnarray}
 \lambda_\pm =\frac{1+\mathrm{i}}{L_\eta} \sqrt{\frac{1+P_m -\mathrm{i} \Lambda_l \pm \mathrm{i} \sqrt{[\Lambda_l+\mathrm{i}(1+P_m)]^2+4P_m}}{2 P_m} }\label{eq:STBU} ,
\end{eqnarray}
which can be written symmetrically for the diffusivities (\S\ref{sec:symStmL}). The non-magnetic purely viscous limit is detailed in appendix~\ref{sec:stoBL}. 

If we consider the steady limit $\omega=0$, we obtain that ${\lambda}_+$ vanishes as  ${\lambda}_+ =\mathrm{i} L_\eta^{-1} (2/\Lambda_l)^{1/2} \propto \omega$ and ${\lambda}_-^{-1}=L_\nu/\sqrt{2 \Lambda_l}=\tilde{B}_{0r}^{-1}\sqrt{\rho_f \nu\eta_f \mu_f}$,
with the viscous skin depth $L_\nu=L_\eta P_m^{1/2}$. In this limit, ${\lambda}_-^{-1}$ is independent of $\omega$ and is exactly the usual Hartmann layer thickness \cite[$\lambda_- D$ being then the Hartmann number, see e.g.][]{dormy2007mathematical}. Such layers have thus also been termed Stokes-Hartmann layers \cite[e.g.][]{cuevas2002magnetic}.

In the strong-field limit $\Lambda_l\gg1$, diffusive effects modify Alfv\'en-wave propagation (ideally at speed $V_A=\tilde{B}_{0r}/\sqrt{\rho_f\mu_f}$). One has $\lambda_- \approx (2\Lambda_l/P_m)^{1/2}/L_\eta \to \infty$, while $\lambda_+$ gives 
\begin{eqnarray}
\frac{ \omega}{\Imag(\lambda_+)}=V_A \, \left(1-5P_m \omega^4 L_\eta^4 V_A^{-4}/16\right)^{-1} ,
\end{eqnarray}
which is the diffusion-modified phase velocity, and attenuation length
\begin{eqnarray} \label{eq:fine}
L_A\equiv\Real(\lambda_+)^{-1}=2^{1/2} (1+P_m)^{-1}L_\eta\Lambda_l^{3/2},
\end{eqnarray}
reduced by the factor $1+P_m$ owing to viscous diffusion (compared with BG95). At a given latitude, the finite-cavity requirement $L_A\ll D$ sets an upper bound for $\Lambda_l$, given by
\begin{eqnarray}
\Lambda_l \ll (2^{-1/2}(1+P_m)L_\eta^{-1} D)^{2/3},
\end{eqnarray}
where waves are attenuated before reaching the opposite boundary, and the inner- and outer-boundary-layer constructions remain independent. When $L_A=\mathcal{O}(D)$, reflected waves couple the two boundaries and may generate global cavity resonances, for which the local radiative drag law no longer describes the finite-cavity response. This possibility was demonstrated experimentally by \citet{motz1966wave}, who generated Alfv\'en waves in a mercury-filled spherical shell and determined their propagation speed from a standing-wave resonance.

In the geophysically relevant limit $P_m \ll 1$, equation~(\ref{eq:STBU}) gives
\begin{subequations}
\label{syst:lplmc}
\begin{equation}
 \lambda_+  \to (1+\mathrm{i})L_\eta^{-1 } (1-\mathrm{i} \Lambda_l)^{-1/2} =\lambda \quad , \qquad \lambda_- \to (1+\mathrm{i}) L_\nu^{-1}  \sqrt{1-\mathrm{i} \Lambda_l} ,
\tag{\theequation~\emph{a,b}}
\end{equation}
\end{subequations}
where $\lambda_+$ recovers the inviscid MHD boundary-layer wavenumber (\S\ref{sec:BGinv}), and $\lambda_-^{-1}\propto L_\nu=L_\eta P_m^{1/2}$ vanishes with $P_m$, reducing to the viscous Stokes layer for $\Lambda_l\ll1$. This formulation provides a continuous transition between regimes (figure~\ref{fig:PmlambdaLetaFull}). For an axial field, $\Lambda_l=\Lambda_z\cos^2\theta$, the viscomagnetic oscillatory boundary layer reduces to a purely viscous skin layer at the equator. In the opposite limit $P_m\gg1$,
\begin{subequations}
\label{syst:hydrovisL}
\begin{equation}
 {\lambda}_+ \to \frac{1+\mathrm{i}}{L_\nu}  -\frac{(1-\mathrm{i})\Lambda_l}{2L_\nu P_m}  +\mathcal{O}(P_m^{-2}), \quad  {\lambda}_- \to \frac{1+\mathrm{i}}{L_\eta}  +\frac{(1-\mathrm{i})\Lambda_l}{2L_\eta P_m}  +\mathcal{O}(P_m^{-2}) ,
\tag{\theequation~\emph{a,b}}
\end{equation}
\end{subequations}
so that, at leading order, $\lambda_+$ recovers the viscous Stokes boundary layer, while $\lambda_-$ corresponds to the magnetic skin depth wavenumber in the absence of flow, the fluid behaving as a solid.

\begin{figure}
\centering
\begin{subfigure}[b]{0.8\textwidth}
\centering
\includegraphics[width=0.8\linewidth]{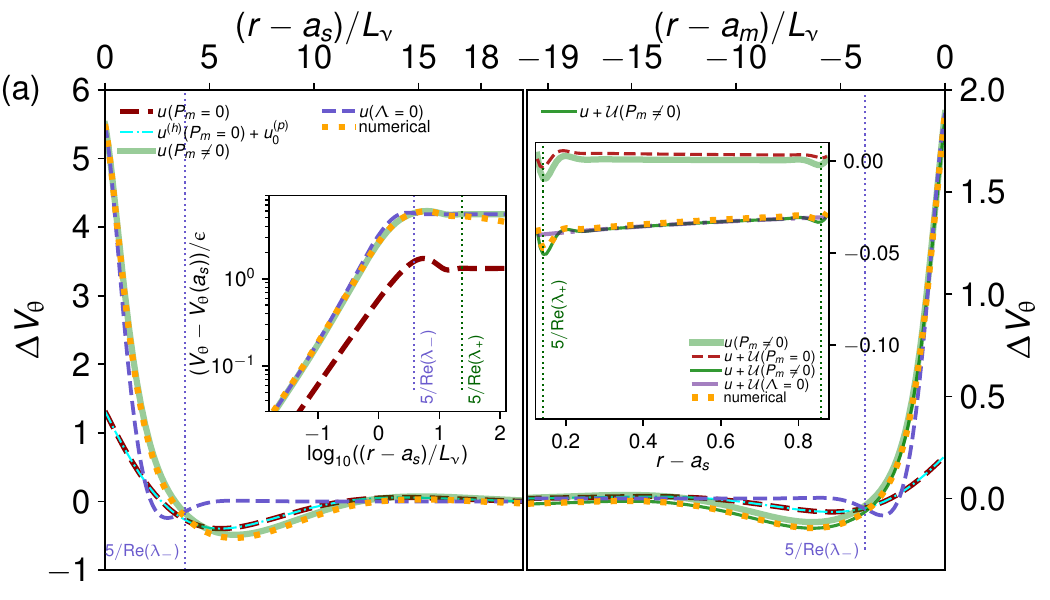}
\end{subfigure}
\begin{subfigure}[b]{0.9\textwidth}
\centering
\includegraphics[width=0.8\linewidth]{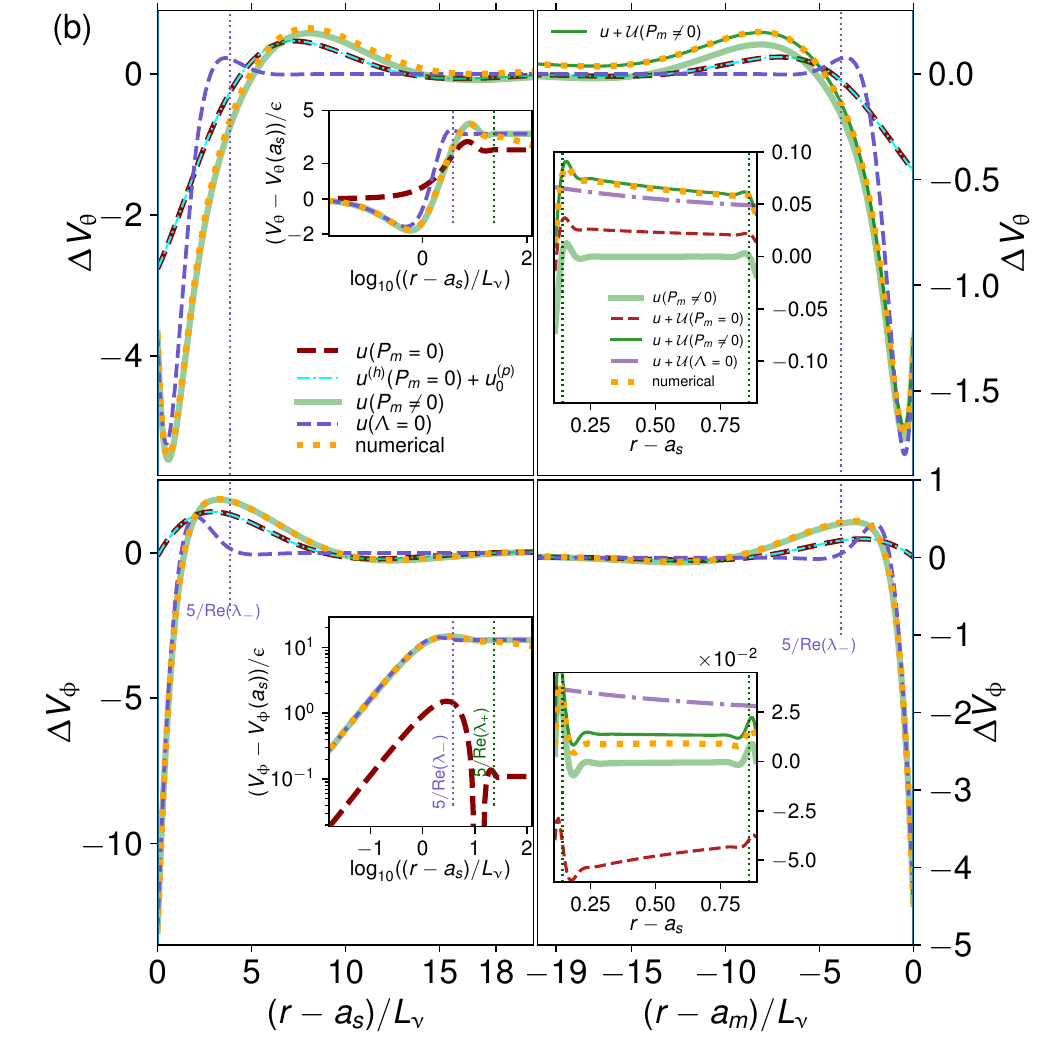} 
\end{subfigure}
\caption{Tangential components of the flow perturbation near the inner and outer boundaries. (a) Polar and (b) equatorial forcing. \textsc{xshells} (orange-dotted), inviscid MHD (dark-red dashed), inviscid MHD using diffusionless particular solution (cyan dash-dotted), viscomagnetic oscillatory (thick green), and purely viscous (purple dashed). Insets: left, zoom on $r=a_s$; right, bulk profiles vs. sum of boundary-layer and secondary flows from the inviscid MHD (dark-red dashed), viscomagnetic (dark-green), and purely viscous (purple dash-dotted). Parameters: as in figure~\ref{fig:xRadialMagneticBoundaryLayer}, but $\gamma=0$.} 
   \label{fig:xRyThetaPhiVelocityBLViscoMag}
\end{figure} 
\begin{figure}
\centering
\begin{subfigure}[b]{0.8\textwidth}
\centering
\includegraphics[width=0.8\linewidth]{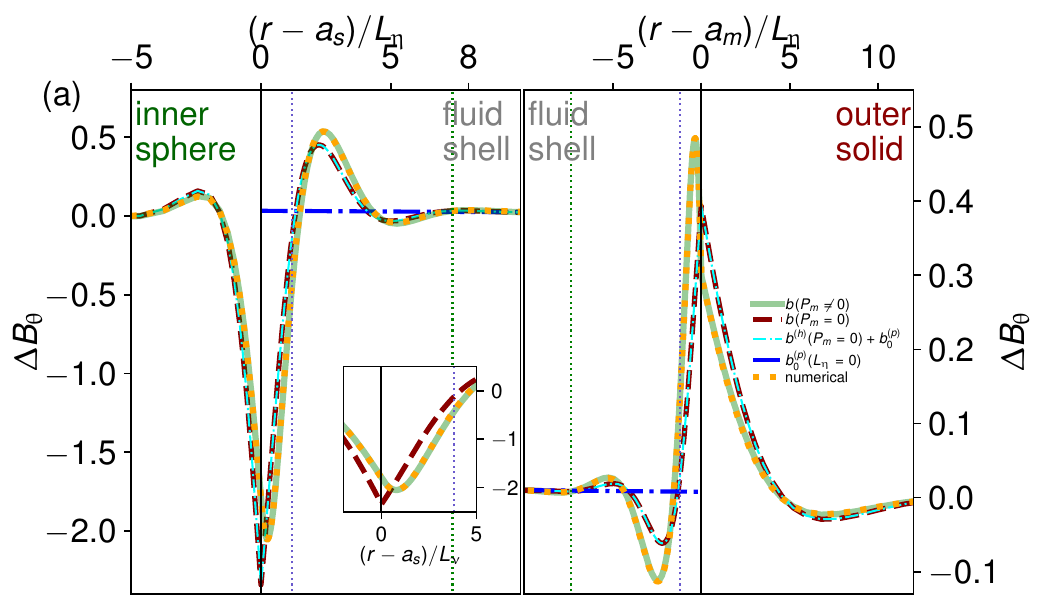}
\end{subfigure}
\begin{subfigure}[b]{0.9\textwidth}
\centering
\includegraphics[width=0.8\linewidth]{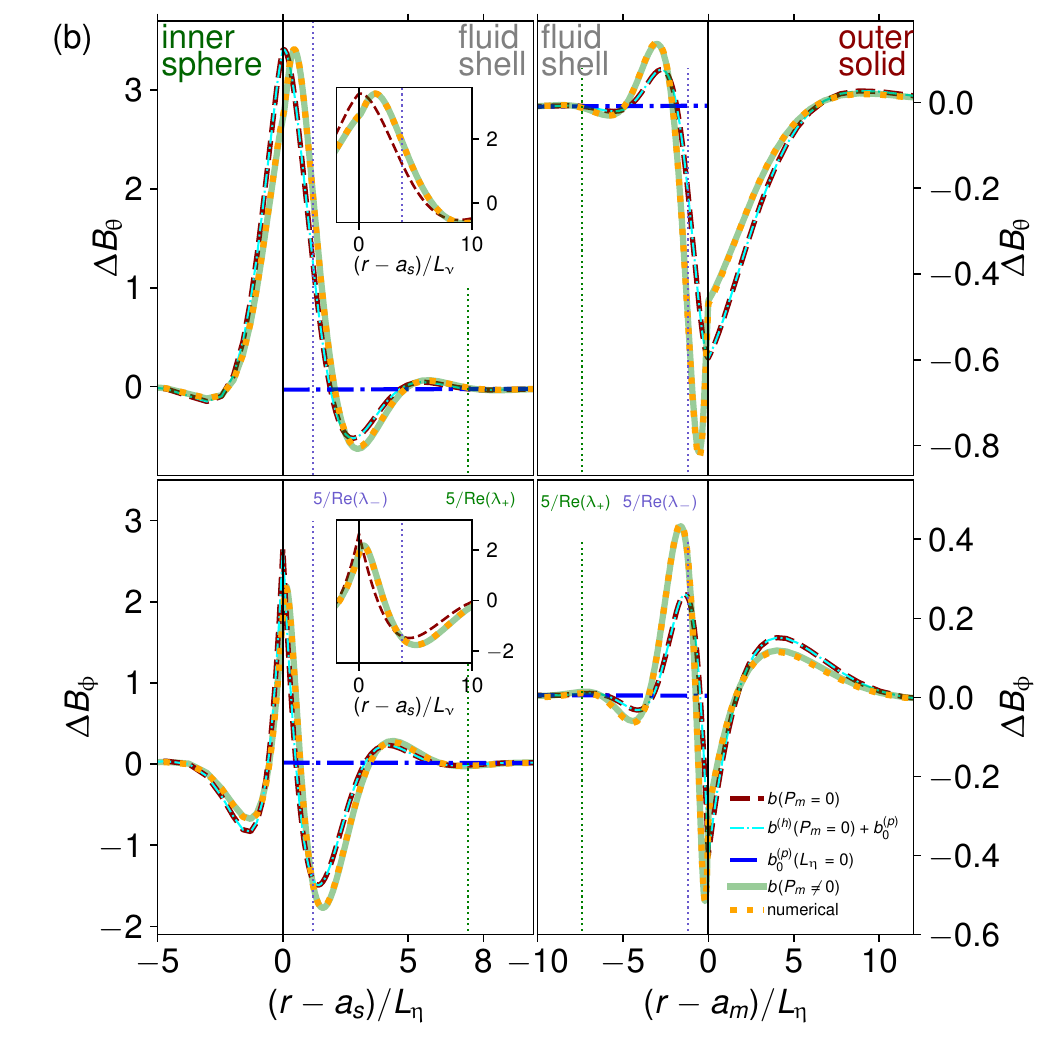} 
\end{subfigure}
\caption{Tangential components of the magnetic perturbation near the inner and outer boundaries. (a) Polar and (b) equatorial forcing. \textsc{xshells} (orange-dotted), inviscid MHD (dark-red dashed), inviscid MHD with diffusionless correction (cyan dash-dotted), viscomagnetic oscillatory (thick green),  purely viscous (purple dash-dotted), and diffusionless bulk solution (blue dash-dotted,~\ref{eq:brBdn1}). Insets show zooms near the interfaces. Parameters: as in figure~\ref{fig:xRadialMagneticBoundaryLayer}, but $\gamma=0$.} 
   \label{fig:xRyThetaPhiMagneticBLViscoMag}
\end{figure} 

The boundary-layer solution (\ref{eq:viscmag1})-(\ref{eq:viscmag2}) is completed by determining the prefactors from the boundary conditions (appendix~\ref{sec:CFbvis}), whose
complete analytical expressions are provided in the accompanying Zenodo repository \cite{zenodo}. The resulting oscillating viscomagnetic solution (thick dark-green curves in figures~\ref{fig:xRyThetaPhiVelocityBLViscoMag} and~\ref{fig:xRyThetaPhiMagneticBLViscoMag}) reproduces the \textsc{dns} throughout the boundary layer. Within one viscous skin depth $L_\nu$ of the solid boundaries, viscous effects dominate and the solution reduces to the purely viscous Stokes layer (appendix~\ref{sec:stoBL}), in agreement with the \textsc{dns} (left inset of figure~\ref{fig:xRyThetaPhiVelocityBLViscoMag}); this remains true even under strong magnetic forcing (figure~\ref{fig:xRyThetaPhiMagneticBLViscoMagWaves}, $\Lambda_l=2$). However, for $L_\nu<(r-a_s)<2L_\eta$, the purely viscous solution agrees with the \textsc{dns} only for very weak local magnetic fields (figure~\ref{fig:xRyViscoMagRotVThetaPhiSmallLambda}, $\Lambda_l=5\times10^{-3}$), whereas the viscomagnetic solution remains accurate from the bulk-diffusive regime to Alfv\'en waves emission (figures~\ref{fig:xRyThetaPhiVelocityBLViscoMagWaves} and~\ref{fig:xRyThetaPhiMagneticBLViscoMagWaves}).

For $P_m \ll 1$, magnetic coupling increases the penetration depth of the velocity boundary layer. For the polar mode, the viscomagnetic solution accurately reproduces the meridional velocity and magnetic-field components under weak rotation up to $\gamma=0.1$ (figures~\ref{fig:xRyViscoMagRotBThetaPhiGamma1em1} and~\ref{fig:xRyViscoMagRotVRThetaPhiPolarGamma1em1}). Similar agreement is obtained for equatorial modes (figures~\ref{fig:xRyViscoMagRotBThetaPhiEquatorialGamma1em1} and~\ref{fig:xRyViscoMagRotVRThetaPhiEquatorialGamma1em1}), where oscillatory forcing dominates the dynamics and masks rotational effects for $\gamma<0.1$.

Viscous effects near the boundaries are also visible in radial profiles of the tangential magnetic field (figure~\ref{fig:xRyThetaPhiMagneticBLViscoMag} and right insets), where extrema are slightly shifted from the interfaces—more strongly at the outer boundary—due to enhanced Ohmic dissipation. In the bulk (right insets of figure~\ref{fig:xRyThetaPhiVelocityBLViscoMag}), the boundary-layer solution (thick light green) alone cannot reproduce the velocity perturbation because it vanishes outside the layer; adding the higher-order viscomagnetic correction $\mathcal{U}$ (dark-green) restores agreement with the \textsc{dns} (orange-dotted). Radial plots also display vertical dotted lines at five times the characteristic penetration lengths $\Real(\lambda_\pm)^{-1}$ defined in (\ref{eq:STBU}), providing estimates of the viscously dominated region, $5\Real(\lambda_-)^{-1}$, and of the full velocity or magnetic boundary layer, $5\Real(\lambda_+)^{-1}$, even in the presence of waves (e.g. figure~\ref{fig:xRyThetaPhiVelocityBLViscoMag} and its right insets).

\subsection{Tangential-stress drag from the viscomagnetic oscillatory layers} \label{sec:fovisma}
 
\begin{figure}
\centering
\begin{subfigure}[b]{1\textwidth}
\centering
\includegraphics[width=1.0\linewidth]{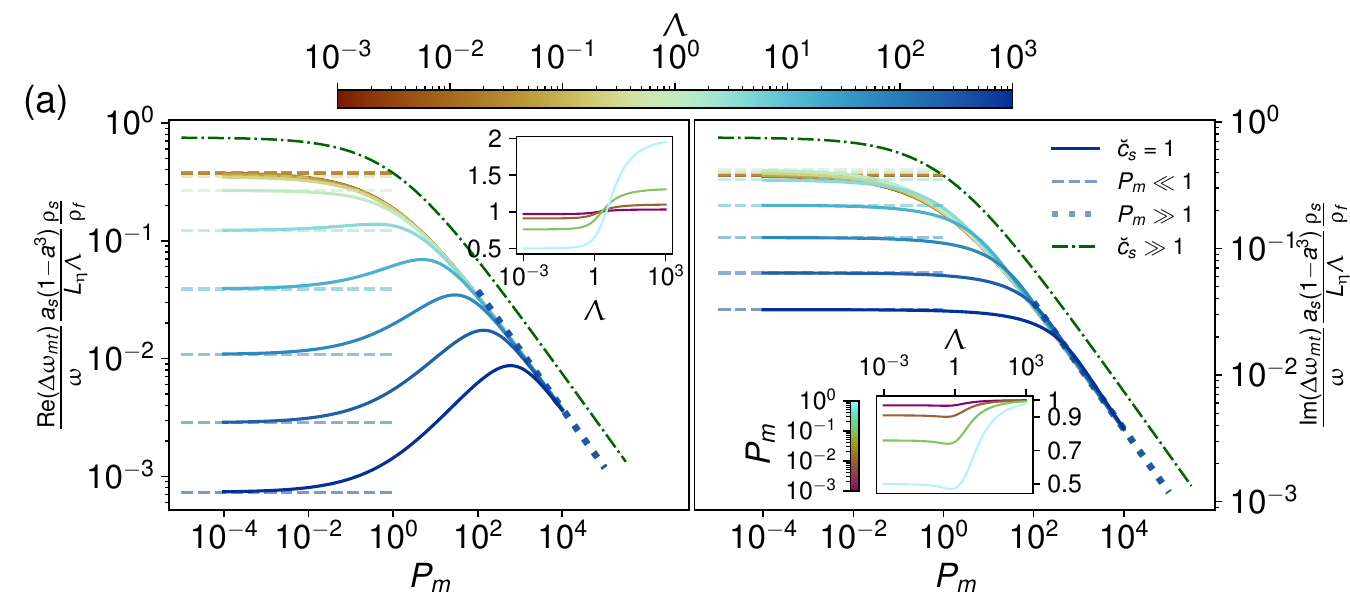}
\end{subfigure}
\begin{subfigure}[b]{1\textwidth}
\centering
\includegraphics[width=1\linewidth]{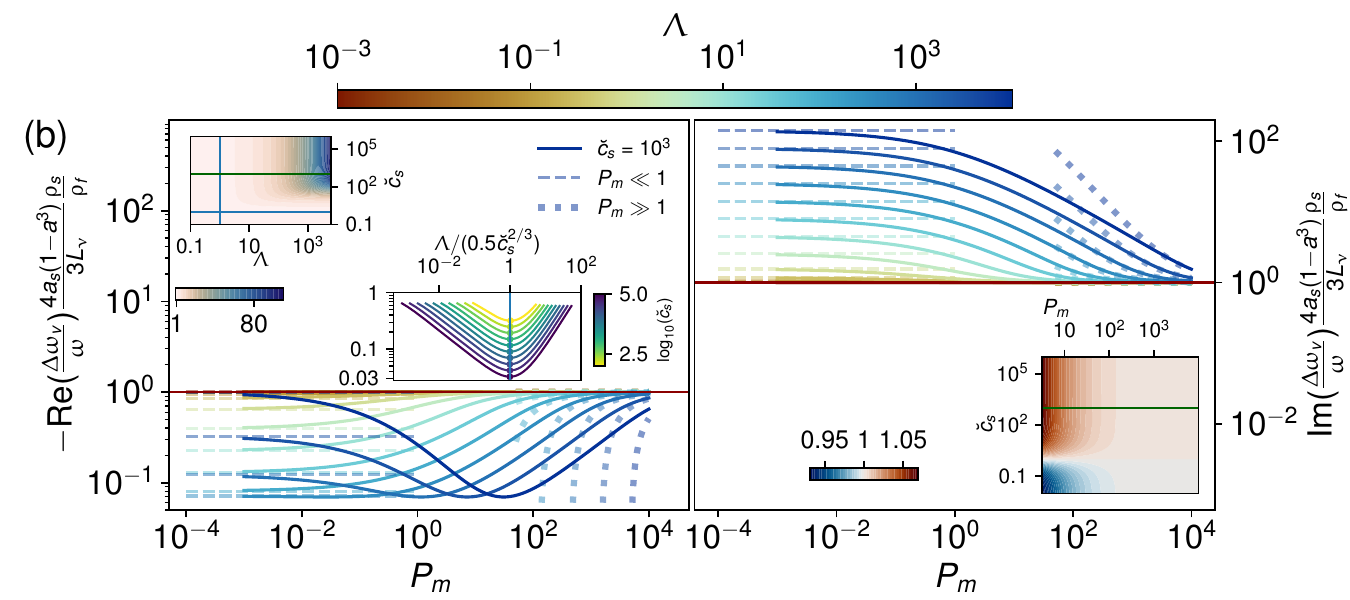}
\end{subfigure}
\caption{(a) Normalised magnetic-stress detuning assuming $\breve{c}_s = 1$; insets show the ratio of equation~(\ref{eq:extBuff}) to equation~(\ref{eq:Bufextde}b). Dashed and dotted lines denote the low-$P_m$ (equation~(\ref{eq:Bufextde}b)) and large-$P_m$ limits (equation~(\ref{eq:dmnd})). (b) Viscous-stress detuning, equation~(\ref{eq:extBuffvisc}), normalised by its non-magnetic limit (\ref{eq:batch}), evaluated at $\breve{c}_s = 10^3$; $P_m\ll1$ (dashed) and $P_m\gg1$ (dotted) correspond to equations (\ref{eq:disSh1}) and (\ref{eq:disSh2}). Left ($P_m = 10^{-3}$) and right ($\Lambda = 10$) insets show maps of equation~(\ref{eq:extBuffvisc}), green horizontal line is at $\breve{c}_s = 10^3$; the central inset displays the normalised real part of (\ref{eq:disSh1}).}
  \label{fig:PmLambdaMagneicViscoDetuning}
\end{figure}

The forces $\boldsymbol{F}_{mt}$ and $\boldsymbol{F}_v$ can now be calculated, including the contribution of the particular solution. However, for thin enough boundary layers, the particular solution can be discarded. One can then use the \textit{huBL} approach to obtain estimates of the forces $\boldsymbol{F}_{mt}$ and $\boldsymbol{F}_v$. Using $\tilde{\lambda}_\pm=\lambda_\pm L_\eta$ with a constant radial magnetic field at the boundary, the polar mode leads to (for the inner boundary)
\begin{eqnarray}
  \left.  \frac{\Delta \omega_{mt}}{\omega} \right|_{r=a_s} &=&\frac{3}{4} \Lambda  \frac{\rho_f}{\rho_s} \frac{ L_\eta}{ a_s } \frac{1+\mathrm{i}}{\breve{c}_s (2\mathrm{i}+\tilde{\lambda}_+ \tilde{\lambda}_-) +(1+\mathrm{i})(\tilde{\lambda}_++\tilde{\lambda}_-) } \frac{\tilde{\lambda}_+ \tilde{\lambda}_-}{1-a^3} ,  \label{eq:extBuff}  \\
      \left.  \frac{\Delta \omega_v}{\omega} \right|_{r=a_s}&=&\frac{3}{8} P_m \frac{\rho_f}{\rho_s} \frac{L_\eta}{a_s} \frac{ 2\mathrm{i} (\tilde{\lambda}_+ +\tilde{\lambda}_-)\breve{c}_s+\tilde{\lambda}_+\tilde{\lambda}_-(1+\mathrm{i})+2(\mathrm{i}-1)}{ \breve{c}_s(2 \mathrm{i}+\tilde{\lambda}_+\tilde{\lambda}_-)+(1+\mathrm{i})(\tilde{\lambda}_++\tilde{\lambda}_-)} \frac{\tilde{\lambda}_+\tilde{\lambda}_-}{1-a^3} , \, \label{eq:extBuffvisc} 
\end{eqnarray}
for no-slip boundary conditions, and (noting  $\xi=4+2\mathrm{i}(\tilde{\lambda}_+^2+\tilde{\lambda}_-^2)=-4(1-\mathrm{i} \Lambda )/P_m$)
\begin{eqnarray}
   \left. \frac{\Delta \omega_{mt}}{\omega} \right|_{r=a_s}&=&\frac{3}{2} \Lambda  \frac{\rho_f}{\rho_s} \frac{ L_\eta}{ a_s } \frac{\mathrm{i} \tilde{\lambda}_+ \tilde{\lambda}_-(\tilde{\lambda}_++\tilde{\lambda}_-)}{  \breve{c}_s (1+\mathrm{i})\tilde{\lambda}_+ \tilde{\lambda}_-(\tilde{\lambda}_++ \tilde{\lambda}_-)  +\xi+2\mathrm{i}\tilde{\lambda}_+ \tilde{\lambda}_- } \frac{1}{1-a^3} ,  \label{eq:extBuff2}  
\end{eqnarray}
for stress-free boundary conditions, shown in figures~\ref{fig:PmLambdaMagneicViscoDetuning} and~\ref{fig:PmLambdaMagneticDetuningStressFree} as functions of $P_m$ and $\Lambda$. In the stress-free case, $\tilde{\lambda}_+\tilde{\lambda}_- = 2 \mathrm{i} P_m^{-1/2}$ and the viscous stress at the boundary vanishes, so that $\Delta \omega_v=0$. As in \S\ref{sec:BGinv}, the total magnetic-tension dissipation includes the outer-boundary contribution, obtained by multiplying the above expressions by $a$ and replacing $(\breve{\mu}_s,\breve{\eta}_s)$ with $(\breve{\mu}_m,\breve{\eta}_m)$. The total magnetic-tension dissipation from both boundaries then follows as in (\ref{eq:Im_maT}). Equations (\ref{eq:extBuff})-(\ref{eq:extBuff2}) should further be multiplied by $1+\alpha \gamma^2$, with $\alpha=-6a^3(1+a)(1+a+a^2)^{-1}/5$, if the flow $\boldsymbol{U}_2$ of (\ref{eq:Busse_ordre4}) is used instead of $\boldsymbol{U}_1$.

Consider first the limit $\breve{c}_s \to \infty$, reached either for $\sigma_s \ll \sigma_f$ (insulating inner sphere) or $\mu_s \gg \mu_f$ (high-permeability inner sphere). Equation~(\ref{eq:extBuff}) then gives a vanishing detuning
\begin{eqnarray}
   \left.  \frac{\Delta \omega_{mt}}{\omega} \right|_{r=a_s} &=&\frac{3}{4}  \frac{\Lambda}{\breve{c}_s} \frac{\rho_f}{\rho_s} \frac{ L_\eta}{ a_s }  \frac{1+\mathrm{i}}{1-a^3} \frac{1 }{1+\sqrt{P_m}} , 
   \label{eq:magneticDetuningMagnetoVisco}
\end{eqnarray}
for a no-slip sphere. For a stress-free boundary, (\ref{eq:extBuff2}) leads to (\ref{eq:magneticDetuningMagnetoVisco}) multiplied by $1+\sqrt{P_m}$, yielding a $P_m$-independent detuning.

In the limit $P_m \ll 1$,  the scaled magnetic tension (\ref{eq:extBuff}) and (\ref{eq:extBuff2}) reduce to the inviscid case (\ref{eq:Bufextde}), and the  viscous stress detuning simplifies to 
\begin{eqnarray} \label{eq:disSh1}
          \left.  \frac{\Delta \omega_v}{\omega} \right|_{r=a_s}&\approx&-\frac{3}{4}  \frac{\rho_f}{\rho_s} \frac{L_\nu}{a_s} \frac{1-\mathrm{i}}{1-a^3} \left( \frac{1+\breve{c}_s\sqrt{1-\mathrm{i} \Lambda} }{\breve{c}_s+\sqrt{1-\mathrm{i} \Lambda}} \right),
\end{eqnarray}
as shown by dashed curves in figure~\ref{fig:PmLambdaMagneicViscoDetuning}b for $\breve{c}_s=10^3$ (approximately insulating solid sphere). In the limit $\Lambda \ll1$ and $\gamma\ll1$,  this gives $\Imag(\Delta \omega_{mt})>\Imag(\Delta \omega_v)$ when $\Lambda>(1+\breve{c}_s) \sqrt{P_m}$.

Equation~(\ref{eq:disSh1}) shows that magnetic effects vanish for $\breve{c}_s=1$, recovering the hydrodynamic result (\ref{eq:batch}) ; see the red line at $1$ in figure~\ref{fig:PmLambdaMagneicViscoDetuning}b. Its expansion for $\breve{c}_s \gg 1$ gives a minimum of $-\Real(\Delta \omega_v)$ at $\Lambda=\breve{c}_s^{2/3}/2$ for $\Lambda \gg 1$.
A minimum appears in the central inset of figure~\ref{fig:PmLambdaMagneicViscoDetuning}b. In the small-$P_m$ limit ($10^{-3}$, top-left inset), the magnetic correction in parentheses matches (\ref{eq:extBuffvisc}) normalised by the purely viscous detuning. This regime is particularly relevant to planetary interiors. In P26, the time-averaged viscous dissipation was shown numerically to agree with the power-based prediction, including confinement effects, and to recover the expected factor of two between the mean viscous dissipation of the polar and equatorial modes (see §\ref{sec:dragduissp2}). The top-left inset in figure~\ref{fig:PmLambdaMagneicViscoDetuning}b shows the imaginary part of this normalised quantity, indicating that magnetic corrections become significant only for large $\Lambda$ and $\breve{c}_s$, as for a high-permeability inner sphere (e.g.\ $\breve{\mu}_s=800$ for an iron sphere in Galinstan) oscillating in a sufficiently strong magnetic field.

In the opposite limit $P_m \gg 1$, equations (\ref{eq:extBuff})-(\ref{eq:extBuffvisc}) give
\begin{eqnarray}
  \left.  \frac{\Delta \omega_{mt}}{\omega} \right|_{r=a_s} &=&\frac{3}{4} \Lambda  \frac{\rho_f}{\rho_s} \frac{ L_\eta}{ a_s }  \frac{1+\mathrm{i}}{ (\breve{c}_s+1)(1-a^3) \sqrt{P_m}} , \label{eq:dmnd}  \\
          \left.  \frac{\Delta \omega_v}{\omega} \right|_{r=a_s}&\approx&-\frac{3}{4} \frac{\rho_f}{\rho_s} \frac{L_\nu}{a_s} \left[ \frac{1-\mathrm{i}}{1-a^3} + \frac{1+\mathrm{i}}{1-a^3} \frac{\breve{c}_s-1}{\breve{c}_s+1} (P_m^{-3/2}-P_m^{-1}/2) \Lambda \right],  \label{eq:disSh2}
\end{eqnarray}
while (\ref{eq:extBuff2}) yields $\sqrt{P_m}$ times (\ref{eq:dmnd}), leading to a $P_m$-independent detuning. For any $P_m$, magnetic corrections to $\Delta \omega_v$ vanish when $\breve{c}_s=1$, in which case (\ref{eq:extBuffvisc}) reduces to the non-magnetic limit given by the first term of (\ref{eq:disSh2}), consistent with (\ref{eq:batch}) and shown by the red line at 1 in figure~\ref{fig:PmLambdaMagneicViscoDetuning}b. In this large $P_m$ limit, $\Lambda=(1+\breve{c}_s)P_m$ marks the transition between magnetic- and viscous-drag-dominated regimes, where the leading-order terms of equations (\ref{eq:dmnd}) and (\ref{eq:disSh2}) have comparable real and imaginary components.

Rotation is absent from (\ref{eq:disSh1}) since the leading $\gamma$-correction is azimuthal; however, viscous pressure corrections reintroduce rotational effects in the total force $\boldsymbol{F}_\nu$. As shown in appendix~\ref{sec:appSmyli}, the results of \cite{smylie1998viscous,smylie2000the} are recovered from (\ref{eq:disSh2}) at order $\gamma$, with higher-order corrections modifying $\Delta \omega_v$. Appendix~\ref{sec:stoBL} further recovers the Stokes result and the detuning from a no-slip outer boundary by multiplying (\ref{eq:disSh1}) by $a$. 

Discarding magnetic and rotational effects, the above result extends directly to the equatorial modes. Restoring the azimuthal dependence $\mathrm{e}^{\mathrm{i}\phi}$ and expressing the force $\boldsymbol{F}_{v}$ in the Cartesian basis, the force components along $\hat{\boldsymbol{x}}$ and $\hat{\boldsymbol{y}}$ are $F_{vz}$ and $\mathrm{i} m F_{vz}$, where $\boldsymbol{F}_{v}=F_{vz} \hat{\boldsymbol{z}}=A_{vz} \epsilon \mathrm{e}^{\mathrm{i}\omega t} \hat{\boldsymbol{z}}$ denotes the corresponding polar-mode force (such that the force is $A_{vz} \boldsymbol{\delta}$). All the previous results, notably related to $\Delta \omega_{v}$, thus hold for the equatorial mode, agreeing with \cite{smylie2000the}, who obtained the same force for the polar and the equatorial modes in the limit of vanishing rotation effects and outer boundary displacement.

\subsection{Viscous and Ohmic dissipation in viscomagnetic oscillatory layers} \label{sec:RelDisMKL2}

To obtain the total dissipation, one can now wonder how pressure effects modify the results of \S\ref{sec:fovisma}. Pressure contributions from the bulk flow $\boldsymbol{\mathcal{U}}$
are required to recover the complete viscous force. When only mean
dissipation is required, $\widehat{\mathcal D}_\nu$ follows directly from (\ref{eq:totPowoh0}). Both approaches are detailed in appendix~\ref{sec:stoBL}, where the leading-order Stokes limits (\ref{eq:Stob1})-(\ref{eq:Stob2}) are recovered and extended to weak rotation.

Performing the volume integration of (\ref{eq:totPowoh0}) at leading order in $L_\nu/a_s \ll 1$ yields the dissipative viscous force at the inner boundary
\begin{eqnarray} \label{eq:Fv_pola}
{F}_\nu^{d,\mathrm{pol}} &=&  6 \pi  \frac{P_m  a_s }{(1-a^3)^2} \frac{a_s}{L_\nu} \left(1+\frac{\gamma^2}{5} \right) \epsilon \omega , \quad \widehat{\mathcal D}_\nu^{\mathrm{pol}} = \frac{9}{8} \frac{\rho_f}{\rho_s}\frac{L_\nu}{ a_s} \frac{1+\gamma^2/5}{(1-a^3)^2} , 
\end{eqnarray}
where $P_m$ arises from the chosen time scale (equivalent to $\nu$ in dimensional form). Equation~(\ref{eq:Fv_pola}) follows from the $\mathcal{O}(\gamma^4)$ expansion (\ref{eq:Busse_ordre2}) of the B74 bulk flow $\boldsymbol{U}$ (appendix~\ref{sec:Bus22}). Equations (\ref{eq:disSh2})-(\ref{eq:Fv_pola}) show that the viscous dissipation exceeds the tangential-stress contribution by $3(1+\gamma^2/5)/[2(1-a^3)]$; neglecting hydrodynamic or magnetic pressure therefore omits rotation and geometric effects, modifying the scaling for $a\sim1$. 

For a no-slip outer boundary, the corresponding force and scaled mean dissipation are obtained by multiplying (\ref{eq:Fv_pola}) by $a^4$. The total dissipative force is thus $(1+a^4)$ times (\ref{eq:Fv_pola}), recovering the leading-order term of (\ref{eq:Stob2}) in the rapid-oscillation limit $L_\nu\to0$ (appendix~\ref{sec:stoBL}). Comparing with the bulk dissipation (\ref{eq:sfPoJ}), the boundary layer dissipation is thus found to dominate when $L_\nu<a_s/2$ in the non-rotating unbounded limit $\gamma =a=0$, similarly to the transition $L_\eta<a_s/2$ found in \S\ref{sec:RelDisMKL} for the Ohmic dissipation. Balancing the two leading terms of the Stokes' viscous force expansion (\ref{eq:Stob2}) gives a similar $a$-dependent transition that evolves from $L_\nu \approx a_s$ for $a\to0$ to $L_\nu \approx (a_m-a_s)/2 $ for $a \to 1$.

For the equatorial modes, performing the volume integration of (\ref{eq:totPowoh0}) at leading order in $L_\nu/a_s \ll 1$ gives (at $r=a_s$)
\begin{eqnarray} \label{eq:denueE}
{F}_\nu^{d,\mathrm{eq}} &=&  6 \pi \frac{P_m  a_s }{(1-a^3)^2} \frac{a_s}{L_\nu}  \, \epsilon \omega \quad , \quad
\widehat{\mathcal D}_\nu^{\mathrm{eq}} = \frac{9}{4}  \frac{\rho_f}{\rho_s} \frac{L_\nu}{ a_s}  \frac{ 1 }{(1-a^3)^2} , \label{eq:denueE2}
\end{eqnarray}
and ${F}_\nu^{d}$ is thus identical for all three modes for $\gamma=0$, as confirmed by direct calculation (appendix~\ref{sec:stoBL}). At $\gamma=0$, identical dissipative forces therefore imply identical modal damping rates; the factor of two between (\ref{eq:Fv_pola}) and (\ref{eq:denueE2}) reflects only
$\mathcal{E}_{\mathrm{eq}}/\mathcal{E}_{\mathrm{pol}}=2$. Extending the integration to the order $L_\nu^2/a_s^2 \ll 1$ multiplies (\ref{eq:Fv_pola})-(\ref{eq:denueE}) by $1+L_\nu/a_s$, in agreement with the Stokes result (\ref{eq:Stob2}) for the unbounded polar mode.

Even within the \textit{huBL} approximation, no closed-form expression can be derived for the total Ohmic dissipation in viscomagnetic oscillatory layers. Nevertheless, the analytical results obtained for inviscid layers in \S\ref{sec:RelDisMKL}, together with the viscous-dissipation results derived above, strongly suggest that the total Ohmic dissipation can be estimated by multiplying (\ref{eq:extBuff}) or (\ref{eq:extBuff2}) by the correction factor $3 \mathcal{I}_{mt}^{\mathrm{pol}}/(2(1-a^3))$.

\begin{figure}
\centering
\begin{subfigure}[b]{0.8\textwidth}
\centering
\includegraphics[width=0.9\linewidth]{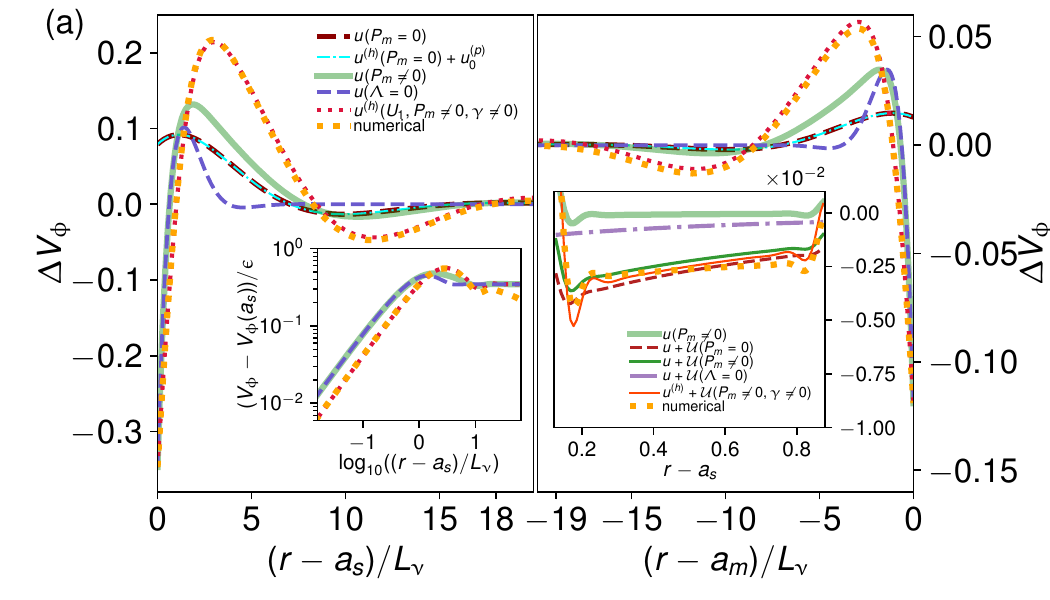}
\end{subfigure}
\begin{subfigure}[b]{0.8\textwidth}
\centering
\includegraphics[width=0.9\linewidth]{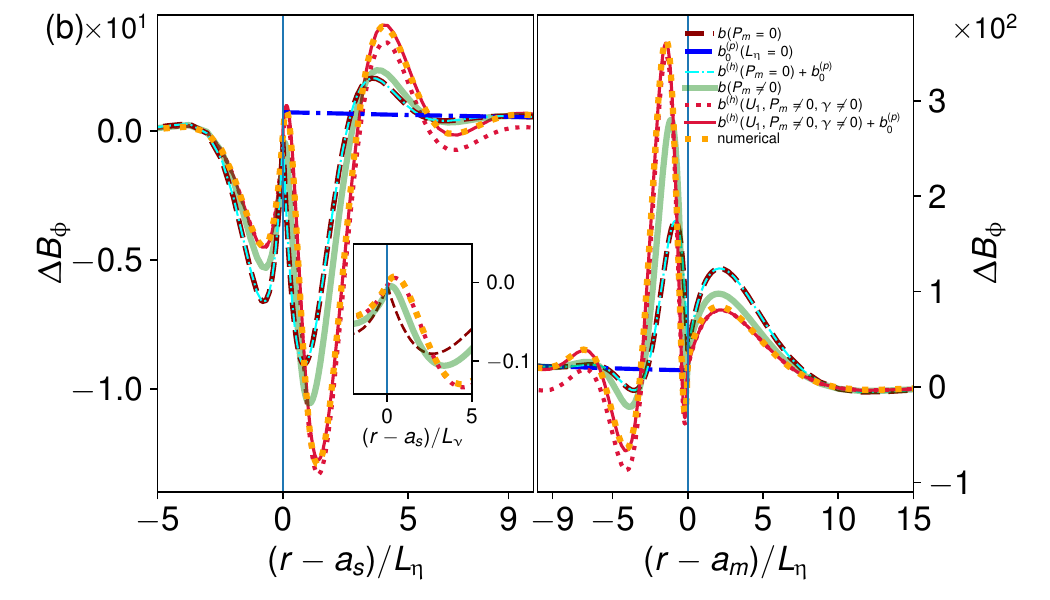}
\end{subfigure}
\caption{Azimuthal velocity (a) and magnetic (b) perturbations close to the inner (left) and outer (right) boundaries for the polar mode. \textsc{xshells} (orange-dotted), inviscid MHD (dark-red dashed), inviscid MHD with diffusionless correction (cyan dash-dotted), viscomagnetic oscillatory (thick green), purely viscous (purple dashed), homogeneous MHD Stokes--Ekman (red dotted), MHD Stokes--Ekman with diffusionless correction (red solid). Insets: left, zoom on $r=a_s$; right, bulk profiles vs. sum of boundary-layer and secondary flows from the inviscid MHD (dark-red dashed), viscomagnetic (dark-green), and purely viscous (purple dash-dotted), MHD Stokes--Ekman (red solid) solutions. Parameters: as in figure~\ref{fig:xRadialMagneticBoundaryLayer}. See \ref{fig:xRyViscoMagRotBThetaPhiGamma1em1} and \ref{fig:xRyViscoMagRotVRThetaPhiPolarGamma1em1} for the other components.} 
   \label{fig:xRyViscoMagRotVPhiGamma01}
\end{figure}

\subsection{Magnetohydrodynamic Stokes–Ekman boundary layer} \label{sec:rotvisc}
Extension to rotating viscous flows with finite $\gamma$ is achieved by reintroducing the Coriolis term, yielding MHD Stokes--Ekman boundary layers. As before, the general boundary-layer solution is determined in terms of a given basic flow $\boldsymbol{U}$ and particular solution. Integration of the homogeneous part of (\ref{eq:dc3})-(\ref{eq:dc2}) then gives, at both boundaries,
\begin{eqnarray} \label{eq:rotvisc1}
 \boldsymbol{b}_\parallel^{(h)} &=& (A_{\pm 1} \mathrm{e}^{\pm {\iota}_+ \tilde{r}_\pm}+A_{\pm 2} \mathrm{e}^{\pm {\kappa}_+ \tilde{r}_\pm}) (\mathrm{i} \hat{\boldsymbol{\theta}} +\hat{\boldsymbol{\phi}})+(B_{\pm 1} \mathrm{e}^{\pm{\iota}_- \tilde{r}_\pm}+B_{\pm 2} \mathrm{e}^{\pm {\kappa}_- \tilde{r}_\pm}) (-\mathrm{i} \hat{\boldsymbol{\theta}} +\hat{\boldsymbol{\phi}}), \qquad
\end{eqnarray}
where $\boldsymbol{u}_\parallel^{(h)}$ can be retrieved from the induction equation~$  B_{0r}  \boldsymbol{u}_\parallel^{(h)}=- \partial_r \boldsymbol{b}_\parallel^{(h)} + \mathrm{i} \omega \int \boldsymbol{b}_\parallel^{(h)} \mathrm{d}r$, 
\begin{eqnarray}
  \frac{2 B_{0r}}{\omega}\boldsymbol{u}_\parallel^{(h)}  &=& \pm (  ({\iota}_+ L_\eta^2-2 \mathrm{i} /{\iota}_+) A_{\pm 1} \mathrm{e}^{\pm {\iota}_+ \tilde{r}_\pm} + ({\kappa}_+ L_\eta^2-2 \mathrm{i} /{\kappa}_+)A_{\pm 2} \mathrm{e}^{\pm {\kappa}_+ \tilde{r}_\pm}) (\mathrm{i} \hat{\boldsymbol{\theta}} +\hat{\boldsymbol{\phi}}) \nonumber \\ & & \pm ( ({\iota}_- L_\eta^2-2 \mathrm{i} /{\iota}_-)  B_{\pm 1} \mathrm{e}^{\pm{\iota}_- \tilde{r}_\pm}+ ({\kappa}_- L_\eta^2-2 \mathrm{i} /{\kappa}_-)B_{\pm 2} \mathrm{e}^{\pm {\kappa}_- \tilde{r}_\pm}) (-\mathrm{i} \hat{\boldsymbol{\theta}} +\hat{\boldsymbol{\phi}}), \qquad \label{eq:rotvisc2}
\end{eqnarray}
with the four radial wavenumbers
\begin{eqnarray} 
  \underline{\iota}_\pm = \frac{1+ \mathrm{i}}{L_\eta} \sqrt{\frac{1+P_m - \mathrm{i} \Lambda_l \pm( \gamma  \cos \theta + \mathrm{i} \sqrt{ \varpi_\pm} )}{2 P_m} } , \label{eq:lrovi1} 
\end{eqnarray}
\begin{eqnarray}
  \underline{\kappa}_\pm  = \frac{1+ \mathrm{i}}{L_\eta}  \sqrt{ \frac{1+P_m - \mathrm{i} \Lambda_l \pm( \gamma  \cos \theta - \mathrm{i} \sqrt{ \varpi_\pm} )}{2 P_m}} , \label{eq:lrovi2}
\end{eqnarray}
where
\begin{eqnarray}
 \varpi_\pm=[\Lambda_l+\mathrm{i}(1+P_m)]^2+4P_m \pm 2(P_m+\mathrm{i}\Lambda_l-1)\gamma\cos\theta-\gamma^2\cos^2\theta .
\end{eqnarray}
Underlined quantities in (\ref{eq:lrovi1})-(\ref{eq:lrovi2}) relate to those in (\ref{eq:rotvisc1})-(\ref{eq:rotvisc2}) via $X=\underline{X}\,\mathrm{sign}[\Real(\underline{X})]$, so that $\Real(X)>0$ and the boundary-layer solution decays away from the boundary. As reviewed by \cite{hide1972hydromagnetic} or \cite{acheson1973hydromagnetics}, such MHD viscous layers from combined oscillation and rotation have been studied previously \cite[e.g.][]{hide1960hydromagnetic}. For instance, the four boundary layers (\ref{eq:lrovi1})-(\ref{eq:lrovi2}) are given by \cite{gans1970hydromagnetic,gans1971hydromagnetic}. These layers have also been examined near electrically conducting solids for MHD inertial waves \citep{kerswell1994tidal} and precession-nutation flows \citep{mathews2005viscoelectromagnetic,deleplace2006viscomagnetic}. Asymptotic limits are reviewed in \S\ref{sec:appBLHa} of the Supplementary Material.

The prefactors $(A_j,B_j)$ in (\ref{eq:rotvisc1})-(\ref{eq:rotvisc2}) follow from the boundary conditions (appendix~\ref{sec:CFbvisrot}) and are additionally provided in the accompanying Zenodo repository
\cite{zenodo}. With rotation, polar forcing generates finite azimuthal velocity and magnetic components (figure~\ref{fig:xRyViscoMagRotVPhiGamma01}). For weak rotation (e.g.\ $\gamma=0.1$), \textsc{dns} results are well reproduced by the MHD Stokes--Ekman model associated with the rotation-corrected potential base flow (\ref{syst:Lamb2}) (red dotted). Rotation also deepens the penetration of boundary-layer perturbations relative to the purely viscous skin layer (purple dashed) and the viscomagnetic oscillatory solution (dark-green), as seen near the inner boundary in the left insets of figure~\ref{fig:xRyViscoMagRotVPhiGamma01}. Adding the diffusionless particular solution to the homogeneous MHD Stokes--Ekman model (solid red) yields an accurate prediction of the azimuthal magnetic perturbation (figure~\ref{fig:xRyViscoMagRotVPhiGamma01}b).
\begin{figure}
\centering
\begin{subfigure}[b]{0.8\textwidth}
\centering
\includegraphics[width=1\linewidth]{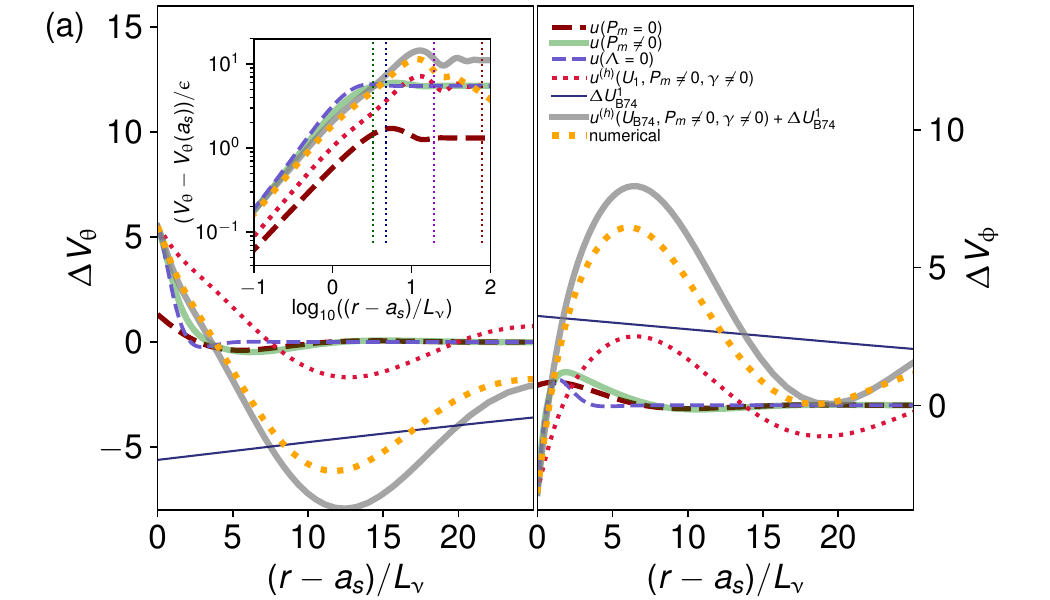}
\end{subfigure}
\begin{subfigure}[b]{0.8\textwidth}
\centering
\includegraphics[width=1\linewidth]{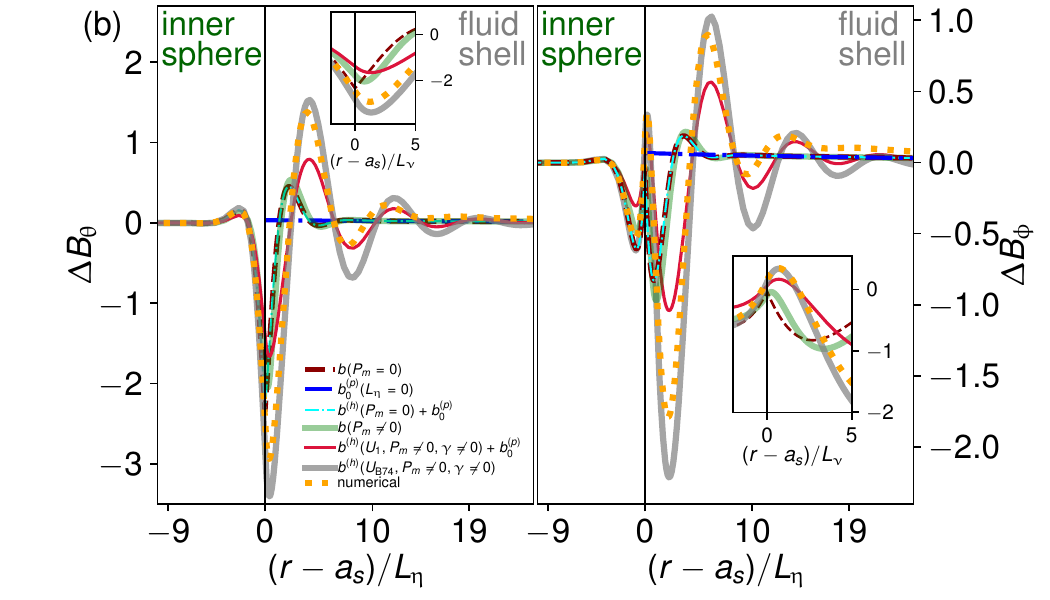}
\end{subfigure}
\caption{Meridional (left) and azimuthal (right) velocity (a) and magnetic (b) perturbations near $r=a_s$ (polar mode). \textsc{xshells} (orange-dotted), inviscid MHD (dark-red dashed), inviscid MHD with diffusionless correction (cyan dash-dotted), viscomagnetic oscillatory (thick green), purely viscous (purple dashed), homogeneous MHD Stokes--Ekman based on $\boldsymbol{U}_1$ (red dotted), MHD Stokes--Ekman based on $\boldsymbol{U}_1$ with diffusionless correction (red solid), MHD Stokes--Ekman based on $\boldsymbol{U}_\mathrm{B74}$ (grey) and diffusionless bulk solution (blue dash-dotted,~\ref{eq:brBdn1}). Insets show zooms near the interfaces. Parameters: as in figure~\ref{fig:xRadialMagneticBoundaryLayer}, but $\gamma=0.92$.} 
   \label{fig:xRyViscoMagRotVPhiGamma092}
\end{figure} 
At $\gamma=0.1$, the equatorial modes are accurately captured by the viscomagnetic oscillatory model, which reproduces both polar and azimuthal velocity and magnetic perturbations (dark-green lines in~\ref{fig:xRyViscoMagRotVRThetaPhiEquatorialGamma1em1}).
At larger $\gamma$, the model based on the base flow (\ref{eq:LambEquatorial}) fails to reproduce the \textsc{dns} for equatorial modes, see figure~\ref{fig:xRyVthetaVphiGammaEqua033} for $\gamma=0.33$.

For the polar mode, the Stokes--Ekman solution based on the rotation-corrected potential base flow (\ref{syst:Lamb2}) loses accuracy at larger $\gamma$, e.g.\ $\gamma=0.92$, as shown by the red curves in figure~\ref{fig:xRyViscoMagRotVPhiGamma092}. The discrepancy between the rotation-corrected potential base flow and B74 (dark blue line) is of the same order as the boundary-layer perturbation. Using $\boldsymbol{U}_\mathrm{B74}$ in the MHD Stokes--Ekman solution (grey in figure~\ref{fig:xRyViscoMagRotVPhiGamma092}) restores accurate estimates of both polar and azimuthal velocity and magnetic perturbations, notably near $r=a_s$.

One can also calculate higher-order flows. While $\mathfrak{u}_r$ and $\mathfrak{u}_r^{(o)}$ can be calculated using the flow (\ref{syst:psi}) and the particular solution of appendix~\ref{sec:diffuWekL}, calculating the bulk flows $\mathcal{U}$ and $\mathcal{U}^{(o)}$ is only tractable by discarding the particular solution and by using the simpler rotation-perturbed flow (\ref{syst:Lamb2}). Assuming that they are rotation-perturbed potential flows (as we did for $\boldsymbol{U}_1$ in \S\ref{sec:InvBulk}), we obtain explicit analytical expressions for them. For instance, $ \mathcal{U}_\theta$ can be simplified in the limit $\Lambda_l \gg 1$, reading
\begin{eqnarray}
 \mathcal{U}_\theta =\frac{3 }{8} \frac{L_\eta \omega}{a_s r^3} \frac{(a_s^3+2a^3 r^3) \sqrt{2 \Lambda_l}}{(1-a^3)^2} \frac{\tilde{\gamma}_+(1-3 \gamma \cos \theta)+6 \mathrm{i} \gamma \cos(\theta) \sqrt{\gamma \cos \theta +1} }{\sqrt{(\gamma \cos \theta +1)(\gamma \cos \theta -1)}} \sin \theta , \qquad 
\end{eqnarray}
with $\tilde{\gamma}_+=\sqrt{(\gamma \cos \theta -1)}+ \sqrt{- (\gamma \cos \theta +1)}$. We found good agreement between the bulk numerical results (orange-dotted) and these higher-order flows (light red) for the azimuthal component, as shown in the right inset of figure~\ref{fig:xRyViscoMagRotVPhiGamma01}a.

Direct evaluation of forces from the flow (\ref{syst:psi}) is analytically intractable. For $\gamma \ll 1$, \textsc{dns} agree with the simpler flow (\ref{syst:Lamb2}), which can therefore be used to compute magnetic-tension and viscous tangential stresses. The \textit{huBL} approach, previously based on the mean value of $B_{0r}^2$, becomes less straightforward for terms involving $\gamma\cos\theta$. To illustrate this, the surface integration is simplified in the limit $\Lambda_l\gg1$. For a monopole field at $r=a_s$, $\Lambda_l=\Lambda$ is constant, and with no-slip conditions one obtains (for $\Lambda_l\gg1$, $\breve{\mu}_s=\breve{\eta}_s=1$)
\begin{eqnarray}
  \left.  \frac{\Delta \omega_{mt}}{\omega} \right|_{r=a_s} &=&\frac{3}{4}  \frac{\rho_f}{\rho_s} \frac{ L_\eta}{ a_s } \frac{(1-\mathrm{i})(1-\gamma^2/5) + (\mathrm{i} \sqrt{2 \Lambda}+(1-\mathrm{i}) \sqrt{P_m} )(1-\gamma^2/8) }{1-a^3}  , \label{eq:gis01}\\
      \left.  \frac{\Delta \omega_v}{\omega} \right|_{r=a_s}&=& - \frac{3}{4} \frac{\rho_f}{\rho_s} \frac{L_\nu }{a_s }  \frac{1-\mathrm{i}}{1-a^3} (1-\gamma^2/8) , \label{eq:gis02}
\end{eqnarray}
where (\ref{eq:gis01}) generalises (\ref{eq:asymBG}) by including viscous and rotational effects, and (\ref{eq:gis02}) extends the hydrodynamic viscous detuning to rotation (e.g. equation~\ref{eq:disSh2}). Equation~(\ref{eq:gis01}) therefore gives the magnetic-tension force including confinement, viscosity, magnetic field and rotation. In the unbounded limit, equation~(\ref{eq:gis02}) recovers the result (\ref{eq:rotcorr1}) of \cite{smylie2000the}, including its rotational dependence. Note that the rotational dependence $1-\gamma^2/8$ of the leading-order term of (\ref{eq:gis01})-(\ref{eq:gis02}) becomes $1-6\gamma^2 a^3(1+a)  (1+a+a^2)^{-1}/5$ when using the flow $\boldsymbol{U}_2$ given by equation~(\ref{eq:Busse_ordre4}). Rotation decreases the magnetic tension and the tangential viscous stress (and increases Ohmic and viscous dissipation).

For a uniform axial field, the integral can be evaluated exactly using an integrand simplified in the limit $\Lambda_l \gg 1$. With $\Lambda = \langle \Lambda_l \rangle = \Lambda_z/3$, analogous expressions are obtained, where, for instance, the leading-order term $\sqrt{2 \Lambda}$ in (\ref{eq:gis01}) is replaced by $3 \sqrt{2 \Lambda_z}/8$, thereby recovering the ratio between (\ref{syst:exint}) and (\ref{eq:exact}) in this limit. As before, the total dissipation can then be estimated by multiplying (\ref{eq:gis01}) and (\ref{eq:gis02}) by $3 \mathcal{I}_{mt}^{\mathrm{pol}} /(2(1-a^3)) $ and $3/(2(1-a^3))$, respectively. Performing the same exact integration with the simplified integrand in the limit $\Lambda_l \gg 1$ further shows that both viscous and magnetic torques vanish. Replacing $\gamma\cos\theta$ by $\gamma$, the \textit{huBL} approach reproduces equations~(\ref{eq:gis01})--(\ref{eq:gis02}) and therefore provides accurate stress estimates in this limit, but not the torques (\S\ref{sec:huBL_sMat}).

\section{Conclusion} \label{sec:discon}
\subsection{Summary}

\begin{figure}
    \centering
    \includegraphics[trim=0 0 0 0,clip,width=0.75\linewidth]{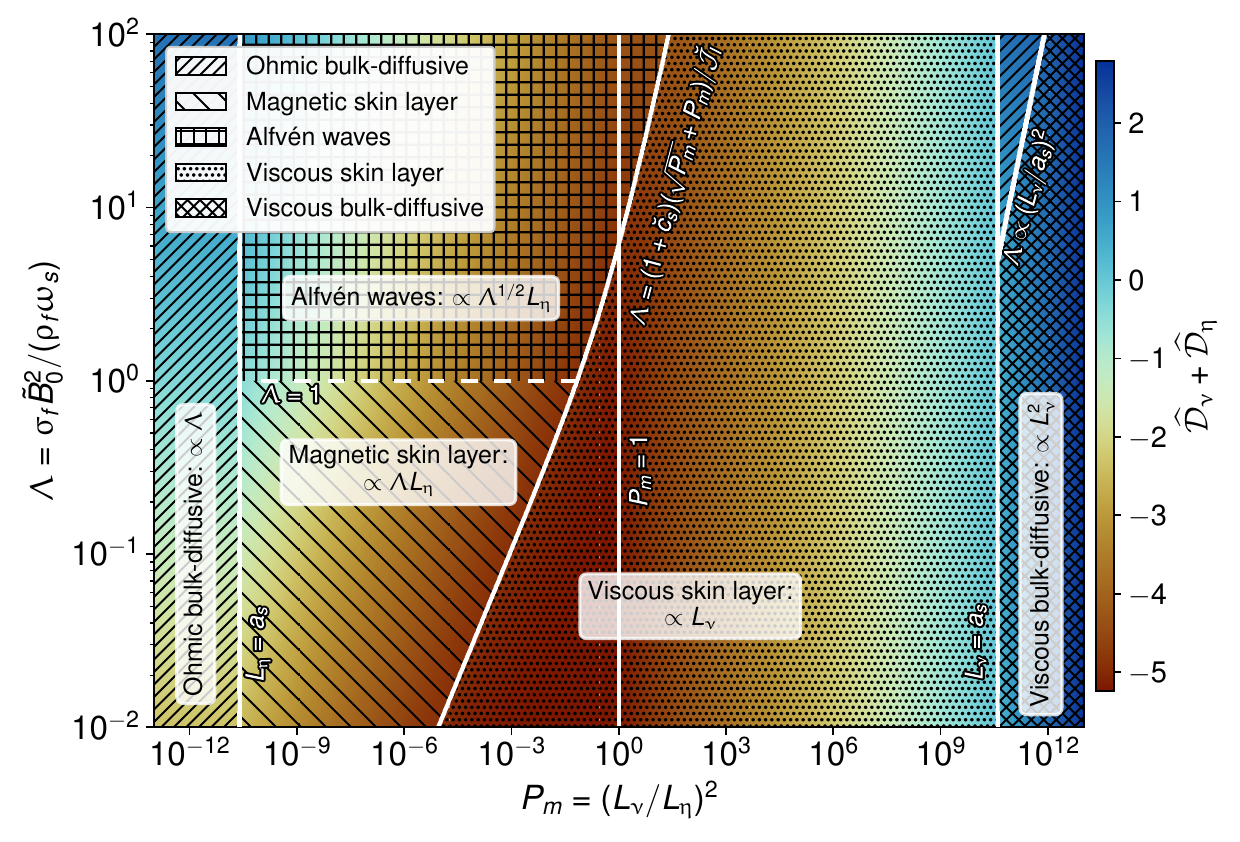}
\caption{Schematic diagram. For $P_m<1$ ($P_m>1$), $L_\eta$ ($L_\nu$) varies at fixed $L_\nu=10^{-6}$ ($L_\eta=10^{-6}$). Colour (log$_{10}$): sum of (\ref{eq:sfPoJ}) and (\ref{eq:imB1}), with formal continuation for $\Lambda_z \gg 1$, in bulk-diffusive regimes ($L_\nu\gtrsim a_s$ or $L_\eta\gtrsim a_s$ in the large-gap limit), and imaginary parts of (\ref{eq:extBuff}) and (\ref{eq:extBuffvisc}), weighted by $3\mathcal{I}_{mt}^{\mathrm{pol}}/[2(1-a^3)]$ and $3/[2(1-a^3)]$. Hatching: the dominant mechanism. $\Lambda\approx1$ illustrates the transition from the magnetic-skin-layer to the Alfv\'en-wave regime within the local range $L_A\ll D$; reflections and global cavity resonances are excluded. For $a_s=0.2$, $a_m=1.2$, $\gamma=0$, $\rho_f=\rho_s$, $\breve{\mu}_s=\breve{\eta}_s=f_\Lambda=1$, $\mathcal{I}_{mt}^{\mathrm{pol}}=3/5$.}
  \label{fig:reg}
\end{figure}

The fluid response separates into in-phase and out-of-phase components, corresponding to the real and imaginary parts of $\boldsymbol{\Delta\omega}$: the former defines the added-mass coefficient, whereas the latter quantifies viscous and magnetic drag. The principal analytical expressions derived throughout this work and their domains of validity are summarised in the lower part of table~\ref{tab:eqcomp}, alongside the classical theories listed in its upper part. Figure~\ref{fig:reg} synthesises the dominant dissipation mechanisms in the $(P_m,\Lambda)$ plane, and reveals the
transition curve $ \Lambda =(1+\breve{c}_s)(\sqrt{P_m}+P_m)/ \mathcal{I}_{mt}^{\mathrm{pol}}$, obtained by balancing the leading viscous and magnetic boundary-layer contributions, which separates viscously and magnetically dominated energy loss for a nearly diffusionless bulk. Because inertial and dissipative effects are encoded in the real and imaginary parts of the same vector $\Delta\boldsymbol{\omega}$, force measurements and energetic losses admit a unified interpretation. Within their weak-Lorentz domain, the quasi-static MHD asymptotics derived in appendix~\ref{appA} complete the large-diffusivity branches relevant to shallow conducting subsurface oceans (figure~\ref{fig:planets}).

\begin{table}
  \begin{center}
\def~{\hphantom{0}}
  \begin{tabular}{lcccccc}
    & Equations &  Viscous  & Confinement  & Rotation & Magnetic & Equatorial \\ 
 \cite{stokes1851effect}: $C_a$, $F^d_\nu$ & (\ref{eq:StokE59})   & $\checkmark$   & $\checkmark$  & $\times$ & $\times$  & $\times$ \\
 \cite{singh1965drag}: $C_a$, $F^d_\nu$, $F^d_\eta$ & (\ref{eq:FStomag})-(\ref{eq:FStomag2})   & $\checkmark$   & $\times$  & $\gamma \ll 1$ & $L_\eta \gg a_s$  & $\times$ \\
   \cite{busse1974free}: $C_a^i$ & (\ref{eq:Ca_Busse}) & $\times$   & $\checkmark$ & $\gamma < 1$ & $\times$ & $\times$\\    
     \cite{buffett1995magnetic}: $\Delta \omega_{mt}$ & (\ref{eq:Bufextde})$^a$ & $\times$   & $\times$ & $\times$ & $L_\eta \ll a_s$ & $\times$ \\ 
    SM98$^b$: $C_a^i$, $\Delta  \omega_{v}$ & (\ref{eq:smh})-(\ref{eq:disSh1Smylie2})  & $L_\nu \ll a_s$   & $\checkmark$ & $\gamma < 1$  & $\times$ & $\checkmark$  \\ \hline
  $C_a^i$ & (\ref{eq:OURCa}) & $\times$    & $\checkmark$ & $\gamma \ll 1$  & $\times$ & $\checkmark$   \\ 
  $\mathcal{D}_\nu$ of the bulk & (\ref{eq:sfPoJ0})   & stress-free  & $\checkmark$& $\gamma \ll 1$  & $\times$ & $\checkmark$  \\
   $\Delta  \boldsymbol{\omega}_{mt}$ & (\ref{eq:Bufextde}) & $\times$   & $\checkmark$   & $\gamma \ll 1$ & $L_\eta \ll a_s$ & $\checkmark$  \\ 
   $\mathcal{D}_\eta$ at $L_\eta \ll a_s$ & (\ref{eq:pmBatch21}) $^c$ & $\times$   & $\checkmark$ &  $\gamma \ll 1$   &  $L_\eta \ll a_s$ & $\checkmark$ \\  
   $\Delta  \boldsymbol{\omega}_{mt}$ \& $\Delta  \boldsymbol{\omega}_{v}$ & (\ref{eq:extBuff}) - (\ref{eq:extBuffvisc})& $L_\nu \ll a_s$   & $\checkmark$  &  $\gamma \ll 1$  &  $L_\eta \ll a_s$  & $\checkmark$  \\  
  $\mathcal{D}_\nu$ at $L_\nu \ll a_s$  &(\ref{eq:Fv_pola})-(\ref{eq:denueE}) & $L_\nu \ll a_s$   & $\checkmark$ & $\gamma \ll 1$ & $\times$ & $\checkmark$ \\ 
 $\mathcal{D}_\eta$ at $L_\eta \gg a_s$ & (\ref{eq:imB1}) $^d$ & $\times$   & $\checkmark$  & $\gamma \ll 1$  & $L_\eta \gg a_s$ & $\checkmark$ \\
  \end{tabular}
  \caption{First five rows: previous theories. Others: our expressions, with their validity domains. For the added-mass force ($C_a$, $C_a^i$), magnetic-tension and viscous-stress forces ($\Delta\boldsymbol{\omega}_{mt}$, $\Delta\boldsymbol{\omega}_{v}$), and the associated dissipation ($\mathcal{D}_\nu$, $\mathcal{D}_\eta$). Equatorial modes: $\gamma=0$ here, but not for SM98. $^a$At $a=\gamma=0$ and $\breve{c}=1$. $^b$From \cite{smylie1998viscous,smylie2000the}. $^c$ For $\Lambda_z \leq \mathcal{O}(1)$. $^d$ Exact within the quasi-static, weak-Lorentz-force limit for the non-rotating polar mode; rotation-dependent and equatorial terms are estimates.
  } 
  \label{tab:eqcomp}
  \end{center}
\end{table}

In the large magnetic-diffusivity limit, the present theory provides the confined counterpart of the classical unbounded oscillatory solution of \citet{singh1965drag}, and complements the finite-field inductionless extension of \citet{motz1966wave,motz1966magnetohydrodynamic}. The large-diffusivity branch of figure~\ref{fig:reg} should therefore be interpreted as the continuation of the inductionless asymptotic theory. This regime should not be confused with the strong-field studies of \citet{yonas1967measurements,maxworthy1968experimental,sekhar2005magnetohydrodynamic}, for a steadily translating sphere, in which the Lorentz force modifies the global flow and pressure distribution, producing a drag proportional to $\Lambda^{1/2} \propto \tilde{B}_0$. Our theoretical framework also predicts such a $\Lambda^{1/2}$ scaling within the local Alfv\'en-wave regime ($\Lambda\gg1$ with $L_A\ll D$), where the drag scales as $\Lambda^{1/2}L_\eta$, in agreement with the \textsc{dns} of P26. In this limit, the drag asymptotically recovers the Alfv\'en-wing scaling for a conducting obstacle moving through a magnetised plasma \citep{drell1965drag,neubauer1980nonlinear}, also recently retrieved in global simulations of \citep{corso2026simulations}. Moreover, when normalised by the magnetic energy density, this asymptotic drag law recovers the mathematical structure of the classical Stokes law, with the Alfv\'en number replacing the radius-based Reynolds number. This correspondence highlights the analogous roles played by Alfv\'en-wave radiation and viscous diffusion as momentum-transport mechanisms in ideal MHD and low-Reynolds-number hydrodynamics, respectively. In the opposite non-magnetic limit ($\Lambda\ll1$), the classical analysis of \citet{stokes1851effect} continuously connects the viscous skin-layer and bulk-diffusive regimes (appendix~\ref{app:revisit_stu}).

\subsection{Perspectives}
A first limitation concerns equatorial modes at non-negligible rotation, for which no analytical base flow is available. Further analytical developments are therefore required to obtain a tractable description of rotation-modified equatorial modes. A promising route is to build upon the rotation-corrected basic flow of \cite{smylie1998viscous}, including rotational, magnetic and viscous effects. Their formulation, close to an inertial-mode basis, would provide a route to exploring strong-rotation regimes ($\gamma>1$), although inviscid inertial modes in a spherical shell are singular and only regularised by viscosity. 

Another limitation arises from the small-amplitude assumption underlying both the theory and the approximate boundary condition implemented in \textsc{xshells}. Although well suited to planetary applications, this assumption becomes restrictive for laboratory experiments, where the approximate boundary condition overestimates the Ohmic dissipation at larger amplitudes (see P26). Additional \textsc{comsol} simulations, using an Arbitrary-Lagrangian-Eulerian (ALE) formulation to model the sphere motion exactly, nevertheless show good agreement for modest deformations. Extending the present framework to finite amplitudes, particularly for the Ohmic dissipation, remains an important direction for future work.

For planetary applications, a final point concerns translation of the outer solid. The present problem prescribes the motion of the inner sphere relative to a fixed outer boundary. For Slichter modes, this corresponds to the fixed-mantle approximation of B74 and \cite{rieutord2002slichter}, whereas a free planetary mode also involves mantle translation. In the idealised non-rotating three-layer model of \cite{grinfeld2005motion}, however, including this motion changes the Slichter period by less than $0.01\,\%$ for the Earth and only about $0.1\,\%$ for Mercury, despite a mantle-displacement amplitude of about $10\,\%$ of the inner-core displacement in the latter case. Thus, within that model, the fixed-mantle approximation is excellent for the Slichter period, particularly for the Earth. More generally, including subsurface-ocean configurations in which the solid layers bounding the liquid may move relative to one another, the force laws derived here provide the local response to the prescribed relative motion. For a complete planetary mode, the associated pressure forces must be coupled to the appropriate global dynamics of these solid layers \citep[for the terrestrial Slichter problem,][]{grinfeld2005motion,grefflefftz2007fluid}.

Experimental tests of the predicted drag appear feasible. Table~\ref{tab:em_contrasts} indicates that a $50$~cm-scale Galinstan filled cavity containing an oscillating iron or titanium sphere could reach $\Lambda=\mathcal{O}(1)$ and probe the onset of wave-like magnetic penetration, while providing moderate to strong electromagnetic contrasts. For the tabulated $L_\eta=9$~cm and $P_m\ll1$, the full wavenumber (\ref{eq:STBU}) gives $L_A\simeq0.20$-$0.41$~m over $\Lambda_l=1$-$2$, comparable to the apparatus scale. Such an experiment would be naturally suited to probing the crossover and the onset of reflected wave and global cavity effects. A clean test of the local radiative drag law would additionally require the material dependent drag crossover to be reached and the actual fluid gap to satisfy $D\gg L_A$. Varying the field strength, gap width and material properties would allow these regimes to be explored \citep[e.g.][]{motz1966wave,alboussiere2011experimental}.

\begin{table}
\centering
\begin{tabular}{lcccccc}
 & Core & Icy ocean & Iron/sea & Iron/GaInSn & MnZn-like/GaInSn & Ti/GaInSn \\
   $T_s=2\pi/\omega_s$
 & $4$~h
& $10$~h
& $0.1\,\mathrm{s}$
& $ 0.1\,\mathrm{s}$
& $ 0.1\,\mathrm{s}$
& $ 0.1\,\mathrm{s}$ \\
  $\rho_f \, (\mathrm{kg\,m^{-3}})$
 & $10^{4}$
& $10^{3}$
& $10^{3}$
& $6.4\times10^{3}$
& $6.4\times10^{3}$
& $6.4\times10^{3}$ \\ 
$\sigma_f \, (\mathrm{S\,m^{-1}})$ 
& $10^{6}$
& $4$
& $4$
& $3.46\times10^{6}$
& $3.46\times10^{6}$ 
& $3.46\times10^{6}$ \\
$\sigma_s \, (\mathrm{S\,m^{-1}})$ 
& $10^{6}$
& $4\times 10^{-3}$
& $10^{7}$
& $10^{7}$
& $10^{2}$ 
& $5.8\times10^{5}$ \\
$\tilde{B}_0 (\mathrm{mT})$
 & $2$
& $<10^{-3}$
& $0$--$500$
& $0$--$500$
& $0$--$500$
& $0$--$500$ \\ 
$\sigma_s/\sigma_f$
& $1$
& $10^{-3}$
& $2.5 \times 10^{6}$
& $2.9$
& $2.9 \times 10^{-5}$
& $0.17$ \\
$\breve{\mu}_s= \mu_s/\mu_f$
& $1$
& $1$
& $800$
& $800$
& $80$
& $1$ \\
$\breve{\eta}_s=\eta_s/\eta_f$
& $1$
& $10^{3}$
& $5 \times 10^{-10}$
& $4 \times 10^{-4}$
& $4 \times 10^{2}$
& $6$ \\
$\Lambda=\sigma_f\tilde{B}_0^2/(\rho_f\omega_s)$
& $1$
& $<10^{-10}$
& 0--$2\times10^{-5}$
& 0--$2$
& 0--$2$
& 0--$2$ \\
$L_\eta=\sqrt{2 \eta_f/\omega_s}$ 
& $60$~m
& $50$~km
& $80$~m
& $9$~cm
& $9$~cm
& $9$~cm \\
\end{tabular}
\caption{Representative parameters for planetary cores, icy oceans and laboratory setups combining seawater (sea) or room-temperature GaInSn eutectic (Galinstan) with pure-iron, MnZn-ferrite-like or Ti-6Al-4V spheres. Iron and MnZn-like represent high-permeability solids with high and low conductivity, respectively; all three are idealised linear models, rather than manufacturer specifications, with constant effective permeability ratios $\breve{\mu}_s=800$, $80$ and $1$. Laboratory fields span $\tilde{B}_0=0$--$500$~mT; field dependence, hysteresis and saturation are neglected.}

\label{tab:em_contrasts}
\end{table}

Further extensions include non-spherical inner bodies, beginning with spheroids using the potential flow of \cite{lai1972stokes}, and boundary layers along spheroidal interfaces \cite[e.g.][]{cebron2021mean}. Incorporating density stratification and gravity-wave coupling \citep{voisin2024added} would also be relevant for planetary interiors such as Mercury's stratified core \citep{seuren2023effects}. Finally, the present Alfv\'en-wave drag law is local and assumes an attenuation length much smaller than the fluid gap. When this condition fails, wave reflections and global cavity resonances, demonstrated experimentally by \citet{motz1966wave}, fall outside the model; their amplitude and absorption also depend on the electromagnetic properties of the bounding solids \citep{schaeffer2016electrical}. Wave--wave interactions are likewise neglected. Including these effects would provide a global description of wave-mediated energy transport.

The present linear theory also predicts zero torques, whereas simulations exhibit finite mean torques in the presence of rotation effects, indicating higher-order effects in oscillation amplitude. Rapid oscillations generate steady streaming at order $\mathcal{O}(\epsilon^2)$, producing mean flows and modified dissipation \cite[e.g.][]{riley2001steady,rednikov2011acoustic,cebron2021mean}. Extending this study to include such nonlinear corrections would clarify the role of oscillations and rotation-induced mean secondary flows.

To facilitate reuse of the theory, the accompanying Zenodo repository provides machine-readable implementations of all analytical expressions derived in this work, enabling their direct evaluation for parameter values within their stated asymptotic domains and comparison with future experiments and numerical simulations. More broadly, this work unifies previously disconnected asymptotic theories of oscillatory magnetohydrodynamic drag into a single analytical description of the confinement, rotation and magnetic-coupling regimes relevant to planetary interiors, icy-moon oceans and liquid-metal experiments, within their stated asymptotic domains.

\backsection[Supplementary data]{\label{SupMat}Supplementary material is available as a pdf file.}

\backsection[Acknowledgements]{The authors thank Nathana\"el Schaeffer for his assistance with the \textsc{xshells} code, and Nathana\"el Schaeffer, Dominique Jault and Mioara Mandea for fruitful discussions. They also thank Romain Claveau for his careful reading of the manuscript.}

\backsection[Funding]{
PP and DC received funding from the ERC under the European Union’s Horizon 2020 research and innovation program via the THEIA grant no. 847433. PP received funding from the European Research Council (ERC) under the European Union Horizon 2020 research and innovation programme via GRACEFUL Synergy Grant No. 855677. Most of the computations were performed using the GRICAD infrastructure (\url{https://gricad.univ-grenoble-alpes.fr}), and HPC resources (Jean Zay V100) of IDRIS under
allocation AD010413621 attributed by GENCI (Grand Équipement National de Calcul Intensif).}

\backsection[Declaration of interests]{The authors report no conflict of interest.}

\backsection[Data availability statement]{
Simulations were performed using the \textsc{XSHELLS} code, which is
freely available at \url{https://nschaeff.bitbucket.io/xshells/}. The
accompanying Zenodo repository \citep[\href{https://doi.org/10.5281/zenodo.20928186}{doi:10.5281/zenodo.20928186}]{zenodo} contains the numerical data underlying the figures, the Python scripts used to reproduce all
figures, and implementations of all analytical expressions derived in this article, including the complete boundary-condition prefactors, asymptotic coefficients and closed-form solutions omitted from the manuscript for conciseness. The repository therefore enables complete reproduction and direct reuse of both the analytical and numerical results presented here.
}

\backsection[Author ORCIDs]{D. Cébron, \url{https://orcid.org/0000-0002-3579-8281}; P. Personnettaz, \url{https://orcid.org/0000-0001-8990-0643}}

\backsection[Author contributions]{DC conceived the study, secured the computational resources, and developed the theoretical framework. PP performed the \textsc{dns}, compared them to the theory, and produced the figures. Both authors interpreted the results, contributed to the manuscript, and approved the final version.}

\appendix

\section{Revisiting previous studies} \label{app:revisit_stu}

This appendix revisits several classical analytical results and derives explicit solutions and asymptotic expansions used throughout the paper, most notably a closed-form solution for the bounded oscillatory Stokes flow. It also provides complementary bulk-diffusive asymptotic solutions, for which viscous or magnetic diffusion extends across the relevant system scale, of the order $\ell=\min(a_s,D)$, rather than remaining confined to thin boundary layers.

\subsection{Explicit closed-form solution for the bounded oscillatory Stokes flow} \label{sec:ExpStoU}

\begin{figure}
    \centering
    \includegraphics[trim=10 15 10 29,clip,width=0.85\linewidth]{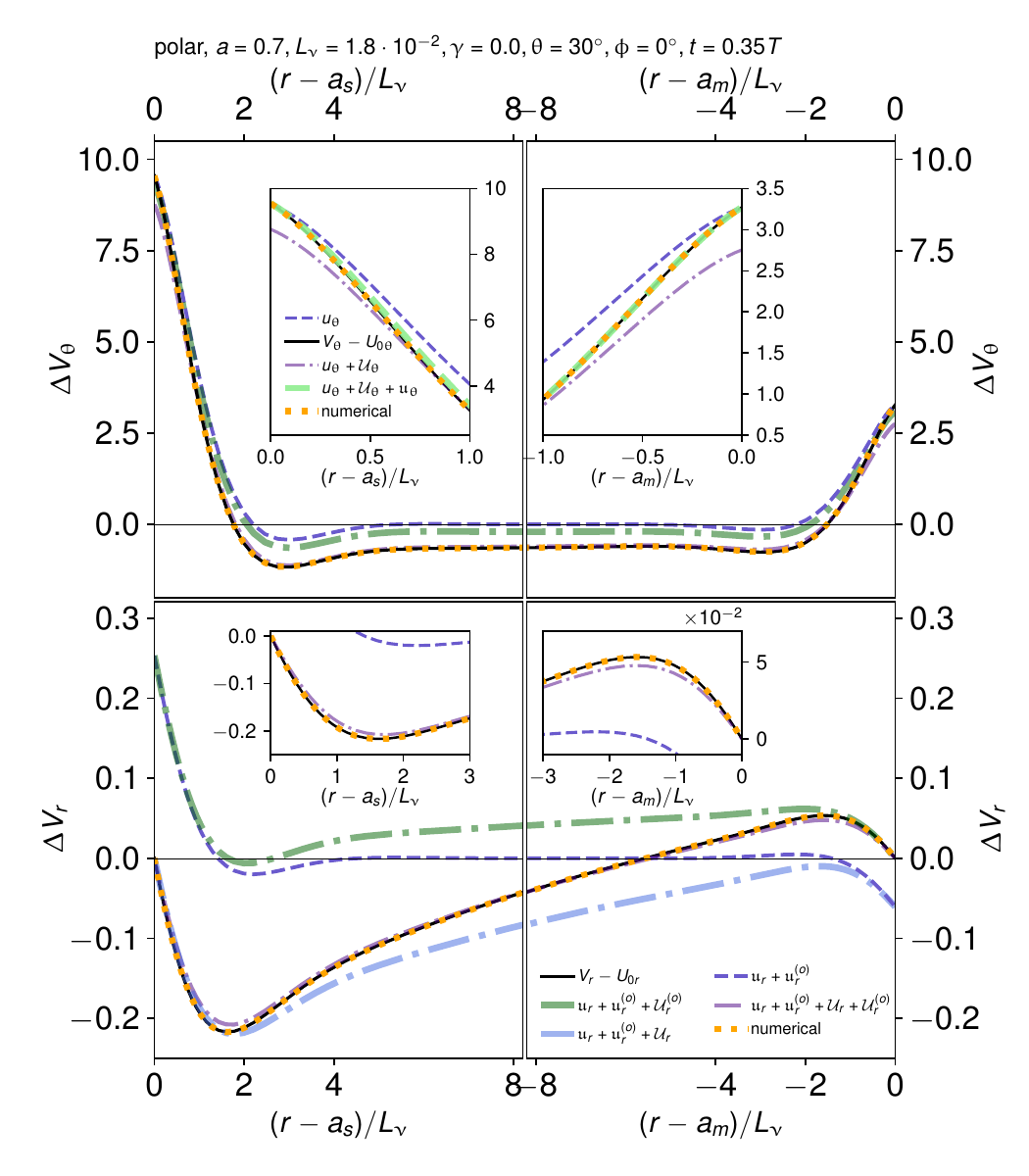}
    \caption{Meridional $\Delta V_\theta$ (top) and radial $\Delta V_r$ (bottom)  velocity perturbation profiles at $\theta=30^{\circ}$, $\phi=0$, $L_\nu=0.02$, $\gamma=0$.  
    \textsc{xshells} (orange-dotted), Stokes solution (black), viscous BLT equation~(\ref{eq:visBLst}) (purple dashed), viscous boundary layer and inner secondary bulk flow (light-blue dash-dotted), boundary-layer solutions and outer-secondary bulk flow (dark-green dash-dotted), boundary-layer solutions and inner- and outer-secondary bulk flows (purple dash-dotted).}
  \label{fig:Stokes}
\end{figure}

Considering the non-rotating problem, \cite{stokes1851effect} derived the bounded oscillatory solution through an implicit system (his equations 54--57), but did not 
provide an explicit closed-form solution. We integrate Stokes' implicit system analytically to obtain an explicit velocity field valid for arbitrary gap width and oscillation frequency. This exact viscous solution is thus continuously valid from the thin-gap limit $a_m-a_s \ll L_\nu$ to the opposite large gap limit, as well as from the steady limit $\omega_s \to 0$ of large $L_\nu \gg a_m>a_s$ to the rapidly oscillating limit of small $L_\nu \ll a_s<a_m$. Using $U_0=\mathrm{i} \omega_s \epsilon$, such that the boundary velocity is $ \mathrm{d}_t \boldsymbol{\delta}=U_0 \mathrm{e}^{\mathrm{i} \omega_s t} \hat{\boldsymbol{z}} $, we obtain the (dimensional) velocity field
\begin{eqnarray}
[V_r,V_\theta]&=& \frac{ 3 U_0 a_s}{\lambda_\nu^2r^3} \left[ \frac{ \mathcal{N}_r+\tilde{\mathcal{N}}_r }{\breve{\mathcal{D}}+\tilde{\mathcal{D}}} \cos \theta,  \frac{ \mathcal{N}_\theta +\tilde{\mathcal{N}}_\theta }{\breve{\mathcal{D}}+\tilde{\mathcal{D}}} \frac{\sin \theta}{2} \right]   \mathrm{e}^{\mathrm{i} \omega_s t} , \label{eq:VnRSTO1}
\end{eqnarray}
with $\lambda_\nu=(1+\mathrm{i})/L_\nu$, where the tilde indicates terms that are (exponentially) negligible in the large gap limit  $a_m-a_s \gg L_\nu$ (e.g.\ discarded by Stokes in \S22, p.33), and noting
\begin{eqnarray}
\tilde{\mathcal{N}}_r&=&\mathcal{M}_s^+ \mathrm{e}^{\lambda_\nu(2 a_s-a_m-r)} - \mathcal{M}_m^- \mathrm{e}^{\lambda_\nu( a_s-2a_m+r)} - \mathcal{K}^+ \mathrm{e}^{2 \lambda_\nu ( a_s-a_m)} +2 a_m \lambda_\nu^3 r^3 \mathrm{e}^{\lambda_\nu(a_s-a_m)} , \qquad \\
\mathcal{N}_r &=& \mathcal{M}_s^- \mathrm{e}^{\lambda_\nu(r-a_m)}- \mathcal{M}_m^+ \mathrm{e}^{\lambda_\nu(a_s-r)} + \mathcal{K}^- , \\
\tilde{\mathcal{D}}&=& [(a_s^3-a_m^3) \lambda_\nu^2-3(a_s^2+a_m^2)\lambda_\nu+3(a_s-a_m)] \mathrm{e}^{2 \lambda_\nu (a_s-a_m)}+12 \lambda_\nu a_s a_m \mathrm{e}^{\lambda_\nu(a_s-a_m)} , \\
\breve{\mathcal{D}}&=& (a_m^3-a_s^3) \lambda_\nu^2-3(a_s^2+a_m^2) \lambda_\nu+3(a_m-a_s)
\end{eqnarray}
for the radial flow $V_r$, and
\begin{eqnarray}
\tilde{\mathcal{N}}_\theta&=& \breve{\mathcal{M}}_s^+ \mathrm{e}^{\lambda_\nu (2 a_s-a_m-r)} + \breve{\mathcal{M}}_m^- \mathrm{e}^{\lambda_\nu ( a_s-2a_m+r)} -\breve{\mathcal{K}}^+ \mathrm{e}^{2 \lambda_\nu (a_s-a_m)} -4 a_m \lambda_\nu^3 r^3 \mathrm{e}^{\lambda_\nu (a_s-a_m)}  , \qquad \\
\mathcal{N}_\theta &=&  - \breve{\mathcal{M}}_s^- \mathrm{e}^{\lambda_\nu (r-a_m)}- \breve{\mathcal{M}}_m^+ \mathrm{e}^{\lambda_\nu (a_s-r)} + \breve{\mathcal{K}}^-
\end{eqnarray}
for the meridional flow $V_\theta$, where
\begin{eqnarray}
\mathcal{M}_j^{\pm} &=& a_m (\lambda_\nu^2 a_j^2 \mp 3 \lambda_\nu a_j +3)(\lambda_\nu r \pm 1) ,   \\
\breve{\mathcal{M}}_j^{\pm} &=& a_m (\lambda_\nu^2 a_j^2 \mp 3 \lambda_\nu a_j +3)(\lambda_\nu^2r^2\pm \lambda_\nu r + 1) ,   \\
\mathcal{K}^{\pm} &=& \frac{\lambda_\nu^2 a_s^2 \mp 3 \lambda_\nu a_s+3}{3}[\lambda_\nu^2(a_m^3-r^3)+3 a_m(1 \pm \lambda_\nu a_m)], \\
\breve{\mathcal{K}}^{\pm} &=& \frac{\lambda_\nu^2 a_s^2 \mp 3 \lambda_\nu a_s+3}{3}[\lambda_\nu^2(a_m^3+2r^3)+3 a_m(1 \pm \lambda_\nu a_m)] , \label{eq:stoENG}
\end{eqnarray}
such that $(\mathcal{M}_j^{\pm},\mathcal{K}^{\pm})$ are only used for $V_r$, and $(\breve{\mathcal{M}}_j^{\pm},\breve{\mathcal{K}}^{\pm})$ are only used for $V_\theta$. Equations~(\ref{eq:VnRSTO1})-(\ref{eq:stoENG}) provide explicitly the complete bounded Stokes field. This solution, shown as the black solid line in figure~\ref{fig:Stokes}, agrees perfectly with \textsc{xshells} \textsc{dns} (orange-dotted curves) across the full range of radii, for both the meridional and radial components.

\subsection{Stokes force on a bounded oscillating no-slip sphere} \label{sec:StokEq59}
\begin{figure}
\centering
\begin{subfigure}[b]{0.52\textwidth}
\centering
\includegraphics[trim=10 15 10 10,clip,width=1\linewidth]{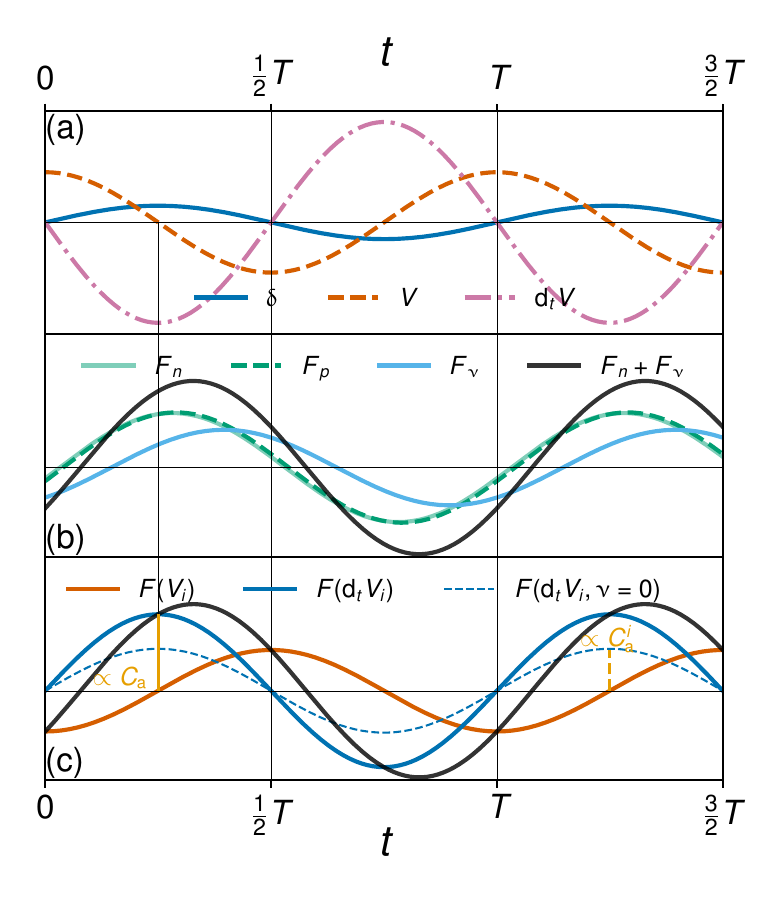}
\end{subfigure}
\caption{Time evolution of the (a) harmonic displacement (dark blue) and its two first time derivatives (orange dashed and pink dash-dotted), (b) normal force (light green), pressure force (green dashed) and viscous (cyan) forces and total force (black); and (c) velocity-dependent (orange) and acceleration-dependent (blue) parts of the total force. The  total (and inviscid part) of the acceleration contribution is shown as a blue solid (and dashed) curve. Yellow segments mark quantities proportional to $C_a$ and $C_a^i$.}
\label{fig:sketchFSto}
\end{figure}

In the absence of rotation and magnetic effects, the viscous flow driven by the polar mode with no-slip conditions was derived by \cite{stokes1851effect}, together with the associated forces. In the unbounded limit, his equation~(51) yields the dimensional drag force ${\boldsymbol{F}}_n+{\boldsymbol{F}}_v$ with
\begin{eqnarray}
   {\boldsymbol{F}}_n+{\boldsymbol{F}}_v =- {m}_{fs} \left[ \frac{1}{2}+\frac{9}{4} \frac{{L}_\nu}{{a}_s} \right] \mathrm{d}_t {\boldsymbol{V}}_s-6 \pi {a}_s \rho_f \nu \left( 1+\frac{{a}_s }{{L}_\nu} \right)  {\boldsymbol{V}}_s ,  \label{eq:FSto}
\end{eqnarray}
where the normal component ${\boldsymbol{F}}_n$ and tangential component ${\boldsymbol{F}}_v$ are respectively given by
\begin{eqnarray}
    [{\boldsymbol{F}}_n,{\boldsymbol{F}}_v] &=& - {m}_{fs} \left[ \frac{1}{2}+\frac{3}{4} \frac{{L}_\nu}{{a}_s},\frac{3}{2} \frac{{L}_\nu}{{a}_s} \right] \mathrm{d}_t {\boldsymbol{V}}_s-2 \pi {a}_s \rho_f \nu \left( \left[1-\frac{4}{3},\frac{10}{3} \right]+\frac{{a}_s }{{L}_\nu} [1,2] \right)  {\boldsymbol{V}}_s  , \qquad \, \label{eq:FStoo}
\end{eqnarray}
with ${\boldsymbol{F}}_n={\boldsymbol{F}}_p+{\boldsymbol{F}}_{vn}$, where the force contribution ${\boldsymbol{F}}_v$ only arises from viscous stresses, and where ${\boldsymbol{V}}_s=\mathrm{d}_t {\boldsymbol{\delta}}$ is the uniform inner-boundary velocity, ${L}_\nu=\sqrt{2 \nu/{\omega_s}}$ the viscous skin depth, and the fluid mass ${m}_{fs}= 4 \pi a_s^3 \rho_f/3$ displaced by the sphere. The viscous normal contribution ${\boldsymbol{F}}_{vn}$ enters only through the coefficient $-4/3$ in the term proportional to ${\boldsymbol{V}}_s$, and is therefore negligible in the rapid oscillation limit $L_\nu \to 0$, for which ${\boldsymbol{F}}_n \approx {\boldsymbol{F}}_p$. This behaviour is illustrated by the two green curves in the middle panel of figure~\ref{fig:sketchFSto}b, showing the temporal evolution of the force components.

In (\ref{eq:FSto}), the prefactor of $- {m}_{fs} \mathrm{d}_t {\boldsymbol{V}}_s$ defines the added-mass coefficient $C_a$. Regarding its viscous contribution, $2/3$ arises from $\boldsymbol{F}_v$, while the inviscid part $C_a^i=1/2$ fully determines the force ${\boldsymbol{F}}_p=-C_a^i {m}_{fs} \mathrm{d}_t {\boldsymbol{V}}_s$ in the inviscid limit. The corresponding acceleration force is shown by the blue-solid curve in figure~\ref{fig:sketchFSto}(c), with its inviscid component indicated by dashed blue lines. For a viscous fluid, one recovers $C_a=1/2+(9/4)\,{L}_\nu/{a}_s$ and the drag $-6 \pi {a}_s \rho_f \nu (1+{L}_\nu/{a}_s ) {\boldsymbol{V}}_s$ from \cite{stokes1851effect}, later extended to spheroids by \cite{lai1972stokes}. As emphasised therein, the fluid force decomposes into added-mass, steady drag, and history-force contributions, the latter associated with the $L_\nu$-dependent terms in (\ref{eq:FSto}).

\begin{figure}
\centering
\includegraphics[trim=0 15 10 10,clip,width=0.9\linewidth]{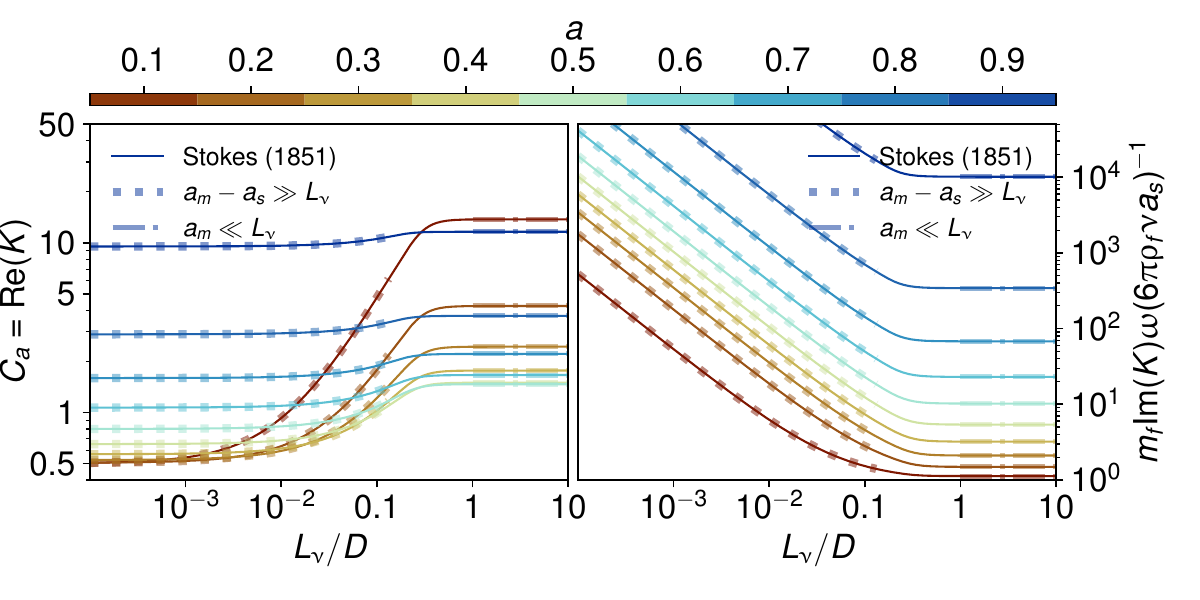}
    \caption{Approximations of the added-mass coefficient (left) and viscous contribution $\Imag(K)$ (right), as function of the viscous skin depth $L_\nu$ (and bounding effect $a$).
    Solid line: equation~(\ref{eq:StokE59}); dotted lines: equations (\ref{eq:Stob1}) and (\ref{eq:Stob2}); dot-dashed lines: equations (\ref{eq:dc_Caa7a}) and (\ref{eq:dc_Caa7b}), respectively for left and right images.}
    \label{fig:EqA}
\end{figure}

While lesser known \cite[][]{smylie1998viscous}, the bounding effect due to the spherical outer boundary has been obtained by \cite{stokes1851effect}, and his equation~(59) gives the complex constant $K$ related to the total force through ${\boldsymbol{F}}_n+{\boldsymbol{F}}_v={m}_{fs}[\Real(K) \mathrm{d}_t {\boldsymbol{V}}_s-\Imag(K) \omega_s {\boldsymbol{V}}_s]$, where 
\begin{eqnarray}
K&=&1-\frac{3 a_m}{2\lambda_\nu^2 a_s^2} \frac{\varkappa{(a_s,a_m)}-\varkappa{(a_m,a_s)}}{12 \lambda_\nu a_s a_m + \zeta{(a_s,a_m)}+\zeta{(a_m,a_s)}} , \label{eq:StokE59} \\
    \varkappa{(a_i,a_j)}&=&(\lambda_\nu^2 a_i^2+3\lambda_\nu a_i +3)(\lambda_\nu^2 a_j^2-3\lambda_\nu a_j+3) \mathrm{e}^{\lambda_\nu(a_j-a_i)} , \\
    \zeta{(a_i,a_j)}&=&[a_j(\lambda_\nu^2 a_j^2-3 \lambda_\nu a_j +3)-a_i(\lambda_\nu^2 a_i^2+3\lambda_\nu a_i+3)]\mathrm{e}^{\lambda_\nu(a_j-a_i)} ,
\end{eqnarray}
noting $\lambda_\nu=(1+\mathrm{i}) L_\nu^{-1}$, yielding $C_a=-\Real(K)$ and the viscous contribution from ${m}_{fs} \omega_s \Imag(K)=(8 \pi/3)\rho_f a_s^3 \nu L_\nu^{-2} \Imag(K)$. These are shown as solid lines in figure~\ref{fig:EqA}. In the limit $L_\nu \ll D$, (\ref{eq:StokE59}) reduces to equation~(61) of \cite{stokes1851effect}. 

Considering rapid oscillations and confinement effects, (\ref{eq:StokE59}) modifies (\ref{eq:FSto}) into
\begin{eqnarray}
    C_a&=&\frac{1}{2} \frac{1+2a^3}{1-a^3}+\frac{9}{4} \frac{1+a^4}{(1-a^3)^2} \frac{L_\nu}{a_s}-\frac{27a^3(1+a)^2}{8(1-a^3)^{4}}  \left[f_1 \frac{L_\nu^3}{a_s^3} +\frac{f_2}{1-a^3}\frac{L_\nu^4}{a_s^4} \right], \qquad \label{eq:Stob1} \\
   \frac{{m}_{fs} \Imag(K) \omega_s}{6\pi \rho_f \nu a_s}  &=&   \frac{1+a^4 }{(1-a^3)^2} \frac{a_s}{L_\nu}+\frac{1+2(a^3+a^5)+6a^4+a^8}{(1-a^3)^3}  + \frac{3a^3(1+a)^2}{2(1-a^3)^{4}}  f_1 \frac{L_\nu}{a_s} , \label{eq:Stob2}
\end{eqnarray}
with $f_1=2(1+a^4)+a+a^3$ and $f_2=1+4(a+a^7)+a^2+7(a^3+a^5)+10a^4+a^6+a^8$. In the unbounded limit $a=0$, these reduce to $1/2+9 L_\nu/(4 a_s)$ and $a_s/L_\nu+1$. In (\ref{eq:Stob1})-(\ref{eq:Stob2}), the leading viscous term (in $1+a^4$) arises solely from boundary layers, with $1+a^4$ reflecting the additive contributions of the inner and outer layers. For a stress-free outer boundary, $1+a^4$ should be replaced by $1$. Equations~(\ref{eq:Stob1})-(\ref{eq:Stob2}) accurately capture this regime, as illustrated by the dotted curves in figure~\ref{fig:EqA}. The out-of-phase component $\Imag(K)$ increases monotonically with confinement and decreasing viscous skin depth.

In the opposite slow oscillation limit $L_\nu \gg a_m $, equation~(\ref{eq:StokE59}) yields
\begin{eqnarray} \label{eq:dc_Caa7}
C_a&=&  \frac{81+119a+41a^2+56a^3+256a^4+280a^5+112a^6}{112\, (1+(7/4)a+a^2)^2 a (1-a)} , \label{eq:dc_Caa7a}\\
 \frac{{m}_{fs} \Imag(K) \omega_s}{6\pi \rho_f \nu a_s}   &=&   \frac{1+a+a^2+a^3+a^4}{(1-a)^3(1+(7/4)a+a^2)}  , \label{eq:dc_Caa7b}
\end{eqnarray}
extending the unbounded slow-oscillation limit of (\ref{eq:FSto}) to spherical confinement. Although obtained in the formal limit $L_\nu \gg a_m $, these $L_\nu$-independent expressions already approximate the full solution accurately for $L_\nu\gtrsim D$ in the cases shown in figure~\ref{fig:EqA} (dot-dashed curves); the exact expression~(\ref{eq:StokE59}) itself remains valid for arbitrary $L_\nu$ within the linear problem. For the friction term $\Imag(K)$, this regime implies linear scaling with viscosity, characteristic of a purely bulk dissipation (see section~\ref{sec:bulkrot}). In the limit $a_m \ll L_\nu$, $C_a$ attains a minimum $C_a \approx 1.46$ at $a=a_{min}$ (figure~\ref{fig:EqA}). Solving $\partial_a C_a=0$ from (\ref{eq:dc_Caa7}) yields a cumbersome exact expression, but a second-order Taylor expansion about $a\approx 1/2$ provides the accurate estimate $a_{min}=87133/169336 \approx 0.515$ (relative error $<10^{-4}$).

\subsection{Viscous force on oscillating stress-free spheres} \label{app:steLeal}
For a stress-free sphere, ${\boldsymbol{F}}_v=\boldsymbol{0}$ and ${\boldsymbol{F}}_n$ reads \citep{yang1991note,zhang2023drag}
\begin{eqnarray}
    {\boldsymbol{F}}_n=- C_a {m}_{fs}  \mathrm{d}_t {\boldsymbol{V}}_s- {a}_s \rho_f \nu  \, (4 \pi {\boldsymbol{V}}_s + \boldsymbol{F}_i)  , \label{eq:FSto2}
\end{eqnarray}
where $C_a=1/2$. The viscous part of ${\boldsymbol{F}}_n$ contains a non-Basset memory integral contribution $\boldsymbol{F}_i$, that vanishes for no-slip spheres and that is generally not in phase with ${\boldsymbol{V}}_s$ \citep{yang1991note}. Equation~(2.13) of \cite{zhang2023drag} provides a general expression for $\boldsymbol{F}_i$, which shows that $\boldsymbol{F}_i$ simplifies to $8 \pi {\boldsymbol{V}}$ and $\boldsymbol{0}$ in the respective limits of rapid and slow oscillations. The respective viscous force then reduces to the Levich drag force with the numerical prefactor $12 \pi$ \citep{levich1949motion,joseph2004dissipation},  and to the Stokes (Hadamard-Rybzynski) drag force with the numerical prefactor $4 \pi$ \citep{yang1991note}. In both cases, $2/3$ comes from normal viscous stress and $1/3$ from the viscously-modified pressure field.

\subsection{Revisiting the results of \cite{smylie1998viscous,smylie2000the}} \label{sec:appSmyli}
\cite{smylie1998viscous,smylie2000the} compute the flows associated with polar and equatorial translational modes, neglecting magnetic effects but retaining rotation for $\gamma<1$. Although inertial modes are not excited in this regime, they seek solutions of analogous structure (products of Legendre functions) and derive the flow driven by oscillations of a spheroidal inner boundary. While the rotation-perturbed flow (Eq.~\ref{syst:Lamb2}) is restricted to leading order in $\gamma$, their formulation includes higher-order corrections. To simplify the enforcement of boundary conditions, they neglect the small eccentricity of the inner boundary, but account for a mobile outer spherical boundary. The fluid is thus enclosed within a solid shell (of mass $M_s$ in their notation) whose motion ensures conservation of total linear momentum across the three layers. Viscous effects at the inner boundary are incorporated through an oscillatory Stokes--Ekman layer, assuming $L_\nu \ll a_s$. However, the viscous layer at the outer boundary is neglected. Although they derive the viscously-modified pressure $\chi'$, driven by the boundary-layer radial flow $u_r'$, they discard it relative to the inviscid contribution \cite[despite its central role in the total viscous force $\boldsymbol{F}_\nu$, see p.~355 of][]{batchelor1967introduction}. Thus, their pressure force recovers only the inviscid contribution $C_a^{i}$ to the added-mass, while their viscous force reduces to the tangential stress component $\boldsymbol{F}_v$ \cite[i.e.\ $2/3$ of $\boldsymbol{F}_\nu$, see][]{stokes1851effect}.

In the present formulation, the outer boundary is at rest in our working frame of reference; comparison with \cite{smylie1998viscous,smylie2000the}, who allow the outer solid shell to translate under total linear-momentum conservation, therefore corresponds to their fixed-shell limit $M_s\rightarrow\infty$, whose planetary interpretation is discussed in \S\ref{sec:phymodel}. Considering the three Slichter modes, \cite{smylie2000the} give the pressure force (equations 35-39), from which the viscous force can be obtained (equations~25-26 for the polar mode, and 27-28 for the two equatorial modes). In the limit of infinite mass $M_s$, these equations give then 
\begin{subequations}
\label{eq:smh}
\begin{equation}
    C_a^i=\frac{1}{2} \frac{1+2a^3}{1-a^3}-\frac{3\gamma^2}{20} \frac{2-5a^3}{(1-a^3)^2}  , \,  C_a^i =\frac{1}{2} \frac{1+2a^3}{1-a^3}+ \frac{\gamma}{4}\frac{1-4a^3}{1-a^3}-\frac{9\gamma^2 }{40} \frac{(1+5a^3)}{(1-a^3)^2}    ,
\tag{\theequation~\emph{a,b}}
\end{equation}
\end{subequations}
for the polar and equatorial modes, respectively. For the polar mode, the use of series expansion in $\gamma  \ll 1$ shows that the B74 three estimates of $\zeta_m$ allow us to retrieve exactly the first term of (\ref{eq:smh}a), but none of them recovers exactly the form of the $\gamma^2$ term. However, the inner bound $\zeta_m=\zeta_m^{(i)}$ of equation~(\ref{syst:zet_bd}) recovers equation~(\ref{eq:smh}a) where $2-5a^3$ is replaced by $2-5a^3+3a^5$, and thus both give this term as $-3(1-a^3/2) \gamma^2/10 +\mathcal{O}(a^5 \gamma^2) $ for $a\ll1$. Note that the outer bound $\zeta_m=\zeta_m^{(o)}$ of equation~(\ref{syst:zet_bd}) recovers equation~(\ref{eq:smh}a) where $2-5a^3$ is replaced by $2+10a^3-12a^5$.

\begin{figure} 
    \centering
    \begin{subfigure}[b]{0.48\textwidth}
    \centering
    \includegraphics[trim=10 10 10 10,clip,width=0.9\linewidth]{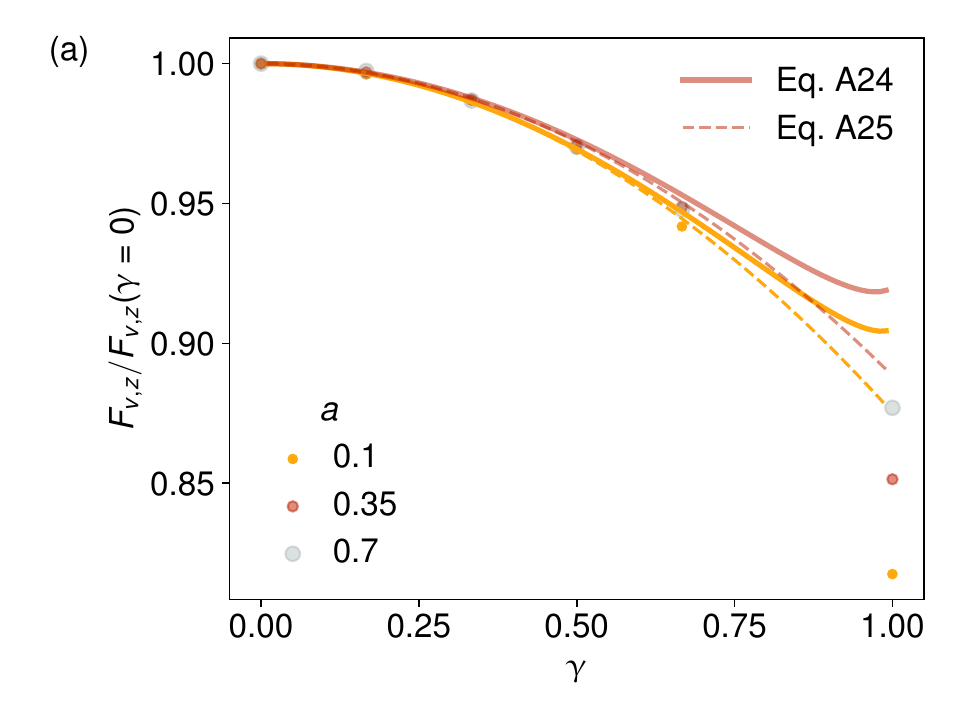}
    \end{subfigure}
    \begin{subfigure}[b]{0.48\textwidth}
    \centering
    \includegraphics[trim=10 10 10 10,clip,width=0.9\linewidth]{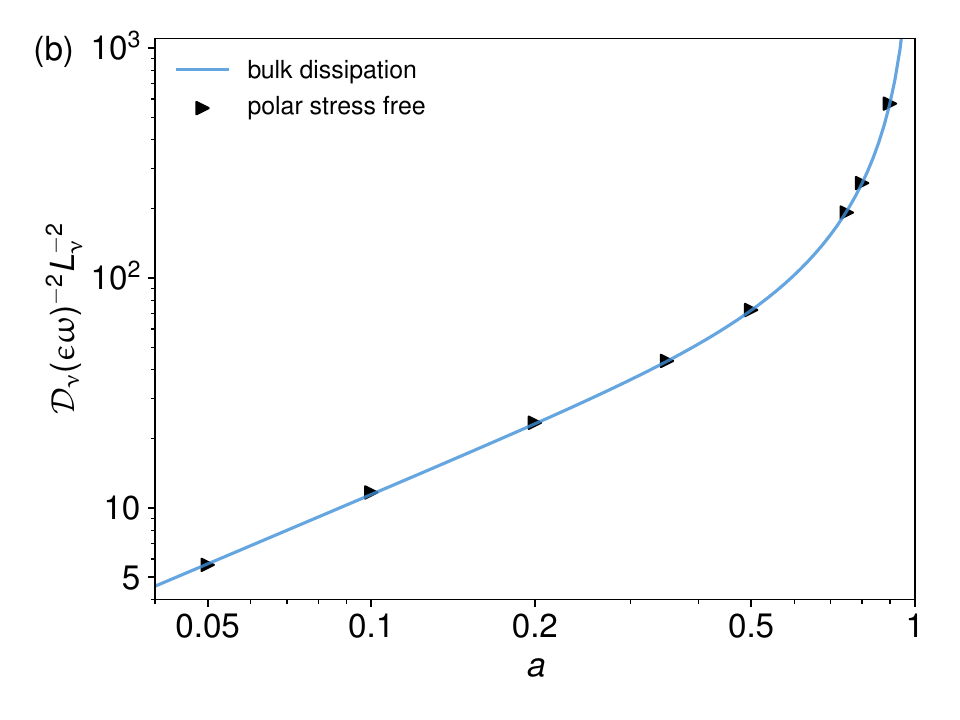}
    \end{subfigure}
        \caption{(a) Rotation effect on the axial force from tangential viscous stress for the polar mode without magnetic effects. Equations (\ref{eq:rotcorr})–(\ref{eq:rotcorr1}) are compared with \textsc{comsol} \textsc{dns} (colored dots) for a stress-free outer boundary: $L_\nu \gamma^{-1/2}/a_m=10^{-2}$, $\epsilon/L_\nu=0.1$. (b) Viscous dissipation for the polar mode with stress-free conditions at both boundaries: blue line (equation~\ref{eq:sfPoJ0}), \textsc{comsol} \textsc{dns} (black triangles).}
    \label{fig:x_Gamma_y_Fvratio}
\end{figure}

The viscous force on the inner boundary gives the detuning 
\begin{eqnarray} \label{eq:disSh1Smylie}
          \left.  \frac{\Delta \omega_v}{\omega} \right|_{r=a_s}&=&-\frac{3}{4} \frac{\rho_f}{\rho_s} \frac{L_\nu}{a_s} \frac{1-\mathrm{i}}{1-a^3} (1 -k_\gamma) 
\end{eqnarray}
with the rotation correction $k_\gamma$, given by 
\begin{eqnarray}
\label{eq:rotcorr}
k_\gamma &=& 1 + \frac{[10(1-a^3)-\gamma^2(2-5a^3)][(2+\gamma)(1-\gamma_2)+\gamma^2(3 \gamma_2 -4) -3 \gamma^3]}{25(1-a^3) \gamma^3 (1+\gamma)^{1/2}} \\ & \approx & \left( \frac{1}{8}-\frac{17}{40}a^3 \right) \frac{\gamma^2}{1-a^3} +\mathcal{O}(\gamma^4)
\label{eq:rotcorr1}
\end{eqnarray}
for the polar mode, and
\begin{eqnarray} 
k_\gamma &=& 1 + \frac{(1+3\gamma+7\gamma^2/2)\gamma_1-(1+\gamma)^{5/2}}{10 \gamma^3} \frac{2(1-a^3)(2-\gamma)-3\gamma^2(a^3+1/5)}{1-a^3} \, \, \\
&\approx & \frac{\gamma}{4}+\left( \frac{1}{8}-\frac{31}{40}a^3 \right) \frac{\gamma^2}{1-a^3} +\mathcal{O}(\gamma^3) \label{eq:disSh1Smylie2}
\end{eqnarray}
for both equatorial modes (with $\gamma_1 = \sqrt{1-\gamma}$ and $\gamma_2 = \sqrt{1-\gamma^{2}}$). From \cite{smylie2000the}, the detuning associated with tangential viscous stresses is thus identical for prograde and retrograde modes. In their notation, this follows in the limit $M_s \to \infty$ and $s=\gamma^{-1}$, where the factor $(1\mp \mathrm{i})f^{e}$ in their equation~(27) equals twice $(1-\mathrm{i})f^a$ in their equation~(25). Their detuning (\ref{eq:disSh1Smylie}) is thus identical for all modes when rotation is neglected, a result that we also retrieve (see \S\ref{sec:fovisma}). Rotation reduces the tangential viscous force while enhancing dissipation, implying an increase in the dissipative component of the viscously-induced pressure. Hence, the approach of \cite{smylie2000the} is not suitable for estimating viscous dissipation.

For $\gamma \ll 1$, equation~(\ref{eq:Busse_ordre2_Ca}) gives $C_a^i-1/2 \approx -3 \gamma^2/10-12 \gamma^4/175 +\mathcal{O}(\gamma^6)$, with the leading term recoverable from (\ref{eq:OURCa}) or from (\ref{eq:smh}a) in the limit $a \ll 1$. Bounding effects are incorporated by adding $3a^3/2+k_a a^3 \gamma^2$, where $k_a=3/20$ for $\zeta_m=\zeta_m^{(i)}$ and (\ref{eq:smh}a) of \cite{smylie1998viscous,smylie2000the}, but $k_a=-3/5$ and $k_a=-21/10$ for $\zeta_m=\zeta_m^{(e)}$ and $\zeta_m=\zeta_m^{(o)}$, respectively. By comparing the magnitude of the tangential part of the viscous force with and without rotation, in figure~\ref{fig:x_Gamma_y_Fvratio}a we tested these analytical formulas for the polar mode.
The correction of \cite{smylie2000the} presented in equation~(\ref{eq:rotcorr}) and in equation~(\ref{eq:rotcorr1}) is accurate for rapidly oscillating spheres,  $\gamma < 0.6$, and in the presence of a large gap $a<0.4$. 

\subsection{Higher-order rotational corrections from the basic flow of B74} \label{sec:Bus22}
In the limit $\gamma \ll 1$, equation~(\ref{syst:psi}) and $\zeta_m=\zeta_m^{(o)}$ give 
\begin{eqnarray}
    \boldsymbol{U}_2&=& \boldsymbol{U}_1+ \frac{3  \omega}{5} \frac{\gamma^2 }{r^5(1-a)\breve{a}^2}  \, \left(\mathrm{i}  \tilde{f}_r (r-a_s) \cos \theta \hat{\boldsymbol{r}}+ \mathrm{i}  \tilde{f}_\theta \sin \theta\hat{\boldsymbol{\theta}}+  \gamma a_s^3 \tilde{f}_\phi \sin (2\theta) \hat{\boldsymbol{\phi}} \right)  , \qquad \label{eq:Busse_ordre4} 
\end{eqnarray}
at order $\mathcal{O}(\gamma^4)$ , with $\breve{a}=1+a+a^2$ and
\begin{eqnarray}
\tilde{f}_r &=& -5 a_s^{3} \breve{a} (r +a_s ) \cos^2 \theta+3 \breve{a} \,a_s^{4}+3 r \breve{a} \,a_s^{3}+ a^{3} (a +1)(2  r^{2}  a_s^{2}+2r^{3} a_s +2 r^{4}), \qquad \\
\tilde{f}_\theta &=& -\frac{5 a_s^{3} (-3 a_s^{2}+r^{2}) \breve{a} \cos^2 \theta}{4}-\frac{3 \breve{a} \,a_s^{5}}{4}-r^{2} \left(a^{4}+a^{3}-\frac{1}{4} \breve{a} \right) a_s^{3}-2 a^{3} r^{5} (a +1), \\
\tilde{f}_\phi &=& \frac{(-35 a_s^{2}+25 r^{2}) \breve{a} \cos^2 \theta}{8}+\frac{(15 a_s^{2}-13 r^{2}) \breve{a}}{8}+\frac{3 a^{3} r^{2} (a +1)}{2}.
\end{eqnarray}

\subsection{Inductionless viscomagnetic drag on an oscillating sphere} \label{sec:singh}
For an unbounded conducting fluid permeated by a uniform axial magnetic field of moderate strength $B_0$, \cite{singh1965drag} derived the leading viscomagnetic correction to Stokes drag (\ref{eq:FSto}) in the quasi-static (Ohmic bulk-diffusive) MHD limit by expanding in $P_m \mathrm{Wo}^2=2a_s^2/L_\eta^2 \ll 1$. Using the notation of \S\ref{sec:StokEq59}, the drag ${\boldsymbol{F}}={\boldsymbol{F}}_n+{\boldsymbol{F}}_v$ is
\begin{eqnarray}
    {\boldsymbol{F}} &=& - {m}_{fs} C_a \mathrm{d}_t {\boldsymbol{V}}_s-\rho_f\left[ 6 \pi {a}_s  \nu \left( 1+\frac{a_s}{L_\nu} \left( 1+ \frac{\Lambda_z}{10} \right) \right) +\frac{2 \pi}{5} \omegas a_s^3  \Lambda_z \right]  {\boldsymbol{V}}_s  , \qquad \,  \label{eq:FStomag} \\
    C_a &=& \frac{1}{2}+\frac{9}{4} \frac{{L}_\nu}{{a}_s} \left(1-\frac{\Lambda_z}{10} \right) , \label{eq:FStomag2}
\end{eqnarray}
for $\Lambda_z<a_s^2/L_\eta^2\ll1$. Magnetic effects reduce the fluid added-mass through the $\mathcal{O}(\Lambda_z)$ correction to $C_a$, while the dissipative term recovers the classical quasi-static magnetic drag $2\pi\sigma_f \tilde{B}_0^2a_s^3/5$ in the inviscid limit. Higher-order corrections in $\Lambda_z$ were subsequently derived by \cite{motz1966magnetohydrodynamic}, who showed that the drag is multiplied by $f_\Lambda=1-1.08\Lambda_z+1.15\Lambda_z^2$ (accounting for the different definitions of the dimensionless parameters between these works). This finite-field correction was obtained for an unbounded,
non-rotating dielectric sphere. It is quoted here to document the
connection with the classical inductionless literature, but is not
used as a large-$\Lambda_z$ asymptote for the confined spherical-shell
problem.

\section{Magnetohydrodynamic particular solution driven by bulk flows}\label{appA}
The particular solution forced by the bulk flow has not been obtained analytically in the general case. Nevertheless, asymptotic solutions can be derived in limits that are directly relevant to the bulk-diffusive sectors of figure~\ref{fig:reg}. We first consider the diffusionless limit with weak Lorentz forces, and then the limit of large magnetic diffusivities.

\subsection{Diffusionless limit with weak Lorentz forces} \label{sec:diffuWekL}
For $\Lambda_l \ll 1$, the Lorentz forces can be neglected at leading order. The bulk flow is then the hydrodynamic basic flow. This flow will force a magnetic field solution $\boldsymbol{b}^{(p)} \approx \boldsymbol{b}^{(p)}_0$, which will then force a particular solution $\boldsymbol{u}^{(p)} \approx \boldsymbol{u}^{(p)}_0$ at next order in $\Lambda_l$. At the order $\epsilon$, the form (\ref{syst:ansstz}) simplifies equation~(\ref{eq:dc1}) into $\mathrm{i} \omega \boldsymbol{b}^{(p)}_0 = \nabla  \times (\boldsymbol{U} \times \boldsymbol{B}_0) $ in the diffusionless limit. This divergence-free solution is valid for arbitrary fields $\boldsymbol{B}_0$. For a uniform axial magnetic field $\boldsymbol{B}_0 = [B_0 \cos \theta,-B_0 \sin \theta, 0]^{\top}$, it reads
\begin{eqnarray}
{b}^{(p)}_{0r}  &=& 
     \frac{\mathrm{i} B_0 }{\omega r} \left[ \sin (\theta) \partial_\theta {U}_r -\frac{U_\theta}{\sin \theta}+  \mathrm{i} m  {U}_\phi \cot \theta +(2 {U}_r +2{U}_\theta \cot \theta+ \partial_\theta {U}_\theta  ) \cos \theta\right] , \qquad  \label{eq:brgenB0q1} \\
{b}^{(p)}_{0\theta} &=& - \mathrm{i} B_0 \omega^{-1} \left[ \mathrm{i} m r^{-1} {U}_\phi+ (r^{-1} {U}_\theta + \partial_r {U}_\theta) \cos (\theta) + (r^{-1} {U}_r+\partial_r {U}_r) \sin \theta \right] , \\
{b}^{(p)}_{0\phi} &=&   \mathrm{i} B_0 \omega^{-1} [r^{-1} \sin( \theta) \partial_\theta {U}_\phi -\cos( \theta) \partial_r {U}_\phi ]  \label{eq:brgenB0q3},
\end{eqnarray}
using the spherical coordinates $(r,\theta,\phi)$ for $\boldsymbol{b}^{(p)}_0=[{b}^{(p)}_{0r},{b}^{(p)}_{0\theta},{b}^{(p)}_{0\phi}]^\top$. In the inner domain, using $\boldsymbol{U}=\mathrm{d}_t \boldsymbol{\delta}$ in equations (\ref{eq:brgenB0q1})-(\ref{eq:brgenB0q3}) gives $\boldsymbol{b}^{(p)}_0=\boldsymbol{0}$ for the three Slichter modes, and thus $\boldsymbol{\mathcal{B}}=\boldsymbol{0}$  as the induction term is $\boldsymbol{0}$ at this order. Considering instead the fluid layer, the polar mode basic flow (\ref{syst:Lamb2}) for $\boldsymbol{U}$ simplifies equations (\ref{eq:brgenB0q1})-(\ref{eq:brgenB0q3}) into
\begin{eqnarray}
 \boldsymbol{b}^{(p)}_{0} &=&-\frac{3}{2} \frac{(a_s/r)^3}{1-a^3}  \frac{ B_0}{r} \left[   (3 \cos^2 \theta -1) ,  \sin (2 \theta) ,  2 \mathrm{i}  \frac{\gamma}{2} (5 \cos^2 \theta-1) \sin \theta  \right] , \label{eq:brBdn1} 
\end{eqnarray}
which is easily recovered from equation~(\ref{eq:brr}) by putting the left-hand side to zero (removing the diffusion term). Besides, equations (\ref{eq:bpfl}) and (\ref{eq:brBdn1}) agree in the limit of small $L_\eta$ and large $r$ (outside the boundary layer). 
For the equatorial modes, the basic flow (\ref{syst:Lamb}) gives
\begin{eqnarray}
 \boldsymbol{b}^{(p)}_{0} &=& \frac{3}{2} \frac{(a_s/r)^3}{1-a^3}  \frac{ B_0}{r} \left[-   \frac{3 }{2} \sin2 \theta ,   \cos 2 \theta , \pm  \mathrm{i} \cos \theta \right] . \label{eq:brBdn2}  
\end{eqnarray}
The diffusionless particular solution $\boldsymbol{u}^{(p)} \approx \boldsymbol{u}^{(p)}_0=[{u}^{(p)}_{0r} ,{u}^{(p)}_{0\theta} ,{u}^{(p)}_{0\phi}]$ can then be obtained from equation~(\ref{eq:Lambi1}), i.e.\ from $ \mathrm{i} \omega \boldsymbol{u}^{(p)}_0=B_{0r} \partial_r \boldsymbol{b}^{(p)}_0$ as the BLT pressure does not vary across the boundary-layer (in the $r$ direction). For instance, equation~(\ref{eq:brBdn1}) gives
\begin{eqnarray}
\boldsymbol{u}^{(p)}_0&=&  \frac{6}{\omega} \left( \frac{a_s}{r} \right)^3 \frac{ B_0^2}{r^2} \frac{\sin 2 \theta }{1-a^3} \left[\frac{1-3 \cos^2 \theta}{2 \sin \theta}  \mathrm{i},- \mathrm{i} \cos \theta , \frac{\gamma}{2} (5 \cos^2 \theta-1) \right] , 
\end{eqnarray}
for the polar mode in a uniform axial magnetic field. Similarly, equation~(\ref{eq:brBdn2}) gives
\begin{eqnarray}
\boldsymbol{u}^{(p)}_0&=& \frac{6}{\omega} \left( \frac{a_s}{r} \right)^3 \frac{ B_0^2}{r^2} \frac{\cos \theta }{1-a^3} \left[-\frac{3}{2} \mathrm{i} \sin 2 \theta,  \mathrm{i} \cos 2 \theta,  \mp  \cos \theta \right] , \label{eq:didy1u} 
\end{eqnarray}
for the equatorial modes. As expected, these velocities are all of the order $\Lambda_l$.

The diffusionless solution accurately captures the perturbation field outside boundary layers, as illustrated by the dot-dashed blue curve in figure~\ref{fig:xRadialMagneticBoundaryLayer}, and also provides an effective approximate particular solution within the boundary-layer problem. As demonstrated in several places (e.g.\ figure~\ref{fig:xRyThetaPhiMagneticBLViscoMagWaves}), it closely reproduces the inviscid MHD particular solution. From a computational standpoint, this approximation is advantageous as it avoids exponential and hyperbolic sine integral functions. This efficiency is particularly valuable when the analytical particular solution is unavailable, as in the case of the Stokes--Ekman boundary layer or when we use the base flow of B74.
In figure~\ref{fig:xRyViscoMagRotVPhiGamma01}b, adding the diffusionless magnetic field solution to the homogeneous Stokes--Ekman boundary layer allows the \textsc{dns} to be perfectly matched. See the difference between the red-dotted line (only homogeneous), and the red-solid line (sum of the two contributions). 

\subsection{Quasi-static MHD limit of large magnetic diffusivity}
\label{app:largeDeta}
For large magnetic diffusivity, neglecting $\mathrm{i}\omega\boldsymbol{b}$ in the induction equation yields the quasi-static Ohm's law $\breve{\eta}\nabla\times\boldsymbol{b}=\boldsymbol{U}\times\boldsymbol{B}_0-\nabla\Phi_e$, where $\Phi_e$ is the appropriately scaled electric potential, determined by charge conservation and the electrical interface conditions. For the non-rotating polar mode ($\gamma=0$) with axial field $\boldsymbol{B}_0=[B_0\cos\theta,-B_0\sin\theta,0]^\top$, both $\boldsymbol{U}$ and $\boldsymbol{B}_0$ are axisymmetric and meridional. Hence, $\boldsymbol{U}\times\boldsymbol{B}_0$ is purely azimuthal, divergence-free and tangent to every spherical interface, so that $\Phi_e$ may be taken constant without further approximation. Moreover, the imposed velocity of the inner sphere is parallel to $\boldsymbol{B}_0$, and the current therefore vanishes within the sphere, irrespective of its conductivity. By contrast, the rotation-dependent part of the polar flow and the equatorial modes generally require a non-uniform electric potential. For these contributions, setting $\nabla\Phi_e=0$ below provides only a simple quasi-static estimate. With this convention, the flow (\ref{syst:Lamb}) gives
\begin{eqnarray}
    \frac{\nabla  \times \boldsymbol{b}_0^{(p)}}{B_0}=-\frac{3  a_s^3  \sin 2 \theta}{2 L_\eta^2 (1-a^3)r^3} [\gamma\sin \theta \hat{\boldsymbol{r}}+\gamma \cos \theta \hat{\boldsymbol{\theta}}+ \mathrm{i}  \hat{\boldsymbol{\phi}} ] , \label{eq:b0diff}
\end{eqnarray}
in the fluid, allowing the calculation of Ohmic dissipation by computing $(\nabla\times\boldsymbol{B})^2$ from the real parts. Since $\Real(\nabla\times\boldsymbol{b})=\breve{\eta}^{-1}\Real(\boldsymbol{U}\times\boldsymbol{B}_0)$, integration over the fluid volume gives
\begin{eqnarray}
 \int_{\Omega_f} \frac{(\nabla \times {\boldsymbol{B}})^2}{B_0^2}\mathrm{d}V = \frac{8 \pi a_s^3 \epsilon^2 }{5 L_\eta^4 } \frac{\sin^2 (\omega t)+\gamma^2 \cos^2 (\omega t)}{1-a^3} , \label{eq:Stoma2}
\end{eqnarray}
whose time average is $(1+\gamma^2)/2$ times the $\sin^2\omega t$ prefactor. This yields
\begin{eqnarray} \label{eq:imB0}
\left.   \widehat{\mathcal D}_\eta \right|_{\Omega_f}^{\mathrm{pol}} =  \frac{3}{20} \frac{\rho_f}{\rho_s} \frac{1+ \gamma^2}{1-a^3} \Lambda_z .
\end{eqnarray}
For $\gamma=0$, equation~(\ref{eq:imB0}) is therefore exact within the present quasi-static, weak-Lorentz-force approximation. It extends the classical magnetic drag derived by \cite{singh1965drag} from an unbounded conducting fluid to confined spherical shells. In the unbounded limit ($a=0$), it exactly recovers Singh's leading-order expression $2 \pi \sigma_f \tilde{B}_0^2 a_s^3/5$ given in \S\ref{sec:singh}, while finite confinement enhances the dissipative drag through the geometrical factor $(1-a^3)^{-1}$. This exact agreement follows because the non-rotating polar forcing requires neither an electric-potential correction nor Ohmic dissipation within the sphere. Equation~(\ref{eq:imB0}) also recovers, up to the factor $3/(1-a^3)$, equation~(7) of \cite{chyba2021magnetic} for a sphere in an oscillating uniform axial magnetic field at $L_\eta\gg a_s$; the remaining prefactor arises solely from the different velocity field $\boldsymbol{U}$. The use of the unperturbed hydrodynamic bulk flow requires $\Lambda_z\ll1$. For $\gamma\neq0$, however, the factor $1+\gamma^2$ should be regarded as an estimate obtained by neglecting the electric potential associated with the rotation-induced flow. For the unbounded, non-rotating configuration,
\citet{motz1966magnetohydrodynamic} obtained the finite-field corrective factor $f_\Lambda=1-1.08\Lambda_z+1.15\Lambda_z^2$.

Considering next the equatorial modes and again setting $\nabla\Phi_e=0$, integration of the resulting bulk current $\nabla\times\boldsymbol{b}_0^{(p)}$ in the fluid gives the estimate
\begin{eqnarray} \label{eq:imB2}
\left. \widehat{\mathcal D}_\eta^{\mathrm{eq}} \right|_{\Omega_f} =   \frac{\rho_f}{\rho_s} \frac{\Lambda_z}{1-a^3} \left[ \frac{7}{20}+a^3 \right] .
\end{eqnarray}
Equation~(\ref{eq:imB2}) can be compared with the dimensional drag derived by \cite{reitz1961the},
\begin{eqnarray} \label{eq:reitz}
\boldsymbol{F}_\eta=- \frac{3 \pi}{5} a_s^3 B_0^2 \sigma_f \frac{1+3\breve{\sigma}_s}{2+\breve{\sigma}_s} \boldsymbol{V}_s,
\end{eqnarray}
for a conducting sphere ($\breve{\mu}_s=1$, $\breve{\sigma}_s=\sigma_s/\sigma_f$) moving steadily through an unbounded inviscid fluid perpendicular to a uniform axial magnetic field (their case~2). Since equation~(\ref{eq:imB2}) accounts only for Ohmic dissipation in the fluid, it should be compared with the insulating limit $\breve{\sigma}_s=0$, for which no dissipation occurs within the sphere. In the unbounded limit, the exact result (\ref{eq:reitz}) then gives the prefactor $9/40$, instead of $7/20$ in equation~(\ref{eq:imB2}). Thus, neglecting the non-uniform electric potential causes equation~(\ref{eq:imB2}) to overestimate the equatorial drag in this reference case by the factor $14/9$, or approximately $56\%$. When dissipation within the sphere is included through $\breve{\sigma}_s\neq0$, the exact result (\ref{eq:reitz}) gives the prefactor $3/5$ for the case $\breve{\sigma}_s=1$ considered by P26, in excellent agreement with their \textsc{dns}. 

The present analysis therefore exactly recovers the quasi-static drag of \cite{singh1965drag} for the non-rotating polar mode and extends it to confined spherical shells. The rotation-dependent contribution in equation~(\ref{eq:imB0}) and the equatorial expression (\ref{eq:imB2}) provide simple estimates when the electric-potential problem is not solved explicitly. The non-rotating polar result and its confinement dependence were validated by the \textsc{dns} of P26, while their equatorial result at $\breve{\sigma}_s=1$ agrees with the conductivity-dependent exact coefficient (\ref{eq:reitz}). Although this quasi-static regime is not relevant to planetary cores, it applies to shallow subsurface oceans (figure~\ref{fig:planets}) and laboratory experiments.

\section{Non-magnetic Stokes boundary-layer solution} \label{sec:stoBL} 
For rapid oscillations of a solid sphere in a viscous fluid, the fluid force is dominated by the flow in the thin boundary layer and the boundary layer theory (BLT) can then be used instead of the exact lengthy calculation of \cite{stokes1851effect} discussed in appendix~\ref{sec:ExpStoU}.

\subsection{Boundary-layer tangential flow and higher-order radial flow}
In the non-magnetic limit, the bulk flow $\epsilon \boldsymbol{U}_1=\epsilon (U_{1\phi} \hat{\boldsymbol{\phi}}+\nabla \varPhi)$ given by equations (\ref{syst:Lamb})-(\ref{syst:Lamb2}) drives Stokes boundary layer flows at both solid interfaces for no-slip conditions. Like \cite{batchelor1967introduction}, we introduce $U_0=\mathrm{i} \omega \epsilon$, such that the boundary velocity is $ \mathrm{d}_t \boldsymbol{\delta}=U_0 \mathrm{e}^{\mathrm{i} \omega t} \hat{\boldsymbol{z}} $. Noting $\lambda_\nu=(1+\mathrm{i})/L_\nu$, the boundary layer equations give
\begin{eqnarray} \label{eq:visBLst}
   \frac{ \epsilon \boldsymbol{u}_\parallel}{\mathrm{e}^{\lambda_\nu (a_s-r)}}=\frac{\epsilon  \boldsymbol{u}_\parallel^{(o)}}{a^3 \, \mathrm{e}^{-\lambda_\nu (a_m-r)}}=-\frac{3 U_0 \sin \theta}{2(1-a^3)}  \left[\hat{\boldsymbol{\theta}}+  \mathrm{i} {\gamma}  \cos \theta  \hat{\boldsymbol{\phi}}  \right] ,
\end{eqnarray}
for the polar mode (for a no-slip outer boundary). Ignoring rotation effects, we also obtain
\begin{eqnarray}
\frac{ \epsilon \boldsymbol{u}_\parallel}{\mathrm{e}^{\lambda_\nu (a_s-r)}} &=&\frac{\epsilon   \boldsymbol{u}_\parallel^{(o)}}{a^3 \, \mathrm{e}^{-\lambda_\nu (a_m-r)}}=\frac{3 U_0 }{2(1-a^3)}  \left[\cos \theta \hat{\boldsymbol{\theta}} \pm  \mathrm{i} \hat{\boldsymbol{\phi}}\right]  , \label{eq:eqvcStL}
\end{eqnarray}
for the equatorial modes. Here, $\boldsymbol{u}_\parallel$ and $\boldsymbol{u}_\parallel^{(o)}$ are the boundary-layer solutions at the inner and outer boundaries, respectively. 
At the top of figure~\ref{fig:Stokes}, the boundary-layer solutions for the inner (left) and outer (right) are presented as dashed purple lines against the orange-dotted \textsc{xshells} solution.
They capture quite well the behaviour at the boundary, matching the value at the solid wall, but they decay to zero in the bulk.

For the polar mode, the dimensional viscous force driven by the tangential stress is then
\begin{eqnarray}
  \boldsymbol{F}_{v} =  \int_S \rho_f \nu \frac{1+\mathrm{i}}{1-a^3} \frac{U_w}{L_\nu} \mathrm{e}^{\mathrm{i} \omega t}  (\hat{\boldsymbol{\theta}} \cdot \hat{\boldsymbol{z}}) \, \hat{\boldsymbol{z}} \, \mathrm{d}S , \label{eq:Fvsh}
\end{eqnarray}
with the basic potential velocity $(3/2) \mathrm{d}_t \boldsymbol{\delta} \sin \theta =U_w \mathrm{e}^{\mathrm{i} \omega t} \hat{\boldsymbol{z}}$ at $r=a_s$, and using the dimensional variables \cite[see also p. 357 of][]{batchelor1967introduction}. Equation~(\ref{eq:Fvsh}) gives then
\begin{eqnarray}
\boldsymbol{F}_{v} &=&  - 4 \pi (1+\mathrm{i}) \frac{\rho_f \nu  a_s U_0}{1-a^3} \frac{a_s}{L_\nu} \mathrm{e}^{\mathrm{i} \omega t} \hat{\boldsymbol{z}}  = - 4 \pi \frac{\rho_f \nu  a_s }{1-a^3} \frac{a_s}{L_\nu} \mathrm{d}_t \boldsymbol{\delta}-\frac{3}{2}  \,  \frac{{m}_{fs}}{1-a^3} \frac{L_\nu}{a_s} \mathrm{d}_t^2 \boldsymbol{\delta} , \label{eq:fghjs} \\ 
   \frac{\Delta \omega_v}{\omega} &=&  \frac{3}{4} \frac{1+\mathrm{i}}{1-a^3} \frac{\rho_f}{\rho_s} \frac{L_\nu}{a_s} \frac{U_0}{\epsilon \omega}  = - \frac{3}{4} \frac{1-\mathrm{i}}{1-a^3} \frac{\rho_f}{\rho_s} \frac{L_\nu}{ a_s}  , \label{eq:batch}
\end{eqnarray}
equations~(\ref{eq:fghjs})-(\ref{eq:batch}) recover the viscous tangential stress contribution of equation~(\ref{eq:FSto}), from \cite{stokes1851effect}, and equation~(25) of \cite{smylie2000the} taken in the limit of negligible rotation effects and outer boundary displacement (infinite shell mass $M_s$ in their notations). In this limit, their results give actually the same $\boldsymbol{F}_{v}$ for equatorial modes (using their equations 27 and 35). Note that the boundary layer of the outer boundary can be tackled similarly, showing that the force generated by the tangential stress at $r=a_m$ can be obtained by multiplying equations (\ref{eq:fghjs}) or (\ref{eq:batch}) by $a$.

For the polar mode, the viscously-modified pressure stress actually leads to a total force that is $3/2$ times higher in an unbounded fluid \cite[e.g.][]{batchelor1967introduction}. To obtain the total force within our boundary-layer approach, we first need to calculate the boundary-layer radial flows $\epsilon \mathfrak{u}_r$ and $ \epsilon \mathfrak{u}_r^{(o)}$ at the inner and outer boundaries, respectively. Equation~(\ref{eq:consmass}) gives 
\begin{eqnarray}
\frac{\epsilon \mathfrak{u}_r}{\mathrm{e}^{\lambda_\nu  (a_s-r)}}&=& \frac{\epsilon \mathfrak{u}_r^{(o)}}{-a^4 \mathrm{e}^{\lambda_\nu  (r-a_m)}}= -\frac{3}{2} \frac{1+\mathrm{i}}{1-a^3} \frac{ L_\nu}{a_s }  \epsilon \omega \mathfrak{c} \label{eq:fcc1} , \label{eq:fcc12}
\end{eqnarray}
with $\mathfrak{c} =\cos \theta$ for the polar mode, and $\mathfrak{c} =\sin \theta$ for equatorial modes. Then, $\epsilon \mathfrak{u}_r$ and $ \epsilon \mathfrak{u}_r^{(o)}$  force bulk flows $\boldsymbol{\mathcal{U}}$ and $\boldsymbol{\mathcal{U}}^{(o)}$ through the non-penetration condition at the inner and outer boundaries. At the bottom of figure~\ref{fig:Stokes}, these radial solutions (dashed purple) for the inner (left) and outer (right) boundary regions are compared to the \textsc{xshells} solution (orange-dotted). 

\subsection{Polar mode: viscous bulk flow and its boundary layer correction}
For the polar mode, equation~(\ref{eq:dc40}) and equation~(\ref{eq:fcc12}) for $\mathfrak{u}_r$ give
\begin{eqnarray}
   \epsilon \varPsi &=& - \frac{3}{4} \frac{1-\mathrm{i}}{(1-a^3)^2} \frac{L_\nu U_0 a_s^2  (1+2 r^3/a_m^3) \cos \theta}{r^2 },  \label{eq:fcc2psi} \\
   \epsilon \boldsymbol{\mathcal{U}} &=& \epsilon \nabla \varPsi = \frac{3}{4} \frac{1-\mathrm{i}}{(1-a^3)^2} \frac{a_s^2 L_\nu}{r^3 } U_0  \left[ 2 \left(1-\frac{r^3}{a_m^3} \right)\cos(\theta) \hat{\boldsymbol{r}}+\left(1+2\frac{r^3}{a_m^3} \right)\sin(\theta) \hat{\boldsymbol{\theta}} \right] , \qquad \label{eq:fcc2}
\end{eqnarray}
in the absence of rotation ($\gamma=0$) and magnetic effects. The pressure $\mathcal{P}=- \mathrm{i} \omega \varPsi$ due to $\boldsymbol{\mathcal{U}}$ gives $\boldsymbol{F}_\nu$, which is half of equation~(\ref{eq:fghjs}), but where $(1-a^3)^{-1}$ is replaced by $(1+2a^3)(1-a^3)^{-2}$. In the unbounded limit, we recover all the $L_\nu$ terms of the Stokes pressure force (\ref{eq:FSto}). As in \S\ref{sec:InvBulk}, a rotation perturbation of the axisymmetric flow $\boldsymbol{\mathcal{U}}$ only generates an azimuthal flow $ \mathcal{U}_\phi$ at leading order. Using $\mathcal{U}_\phi= \mathrm{i} \gamma \mathcal{U}_s$, we obtain
\begin{eqnarray}
 \epsilon \mathcal{U}_\phi &=& \frac{9}{4} \frac{1+\mathrm{i}}{(1-a^3)^2} \frac{\gamma}{2} \frac{a_s^2 L_\nu}{r^3} U_0 \sin 2 \theta, \label{eq:fphiSto1}
\end{eqnarray}

If a no-slip condition is imposed at $r=a_m$, one can also search for an irrotational flow with $\mathfrak{u}_r^{(o)}+\mathcal{U}_r^{(o)}=0$ at the outer boundary, and $\mathcal{U}_r^{(o)}(r=a_s)=0$. Using equation~(\ref{eq:fcc12}), we obtain  the potential $\varPsi^{(o)}$ of the bulk flow $ \boldsymbol{\mathcal{U}}^{(o)} $. It turns out that $\varPsi^{(o)}$ is given by equation~(\ref{eq:fcc2psi}) multiplied by $a^4$, replacing $a_m$ by $a_s$. This gives 
\begin{eqnarray}
   \epsilon  \boldsymbol{\mathcal{U}}^{(o)} &=&  \frac{3}{4} \frac{1-\mathrm{i}}{(1-a^3)^2} \frac{a_s^2 L_\nu}{r^3 } U_0 a^4 \left[ 2 \left(1-\frac{r^3}{a_s^3} \right)\cos(\theta) \hat{\boldsymbol{r}}+\left(1+2\frac{r^3}{a_s^3} \right)\sin(\theta) \hat{\boldsymbol{\theta}} \right] ,   \label{eq:fcc22}
\end{eqnarray}
and equations (\ref{eq:fcc1})-(\ref{eq:fcc22}) are recovered from the exact Stokes' solution (\ref{eq:VnRSTO1}) in the limit $L_\nu \ll a_s$. The rotation-induced flow $\mathcal{U}_\phi^{(o)}= \mathrm{i} \gamma \mathcal{U}_s^{(o)}$ is found to be $a^4$ times equation~(\ref{eq:fphiSto1}).

In figure~\ref{fig:Stokes}, we present these bulk flows at the inner (left) and outer (right) boundary combined with the boundary layer meridional and radial components (equations~\ref{eq:visBLst} and~\ref{eq:fcc1}, respectively).
When considering individual contributions, depicted as dash-dotted lines in cyan (inner flow) and green (outer flow), the results fail to match the \textsc{dns} (orange-dotted) outside the specified boundary region. However, by accounting for both contributions (purple dash-dotted), we achieve a close match with the \textsc{dns}, except for the meridional flow at the boundaries where an additional boundary layer correction is required (considered below). 

Since the bulk flows $\boldsymbol{\mathcal{U}}$ and $ \boldsymbol{\mathcal{U}}^{(o)}$ do not verify the boundary conditions in the $\theta$ and $\phi$ directions, they force their own boundary-layer flow $\boldsymbol{\mathfrak{u}}_\parallel$ to ensure a zero velocity at the boundaries. This additional boundary-layer flow is easily obtained as above, reading
\begin{eqnarray} \label{eq:drf1}
 \epsilon \boldsymbol{\mathfrak{u}}_\parallel&=&-[\mathcal{U}_\parallel(r = a_s) +\mathcal{U}_\parallel^{(o)}(r = a_s) ] \, \, \mathrm{e}^{\lambda_\nu (a_s-r)} , \\
  \epsilon \boldsymbol{\mathfrak{u}}_\parallel^{(o)}&=&-[\mathcal{U}_\parallel(r = a_m) +\mathcal{U}_\parallel^{(o)}(r = a_m) ] \, \, \mathrm{e}^{-\lambda_\nu (a_m-r)} , \label{eq:drf2}
\end{eqnarray}
where $\mathcal{U}_\parallel$ and $\mathcal{U}_\parallel^{(o)}$  are the tangential components of $\mathcal{U}$ and $\mathcal{U}^{(o)}$, respectively, in the $\theta$ and $\phi$ directions. Considering the exact Stokes' solution (\ref{eq:VnRSTO1}) at large $\lambda_\nu $, a series expansion of
\begin{eqnarray}
V_\theta \approx \frac{ 3 U_0 a_s}{\lambda_\nu ^2r^3} \left[  \frac{ \mathcal{N}_\theta }{\breve{\mathcal{D}}} \frac{\sin \theta}{2} \right]   \mathrm{e}^{\mathrm{i} \omega t} \approx -\frac{ 3 U_0 a_s}{\lambda_\nu ^2r^3} \left[  \frac{ \breve{\mathcal{M}}_s^- \mathrm{e}^{\lambda_\nu (r-a_m)}+ \breve{\mathcal{M}}_m^+ \mathrm{e}^{\lambda_\nu (a_s-r)}}{\breve{\mathcal{D}}} \frac{\sin \theta}{2} \right]   \mathrm{e}^{\mathrm{i} \omega t}  , \qquad
\end{eqnarray}
indeed allows equations (\ref{eq:drf1})-(\ref{eq:drf2}) to be recovered, giving for instance
\begin{eqnarray} \label{eq:C14}
\frac{\epsilon \boldsymbol{\mathfrak{u}}_\parallel}{ \mathrm{e}^{\lambda_\nu (a_s-r)} }  &=& - \frac{3}{4} \frac{(1-\mathrm{i}) U_0}{(1-a^3)^2} \frac{L_\nu}{a_s} [(1+2 a^3+3a^4 )  \hat{\boldsymbol{\theta}}+3 \mathrm{i} \gamma (1+a^4) \cos  \theta \hat{\boldsymbol{\phi}}] \sin \theta, \qquad \quad  \\
\frac{\epsilon \boldsymbol{\mathfrak{u}}_\parallel^{(o)}}{a^3 \mathrm{e}^{\lambda_\nu (r-a_m)} }  &=& - \frac{3}{4} \frac{(1-\mathrm{i}) U_0}{(1-a^3)^2} \frac{L_\nu}{a_s} [(3+2 a+a^4 ) \hat{\boldsymbol{\theta}}+3 \mathrm{i} \gamma (1+a^4) \cos  \theta \hat{\boldsymbol{\phi}}] \sin \theta,  \label{eq:C15}
\end{eqnarray}
where the terms in $a^4$, or higher powers of $a$, in (\ref{eq:C14})–(\ref{eq:C15}) arise from the $\mathcal{U}^{(o)}(r=a_s)$ contribution, and the rotation-induced $\phi$ component obtained here is absent from the original Stokes solution. The top insets of figure~\ref{fig:Stokes} show that agreement with the \textsc{dns} (orange-dotted) near the boundary is achieved only when combining the boundary-layer solution $u_\theta$, the higher-order bulk flow $\mathcal{U}_\theta$, and the boundary correction $\mathfrak{u}_\theta$ (green dashed). The sum of these three contributions is consistent with the Stokes solution in black (appendix~\ref{sec:ExpStoU}).

\subsection{Polar mode: viscous forces from tangential viscous stress and pressure}

Considering only the inner boundary layer, the dimensional force $\boldsymbol{F}_\nu$ reads
\begin{eqnarray} \label{eq:FpFv}
\boldsymbol{F}_\nu = \frac{3}{2} \frac{\boldsymbol{F}_{v}}{1-a^3} =- 6 \pi (1+\mathrm{i}) \frac{\rho_f \nu  a_s U_0}{(1-a^3)^2} \frac{a_s}{L_\nu} \mathrm{e}^{\mathrm{i} \omega t} \hat{\boldsymbol{z}}  ,
\end{eqnarray}
i.e.
\begin{eqnarray}  \label{eq:FpFv2}
\boldsymbol{F}_\nu =    - 6 \pi \frac{\rho_f \nu  a_s }{(1-a^3)^2} \frac{a_s}{L_\nu} \mathrm{d}_t \boldsymbol{\delta}-\frac{9}{4}  \,  \frac{{m}_{fs}}{(1-a^3)^2} \frac{L_\nu}{a_s} \mathrm{d}_t^2 \boldsymbol{\delta} .
\end{eqnarray}
using dimensional quantities. Equation~(\ref{eq:FpFv2}) recovers the leading order viscous terms of (\ref{eq:Stob1})-(\ref{eq:Stob2}), but where the factor $1+a^4$ is replaced by $1$ in (\ref{eq:FpFv2}). We have indeed neglected here the dissipation in the outer boundary layer, which gives this additional term in $a^4$ (see below). For this polar mode, the scaled mean dissipation equals the modal damping
rate, and the power balance gives a ratio $3/[2(1-a^3)]$ between the
total and tangential-stress dissipative contributions. Direct force
calculation shows that this ratio also holds for their non-dissipative
components in the case considered here.

The total dissipative force may equivalently be evaluated from (\ref{eq:totPowoh0}). For the polar mode, this yields (\ref{eq:Fv_pola}), recovering (\ref{eq:FpFv}) in the non-rotating limit $\gamma=0$ at leading order in $L_\nu/a_s \ll 1$. To quantify rotational corrections to (\ref{eq:FpFv}), we use the $\mathcal{O}(\gamma^3)$ expansion (\ref{eq:Busse_ordre2}) of the unbounded bulk flow $\boldsymbol{U}$ from B74. In this limit, boundary-layer contributions arise only at the inner boundary and read
\begin{eqnarray} \label{eq:visBLstBuss}
   \frac{ \epsilon [\hat{\boldsymbol{\theta}},\hat{\boldsymbol{\phi}}] \cdot  \boldsymbol{u}_\parallel}{\mathrm{e}^{\lambda_\nu (a_s-r)}}=- \frac{3}{2} \frac{U_0 \sin \theta}{1-a^3} \left(1-\frac{\gamma^2}{5}+\gamma^2 \cos^2 \theta \right) \left[1,  \mathrm{i} {\gamma}  \cos \theta   \right].
\end{eqnarray}
Substitution into (\ref{eq:totPowoh0}) recovers (\ref{eq:Fv_pola}). Extending to $\mathcal{O}(\gamma^4)$ introduces the term $-\gamma^4(1+8/35-5\cos^4 \theta)/5$ within parentheses in (\ref{eq:visBLstBuss}), yielding an additional contribution $24 \gamma^4/175$ to the $\gamma^2/5$ term in (\ref{eq:Fv_pola}).

One can now wonder how the presence of the outer boundary layer modifies this equation. Since the boundary layer flow $\boldsymbol{u}_\parallel$ at the outer boundary is $a^3$ times the one at the inner boundary (equation~\ref{eq:visBLst}), the integrand of equation~(\ref{eq:totPowoh0}) has a multiplying factor of $a^6$. But the volume integration $\int_{S_m} \int_{r=-\infty}^{r= a_m}$ of equation~(\ref{eq:totPowoh0}) leads to a factor $a_m^2$ when considering the outer boundary surface $S_m$. In total, the outer boundary layer obeys the same equation as the inner boundary layer, with a supplementary multiplying factor of $a^6 a_m^2/a_s^2=a^4$. Adding the two contributions allows equation~(\ref{eq:Stob2}) to be fully retrieved.

\subsection{Equatorial modes: viscous bulk flow and associated forces}
For equatorial modes, the potential $\varPsi$ of the viscous bulk flow is obtained by restoring the azimuthal dependence $\mathrm{e}^{\mathrm{i} m \phi}$ in $\mathfrak{u}_r$ and seeking an irrotational solution (\S\ref{sec:InvBulk}). This yields $\varPsi$ from (\ref{eq:fcc2psi}) with $\cos \theta$ replaced by $\sin \theta$. The corresponding pressure and velocity fields are $\mathcal{P}=- \mathrm{i} \omega \varPsi$ and $\epsilon \boldsymbol{\mathcal{U}} \mathrm{e}^{\mathrm{i} m \phi}=\epsilon \nabla \left( \varPsi \mathrm{e}^{ \mathrm{i} m \phi} \right)$, giving
\begin{eqnarray}
   \epsilon \boldsymbol{\mathcal{U}} &=& - \frac{3}{4} \frac{1-\mathrm{i}}{(1-a^3)^2} \frac{a_s^2 L_\nu U_0}{r^3 }   \left[ 2 \left(1-\frac{r^3}{a_m^3} \right)\sin(\theta) \hat{\boldsymbol{r}}+\left(1+2\frac{r^3}{a_m^3} \right) (\cos(\theta) \hat{\boldsymbol{\theta}} + \mathrm{i} m  \hat{\boldsymbol{\phi}}) \right]  . \qquad \, \label{eq:fcc2EQ}
\end{eqnarray}
Imposing no-slip at $r=a_m$, equations (\ref{eq:dc40}) and (\ref{eq:fcc12}) for $\mathfrak{u}_r^{(o)}$ yield the outer potential $\varPsi^{(o)}$ associated with $\boldsymbol{\mathcal{U}}^{(o)}$, which can actually be obtained from (\ref{eq:fcc2psi}) multiplied by $a^4$, with $\cos \theta$ replaced by $\sin \theta$ and $a_m$ by $a_s$. Accordingly, $\boldsymbol{\mathcal{U}}^{(o)}$ follows from (\ref{eq:fcc2EQ}) with the same substitutions and prefactor $a^4$. Boundary-layer corrections $\mathfrak{u}_\parallel$ and $\mathfrak{u}_\parallel^{(o)}$ then follow directly from (\ref{eq:drf1})-(\ref{eq:drf2}). Restoring the azimuthal dependence $\mathrm{e}^{\mathrm{i} m \phi}$ and reverting to Cartesian coordinates, integration over the inner boundary $r=a_s$ yields the total viscous force associated with the pressure (and similarly for the tangential viscous stress from the boundary-layer flow~\ref{eq:eqvcStL}). This recovers the dimensional force of the polar mode,
\begin{eqnarray} \label{eq:FpFveq}
\boldsymbol{F}_\nu = \frac{3}{2} \frac{\boldsymbol{F}_{v}}{1-a^3} = - 6 \pi \frac{\rho_f \nu a_s }{(1-a^3)^2} \frac{a_s}{L_\nu} \mathrm{d}_t \boldsymbol{\delta}-\frac{9}{4}  \,  \frac{{m}_{fs}}{(1-a^3)^2} \frac{L_\nu}{a_s} \mathrm{d}_t^2 \boldsymbol{\delta}  .
\end{eqnarray}
This recovers (\ref{eq:denueE}) and yields the viscous correction $C_a-C_a^i=(9/4)(1-a^3)^{-2} L_\nu/a_s$.

\section{Theoretical prefactors from boundary conditions}

This appendix summarises representative analytical expressions of the
prefactors determined from the boundary conditions for the various
models developed in the main text. For readability, only the compact
closed-form expressions most useful for documenting the derivations are reproduced here. The complete collection of analytical formulas is distributed through the accompanying Zenodo repository \cite{zenodo}.

\subsection{Radial magnetic field induced by the base flow} \label{sec:CFbr}
The interface condition, given by the continuity of $\breve{\mu}^{-1}\partial_r (r^2 b_r)$, reduces to its BLT version $[b_r]_-^+ =[\breve{\mu}^{-1} \partial_r b_r]_-^+=0$ at $r=a_s$ and $r=a_m$ \citep[see \S\ref{sec:BINTCBC}, but also equation~22 of][]{plunian2025three}. The four constants are then
\begin{eqnarray}
[\tilde{A}_+,\tilde{A}_-] &=& \frac{[\breve{c}_s \partial_r \Upsilon_s + [\breve{c}_s,-1] \lambda_{\mathfrak{s}} \Upsilon_s]_{r=a_s} }{\lambda_{\mathfrak{s}} (1+\breve{c}_s)} \label{syst:const1}, \\{} [\tilde{B}_+,\tilde{B}_-] &=& - \frac{[\breve{c}_m \partial_r \Upsilon_m + [1,-\breve{c}_m] \lambda_{\mathfrak{s}} \Upsilon_m]_{r=a_m}}{\lambda_{\mathfrak{s}} (1+\breve{c}_m)} ,  \label{syst:const12}
\end{eqnarray}
where $\Upsilon_j$ denotes the difference between the particular solutions in the fluid and solid at $r=a_j$. Similar calculations remain analytically tractable for equatorial modes but not for the rotating flow (\ref{syst:psi}). The boundary-layer approach is valid provided
$L_\eta\ll\ell$, $L_\eta\ll a_s \breve{\eta}_s^{-1/2}$, and $L_\eta\ll \breve{\eta}_m^{-1/2} \min(a_m,a_e-a_m)$, where $a_e-a_m$ is the thickness of the outer solid domain.

\subsection{Tangential components in the inviscid MHD boundary layer} \label{sec:CFbinv}

The interface conditions $\left[\breve{\eta} \partial_r \boldsymbol{b}_\parallel \right]_-^+ + B_{0r} [\boldsymbol{u}_\parallel]_-^+ = [\breve{\mu}^{-1} \boldsymbol{b}_\parallel  ]_-^+ = \boldsymbol{0}$  must then be enforced (\S\ref{sec:BINTCBC}). The prefactors $[A_\pm,B_\pm]$ are obtained generically as functions of an arbitrary $\boldsymbol{U}$ and $[\boldsymbol{b}_\parallel^{(p)},\boldsymbol{u}_\parallel^{(p)}]$. The expressions below assume $[\boldsymbol{b}_\parallel^{(p)},\boldsymbol{u}_\parallel^{(p)}]=[\boldsymbol{0},\boldsymbol{0}]$ in the solids (valid for uniform $\boldsymbol{B}_0$), but remain general if these quantities represent the jump of the particular solution across the interface. One finds $[A_\pm,B_\pm]=[A_\pm^{(0)}+A_\pm^{(1)},B_\pm^{(0)}+B_\pm^{(1)}]$, where $[A_\pm^{(0)},B_\pm^{(0)}]$ is independent of $[\boldsymbol{b}_\parallel^{(p)},\boldsymbol{u}_\parallel^{(p)}]$ and $[A_\pm^{(1)},B_\pm^{(1)}]$ is linear in it. The homogeneous contribution $[\boldsymbol{b}_\parallel^{(h)},\boldsymbol{u}_\parallel^{(h)}]$ is therefore determined by $[A_\pm^{(0)},B_\pm^{(0)}]$. For the polar mode with $B_{0\phi}=0$, this gives
\begin{eqnarray}
   [ A_\pm^{(0)},B_\pm^{(0)}] &=&\breve{\mu}_j^{-1} C_\pm^{(0)}= \pm \sqrt{2/\mathrm{i}} \, B_{0r} \mathcal{K}_j^{-1}  [\mathrm{i} \mathfrak{h} \sin \theta+\omega ^{-1 }U_\theta,\omega ^{-1 }U_\phi ]_{r=a_j} , \label{eq:A1} \\{} 
  [A_\pm^{(1)},B_\pm^{(1)}]  &=& \pm \sqrt{2/\mathrm{i}}  \mathcal{K}_j^{-1} [  \omega^{-1} B_{0r} \boldsymbol{u}_\parallel^{(p)}+ 2^{-1}(L_\eta^2 \partial_r \boldsymbol{b}_\parallel^{(p)} \mp L_\eta \breve{c}_j \tilde{\lambda}_{\mathfrak{s}}\boldsymbol{b}_\parallel^{(p)} )]_{r=a_j} , \qquad \label{eq:A3}
\end{eqnarray}
where the inner boundary-layer prefactors $(A_-,B_-)$ are obtained with $j=s$ and $\mathfrak{h} =1$, while the outer boundary-layer prefactors $(A_+,B_+)$ are obtained with $j=m$ and by removing the boundary motion with $\mathfrak{h} =0$ (as the outer boundary is at rest). Here we have noted $\boldsymbol{b}^{(p)}_\parallel=[b_\theta^{(p)},b_\phi^{(p)}]$, $\boldsymbol{u}^{(p)}_\parallel=[u_\theta^{(p)},u_\phi^{(p)}]$, and $\mathcal{K}_j= L_\eta(\breve{c}_j \tilde{\lambda}_{\mathfrak{s}} +2\mathrm{i} \tilde{\lambda}^{-1})/\sqrt{2 \mathrm{i}}$, where $\tilde{\lambda}=\lambda L_\eta$, $ \tilde{\lambda}_{\mathfrak{s}}={\lambda}_{\mathfrak{s}} L_\eta=1+\mathrm{i}$, such that $\mathcal{K}_s=L_\eta(\breve{c}_s+\sqrt{1-\mathrm{i} \Lambda_l}) $ ; prefactors of a boundary-layer are deduced from those of the other layer by changing the sign of ${\lambda}_{\mathfrak{s}}$ and $\lambda$.

Following BG95, one can discard the contribution of the particular solution when the boundary layer is thin enough. The boundary-layer solution results thus only from the boundary forcing term $[\boldsymbol{U}_\parallel+\boldsymbol{u}_\parallel]_-^+$, and $[A_\pm,B_\pm]=[A_\pm^{(0)},B_\pm^{(0)}]$. The solution becomes then independent of $B_{0\theta}$ and $B_{0\phi}$. Considering the basic flow (\ref{syst:Lamb})-(\ref{syst:Lamb2}), we obtain $[C_+,D_+]=\breve{\mu}_m [A_+,B_+]$, where equations (\ref{eq:A1})-(\ref{eq:A3}) give
\begin{eqnarray}
[A_-,B_-] \mathcal{K}_s &=& \frac{3}{2} \frac{B_{0r}}{1-a^3} [(1+\mathrm{i}) \sin  \theta , -(1-\mathrm{i}) (\gamma/2) \sin 2 \theta  ] \nonumber \\ &=& -a^{-3}  [A_+,B_+] \mathcal{K}_m= \breve{\mu}_s^{-1}  [C_-,D_-] \mathcal{K}_s=- a^{-3} \breve{\mu}_m^{-1} [C_+,D_+] \mathcal{K}_m 
, \qquad \label{eq:A1pDC} 
\end{eqnarray}
for the polar mode, while equatorial modes lead to (ignoring rotation effects) 
 \begin{eqnarray}
 [A_-,B_-] \mathcal{K}_s &=& \frac{3}{2} \frac{B_{0r}}{1-a^3} \left[-(1+\mathrm{i}) \cos  \theta , \pm (1-\mathrm{i})   \right] , 
 \end{eqnarray}
 and 
 \begin{eqnarray}
    [ A_+ , B_+ ]  \mathcal{K}_m &=& - \frac{  [C_-,D_-] \mathcal{K}_s}{\breve{\mu}_s} = -a^3 \left[ A_-,B_-  \right] \mathcal{K}_s  . 
 \end{eqnarray}

\subsection{Tangential components in the viscomagnetic oscillatory layers} \label{sec:CFbvis}

As in \S\ref{sec:BGinv}, the boundary-layer solution decomposes into two contributions and can be written for a given $\boldsymbol{U}$ and particular solution $[\boldsymbol{b}_\parallel^{(p)},\boldsymbol{u}_\parallel^{(p)}]$. Owing to the length of the general expressions, we restrict attention to the polar-mode flow (\ref{syst:Lamb})-(\ref{syst:Lamb2}). Using viscous interface conditions (\S\ref{sec:BINTCBC}), a no-slip inner boundary yields the prefactors $[A_\pm^{(0)},B_\pm^{(0)}]$ of the homogeneous solution. A no-slip outer boundary yields analogous expressions with $(\breve{\mu}_s,\breve{\eta}_s)$ replaced by $(\breve{\mu}_m,\breve{\eta}_m)$. The $\phi$ components vanish in the non-rotating limit.

\subsection{Magnetohydrodynamic Stokes–Ekman boundary layer} \label{sec:CFbvisrot}

Discarding the particular solution isolates the contribution independent of it (superscript $^{(0)}$, as in~\ref{eq:A1}). For a no-slip condition at $r=a_s$ with $\breve{\mu}_s=\breve{\eta}_s=1$, we obtain
\begin{eqnarray}
 \frac{({\check{\iota}}_++1)({\check{\kappa}}_+-{\check{\iota}}_+)  A_{-1}^{(0)}}{{\check{\iota}}_+ {\check{\kappa}}_+} &=&  - \frac{({\check{\kappa}}_++1)({\check{\kappa}}_+-{\check{\iota}}_+)  A_{-2}^{(0)}}{{\check{\iota}}_+ {\check{\kappa}}_+}= \frac{1-\mathrm{i}}{2} \frac{B_{0r} (\omega \sin \theta - \mathrm{i} U_\theta+U_\phi) }{\omega L_\eta} , \qquad
\end{eqnarray}
and
\begin{eqnarray}
   \frac{({\check{\kappa}}_-+1)({\check{\kappa}}_- - {\check{\iota}}_-) B_{-2}^{(0)} }{{\check{\iota}}_- {\check{\kappa}}_-}&=&  -  \frac{({\check{\iota}}_-+1)({\check{\kappa}}_- - {\check{\iota}}_-) B_{-1}^{(0)}}{{\check{\iota}}_- {\check{\kappa}}_-}= \frac{1-\mathrm{i}}{2} \frac{B_{0r} (\omega \sin \theta - \mathrm{i} U_\theta-U_\phi)}{\omega L_\eta} , \qquad
\end{eqnarray}
where $\omega \sin\theta$ arises from boundary motion (absent at $r=a_m$), and $X=\check{X}(1+\mathrm{i})/L_\eta$ for $\check{\iota}_\pm$ and $\check{\kappa}_\pm$. The $\phi$ components vanish in the non-rotating limit. Further progress requires specifying both $\boldsymbol{U}$ and the particular solution. For the polar mode, $\boldsymbol{U}$ follows from (\ref{syst:psi}), but its complexity precludes an analytical particular solution; a diffusionless approximation is nevertheless obtained for weak Lorentz forces (appendix~\ref{sec:diffuWekL}). 

\bibliographystyle{jfm}
\bibliography{minimal_references}
\include{supplementary_material}

\end{document}

%% file: supplementary_material.tex
\newpage
\setcounter{page}{1}
\section*{SUPPLEMENTARY MATERIAL}
\renewcommand{\thefigure}{S\arabic{figure}}
\setcounter{figure}{0}
\renewcommand{\thesection}{S\arabic{section}}
\setcounter{section}{0}
\renewcommand{\appendixname}{}

\section{COMSOL simulation details} \label{sec:COMSim}

In the \textsc{comsol} finite element model, we use the Arbitrary Lagrangian-Eulerian approach (ALE) to translate the inner sphere and modify the mesh in accordance with its imposed kinematics.
This method allows for arbitrary displacements, and enables stress-free or no-slip boundary conditions at the inner or outer boundaries.
Considering the magnetic field departure $\boldsymbol{B}_1$ from an imposed magnetic field $\boldsymbol{B}_0$, equation (\ref{eq:dc1}) for $\boldsymbol{B}=\boldsymbol{B}_0+\boldsymbol{B}_1$ shows that $\boldsymbol{B}_1=\nabla \times \boldsymbol{A}$ can be obtained from the equation
\begin{eqnarray}
    \partial_t \boldsymbol{A} =\boldsymbol{V} \times (\boldsymbol{B}_0+\boldsymbol{B}_1) + \breve{\eta} \nabla^2 \boldsymbol{A} , \label{eq:magpot}
\end{eqnarray}
 with the magnetic potential vector $\boldsymbol{A}$, and using $\breve{\eta}=1$ in the fluid, $\breve{\eta}=\breve{\eta}_s$ in the inner solid sphere, and $\breve{\eta}=\breve{\eta}_m$ in the outer solid domain. As in \cite{cebron2012magnetohydrodynamic}, the non-linear MHD solution is then obtained by integrating equations (\ref{eq:dc0}) in the fluid, and (\ref{eq:magpot}) everywhere using the appropriate magnetic diffusivity $\breve{\eta}$. To consider an unbounded outer solid domain, a typical radius  $  > 7 a_s $ is used. Because of the gauge used by \textsc{comsol}, all domains should be electrically conducting and we typically use a conductivity $10^3$ times smaller than the fluid one to approximate insulating domains.

While equatorial modes require three-dimensional meshes, the polar mode can be simulated using the 2D meridional plane of an axisymmetric model.
The finite element method coupled with the ALE is computationally intensive and, hence, is limited to large viscosity.
Nevertheless, it allowed us to examine limitations of parametric boundary conditions.
Furthermore, we can compute the forces at both boundaries, as the pressure field is a by-product of the numerical solution.
To compute the added-mass coefficient $C_a$, we record the values of the acceleration part of the axial total force exerted at the boundary, when the velocity of the inner sphere is null (e.g. figure~\ref{fig:sketchFSto}, where the yellow vertical segment is proportional to the added-mass coefficient).
Both \textsc{xshells} and \textsc{comsol} can compute all magnetic forces and dissipation magnitude, including magnetic corrections of the added-mass coefficient. However, only \textsc{comsol} can determine the inviscid added-mass coefficient or its viscous correction (as they require the pressure, not computed by \textsc{xshells}).

\section{Effective permeability and magnetic shielding} \label{sec:B00}
Although the boundary-layer theory developed in this work is valid for an arbitrary $\phi$-independent $\boldsymbol{B}_0$, it is useful to relate this field to an externally imposed field $\boldsymbol{B}_e$ through the magnetic permeability of the different domains. This relation can be obtained analytically for a uniform $\boldsymbol{B}_e$ and a spherical outer solid medium with permeability $\mu_m$ and outer radius $a_e$. To this end, dipolar contributions are added to $\boldsymbol{B}_e$ in each domain, arising from internal and external sources of the respective forms $(k_1 x + k_2 y + k_3 z) r^{-3}$ and $k_4 x + k_5 y + k_6 z$. Assuming, for instance, an external H-field of the form $\boldsymbol{B}_e=\mu_0 \boldsymbol{H}_e= \mu_0 H_e \hat{\boldsymbol{z}}$ imposed far away, the field in the vacuum reads $\boldsymbol{H}= \nabla( k_e z r^{-3} +H_e z)$ and the field in the inner sphere reads $\boldsymbol{H}= \nabla( k_s z)$ to ensure regularity at the centre. In the other two domains, the field reads $\boldsymbol{H}= \nabla( k_i z r^{-3} +k_j z)$. The $6$ coefficients $k_i$ are then determined by imposing continuity of $\boldsymbol{H}_\parallel= \boldsymbol{B}_\parallel/\mu$ and of $B_r$ at each interface.

Recalling that the magnetic field inside a full sphere of permeability $\mu_s$ embedded in an unbounded medium of permeability $\mu_0$ is
\begin{eqnarray}
\boldsymbol{B}(r\leq a_s) = \frac{3}{1+2 \mu_{0}/\mu_s} \boldsymbol{B}_e , \label{eq:inneD}
\end{eqnarray}
the field in the inner sphere of the present configuration follows from (\ref{eq:inneD}) with $\mu_s$ replaced by the effective permeability $\mu_2$,
\begin{eqnarray}
\frac{1}{\mu_2}&=&\frac{1}{\mu_1} + \frac{\mu_{0m}(a^2+a_2^2+a a_2)(a_2-a)}{\mu_0 a^3} \frac{(2 \mu_f +\mu_m)\mu_{fs} a^3-\mu_{fm}(2 \mu_f +\mu_s)}{9 \mu_f \mu_m \mu_s}  , \\
\mu_1 &=& \frac{3 \mu_0 \mu_s}{(1-a^3)(\mu_f+\mu_s \mu_0 \mu_f^{-1}- \mu_s)+ \mu_0(2+a^3)} ,
\end{eqnarray}
with $\mu_{ij}=\mu_i-\mu_j$ and $a_2=a_s/a_e$. Several limiting cases follow. For $\mu_m=\mu_0$, corresponding to a solid sphere surrounded by a shell of permeability $\mu_f$, one obtains $\mu_2=\mu_1$, such that
\begin{eqnarray}
\boldsymbol{B}(r\leq a_s) = \frac{9  \boldsymbol{B}_e }{1+4\mu_0/\mu_s+2a^3(1+\mu_0/\mu_s)+2 (1-a^3)(\mu_f/\mu_s+\mu_0/\mu_f)}   ,
\end{eqnarray}
inside the 'coated' sphere. If we further assume $\mu_f=\mu_0$, then $\mu_1=\mu_s$, recovering (\ref{eq:inneD}); alternatively, setting $\mu_s=\mu_0$ yields
\begin{eqnarray}
\frac{\mu_1}{\mu_0} &=& \frac{3}{1+2a^3+(1-a^3)(\mu_f/\mu_0+ \mu_0/ \mu_f)} ,  \label{eq:hjkz}
\end{eqnarray}
or equivalently
\begin{eqnarray}
\boldsymbol{B}(r\leq a_s) = \frac{9  \boldsymbol{B}_e }{5+4a^3+2 (1-a^3)(\mu_f/\mu_0+\mu_0/\mu_f)}  \approx  \frac{9 \mu_0}{2 (1-a^3)} \frac{\boldsymbol{B}_e}{\mu_f} + \mathcal{O} \left( \frac{1}{\mu_f^2} \right)  , \quad \label{eq:fg22}
\end{eqnarray}
where the exact expression recovers the spherical-shell result \cite[p.~265]{stratton2007electromagnetic}, and the large-$\mu_f$ asymptotic expansion retrieves equation~5.122 of \cite{jackson1977classical}, illustrating an efficient magnetic shielding by a thin high-permeability shell. For $\mu_s=\mu_f=\mu_0$, one obtains $\mu_1=\mu_0$ and an expression for $\mu_2$ formally identical to (\ref{eq:hjkz}) where $\mu_f$ and $a$ are replaced by $\mu_m$ and $a_3=a_m/a_e=a_2/a$, respectively.

\section{Accuracy of the \textit{huBL} approach } \label{sec:huBLaccu}
\begin{figure}
    \centering
    \includegraphics[width=1.0\linewidth]{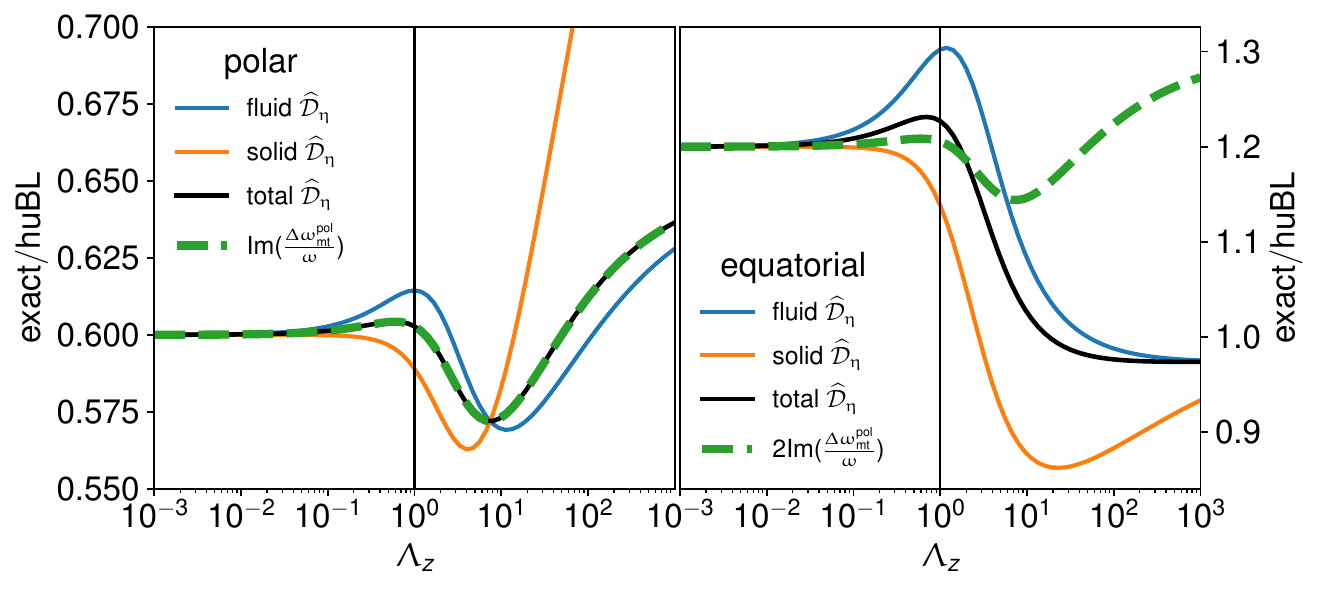}
    \caption{Ratio between the 'exact' dissipation integrals $\widehat{\mathcal D}_\eta$ and the \textit{huBL} counterparts for the polar mode (left) and the equatorial modes (right), in the fluid, in the solid sphere and in total (sum of the two). We also show (left) the corresponding polar mode ratio for $\Imag(\Delta \omega_{mt})$, i.e. $\mathcal{J}_1$ from (\ref{eq:exInt}), and $2 \mathcal{J}_1$ (right). At $\breve{c}=1$ and $\gamma=0$.}
  \label{fig:huBL}
\end{figure}

For a uniform axial field ($\Lambda=\Lambda_z/3$), we calculate the ratio between the exact integrals, retaining the angular dependence of the imposed magnetic field rather than replacing $B_{0r}$ by its surface average, and the corresponding \textit{huBL} counterparts. The results are shown in figure \ref{fig:huBL} for the solid sphere, for the fluid, and for the union of these two domains (named 'total'). This latter ratio is denoted $\mathcal{I}_{\eta}^{\mathrm{pol}}$ and  $\mathcal{I}_{\eta}^{\mathrm{eq}}$ for the polar and equatorial modes, respectively.

At $\breve{c}=1$ and $\gamma=0$, the corresponding magnetic tension ratio $\mathcal{J}_1$ for the polar mode (left panel), analytically given by (\ref{eq:exInt}), reproduces perfectly the total polar mode dissipation ratio. While the respective ratios differs in the fluid and the solid sphere, we observe that $\mathcal{I}_{\eta}^{\mathrm{pol}}$ varies by less than $7\%$ on the range $\Lambda_z \in [0,10^3]$.

For equatorial modes (right figure), $\mathcal{I}_{\eta}^{\mathrm{eq}}$ is exactly $2 \mathcal{J}_1$ for $\Lambda_z \ll 1$, but differs at larger $\Lambda_z$. For $\Lambda_z \in [0,10^3]$, $\mathcal{I}_{\eta}^{\mathrm{eq}}$ only differs from $\mathcal{I}_{\eta}^{\mathrm{eq}} \approx 1.1$ by less than $10\%$, and the three ratios converge towards an asymptotic limit value near $\mathcal{I}_{\eta}^{\mathrm{eq}} \approx 0.97$ for very large $\Lambda_z \gg 1$.

\section{Polar mode: confinement and strong rotation effects on the bulk flow}
\label{Supp:RotationPolar}
\begin{figure}
    \centering
    \begin{subfigure}[b]{0.8\textwidth}
    \centering
    \includegraphics[width=1.0\linewidth]{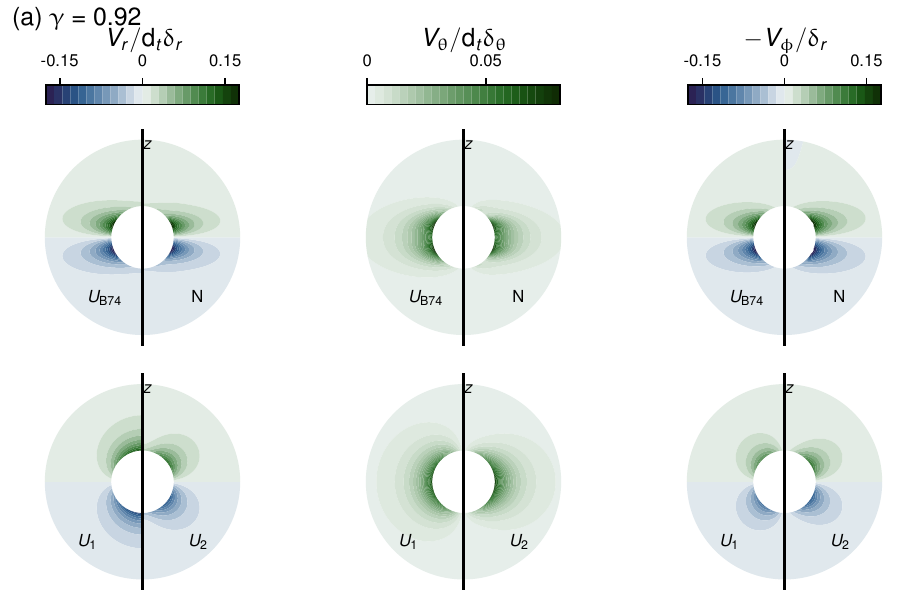}
    \end{subfigure}
    \begin{subfigure}[b]{0.8\textwidth}
    \centering
    \includegraphics[width=1.0\linewidth]{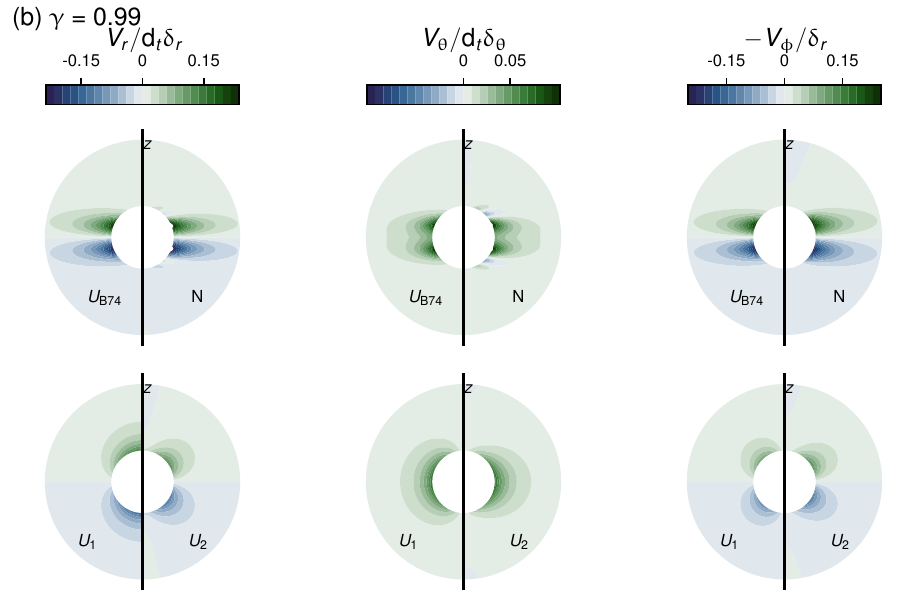}
    \end{subfigure}
    \caption{Surface plots compare B74 velocity field solution ($\boldsymbol{U}_\mathrm{B74}$): $\zeta_m$ from (\ref{syst:zet_est}), rotation-perturbed potential flow $\boldsymbol{U}_1$ (\ref{syst:Lamb2}), and its correction $\boldsymbol{U}_2$ (\ref{eq:Busse_ordre2}) with numerical (N) simulation (meridional plane, zoom up to $r=6a_s$). Parameters: (a) $\gamma=0.92$, (b) $\gamma=0.99$, for the polar mode ($a=0.05$, $L_\nu=0.002$).}
    \label{fig:meridVelocityRotation}
\end{figure}

In figure \ref{fig:meridVelocityRotation}, meridional plots of the three velocity components forced by polar oscillation are shown for (a) $\gamma =$ 0.92 and (b) $\gamma =$ 0.99.
There, we compare B74 solution ($\boldsymbol{U}_\mathrm{B74}$), the rotation-perturbed potential flow $\boldsymbol{U}_1$ (\ref{syst:Lamb2}), and its correction $\boldsymbol{U}_2^{a\to0}$ (\ref{eq:Busse_ordre2}) with numerical simulations (meridional plane, zoom up to $r=6a_s$).
The B74 solution ($\boldsymbol{U}_\mathrm{B74}$) effectively captures the influence of rotation.
Rotation significantly modifies the potential flow $\boldsymbol{U}_1$, which classically propagates in the axial direction.
This modification results in a velocity distribution spreading outward in the equatorial region.

\section{Magnetohydrodynamic boundary layers} \label{sec:stoBL2}

\subsection{Polar mode and uniform axial field: magnetic tension vs. magnetic pressure} \label{sec:magPPOdc}
For the polar mode, the Lorentz force has no axial component: the Lorentz force, $(\nabla\times\boldsymbol{b})\times\boldsymbol{B}_0$, is perpendicular to the imposed axial field $\boldsymbol{B}_0$ at this order and therefore cannot contribute to the $z$-directed drag. This can be verified directly from the BLT. The magnetic-tension contribution is
$\boldsymbol{f}_{mt}\cdot\hat{\boldsymbol{z}}
\simeq B_{0r}\boldsymbol{b}_{\parallel}\cdot\hat{\boldsymbol{z}}
=-B_0b_\theta\sin\theta\cos\theta$,
whereas the magnetic-pressure contribution is
$\boldsymbol{f}_{mp}\cdot\hat{\boldsymbol{z}}
=f_{mp}\,\hat{\boldsymbol{r}}\cdot\hat{\boldsymbol{z}}
=B_0b_\theta\sin\theta\cos\theta$,
since $f_{mp}=-\boldsymbol{B}_0\cdot\boldsymbol{b}$. The two
contributions therefore exactly cancel. Consequently, the drag associated
with the Ohmic dissipation
(\ref{eq:pmBatch21})--(\ref{eq:pmBatch22}) is transmitted to the solid
through the pressure $\mathcal{P}$ of
(\ref{eq:dc40})--(\ref{eq:dc41}), which is of the same order as
$\boldsymbol{b}_{\parallel}$. This closely parallels the classical Stokes
problem, in which the pressure likewise carries a significant fraction of
the viscous drag (Appendix~\ref{sec:stoBL}).

\subsection{Symmetric form of the viscomagnetic oscillatory layers} \label{sec:symStmL}
Using $B_{0r}$ as the magnetic field scale, the Alfv\'en speed $\tilde{V}_A=B_{0r}/\sqrt{\rho_f \mu_f}$ as the velocity scale, and the dimensional $\omegas^{-1}$ as the time scale, such that the length scale is $\tilde{V}_A/\omega_s$, the governing dynamical equations read (still noting $r$ with its dimensionless counterpart)
\begin{eqnarray}
\mathrm{i} [\boldsymbol{b},\boldsymbol{u}]&=& \partial_r^2 [\tilde{\eta} \boldsymbol{b},\tilde{\nu} \boldsymbol{u}] + \partial_r [\boldsymbol{u},\boldsymbol{b}], \label{eq:jault}
\end{eqnarray}
with $(\tilde{\eta},\tilde{\nu})= (\eta,\nu) \omegas/\tilde{V}_A^2$. These equations are symmetric for the magnetic and velocity fields and are often combined using the so-called Elsasser variables $\boldsymbol{u}\pm \boldsymbol{b}$. In these units, equation (\ref{eq:STBU}) adopts the following symmetric form
\begin{eqnarray}
\lambda_\pm=    \frac{1+\mathrm{i}}{2 \sqrt{\tilde{\nu} \tilde{\eta}}} \sqrt{\tilde{\nu}+\tilde{\eta}-\mathrm{i} \mp \sqrt{(\tilde{\nu}+\tilde{\eta}-\mathrm{i})^2-4 \tilde{\nu} \tilde{\eta}}}
\end{eqnarray}
where $\tilde{\eta}$ and $\tilde{\nu}$ play a symmetric role.

\subsection{Additional figures for viscomagnetic oscillatory layers}

\begin{figure}
\centering
\begin{subfigure}[b]{1.1\textwidth}
\centering
\includegraphics[width=1.0\linewidth]{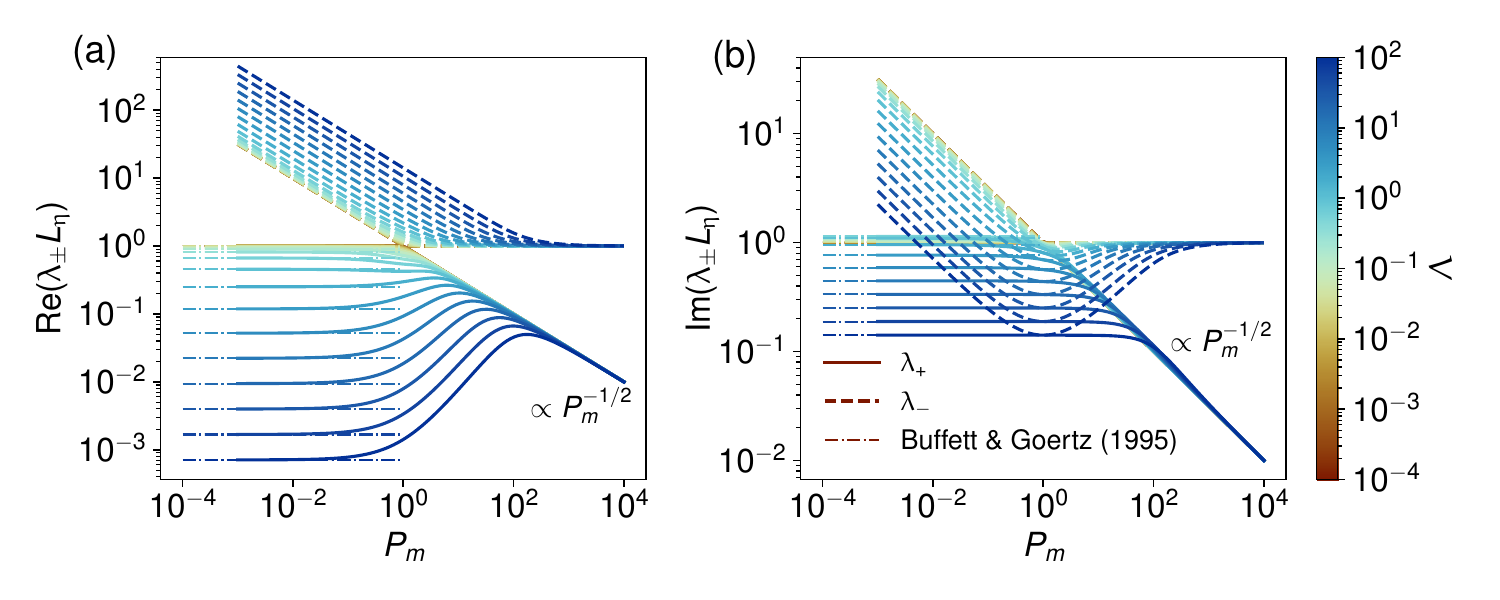}
\end{subfigure}
\begin{subfigure}[b]{\textwidth}
\centering
\includegraphics[width=1\linewidth]{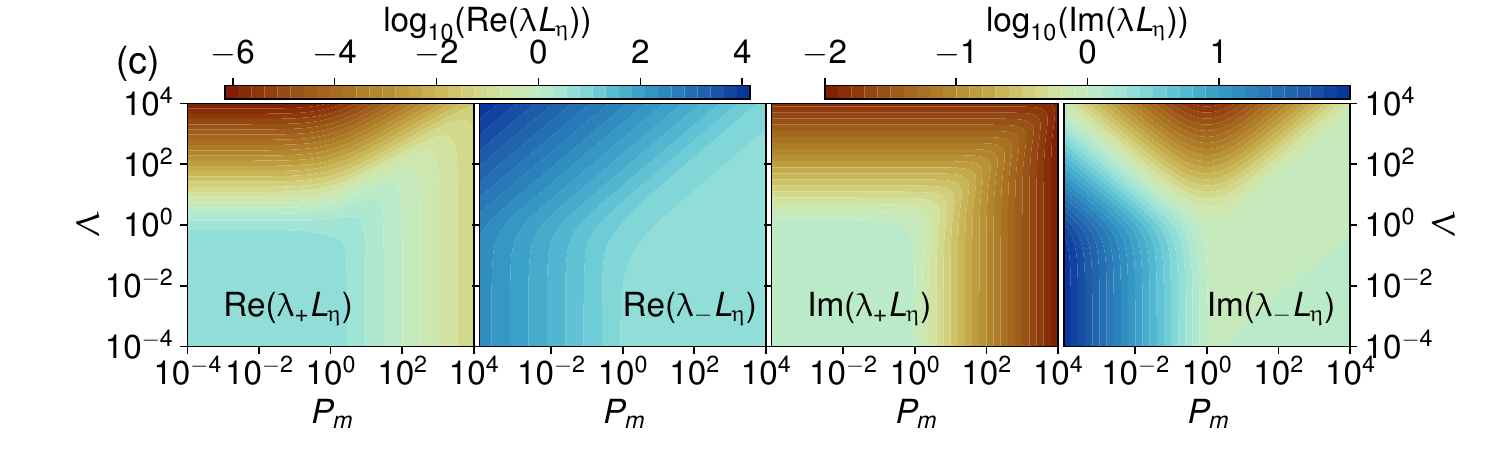}
\end{subfigure}
\caption{Effect of the magnetic Prandtl number $P_m$ and $\Lambda$ on the normalised wavenumber ($\lambda L_\eta$) of the viscomagnetic oscillatory boundary layer, as expressed by equation~(\ref{eq:STBU}). (a) Real and (b) imaginary part. (c) Regime diagram.}
    \label{fig:PmlambdaLetaFull}
\end{figure}

Figure \ref{fig:PmlambdaLetaFull} illustrates the evolution of the two complex wavenumbers of the oscillatory viscomagnetic boundary layer (\ref{eq:STBU}) with $P_m$ and magnetic-field strength $\Lambda$, highlighting the continuous transition between regimes. In particular, the solution recovers the purely viscous Stokes-layer wavenumber $(1+\mathrm{i})/L_\nu$ and the inviscid magnetic skin-depth wavenumber $(1+\mathrm{i})/L_\eta$. For sufficiently large $P_m$, the wavenumbers become independent of $\Lambda$.

\begin{figure}
    \centering
    \includegraphics[width=1.0\linewidth]{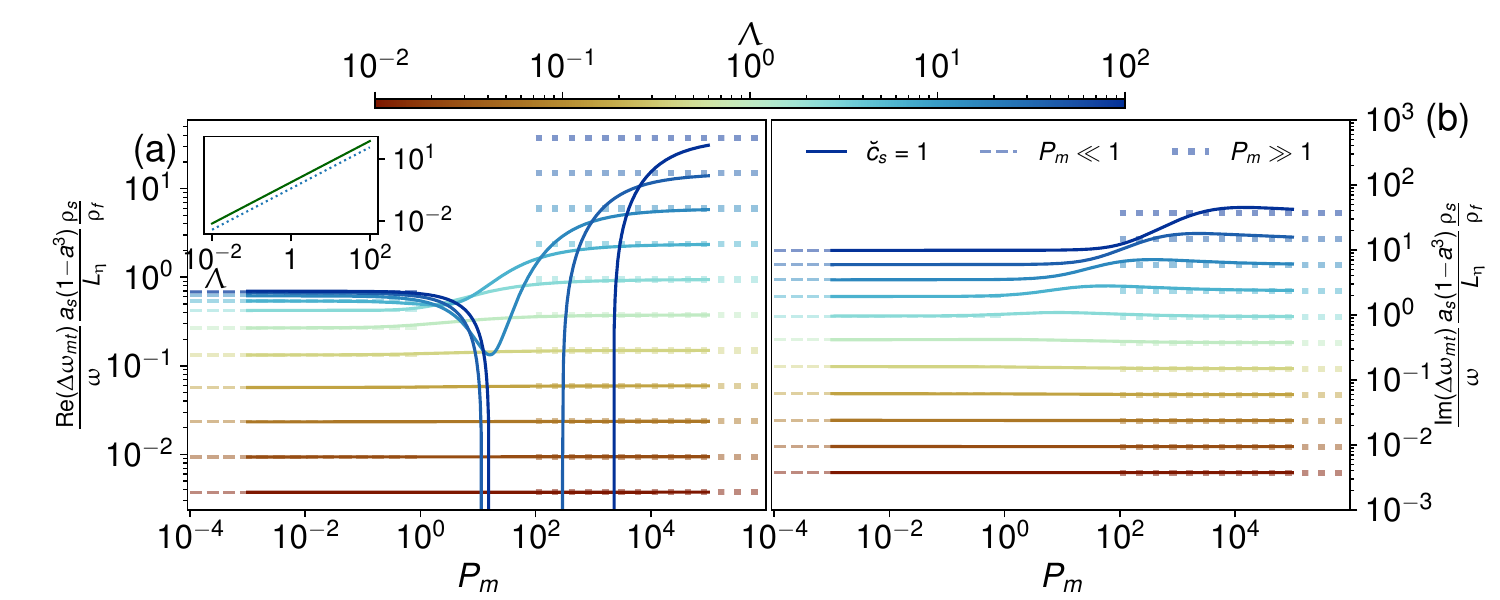}
    \caption{Effect of the magnetic Prandtl number $P_m$ and $\Lambda$ on the normalised magnetic stress detuning of the viscomagnetic oscillatory boundary layer for stress-free boundary conditions, and $\breve{c}_s = 1$. (a) Real and (b) imaginary part. In the inset of (a), magnetic stress detuning for large $P_m$ (dotted line) and for insulating inner sphere (dark green line), as a function of $\Lambda$. The limits $P_m\ll1$ and $P_m\gg1$ correspond to equations (\ref{eq:Bufextde}) and (\ref{eq:dmnd}), respectively.}
  \label{fig:PmLambdaMagneticDetuningStressFree}
\end{figure}

\subsection{Asymptotic limits of MHD Stokes–Ekman layers} \label{sec:appBLHa}

Comparing with $\lambda_\pm $ from equation (\ref{eq:STBU}), a series expansion of equations (\ref{eq:lrovi1})-(\ref{eq:lrovi2}) about the non-rotating limit $\gamma=0$ shows that $\iota_\pm$ tend towards $\lambda_\pm$ whereas $ \kappa_\pm$ tends towards $\lambda_\mp$, with
\begin{eqnarray}
 \frac{\iota_+}{\lambda_+}-1 = 1-\frac{\kappa_-}{\lambda_+}=\frac{1+\mathrm{i} P_m^{-2} \Lambda_l}{2} \gamma \cos \theta , \quad \frac{\iota_-}{\lambda_-}-1 = 1-\frac{\kappa_+}{\lambda_-}= \frac{\mathrm{i} \Lambda_l}{2 P_m^2} \gamma \cos \theta
\end{eqnarray}
for $P_m \gg 1$, and
\begin{eqnarray}
 \frac{\iota_+}{\lambda_+}-1 = 1-\frac{\kappa_-}{\lambda_+}= - \frac{1+2 P_m}{2} \mathrm{i} \Lambda_l \gamma \cos \theta , \quad \frac{\iota_-}{\lambda_-}-1 = 1-\frac{\kappa_+}{\lambda_-}= - \frac{\gamma \cos \theta}{2}
\end{eqnarray}
for $P_m \ll 1$ and $\Lambda_l \ll 1$. In the limit $\Lambda_l, P_m^{-1} \ll 1$, we obtain
\begin{align}
    \iota_+ = \frac{1+\mathrm{i}}{L_\nu} \sqrt{1+ \gamma \cos \theta} , \qquad \kappa_- = \frac{1+\mathrm{i}}{L_\nu} \sqrt{1- \gamma \cos \theta} , \qquad \iota_-=\kappa_+=\frac{1+\mathrm{i}}{L_\eta} ,
\end{align}
recovering the usual expressions for Stokes-Ekman boundary layers \cite[e.g. equation 89 of][]{smylie1998viscous}, e.g. found within rotating spheres in librations \cite[][]{sauret2013libration}. When $\gamma>1$, the boundary layer thickness diverges at the so-called critical latitudes given by $\cos \theta_\pm =\mp 1/\gamma$, which turns out to be also the case for the general expressions (\ref{eq:lrovi1})-(\ref{eq:lrovi2}): $\iota_+$ and $\kappa_-$ vanish for $\theta_+$ and $\theta_-$, respectively, for any values of $P_m$ and $\Lambda_l$ (and thus for any magnetic field geometry). For the uniform axial magnetic field,  $\iota_-$  also vanishes for $\theta_-$, and $\kappa_+$ for $\theta_+$.

In the quasi-steady limit $\omega \to 0$, the solution at $\theta=0$ reduces to the familiar plane Ekman-Hartmann layer from spin-up studies. In particular, \cite{benton1969spin} derived two complex wavenumbers $k$ and $m$ (their equations 30-36). Comparing their steady solution (equations 51-52) with the steady limit of (\ref{eq:lrovi1})-(\ref{eq:lrovi2}) shows that their $m=0$ corresponds to $\iota_+=-\kappa_-=0$, while their $k$ is recovered from $\kappa_+=\iota_-^*$ (with $^*$ denoting complex conjugation). In their scaling, $k=\kappa_+ L_\varOmega=[\Lambda_\varOmega+(\Lambda_\varOmega^2)^{1/2}]^{1/2}$, where $L_\varOmega=L_\eta (P_m/\gamma)^{1/2}$ is the Ekman depth and $\Lambda_\varOmega=\Lambda_l/\gamma+\mathrm{i} \cos \theta$, reducing to $\Lambda_\varOmega=\Lambda_z/\gamma+\mathrm{i}$ at $\theta=0$.

For a uniform magnetic field bounded by an infinite rigid insulating plate, \cite{debnath1974unsteady} analysed oscillatory plane boundary layers (recovered here at $\theta=0$) including viscous, rotational, and magnetic effects. However, the Lorentz force was approximated as $(\nabla \times \boldsymbol{B}) \times \boldsymbol{B}=-\sigma_f \rho_f^{-1} B_0^2 \boldsymbol{V}$ \cite[see equation 2.5 of][]{debnath1972unsteady}, which assumes (i) negligible electric and induced magnetic fields and (ii) velocity predominantly perpendicular to $\boldsymbol{B}_0$ \cite[][]{rossow1958flow}. While valid for variants of Hartmann layers \cite[see also equation 10.16 of][]{jackson1977classical} or specific quasi-static MHD regimes \cite[][]{thess2007transition}, this approximation suppresses magnetic perturbations and yields only two wavenumbers, in contrast with the four in (\ref{eq:lrovi1})-(\ref{eq:lrovi2}). Consequently, the radial wavenumber $\lambda$ of BG95 cannot be recovered as a limiting case. The magnetic Stokes-Ekman layer considered here therefore generalises the Stokes-Ekman-Hartmann layer of \cite{debnath1974unsteady}.

\section{Theory–simulation comparison}
In this section, we propose further comparisons of the boundary layer models with numerical results to cross-validate the theoretical development and the simulations.
\subsection{Boundary layers at small $\Lambda_l$}
\label{S:ProfilesPolar}
\begin{figure}
\centering
\begin{subfigure}[b]{0.80\textwidth}
    \centering
    \includegraphics[width=1\linewidth]{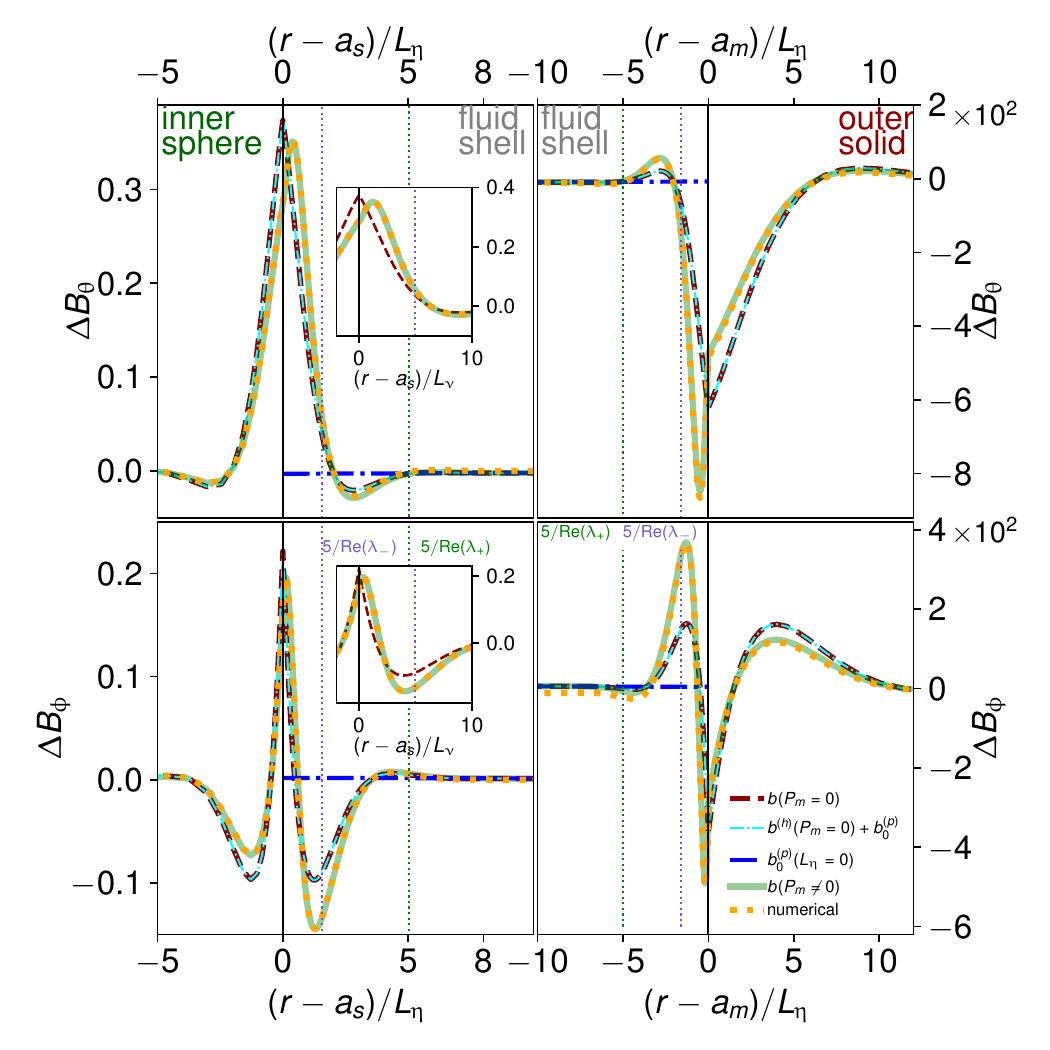}
\end{subfigure}
\caption{Tangential magnetic field perturbations, close to the inner (left) and outer (right) boundaries. Numerical \textsc{xshells} solution (orange-dotted). Boundary layer solutions: inviscid MHD (dark-red dashed), inviscid MHD diffusionless (cyan dash-dotted), viscomagnetic oscillatory (thick green), and diffusionless bulk (dash-dotted blue). Equatorial forcing. Parameters: as in figure~\ref{fig:xRadialMagneticBoundaryLayer}, but  $\Lambda_z=6.67\times10^{-3}$,  $\Lambda_l=5\times10^{-3}$, $\gamma=0$.}
   \label{fig:xRyViscoMagRotMagThetaPhiSmallLambda}
\end{figure}

In the presence of a weak $\Lambda_l$, i.e. a weak magnetic field, the velocity boundary layer follows the viscous skin layer solution (derived in appendix \ref{sec:stoBL}), in purple dashed, in figure
\ref{fig:xRyViscoMagRotVThetaPhiSmallLambda} ($\Lambda_l=5\times10^{-3}$) for equatorial forcing. The inviscid MHD solutions (dark-red dashed and azure dash dotted) result in negligible velocities. The viscomagnetic oscillatory solution (green solid) converges to the purely viscous boundary layer for the velocity.
Including the higher-order viscous bulk flow can help to reconcile the numerical results with the theory, as shown in the right insets of \ref{fig:xRyViscoMagRotVThetaPhiSmallLambda}b.
The numerical radial velocity perturbation, orange dotted in figure \ref{fig:xRyViscoMagRotVThetaPhiSmallLambda}a, is perfectly captured by the higher-order viscomagnetic oscillatory solution and the purely viscous boundary layer.
The viscomagnetic oscillatory solution (green solid) describes perfectly the magnetic field perturbations \ref{fig:xRyViscoMagRotMagThetaPhiSmallLambda}.
Analogous agreements for both velocity and magnetic perturbations were found for polar forcing at small $\Lambda_l$.
\begin{figure}
\centering
\begin{subfigure}[b]{0.8\textwidth}
    \centering
    \includegraphics[width=1\linewidth]{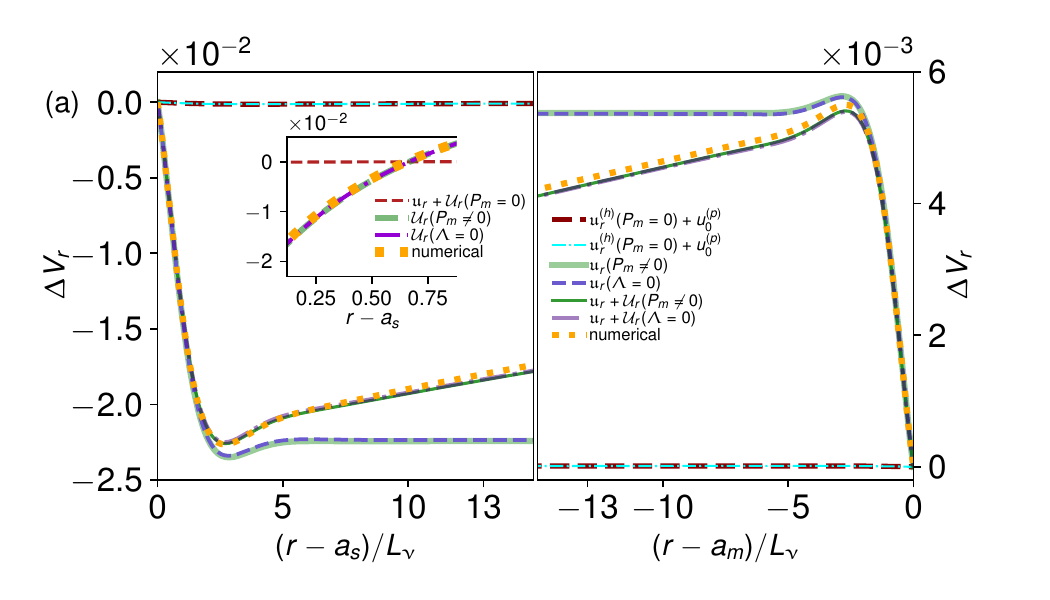}
\end{subfigure}
\begin{subfigure}[b]{0.80\textwidth}
    \centering
    \includegraphics[width=1\linewidth]{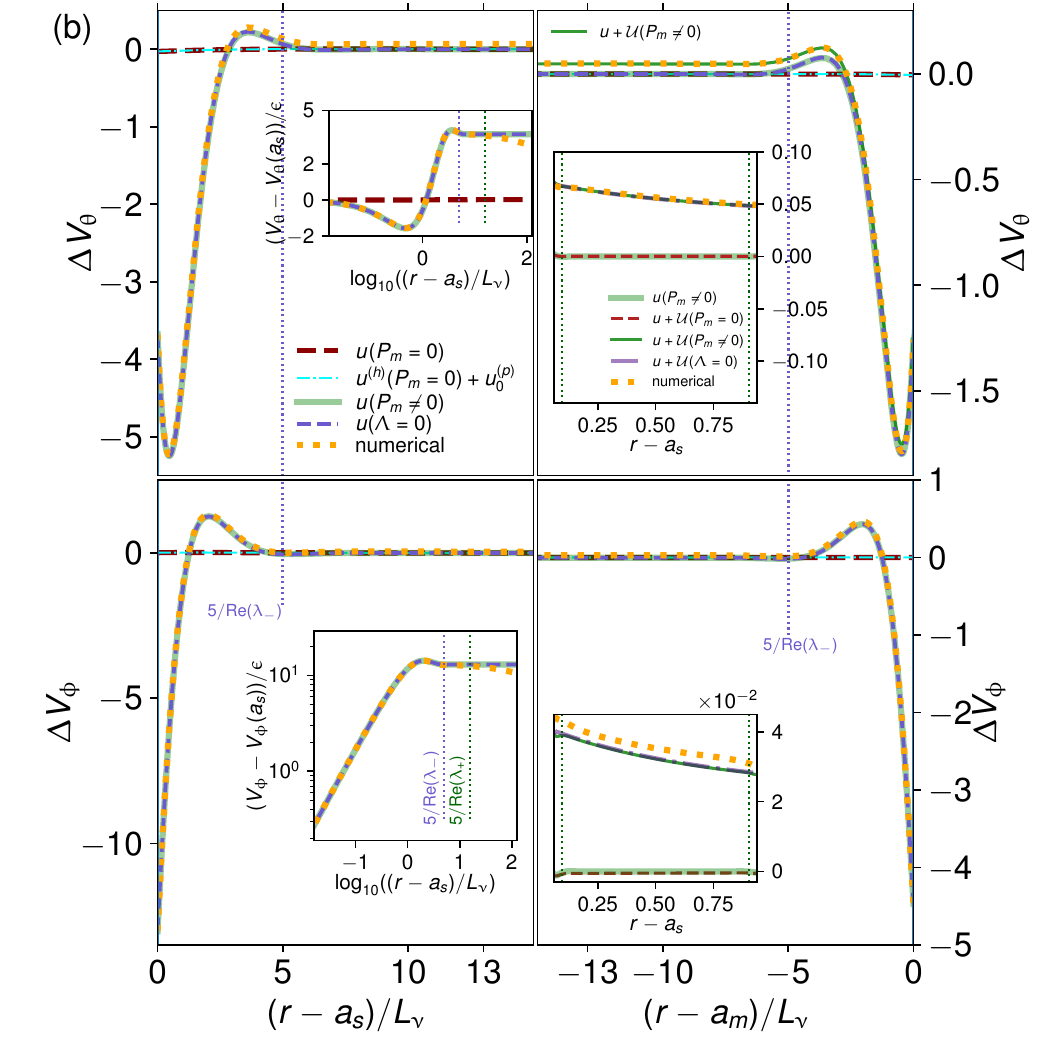}
\end{subfigure}
\caption{Radial $\Delta V_r$ (top) and tangential $\Delta V_\theta$ (middle), $\Delta V_\phi$ (bottom) velocity field perturbations close to the inner (left) and outer (right) boundaries.
     Numerical \textsc{xshells} solution (orange-dotted).
    Boundary layer solutions: inviscid MHD (dark-red dashed), inviscid MHD diffusionless (cyan dash-dotted), viscomagnetic oscillatory (thick green), purely viscous (purple dashed). Insets on the right show the velocity perturbation in the bulk. The homogeneous boundary layer and the secondary bulk flow components are summed:  inviscid MHD (dark-red dashed), viscomagnetic oscillatory (dark-green), purely viscous (purple dash-dotted).
     Equatorial forcing. Parameters: as in figure~\ref{fig:xRadialMagneticBoundaryLayer}, but   $\Lambda_z=6.67\times10^{-3}$, $\Lambda_l=5\times10^{-3}$, $\gamma=0$.}
   \label{fig:xRyViscoMagRotVThetaPhiSmallLambda}
\end{figure}

\subsection{Boundary layers in the weak-rotation regime}

\begin{figure}
\centering
\begin{subfigure}[b]{0.80\textwidth}
    \centering
    \includegraphics[width=1\linewidth]{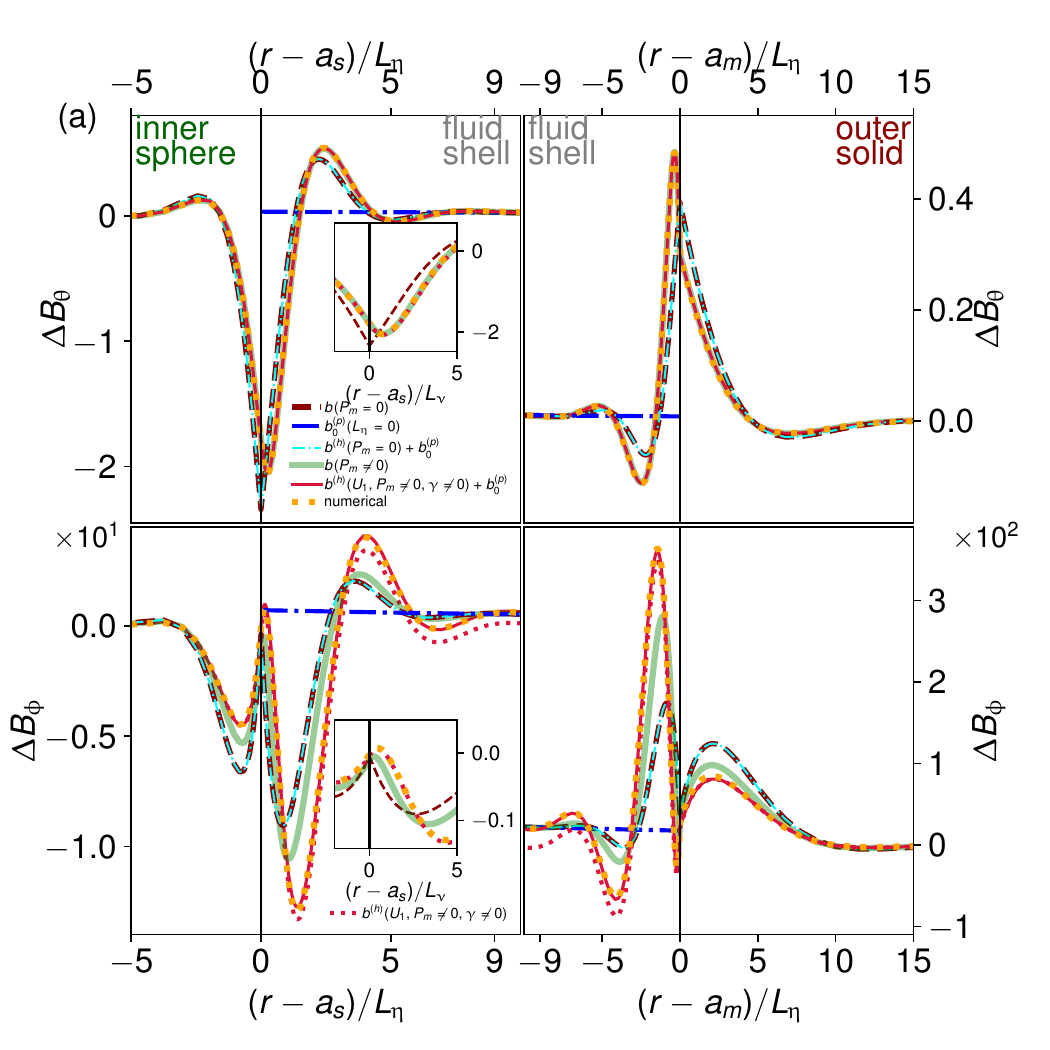}
\end{subfigure}
\caption{
Tangential magnetic field perturbations, close to the inner (left) and outer (right) boundaries.
Numerical \textsc{xshells} solution (orange-dotted).
Boundary layer solutions: inviscid MHD (dark-red dashed), inviscid MHD diffusionless (cyan dash-dotted), viscomagnetic oscillatory (thick green), purely viscous (purple dashed),  MHD Stokes-Ekman (red dotted), and diffusionless bulk (dash-dotted blue).
Polar forcing. Parameters: as in figure~\ref{fig:xRadialMagneticBoundaryLayer}.}
   \label{fig:xRyViscoMagRotBThetaPhiGamma1em1}
\end{figure}

\begin{figure}
\centering
\begin{subfigure}[b]{0.80\textwidth}
\centering
    \includegraphics[width=1\linewidth]{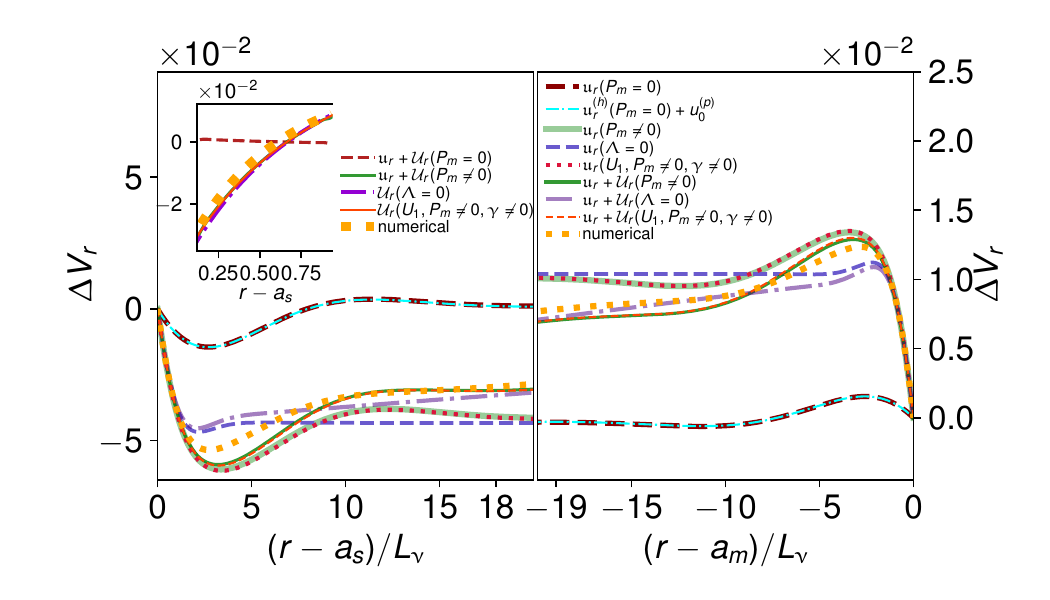}
\end{subfigure}
\begin{subfigure}[b]{0.80\textwidth}
    \centering
    \includegraphics[width=1\linewidth]{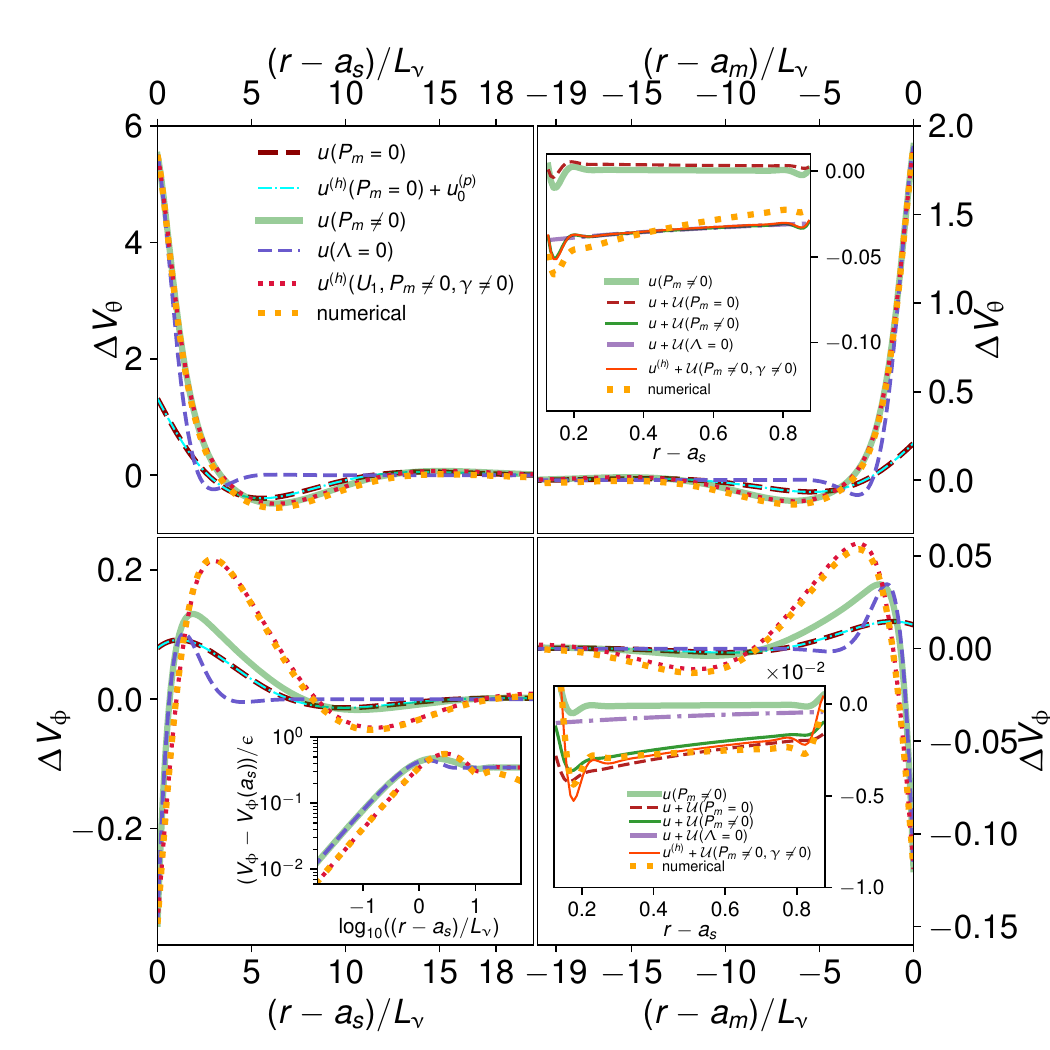}
\end{subfigure}
\caption{
Radial $\Delta V_r$ (top) and tangential velocity field perturbations $\Delta V_\theta$ (middle), $\Delta V_\phi$ (bottom), close to the inner (left) and outer (right) boundaries.
Numerical \textsc{xshells} solution (orange-dotted).
Boundary layer solutions: inviscid MHD (dark-red dashed), inviscid MHD diffusionless (cyan dash-dotted), viscomagnetic oscillatory (thick green), purely viscous (purple dashed).
In the inset: boundary layer solution and secondary bulk flow: inviscid MHD (dark-red dashed), inviscid MHD diffusionless (cyan dash-dotted), viscomagnetic oscillatory (dark green), purely viscous (purple dash-dotted), MHD Stokes-Ekman (red solid).
Polar forcing. Parameters: as in figure~\ref{fig:xRadialMagneticBoundaryLayer}.}
   \label{fig:xRyViscoMagRotVRThetaPhiPolarGamma1em1}
\end{figure}

\begin{figure}
\centering
\begin{subfigure}[b]{0.80\textwidth}
    \centering
    \includegraphics[width=1\linewidth]{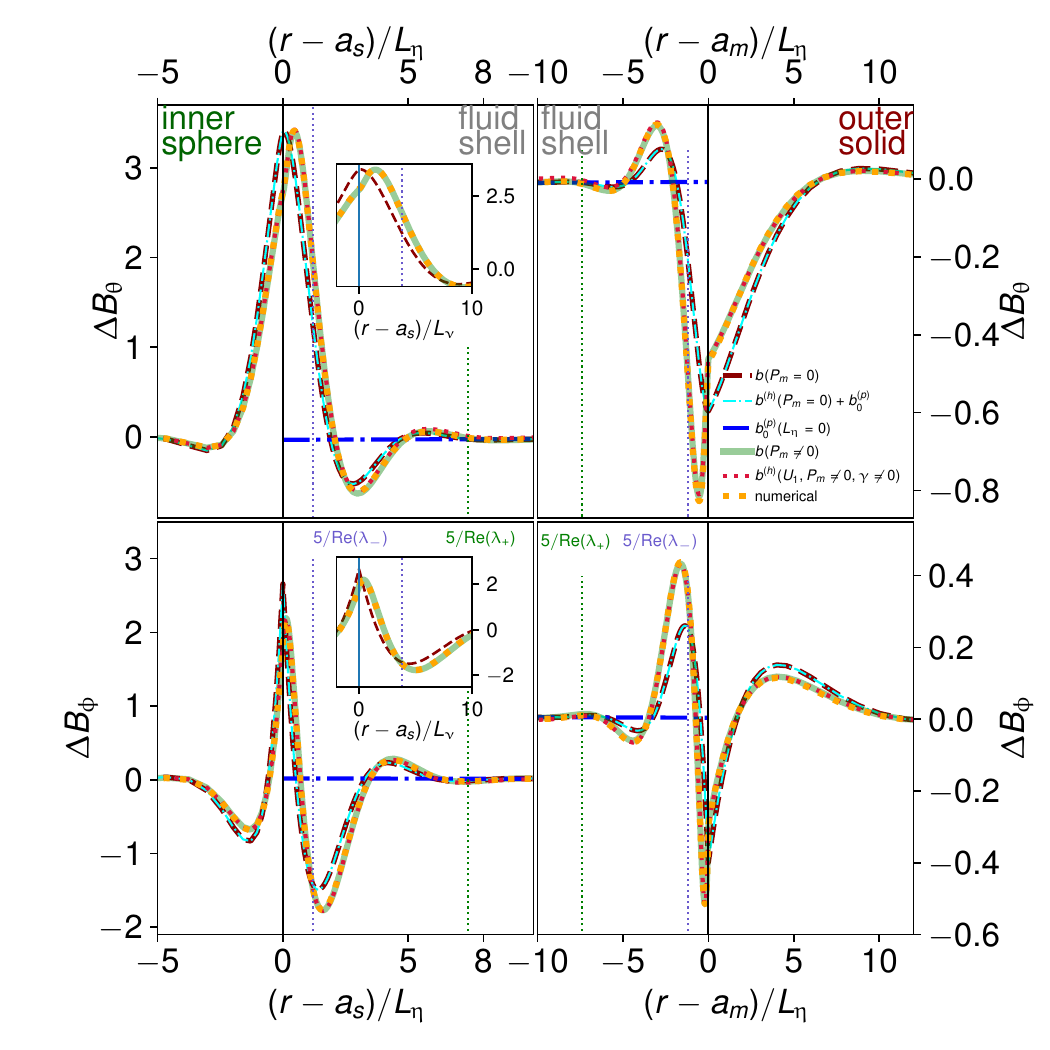}
\end{subfigure}
\caption{
Tangential magnetic field perturbations, close to the inner (left) and outer (right) boundaries.
Numerical \textsc{xshells} solution (orange-dotted).
Boundary layer solutions: inviscid MHD (dark-red dashed), inviscid MHD diffusionless (cyan dash-dotted), viscomagnetic oscillatory (thick green), purely viscous (purple dashed), MHD Stokes-Ekman (red dotted), and diffusionless bulk (dash-dotted blue).
Equatorial forcing. Parameters: as in figure~\ref{fig:xRadialMagneticBoundaryLayer}.}
   \label{fig:xRyViscoMagRotBThetaPhiEquatorialGamma1em1}
\end{figure}

\begin{figure}
\centering
\begin{subfigure}[b]{0.80\textwidth}
    \centering
    \includegraphics[width=1\linewidth]{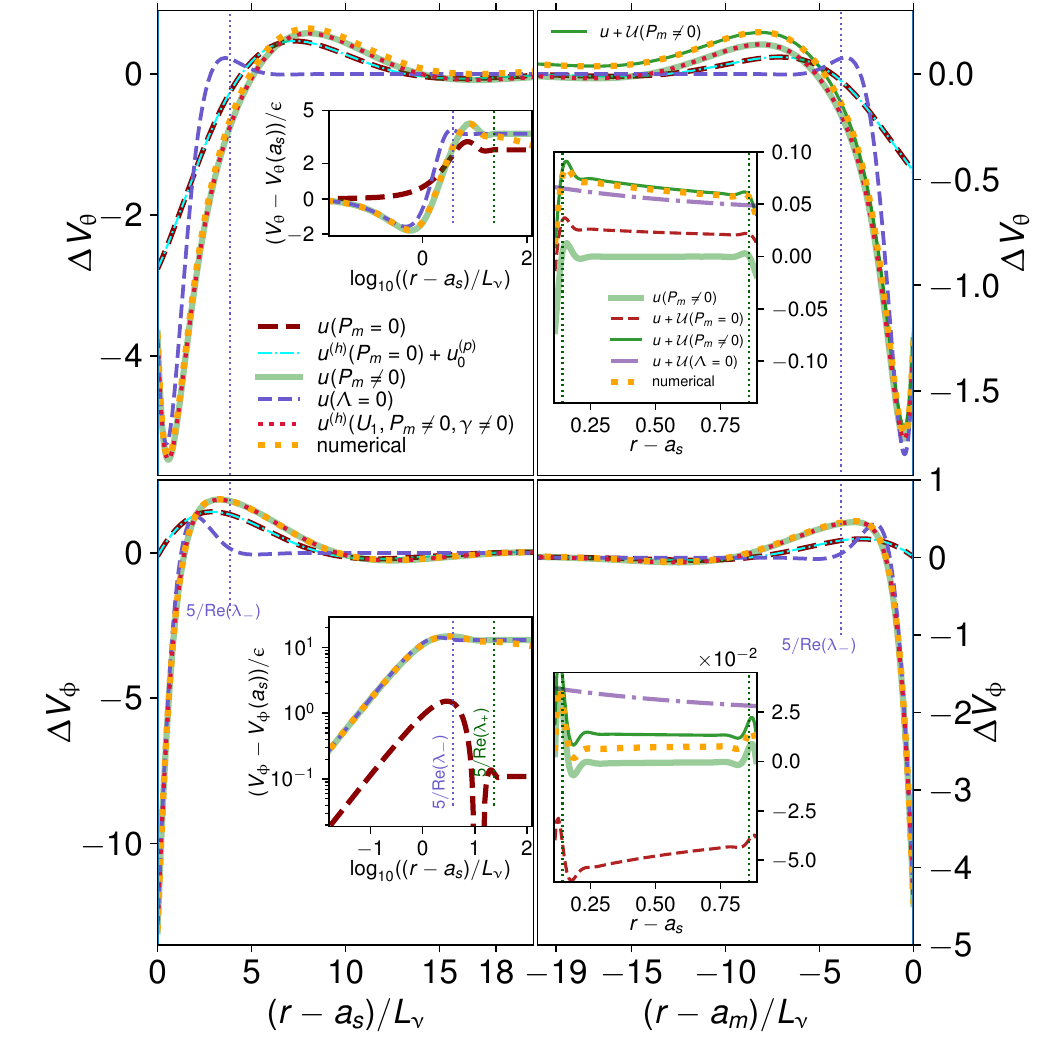}
\end{subfigure}
\caption{
Tangential velocity field perturbations $\Delta V_\theta$ (top), $\Delta V_\phi$ (bottom), close to the inner (left) and outer (right) boundaries.
Numerical \textsc{xshells} solution (orange-dotted).
Boundary layer solutions: inviscid MHD (dark-red dashed), inviscid MHD diffusionless (cyan dash-dotted), viscomagnetic oscillatory (thick green), purely viscous (purple dashed).
In the inset: boundary layer solution and secondary bulk flow: inviscid MHD (dark-red dashed), inviscid MHD diffusionless (cyan dash-dotted), viscomagnetic oscillatory (dark green), purely viscous (purple dash-dotted), MHD Stokes-Ekman (red dotted).
Equatorial forcing. Parameters: as in figure~\ref{fig:xRadialMagneticBoundaryLayer}.}
   \label{fig:xRyViscoMagRotVRThetaPhiEquatorialGamma1em1}
\end{figure}

Figures \ref{fig:xRyViscoMagRotBThetaPhiGamma1em1} and \ref{fig:xRyViscoMagRotVRThetaPhiPolarGamma1em1} show for the polar forcing that the MHD Stokes-Ekman boundary-layer solution based on $\boldsymbol{U}_1$ closely matches the numerical results. For the radial component, the secondary bulk flow $\mathcal{U}_r$ combines with $\mathfrak{u}_r$ to shape the near-boundary profile. In contrast, for tangential components, its contribution is significant only in the bulk where the boundary-layer solution vanishes; there, the MHD Stokes-Ekman secondary flow accurately reproduces the perturbation, as highlighted in the insets. In this weak-rotation regime, velocity effects are predominantly azimuthal. The diffusionless solution $b_0^{(p)}$ provides an accurate approximation of the magnetic particular solution.
For the equatorial forcing in the weak-rotation regime ($\gamma=10^{-3}$), the MHD Stokes-Ekman solution (dotted red) coincides with the simpler MHD oscillatory solution, as shown
in figure \ref{fig:xRyViscoMagRotBThetaPhiEquatorialGamma1em1} and figure \ref{fig:xRyViscoMagRotVRThetaPhiEquatorialGamma1em1}.


\begin{figure}
\centering
\begin{subfigure}[b]{0.80\textwidth}
\centering
\includegraphics[trim=10 0 0 22,clip,width=1\linewidth]{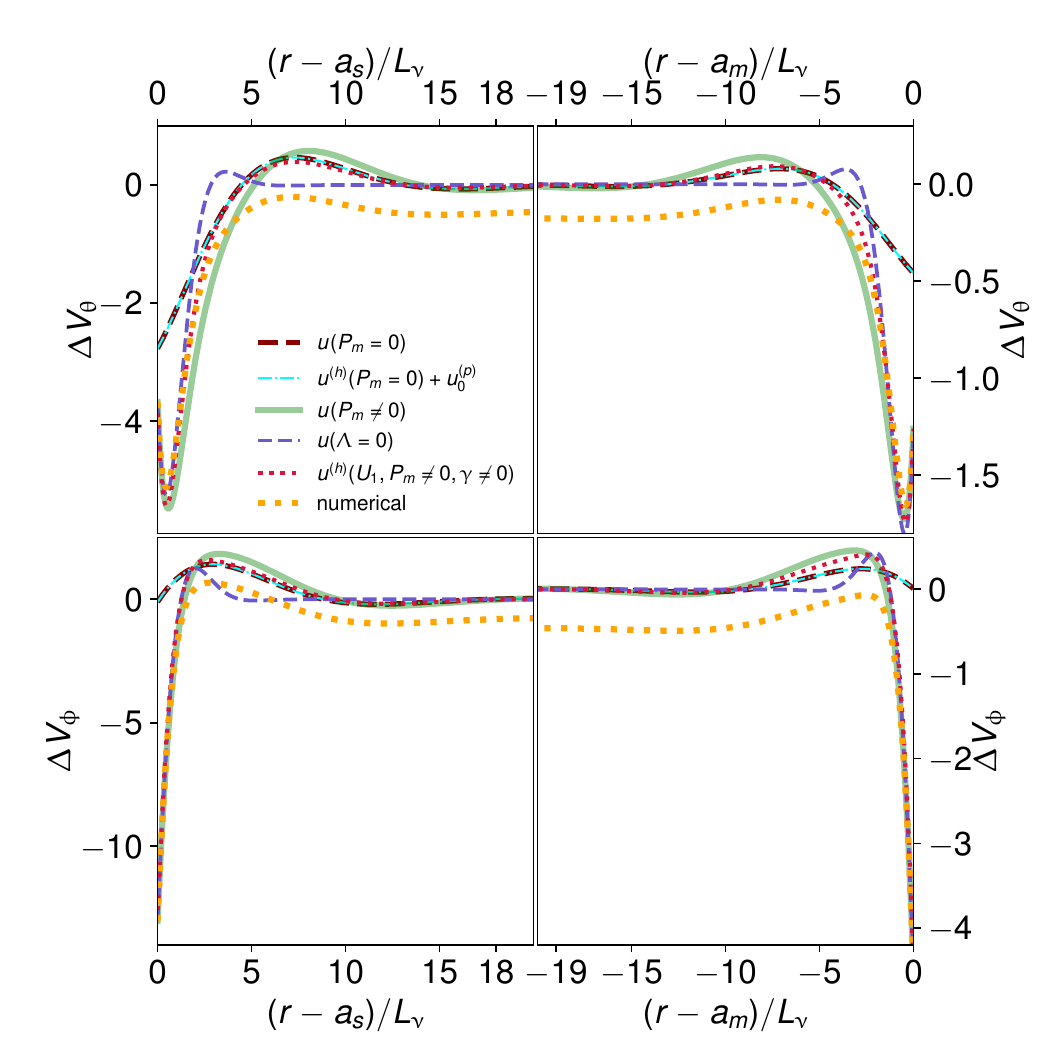}
\end{subfigure}
\caption{Tangential velocity field perturbations $\Delta V_\theta$ (top), $\Delta V_\phi$ (bottom), close to the inner (left) and outer (right) boundaries.
Numerical \textsc{xshells} solution (orange-dotted).
Boundary layer solutions: inviscid MHD (dark-red dashed), inviscid MHD diffusionless (cyan dash-dotted), viscomagnetic oscillatory (thick green), purely viscous (purple dashed), MHD Stokes-Ekman (red dotted).
Equatorial forcing. Parameters: as in figure~\ref{fig:xRadialMagneticBoundaryLayer}, but $\gamma=0.33$.}
   \label{fig:xRyVthetaVphiGammaEqua033}
\end{figure}

In figure \ref{fig:xRyVthetaVphiGammaEqua033}, we present the velocity profiles for an equatorial forcing in the presence of moderate rotation ($\gamma=0.33$). By comparing it with the numerical result in orange dotted, we see that all viscous solutions capture quite well the evolution within a viscous length.
The position of the maximum of the velocity boundary layer profile is well captured by the MHD Stokes-Ekman layer.
The bulk value is not well estimated as we adopt the potential flow solution (\ref{eq:LambEquatorial}), that does not account for rotation effects.

\section{Role of electromagnetic properties in regime transitions} \label{sec:roleEL}

\begin{figure}
\centering
\includegraphics[width=1.0\linewidth]{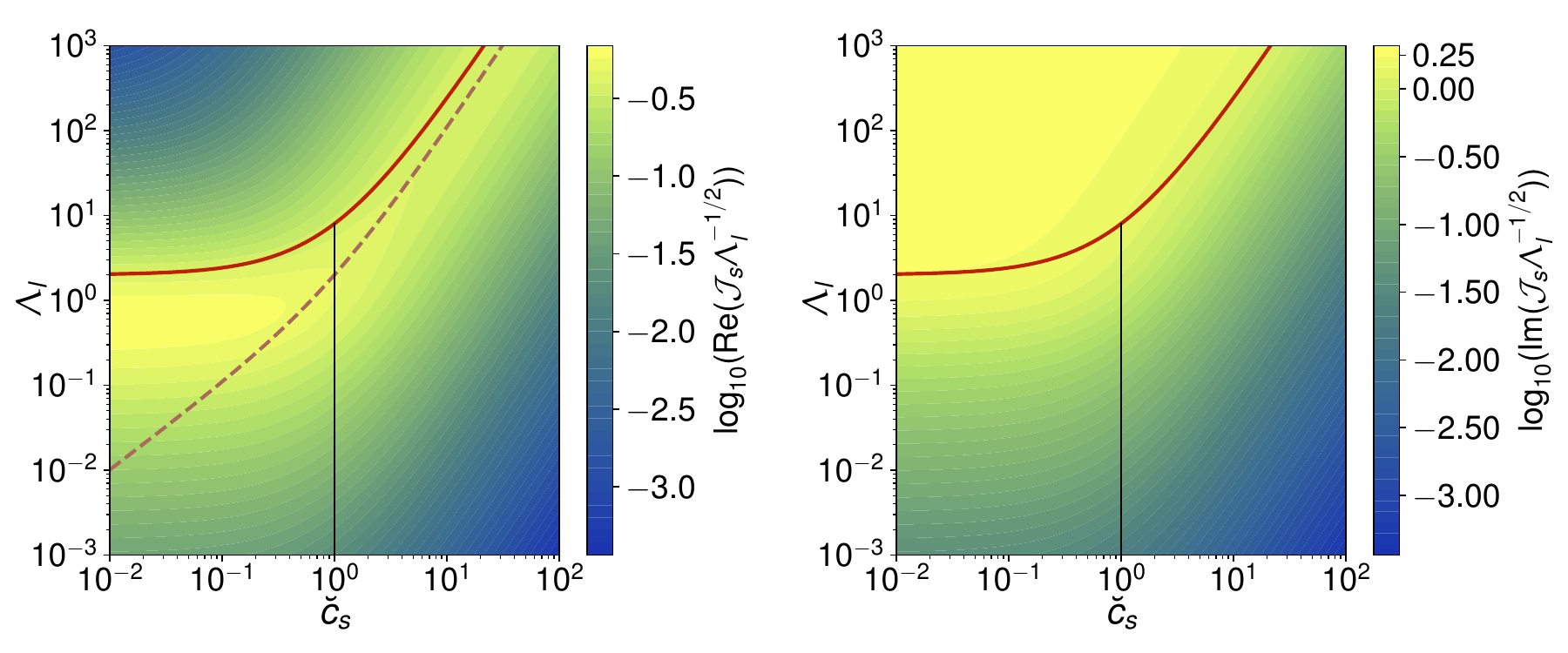}
    \caption{Role of the electromagnetic properties contrast $\breve{c}_s$ and of $\Lambda_l$ on the real (left) and imaginary (right) part of the proxy $\mathcal{J}_s \Lambda_l^{-1/2}$ of the magnetic tension (\ref{eq:Bufextde}). Alfv\'en waves dominated asymptotics occur above the red line $\Lambda_l=2(\breve{c}_s + 1)^2$. Below, the black line $\breve{c}_s=1$ separates the diffusive regime into magnetic-like (left) and hydrodynamic-like (right) regimes.}
    \label{fig:xetamuyLambdacI}
\end{figure}

Figure \ref{fig:xetamuyLambdacI} shows the dependence of the magnetic tension on $\breve{c}_s$ and $\Lambda_l$ through the proxy $\mathcal{J}_s \Lambda_l^{-1/2}$ (eq.~\ref{eq:Bufextde}). Matching its leading-order asymptotic expansions at small and large $\Lambda$, we obtain a transition between the asymptotic behaviours at $\Lambda_l=2(\breve{c}_s + 1)^2$ and $\Lambda_l =\breve{c}_s^2 +\breve{c}_s$ for $\Imag(\mathcal{J}_s/\Lambda_l^{1/2})$ and $\Real(\mathcal{J}_s/\Lambda_l^{1/2})$, respectively. These two transitions are shown by the solid red and black dashed lines in figure \ref{fig:xetamuyLambdacI}, respectively.

The large damping due to Alfv\'en waves radiation is associated to large $\Imag(\mathcal{J}_s/\Lambda_l)$, and is thus obtained above the red lines in figure \ref{fig:xetamuyLambdacI}. It shows that the transition at $\Lambda_l \approx 1$ found by BG95 for $\breve{c}_s=1$ is strongly modified for $\breve{c}_s \gg 1$. Below this transition, the asymptotic formula at $\Lambda_l \ll 1$ shows that the diffusive regime is further divided into two regimes that are separated by the condition $\breve{c}_s=1$ (vertical solid black line in figure \ref{fig:xetamuyLambdacI}) . First, magnetic effects are important for $\breve{c}_s<1$, and the value of $\breve{c}_s$ plays then a negligible role for the magnetic tension. In the opposite case $\breve{c}_s>1$, the value of $\breve{c}_s$ plays an important role as it controls how the magnetic tension vanishes in the hydrodynamic regime (obtained for $\breve{c}_s \gg 1$).

\section{The \textit{huBL} loading from MHD Stokes-Ekman boundary layers } \label{sec:huBL_sMat}

For the polar mode, the forces are along $z$ only. Using the \textit{huBL} approach where ${\check{\iota}}_\pm $ and ${\check{\kappa}}_\pm$ are assumed to be uniform, we obtain (at $\breve{\mu}_s=\breve{\eta}_s=1$)
\begin{eqnarray} \label{eq:rotdet1}
  \left.  \frac{\Delta \omega_{mt}}{\omega} \right|_{r=a_s} &=&\frac{3}{8} \Lambda  \frac{\rho_f}{\rho_s} \frac{ L_\eta}{ a_s } \frac{1+\mathrm{i}}{1-a^3} \frac{2 {\check{\kappa}}_+ {\check{\kappa}}_- {\check{\iota}}_+ {\check{\iota}}_- + \check{\varpi}_+ + {\check{\kappa}}_+ {\check{\iota}}_+ + {\check{\kappa}}_- {\check{\iota}}_-}{({\check{\kappa}}_+ + 1)({\check{\kappa}}_- + 1)({\check{\iota}}_+ + 1)({\check{\iota}}_- + 1)}
\end{eqnarray}
and
\begin{eqnarray}
      \left.  \frac{\Delta \omega_v}{\omega} \right|_{r=a_s}&=&- \frac{3}{8}  \frac{\rho_f}{\rho_s} \frac{L_\nu}{a_s }  \frac{1-\mathrm{i}}{1-a^3} ({\check{\kappa}}_+ {\check{\iota}}_+ +{\check{\kappa}}_- {\check{\iota}}_-) \sqrt{P_m} , \label{eq:rotdet2}
\end{eqnarray}
with no-slip boundary conditions, noting $\check{\varpi}_\pm={\check{\kappa}}_+ {\check{\kappa}}_- {\check{\iota}}_+ \pm {\check{\kappa}}_+ {\check{\kappa}}_- {\check{\iota}}_- + {\check{\kappa}}_+ {\check{\iota}}_+ {\check{\iota}}_- \pm {\check{\kappa}}_- {\check{\iota}}_+ {\check{\iota}}_-$. Simple explicit expressions of equations (\ref{eq:rotdet1})-(\ref{eq:rotdet2}) can be obtained for $\Lambda \gg 1$, which read
\begin{eqnarray}
  \left.  \frac{\Delta \omega_{mt}}{\omega} \right|_{r=a_s} &=&\frac{3}{8}  \frac{\rho_f}{\rho_s} \frac{ L_\eta}{ a_s } \frac{2(1-\mathrm{i}) - ((1+\mathrm{i}) \sqrt{P_m} - \sqrt{2 \Lambda}) \tilde{\gamma}_+ }{1-a^3}  , \label{eq:gis1}
\end{eqnarray}
and
\begin{eqnarray}
      \left.  \frac{\Delta \omega_v}{\omega} \right|_{r=a_s}&=& \frac{3}{8} \frac{\rho_f}{\rho_s} \frac{L_\nu }{a_s }  \frac{1+\mathrm{i}}{1-a^3} \tilde{\gamma}_+ , \label{eq:gis2}
\end{eqnarray}
where rotation effects are given by $\tilde{\gamma}_\pm=\sqrt{\gamma \cos \theta -1}\pm \mathrm{i} \sqrt{\gamma \cos \theta +1}$. Equation (\ref{eq:gis1}) extends equation (\ref{eq:asymBG}) to both viscous and rotation effects, while equation (\ref{eq:gis2}) extends to rotation effects the hydrodynamic viscous detuning (e.g. equation \ref{eq:disSh2}). For stress-free boundary conditions, equation (\ref{eq:gis1}) is retrieved with only the dominant term in $\sqrt{2 \Lambda}$. A typical $\theta$ has naturally to be chosen in these approximations, related to the assumed uniform boundary layer thickness (which is reminiscent of an f-plane approximation).

To ease the comparison with previous formulas for ${\Delta \omega_j}/{\omega}$, we define a formally analogous proxy $\tilde{\Gamma}_{j}=-\boldsymbol{{\Gamma}_{j}} \cdot \hat{\boldsymbol{z}} /(2 a_s m_s \omega^2 \epsilon \mathrm{e}^{\mathrm{i}( \omega t + m \phi)})$. The scaled axial torques are then
\begin{eqnarray}
\tilde{\Gamma}_{mt}&=&  -\frac{3}{8} \Lambda  \frac{\rho_f}{\rho_s} \frac{ L_\eta}{ a_s } \frac{1-\mathrm{i}}{1-a^3} \frac{\check{\varpi}_- + {\check{\kappa}}_+ {\check{\iota}}_+ - {\check{\kappa}}_- {\check{\iota}}_-}{({\check{\kappa}}_+ + 1)({\check{\kappa}}_- + 1)({\check{\iota}}_+ + 1)({\check{\iota}}_- + 1)}
\end{eqnarray}
and
\begin{eqnarray}
\tilde{\Gamma}_{v} &=&- \frac{3}{8}  \frac{\rho_f}{\rho_s} \frac{L_\nu}{a_s }  \frac{1-\mathrm{i}}{1-a^3} ({\check{\kappa}}_+ {\check{\iota}}_+ -{\check{\kappa}}_- {\check{\iota}}_-) \sqrt{P_m} ,
\end{eqnarray}
for the inner solid sphere, leading to, for $\Lambda_l \gg 1$
\begin{eqnarray}
\tilde{\Gamma}_{mt}&=&\frac{3 \mathrm{i}}{8} \frac{\rho_f}{\rho_s} \frac{L_\eta }{a_s }  \frac{2(1-\mathrm{i}) \gamma \cos (\theta) + ((1+\mathrm{i}) \sqrt{P_m} - \sqrt{2 \Lambda}) \tilde{\gamma}_- }{1-a^3} , \label{eq:TorSF1}
\end{eqnarray}
and
\begin{eqnarray}
\tilde{\Gamma}_{v}&=& \frac{3}{8} \frac{\rho_f}{\rho_s} \frac{L_\nu }{a_s }  \frac{1-\mathrm{i}}{1-a^3} \tilde{\gamma}_- , \label{eq:TorSF2}
\end{eqnarray}
whereas an exact integration in the limit $\Lambda_l \gg 1$ gives vanishing torques.

\clearpage

\section{Nomenclature}
\label{sec:nomTab}
\noindent
\begin{longtable}{lp{9cm}}
\endfirsthead
\multicolumn{2}{l}{\textit{Continued}} \\
\endhead

\multicolumn{2}{l}{\textit{Geometry and mass}} \\
$a_s$, $a_m$, $a_e$
  & Inner and outer spherical shell radii, outer solid external radius \\
$a = a_s/a_m$
  & Inner-to-outer radius ratio \\
$D = a_m - a_s$
  & Fluid gap width (length scale) \\
  $M_s$, $M_{fs}$ & Solid mass and displaced fluid mass \\
  [6pt]

\multicolumn{2}{l}{\textit{Oscillation forcing}} \\
$\omega_s$, $T_s$
  & Oscillation angular frequency and period\\
$\omega = \omegas D^2 / \eta_f $
  & Dimensionless oscillation angular frequency \\
$\epsilon_s$, $\boldsymbol{\epsilon}_s$
  & Oscillation amplitude, displacement amplitude vector \\
$\epsilon = \epsilon_s/D$
  & Dimensionless amplitude (small parameter) \\
$\mathcal{V}$ & characteristic translational speed\\
$m$
  & Azimuthal wavenumber ($m=0$: polar; $m=\pm 1$: equatorial) \\
$\mathcal{E}$
  & Kinetic energy prefactor ($\mathcal{E}=1/2$: polar; $\mathcal{E}=1$: equatorial) \\
$\varOmega_o$, $\boldsymbol{\varOmega}_o$
  & Frame rotation rate and vector \\
$\gamma = 2\varOmega_o/\omega_s$
  & Rotation-to-oscillation frequency ratio \\[6pt]

\multicolumn{2}{l}{\textit{Material properties and length scales}} \\
$\rho_f$, $\rho_s$, $\rho_m$
  & Density of the fluid, inner solid, and outer solid \\
$\nu$
  & Kinematic viscosity of the fluid \\
$\sigma_f$, $\sigma_s$, $\sigma_m$
  & Electrical conductivity in the fluid, inner solid, and outer solid \\
$\eta_f$, $\eta_s$, $\eta_m$
  & Magnetic diffusivity in the fluid, inner solid, and outer solid \\
$\breve{\eta}_s = \eta_s/\eta_f$,\;
$\breve{\eta}_m = \eta_m/\eta_f$
  & Magnetic diffusivity ratios \\
$\breve{\mu}_s = \mu_s/\mu_f$,\;
$\breve{\mu}_m = \mu_m/\mu_f$
  & Magnetic permeability ratios \\
$\breve{c}_j = \breve{\eta}_j^{1/2} \breve{\mu}_j$ & Electromagnetic
 properties contrast \\
$L_\nu = \sqrt{2\nu/\omega_s}$
  & Viscous skin depth \\
$L_\eta = \sqrt{2\eta_f/\omega_s}$
  & Magnetic skin depth \\[6pt]

\multicolumn{2}{l}{\textit{Forces, dissipation and frequency correction}} \\
$\boldsymbol{F}$, $\boldsymbol{F}_v$, $\boldsymbol{F}_{vn}$
  & Total fluid force, tangential and normal viscous stress forces \\
 $\boldsymbol{F}_{n}$, $\boldsymbol{F}_p$
  & Total normal force and pressure force \\
$\boldsymbol{F}_\eta = \boldsymbol{F}_{mt} + \boldsymbol{F}_{mp}$
  & Lorentz force (magnetic tension + magnetic pressure) \\
$\boldsymbol{F}_\nu = \boldsymbol{F}_v + \boldsymbol{F}_n - \boldsymbol{F}_n(\nu=0)$
  & Viscous force contributions\\
$\mathcal{D}_\nu$
  & Viscous dissipation \\
$\mathcal{D}_\eta$
  & Ohmic dissipation \\
$C_a$, $C_a^i$
  & Added-mass coefficient and its inviscid limit \\
$C_d$, $C_d^m$
  & Drag coefficient and magnetic drag coeficient \\
$\Delta\boldsymbol{\omega}_j/\omega$
  & Force-based complex frequency correction (detuning) due to force
  $j$; real part: eigenfrequency shift, imaginary part: modal
  damping-rate contribution \\
$\Delta\omega_v$, $\Delta\omega_p$
  & Tangential-viscous-stress and pressure detunings \\
$\Delta\omega_{mt}$, $\Delta\omega_{mp}$
  & Magnetic-tension and magnetic-pressure detunings \\
$\widehat{\mathcal D}_\nu$, $\widehat{\mathcal D}_\eta$
  & Scaled mean viscous and Ohmic dissipations; division by
  $2\mathcal E$ gives the normalised modal damping rates \\
$\mathcal{J}_j$
  & Angular integral proxy for magnetic tension, equation~(5.22b) \\[6pt]

\multicolumn{2}{l}{\textit{Dimensionless parameters}} \\
$P_m = \nu/\eta_f$
  & Magnetic Prandtl number \\
$\Lambda = \sigma_f \tilde{B}_0^2/(\rho_f\omega_s)$
  & Squared Lundquist number based on typical field strength $\tilde{B}_0$ \\
$\Lambda_z = \sigma_f \tilde{B}_{0z}^2/(\rho_f\omega_s)$
  & Same based on axial field amplitude $\tilde{B}_{0z}$ \\
$\Lambda_l = \sigma_f \tilde{B}_{0r}^2/(\rho_f\omega_s)$
  & Same based on local radial field $\tilde{B}_{0r}$ \\
$Lu_\omega = \Lambda^{1/2}$
  & Oscillatory Lundquist number \\
$\mathrm{Wo}=a_s(\omega_s/\nu)^{1/2}$
  & Womersley number \\
$Al = \mathcal{V} /  \tilde{V}_A =  \epsilon_s \omega_s / \tilde{V}_A$
  & Alfv\'en Mach number \\
$N = \sigma_f \tilde{B}_0^2 a_s/(\rho_f \mathcal{V})$
  & Stuart (interaction) parameter \\[6pt]

\multicolumn{2}{l}{\textit{Velocity fields}} \\
$\boldsymbol{\delta}$, $\boldsymbol{V}_s$
  & Relative displacement and velocity of the inner solid \\
$\boldsymbol{V}$, $\boldsymbol{U}$
  & Fluid velocity and basic (potential) flow \\
$\boldsymbol{u}$, $\mathcal{U}$
  & Velocity perturbation,  secondary bulk flow \\
  $\mathfrak{u}_r$, $\boldsymbol{\mathfrak{u}}_\parallel$ &  Higher-order radial and tangential flow \\
$\varPhi$, $\varPsi$
  & Velocity potential and potential for secondary bulk flow \\
$\boldsymbol{\mathcal{W}} = \nabla \times \boldsymbol{V}$
  & Vorticity \\[6pt]

\multicolumn{2}{l}{\textit{Magnetic fields}} \\
 $\tilde{B}_0$ & Characteristic magnetic field amplitude \\
 $\boldsymbol{B}_0$
  & Imposed magnetic field \\
$\tilde{V}_A = \tilde{B}_0/\sqrt{\rho_f \mu_f}$
  & Alfv\'en speed \\
$\boldsymbol{B}$, $\boldsymbol{b}$
  & Total magnetic field and its perturbation \\
  $\boldsymbol{\mathcal{B}}$, $\boldsymbol{\mathfrak{b}}$
  & Secondary bulk and higher-order boundary-layer magnetic field \\
$\boldsymbol{b}^{(p)}$, $\boldsymbol{b}^{(h)}$
  & Particular and homogeneous magnetic solutions \\[6pt]

\multicolumn{2}{l}{\textit{Boundary-layer wavenumbers}} \\
$\lambda_{\mathfrak{s}} = (1+\mathrm{i})/L_\eta$
  & Magnetic skin-depth wavenumber \\
$\lambda = \lambda_{\mathfrak{s}}(1-\mathrm{i}\Lambda_l)^{-1/2}$
  & Inviscid MHD boundary-layer wavenumber \\
$\lambda_\pm$
  & Viscomagnetic oscillatory wavenumbers, equation~(6.3) \\
  $L_A$
  & Alfv\'en-wave attenuation depth \\[6pt]
$\iota_\pm$,\;$\kappa_\pm$
  & MHD Stokes--Ekman wavenumbers, equations~(6.19)--(6.20) \\[6pt]

  \multicolumn{2}{l}{\textit{Abbreviations}} \\
BLT
  & Boundary-layer theory \\
huBL
  & Homogeneous uniform boundary-layer
    approximation \\
B74
  & \cite{busse1974free} \\[4pt]
BG95
  & \cite{buffett1995magnetic} \\
P26
  & \cite{personnettaz2026ohmic} \\

\end{longtable}